\documentclass[12pt]{article}
\usepackage{amsmath}
\usepackage{graphicx,psfrag,epsf}
\usepackage{enumerate}
\usepackage{natbib}
\usepackage{url} 
\usepackage{amsmath}
\usepackage{graphicx}
\usepackage{enumerate}
\usepackage{natbib}
\usepackage{url}
\usepackage{tabularx}
\usepackage{rotating}
\usepackage{algorithm}
\usepackage{float}
\usepackage{subfig}
\usepackage{subfloat}
\usepackage{color}
\usepackage{pgfplots}
\usepackage{tikz}
\usepackage{algorithmic}
\usetikzlibrary{positioning, shapes.geometric}
\usetikzlibrary{shapes}
\usepackage{amsfonts}
\usepackage[bookmarks=true]{hyperref}

\newtheorem{theorem}{Theorem}[section]

\newtheorem{lemma}{Lemma}
\newtheorem{proposition}{Proposition}
\newtheorem{corollary}{Corollary}
\def\beq{\begin{equation}}
\def\eeq{\end{equation}}
\def\beqr{\begin{eqnarray}}
\def\eeqr{\end{eqnarray}}
\def\beqrs{\begin{eqnarray*}}
\def\eeqrs{\end{eqnarray*}}
\def\bet{\begin{theorem}}
\def\eet{\end{theorem}}
\def\bel{\begin{lemma}}
\def\eel{\end{lemma}}
\def\bep{\begin{proposition}}
\def\eep{\end{proposition}}
\def\bec{\begin{corollary}}
\def\eec{\end{corollary}}
\def\n{\nonumber}

\def\mR{\mathbb{R}}
\def\mE{\mathbb{E}}

\def\cov{\mbox{cov}}
\def\vec{\mbox{vec}}

\newcommand{\blind}{1}

\begin{document}

\def\spacingset#1{\renewcommand{\baselinestretch}%
{#1}\small\normalsize} \spacingset{1}
\date{}


\if1\blind
{
  \title{\bf Tucker tensor factor models: matricization and mode-wise PCA estimation}
  \author{Xu Zhang$^1$,
  Guodong Li$^2$,
  Catherine C. Liu$^3$,
  Jianhua Guo$^4$\\
\it{$^1$South China Normal University,
$^2$The University of Hong Kong}\\
\it{$^3$The Hong Kong Polytechnic University}\\
\it{and $^4$Beijing Technology and Business University}}
  \maketitle
} \fi

\if0\blind
{
  \bigskip
  \bigskip
  \bigskip
  \begin{center}
    {\LARGE\bf Tucker tensor factor models: matricization and mode-wise PCA estimation}
\end{center}
  \medskip
} \fi

\bigskip
\begin{abstract}
High-dimensional, higher-order tensor data are gaining prominence in a variety of fields, including but not limited to computer vision and network analysis.
Tensor factor models, induced from noisy versions of tensor {decompositions or factorizations}, are natural potent instruments to study a collection of tensor-variate objects that may be dependent or independent.
However, it is still in the early stage of developing statistical inferential theories for {the} estimation of various low-rank structures, which are customary to play the role of signals of tensor factor models.
{In this paper, we attempt to ``decode" the estimation of a higher-order tensor factor model by leveraging tensor matricization.
Specifically, we recast it into mode-wise traditional high-dimensional vector/fiber factor models, enabling the deployment of conventional principal components analysis (PCA) for estimation.}
Demonstrated by the Tucker tensor factor model (TuTFaM), {which is induced from the noisy version of the widely-used Tucker decomposition}, we summarize that estimations on signal components are essentially mode-wise PCA techniques, and the involvement of projection and iteration will enhance the signal-to-noise ratio to various {extent}.
We establish the inferential theory of the proposed {estimators, conduct rich simulation experiments}, and illustrate how the proposed estimations can work in tensor reconstruction, and clustering for {independent} video and {dependent} economic datasets, respectively.
\end{abstract}

\noindent%
\vfill

\newpage
\spacingset{1.4} 

\section{Introduction}

With the advancements in data acquisition techniques, high-dimensional higher-order tensor data objects are becoming increasingly prevalent in various fields.
Here higher-order is in the sense that the number of modes or ways of the tensor is no less than three \citep{KoldaBader2009SIAM-tensor}.
For example, in the computer vision and medical imaging fields, $3$D images and video sequences are naturally described as $3$rd-order tensors,
where each mode corresponds to different domains, such as spatial dimensions, color channels, and frames;
in the network data analysis fields, transport networks and social networks can be formulated as tensor objects,
where each mode indicates participating entities and types of relations \citep{BiTangYuanZhangQu2021AROSIA-tensors, ChenYangZhang2022JASA-factor, LuPlataniotisVenetsanopoulos2008IEEE-mpca}.
The use of higher-order tensor data objects enables a more comprehensive and expressive representation of complex data, allowing for capturing interactions and dependencies {among} multiple variables or dimensions.
However, it comes into common view that our ability {to acquire and analyze} tensor data has surpassed our capability to develop fundamental mathematical and statistical theories that can justify and rationalize the efficacy of models and methodologies.

In this paper, we attempt to ``decode" {the} estimation of higher-order tensor factor models and develop the corresponding inferential theories.
Specifically, we aim to develop PCA-type estimations, which are the most widely-used techniques for tensor factor analysis with various variants for different low-rank structures \citep{ZareOzdemirIwen2018IEEE-extension}.
We start from a Tucker tensor factor model (TuTFaM), which is a noise version of the widely-used Tucker decomposition \citep{Tucker1966Psychometrika-some}, i.e., the low-rank signal is in the form of a product of a low-dimensional core tensor (representing latent tensor factors) and factor loading matrices across all modes.
In detail, let $\mathcal{X}_{t}\in\mR^{p_{_1}\times \cdots\times p_{_D}}$, where $t \in [T] :=\{1,\cdots, T\}$, denote the $D$th-order tensor observations that may be weakly correlated or uncorrelated.
Then TuTFaM is expressed as
\beqr\label{Model}
\mathcal{X}_{t} &=& \mathcal{S}_t+ \mathcal{E}_{t},\\\nonumber
\mathcal{S}_t &=& \mathcal{F}_{t}\times_{1}\mathbf{A}_{1}\times_{2} \cdots \times_{D} \mathbf{A}_{D},
\eeqr
where $\mathcal{S}_t$ is the unknown low-rank signal part,
$\mathcal{E}_{t}\in\mR^{p_1\times \cdots\times p_D}$ is the noise tensor,
$\mathcal{F}_{t}\in\mR^{k_1\times \cdots\times k_D}$ is the random low-dimensional ($k_d\ll p_d$) latent tensor factor,
$\mathbf{A}_{d}\in\mR^{p_d\times k_d}$ is the unknown and deterministic mode-$d$ loading matrix for $d\in[D]$,
and $\times_d$ is the tensor $d$-mode (matrix) product.
In this context, high dimensionality implies that the sample or time points size $T$ and the size of each mode $p_d$ can be significantly large. Moreover, it is possible for the mode-wise size $p_d$ to surpass the value of $T$.
In addition, model (\ref{Model}) reduces to a bilinear matrix factor model when $D=2$ \citep{WangLiuChen2019JOE-factor, ChenFan2023JASA-statistical, YuHeKongZhang2022JOE-projected}, and a vector factor model when $D=1$ \citep{Forni2000generalized, BaiNg2002Econometrica-determining, Bai2003Econometrica-inferential, LamYaoBathia2011Biometrika-estimation, LamYao2012AOS-factor}.

\subsection{Motivation: matricization and mode-wise PCA estimation}
\label{moPCA}
The proposed estimation procedures are motivated by tensor matricization (also called unfolding or flattening), that is, the elements of an observed tensor object $\mathcal{X}_{t}\in\mR^{p_{_1}\times \cdots\times p_{_D}}$ is reordered into a matrix along each mode $d \in [D]$, where the mode-$d$ fibers are arranged as the columns of the resulting matrix $\mathcal{X}_t^{(d)} \in \mR^{p_{_d}\times p_{_{-d}}}$
with $p_{_{-d}} = \Pi_{_{i \neq d}} ^D p_i$,
and hence the $(i_d,j_d)$th element of $\mathcal{X}_t^{(d)}$ is the $(i_1,\cdots,i_D)$th element of $\mathcal{X}_t$ satisfying $j_d=1+\sum_{d^{\prime}\neq d}^{D}\left(i_{d^{\prime}}-1\right) \prod_{m\neq d}^{d^{\prime}-1} p_{m}$.
The corresponding latent tensor factor is flattened as $\mathcal{F}_{t}^{(d)} \in \mR^{k_d\times k_{{-d}}}$ with $k_{_{-d}} = \Pi_{_{i \neq d}} ^D k_i$.
Therefore, one has the equivalent matricized representation of model (\ref{Model}) as
\beqr
\label{matricized_1}
\mathcal{X}_t^{(d)}=\mathbf{A}_{d}
\mathcal{F}_{t}^{(d)} \mathbf{A}_{[D]/\{d\}}^{\top}
+\mathcal{E}_{t}^{(d)},~{d\in[D]},
\eeqr
where $\mathbf{A}_{[D]/\{d\}}:=\mathbf{A}_D\otimes\cdots\otimes \mathbf{A}_{d+1}\otimes \mathbf{A}_{d-1}\otimes\cdots\otimes \mathbf{A}_1$ {with $\otimes$ being the Kronecker product} and $\mathcal{E}_{t}^{(d)}\in \mR^{p_{_d}\times p_{_{-d}}}$ is the mode-$d$ matricization of the noise tensor $\mathcal{E}_t$.

Let $\mathbf{X}^{(d)}=(\mathcal{X}_{1}^{(d)},\cdots,\mathcal{X}_{T}^{(d)})\in\mR^{p_{_{d}}\times (Tp_{_{-d}})}$ be the mode-$d$ design matrix with dimension $p_{_{d}}$ and sample size $Tp_{_{-d}}$, and $\mathbf{E}^{(d)}=(\mathcal{E}_{1}^{(d)},\cdots,\mathcal{E}_{T}^{(d)})\in\mR^{p_d\times (Tp_{-d})}$ be the noise matrix.
Then $\mathbf{Z}^{(d)}=(\mathcal{Z}_1^{(d)},\cdots,\mathcal{Z}_T^{(d)}) \in\mR^{k_d\times (Tp_{-d})}$ can be regarded as the latent factor matrix of $\mathcal{X}_t^{(d)}$, where $\mathcal{Z}_t^{(d)}=\mathcal{F}_{t}^{(d)} \mathbf{A}_{[D]/\{d\}}^{\top}$.
Hence, based on tensor observations $\{\mathcal{X}_{t}\}_{t=1}^T$ of  interest, model (\ref{matricized_1}) can be reformulated in matrix form as
\beqr
\label{vectorModel}
\mathbf{X}^{(d)}=\mathbf{A}_d\mathbf{Z}^{(d)}+\mathbf{E}^{(d)}, \hspace{4pt} d \in [D].
\eeqr
This model is essential the matrix form of a $p_d$-dim \textit{vector factor model} with sample size $Tp_{-d}$ \citep{Bai2003Econometrica-inferential} in that each column of $\mathbf{X}^{(d)}$ is the mode-$d$ fiber $\mathcal{X}_{t,i_{1}\cdots i_{d-1}: i_{d+1}\cdots i_{D}}$ with the $j$th entry being $x_{t,i_{1}\cdots i_{d-1}j i_{d+1}\cdots i_{D}}$ for $j\in[p_d]$.
Then it is natural to perform the principal components estimation stated in \citet{Bai2003Econometrica-inferential} (also called {the} spectral method in \citet[Remark 2]{ChenFan2023JASA-statistical}) for the mode-wise fiber factor model (1.3).
That is, if the sample variance-covariance matrix ($2$nd sample moment of $\mathbf{X}^{(d)}$) is denoted by
{\setlength\abovedisplayskip{1pt}
\setlength\belowdisplayskip{1pt}
\beqr
\label{Md_hat}
\widehat{\mathbf{M}}_{d}=\frac{1}{T p} \mathbf{X}^{(d)}\mathbf{X}^{{(d)}\top}=\frac{1}{T p}\sum_{t=1}^{T} \mathcal{X}_{t}^{(d)} \mathcal{X}_{t}^{(d)\top},
\eeqr}\noindent
then one obtains the estimator of the deterministic loadings $\mathbf{A}_d$ as follows
\beqr
\label{hat_A_d}
\widehat{\mathbf{A}}_d=\sqrt{p_d}~\mbox{eig}(\widehat{\mathbf{M}}_{d},k_d),~d\in[D],
\eeqr
where {$p = \Pi_{d=1}^D p_d$ and $\mbox{eig}(\mathbf{U}, r)\in\mR^{n\times r}$ is the matrix with the top $r$ eigenvectors of any non-negative definite placeholder matrix $\mathbf{U}\in\mR^{n\times n}$ as columns.}
Next, one can obtain the estimated latent tensor factor and signal part, respectively
\beqr
\label{hat_F}
\widehat{\mathcal{F}}_{t}=\frac{1}{p}\mathcal{X}_{t}\times_{1}\widehat{\mathbf{A}}_1^{\top}
\times_{2} \cdots \times_{D} \widehat{\mathbf{A}}_{D}^{\top} \quad\text{and}\quad
\widehat{\mathcal{S}}_{t}=\widehat{\mathcal{F}}_t\times_1 \widehat{\mathbf{A}}_{1}\times_2\cdots\times_D \widehat{\mathbf{A}}_D.
\eeqr
Such an estimation procedure that is derived from equations \eqref{matricized_1} to \eqref{hat_F} is named mode-wise PCA (moPCA) since one derives estimators of loading matrices  $\mathbf{A}_d$ mode by mode through flattening the observed tensor objects.
\citet{ChenYangZhang2022JASA-factor}'s TOPUP and TIPUP estimation approaches for tensor time series are consistent with the mode-wise spirit of moPCA;
when $D$ reduces to $2$, for the existing works on the bilinear form matrix factor models, their estimation procedures are along row (mode-$1$) and column (mode-$2$) respectively, say, the $\alpha$-PCA in \citet{ChenFan2023JASA-statistical} and the initial estimation in \citet{YuHeKongZhang2022JOE-projected} for dependent or independent observations, and the estimation in \citet{WangLiuChen2019JOE-factor} for matrix time series.

\subsection{Related works}
\label{subsec:literature_review}

\textbf{Two branches of PCA estimations}
PCA estimation is the most commonly used method for high-dimensional factor analysis \citep{ChenFanWang2020-high}.
It is implemented by {the} truncated eigen-decomposition or singular value decomposition on some second-moment-type statistics such as {the} matrix $\widehat{\mathbf{M}}_{d}$ \citep[Remark 2]{ChenFan2023JASA-statistical}.
Existing works apply PCA either on {the} constructed auto-covariance matrix for time series data, such as \citet{ChenYangZhang2022JASA-factor}, \citet{HanChenYangZhang2020arXiv-tensor}, \citet{HanChenZhang2022EJS-rank} (tensor time series), \citet{WangLiuChen2019JOE-factor}, \citet{ChenTsayChen2019JASA-constrained} (matrix time series), and \citet{LamYaoBathia2011Biometrika-estimation}, \citet{LamYao2012AOS-factor} among others (vector time series);
or on {the} constructed contemporaneous variance-covariance (CVC) matrix for weakly correlated or even independent {$D$th-order} tensor observations, such as
\citet{ChenFan2023JASA-statistical}, \citet{YuHeKongZhang2022JOE-projected} (matrix factor models) and \citet{BaiNg2002Econometrica-determining}, \citet{Bai2003Econometrica-inferential} among others (vector factor models);
{there are also several works that combine both auto-covariance and CVC, such as dynamic \citep{ForniHallinLippi2015, ForniHallinLippi2017} and static \citep{ZhangPanYaoZhou2023} vector factor models for vector time series.}
Note that,
CVC matrix not only incorporates contemporaneous information but also includes time-lagged correlation information when stationary dynamic factors are accommodated, though it may not explicitly capture the auto-correlation information of observations.
It inspires us to construct {CVC-based PCA estimations} to fill out the gap for estimation of low-rank structure for uncorrelated or weakly correlated higher-order tensor-variate collection.

\noindent\textbf{Projection and Iteration}
By adhering to the principle of mode-wise thought, moPCA estimation can be further enhanced based on variants of the matricized model \eqref{matricized_1}.
On the one hand, the projection technique \citep{LuPlataniotisVenetsanopoulos2008IEEE-mpca, WangAggarwalAeron2019PR-principal} can be implemented on the raw tensors driven by the practices in modern
machine learning and data science to enhance the signal-to-noise ratio and convergence rate.
This is evidenced by the projected estimation in \citet{YuHeKongZhang2022JOE-projected} that extends the mode-wise estimation in \citet{ChenFan2023JASA-statistical} for $2$nd-order tensor (matrix) observations.
On the other hand, the iteration technique \citep{ZhangXia2018IEEE-tensor, YuanZhang2016FCM-tensor} can be applied to further enhance the signal-to-noise ratio and convergence rate.
This is evidenced by the iterative projected estimations in \citet{HanChenYangZhang2020arXiv-tensor} that extend the mode-wise estimation in \citet{ChenYangZhang2022JASA-factor} for tensor time series.
These results motivate us to develop the projected mode-wise PCA estimation and the iterative projected mode-wise PCA estimation for weakly correlated or uncorrelated tensor observations in Sections \ref{sec:PmoPCA} and \ref{sec:IPmoPCA}, respectively.

\noindent\textbf{Other works}
CP decomposition is another widely used tensor decomposition method, the induced CP tensor factor model has been studied by \citet{HanZhangChen2021arXiv-CP} along the line of auto-covariance type estimation.
Besides tensor factor models, there are limited statistical inference for higher-order tensors on topics such as
tensor completion \citep[among others]{YuanZhang2016FCM-tensor, YuanZhang2017IEEE-incoherent, Zhang2019AOS-cross, XiaYuan2019FCM-polynomial, XiaYuanZhang2021AOS-statistically},
tensor decomposition \citep[among others]{SunLuLiu2017JRSSB-provable, TangBiQu2020JASA-individualized, ZhangXia2018IEEE-tensor, ZhangHan2019JASA-optimal, ZhouZhangZheng2022IEEE-optimal},
supervised learning \citep[among others]{ZhouLiZhu2013JASA-tensor, JohndrowBhattacharyaDunson2017AOS-tensor, ChenRaskuttiYuan2019JMLR-non, ZhouSunZhangLi2021JASA-partially, LiZhang2017JASA-parsimonious}, and etc.

\subsection{Organization}
\label{subsec:organization}

The remainder of this article is organized as follows.
Section 2 develops a set of tensor PCA that involves the projection technique, called PmoPCA, and elucidates the rationale of matricization and mode-wise thought from the perspective of tensor subspace learning and optimization of tensor decomposition.
Section 3 develops the other set of tensor PCA techniques that involve both iteration and projection for the purpose of better signal-to-noise ratio and convergence rate.
Section 4 establishes asymptotic properties of the proposed estimators including convergence rates and asymptotic distributions.
Section 5 evaluates the finite sample performance of {the proposed estimations}.
Section 6 analyzes a moving MNIST video {dataset} and an international import-export transport network dataset, which contains temporally uncorrelated and correlated tensor observations, respectively.
{Section 7 concludes the paper with a short summary}.
All the technique proofs are left in the \textit{Supplementary}.

\section{Projected mode-wise PCA estimation and tensor subspace learning}
\label{sec:PmoPCA}

The tensor factor model TuTFaM, like vector or matrix factor models \citep{LamYaoBathia2011Biometrika-estimation, WangLiuChen2019JOE-factor, ChenFan2023JASA-statistical}, suffers the identification problem.
That is, two sets of $(D+1)$-tuples $\{\mathbf{A}_1,\cdots,\mathbf{A}_D,\mathcal{F}_{t}\}$ and
$\{\mathbf{A}_1\mathbf{H}_1,\cdots,\mathbf{A}_D\mathbf{H}_D, \mathcal{F}_{t}\times_{1}\mathbf{H}_{1}^{-1}\times_{2} \cdots \times_{D} \mathbf{H}_{D}^{-1} \}$
lead to {the} identical model, for any nonsingular matrices $\{\mathbf{H}_d\in\mR^{k_d\times k_d}:d\in[D]\}$.
However, the mode-wise factor loading space $\mathcal{M}(\mathbf{A}_d)$, which is spanned by the columns of $\mathbf{A}_d$, is uniquely defined in the sense that $\mathcal{M}(\mathbf{A}_d)$ and $\mathcal{M}(\mathbf{A}_d\mathbf{H}_d)$ are identical for $d\in[D]$.
Note that, the lack of uniqueness of $\{\mathbf{A}_d\}_{d=1}^D$ brings the advantage of flexible rotation of an estimated factor loading matrix.
To facilitate our estimation, we assume
${p_d^{-1}}\mathbf{A}_d^\top \mathbf{A}_d$ approximates to the identity matrix
$\mathbf{I}_{k_d}$ of size $k_d$ when $p_d$ is quite large (Assumption 3).
This restriction confines the invertible transformation matrix to {the orthogonal one, which aids in simplifying the structure of the estimated loading matrix.}
Once such $\{\mathbf{A}_d\}_{d=1}^D$ is specified, the induced latent tensor factor $\mathcal{F}_t$ is determined. Therefore, the $D$+1 tuple $\{\mathbf{A}_1,\cdots,\mathbf{A}_D,\mathcal{F}_{t}\}$ in TuTFaM is estimable.
Hence, for our proposed estimations and inferential theories, we will focus on the estimation of mode-wise factor loading spaces $\mathcal{M}(\mathbf{A}_d)$ under the orthogonal rotation restriction.

\subsection{Projected mode-wise PCA estimation}
\label{PmoPCA}

The framework of moPCA estimation in the Subsection \ref{moPCA} is based on purely marginal mode information.
Nevertheless, it is more informative to estimate each loading by integrating data information from all modes so that the induced loading estimators can incorporate joint influence among all modes and the signal of the low-rank approximation can be strengthened.
This insight inspires us to give a more delicate consideration to the matricized factor model (\ref{matricized_1}).

{In the estimation procedure of moPCA}, we treat $\mathcal{F}_t^{(d)}$ and $\mathbf{A}_{[D]/\{d\}}$ together and notate their product as $\mathcal{Z}_t^{(d)}$, which acts as the latent factor in the fiber factor model (\ref{vectorModel}).
In essence, this is equivalent to replace each unknown $\mathbf{A}_{d^{\prime}}$ ($ d^{\prime}\in [D]/\{d\}$) with an \textit{identity matrix} in the slice matrix factor model (\ref{matricized_1}).
Such treatment is intuitive and expedient but at the cost of losing all \textit{interrelation information among modes except mode-$d$}.
Instead, one may {want} to incorporate the interaction influence of non-$d$ modes.
We address the problem by projecting $\mathcal{X}_t^{(d)}$ into {non-$d$ modes} hyperspace, that is, projecting the rows of $\mathcal{X}_t^{(d)}$ onto the column space of $\mathbf{A}_{[D]/\{d\}}$.
This is implemented by multiplying $\mathbf{A}_{[D]/\{d\}}$ from right on both sides of equation (\ref{matricized_1}), yielding the following matricized model
\beqr
\label{matricized_2}
\mathcal{Y}_{t,d}^{(d)}:=\frac{1}{p_{-d}}\mathcal{X}_{t}^{(d)}\mathbf{A}_{[D]/\{d\}}
\approx\mathbf{A}_d\mathcal{F}_{t}^{(d)}+\widetilde{\mathcal{E}}_t^{(d)},
\eeqr
where $\widetilde{\mathcal{E}}_t^{(d)}=\mathcal{E}_{t}^{(d)}\mathbf{A}_{[D]/\{d\}}/p_{-d}$ is the noise matrix, and the approximation is induced from the identification conditions.
This new model for a fixed tensor observation incorporates the parameters $\mathbf{A}_D, \cdots\mathbf{A}_{d+1}$, $\mathbf{A}_{d-1}\cdots\mathbf{A}_1$ explicitly, and plays a role in the induced sample variance-covariance matrix later, impacting the estimation of $\mathbf{A}_d$.

Let $\mathbf{Y}^{(d)}=(\mathcal{Y}_{1,d}^{(d)},\cdots,\mathcal{Y}_{T,d}^{(d)})\in\mR^{p_d\times (Tk_{-d})}$
be the observation of sample size $Tk_{-d}$ and dimension $p_d$, and
$\mathbf{F}^{(d)}=(\mathcal{F}_{1}^{(d)},\cdots,\mathcal{F}_{T}^{(d)})\in\mR^{k_d\times (Tk_{-d})}$ be  the latent factor matrix.
By concatenating $\mathcal{Y}_{t,d}^{(d)}$ for $t\in[T]$, we have the following vector factor model
\beqr
\label{vectorModel2}
\mathbf{Y}^{(d)}\approx\mathbf{A}_d\mathbf{F}^{(d)}+\widetilde{\mathbf{E}}^{(d)},
\eeqr
where  $\widetilde{\mathbf{E}}^{(d)}=(\widetilde{\mathcal{E}}_{1}^{(d)},\cdots,\widetilde{\mathcal{E}}_{T}^{(d)})\in\mR^{p_d\times (Tk_{-d})}$ is the noise matrix.
This new projected model is again a $p_d$-dim vector factor model while with sample size $Tk_{-d}$ and each column of $\mathbf{Y}^{(d)}$ is a \textit{projected fiber},
{\setlength\abovedisplayskip{3pt}
\setlength\belowdisplayskip{3pt}
\beqrs
\sum_{{i_1, \cdots, i_{d-1}, i_{d+1}, \cdots, i_D}}\mathcal{X}_{t,i_{1}\cdots i_{d-1}: i_{d+1}\cdots i_{D}}a_{1,i_1j_1}\cdots a_{d-1,i_{d-1}j_{d-1}}a_{d+1,i_{d+1}j_{d+1}}\cdots a_{D,i_{
D}j_{D}},
\eeqrs}\noindent
which is more complicated than the column fiber vector in the vector factor model (\ref{vectorModel}) and apparently contains the information of non-$d$ modes.

Again we perform the method of principal components for the mode-wise projected-fiber vector factor model (\ref{vectorModel2}).
However{,} we can not work on $(Tp_d)^{-1}\sum_{t=1}^{T}  {\mathcal{Y}}_{t,d}^{(d)}{\mathcal{Y}}_{t,d}^{(d)\top}$
directly like on equation (\ref{Md_hat})
because $\mathcal{Y}_{t,d}^{(d)}$ in representation  (\ref{matricized_2}) contains unknown matrix parameters $\mathbf{A}_D, \cdots\mathbf{A}_{d+1},\mathbf{A}_{d-1}\cdots\mathbf{A}_1$.
Therefore we consider substituting the unknown $\mathcal{Y}_{t,d}^{(d)}$ by a reasonable estimator below,
\beqr
\label{Y_td1}
\widehat{\mathcal{Y}}_{t,d}^{(d)}:=\frac{1}{p_{-d}}\mathcal{X}_{t}^{(d)}\breve{\mathbf{A}}_{[D]/\{d\}},
\eeqr
where $\breve{\mathbf{A}}_{[D]/\{d\}}=\breve{\mathbf{A}}_D
\otimes\cdots\otimes\breve{\mathbf{A}}_{d+1}
\otimes\breve{\mathbf{A}}_{d-1}\otimes\cdots\otimes\breve{\mathbf{A}}_1$ and $\{\breve{\mathbf{A}}_{d^{\prime}}\}_{d^{\prime} \in [D]/\{d\}}$ are some estimators that behave as the projected matrices.
The moPCA estimators $\{\widehat{\mathbf{A}}_{d^{\prime}}\}_{d^{\prime} \in [D]/\{d\}}$ in (\ref{hat_A_d}) have the natural priority in playing the role of  $\{\breve{\mathbf{A}}_{d^{\prime}}\}$ in equation (\ref{Y_td1}).
Consequently, based on the following sample variance-covariance matrix
{\setlength\abovedisplayskip{3pt}
\setlength\belowdisplayskip{3pt}
\beqr
\label{Md_tilde}
\widetilde{\mathbf{M}}_{d}=\frac{1}{Tp_d}\sum_{t=1}^{T} \widehat{\mathcal{Y}}_{t,d}^{(d)}\widehat{\mathcal{Y}}_{t,d}^{(d)\top},~d\in[D],
\eeqr}\noindent
by the method of principal components (spectral method),
we obtain estimators of loading matrices, and the resulting estimated latent factors and signal parts as
{\setlength\abovedisplayskip{1pt}
\setlength\belowdisplayskip{1pt}
\beqr\n
\widetilde{\mathbf{A}}_d&=&\sqrt{p_d}~\mbox{eig}(\widetilde{\mathbf{M}}_d,k_d),~d\in[D],\\
\widetilde{\mathcal{F}}_{t}&=&\frac{1}{p}\mathcal{X}_{t}\times_{1}\widetilde{\mathbf{A}}_1^{\top}
\times_{2} \cdots \times_{D} \widetilde{\mathbf{A}}_{D}^{\top},~t\in[T],\\\n
\widetilde{\mathcal{S}}_{t}&=&\widetilde{\mathcal{F}}_t\times_1 \widetilde{\mathbf{A}}_{1}\times_2\cdots\times_D \widetilde{\mathbf{A}}_D,~t\in[T].
\eeqr}\noindent
Such an estimation procedure is named projected mode-wise PCA (PmoPCA) since it implements projection on the mode-wise matricized factor model.
Such projection technique is widely used in tensor/matrix factor model estimations, such as \citet{HanChenYangZhang2020arXiv-tensor}'s iTOPUP and iTIPUP estimation approaches for tensor time series, and
\citet{YuHeKongZhang2022JOE-projected}'s projected estimation for matrix observations ($D=2$).

PmoPCA is superior to moPCA in terms of signal-to-noise ratio under TuTFaM in the sense that the order of the smallest nonzero eigenvalues of $\widetilde{\mathbf{M}}_{d}$ in equation (\ref{Md_tilde}) becomes higher than the zero ones compared to $\widehat{\mathbf{M}}_{d}$ in equation (\ref{Md_hat});
refer to Proposition 1 and Corollary 1 in the \textit{Supplementary}.

\subsection{Tensor subspace learning}
In this subsection, we demonstrate the philosophy of matricization and mode-wise thought from the perspective of tensor subspace learning and optimization on decomposition models.

Tensor subspace learning is one popular way for feature extraction to conduct tasks such as recognition, classification, and reconstruction in SIAM and data science \citep{He2005tensor, LuPlataniotisVenetsanopoulos2008IEEE-mpca}.
For the trivial low-rank approximation and specific Tucker decomposition, the latter of which is used to represent the signal part in model (\ref{Model}), one may denote their tensor subspaces
\beqrs
&&\mathcal{S}_{Trivial}=\left\{\sum_{j=1}^{r} f_{j} \mathcal{A}_{j} | \mathbf{f}=(f_1,\cdots,f_r)^{\top} \in \mR^{r}\right\},\\
&&\mathcal{S}_{Tucker}=\left\{\sum_{i_{1}=1}^{k_{1}} \cdots \sum_{i_{D}=1}^{k_{D}} f_{i_{1} \cdots i_{D}}\left(\mathbf{A}_{1,i_{1}} \circ \cdots \circ \mathbf{A}_{D,i_{D}}\right)|\mathcal{F}\in\mR^{k_1\times\cdots\times k_D}\right\},
\eeqrs
respectively, where the former is spanned by the tensor basis or eigenfaces $\{\mathcal{A}_j\in \mR^{p_{1} \times \cdots \times p_{D}}:j\in[r]\}$ with {entries of $\mathbf{f}$ being factor coefficients},
and the latter is spanned by the {rank-one} tensor basis or eigenfaces $\{\mathbf{A}_{1,i_{1}} \circ \cdots \circ \mathbf{A}_{D,i_{D}}\in \mR^{p_{1} \times \cdots \times p_{D}}:i_d\in[k_d],d\in[D]\}$ with {entries of $\mathcal{F}=(f_{i_1\cdots i_D})_{k_1\times\cdots\times k_{_D}}$ being factor coefficients}, $\mathbf{A}_{d,i_{d}}\in\mR^{p_d}$ being the $i_d$th column of $\mathbf{A}_{d}$, {and} $\circ$ being the outer product.
Compared to eigenfaces $\mathcal{A}_j$ of $\mathcal{S}_{Trivial}$, eigenfaces of $\mathcal{S}_{Tucker}$ have {the} specific \textit{mode-wise outer product {forms}}, which {extract} more structural information from the tensor data.
Tensor subspace is relatively low-dimensional and yet preserves sufficient variability in an input tensor object.
In a geometrical perspective, one may look {on} the tensor subspace analysis as projecting the raw high-dimensional tensor $\mathcal{X}_t$ onto the low-dimensional tensor subspace; and coordinates of the tensor basis once estimated, {act} as extracted latent features.
Estimation of the tensor subspace is achieved by minimizing the distance between the original tensor  {objects and the} tensor subspace, that is, to minimize the {following squared error with respect to the tensor basis and coefficients}
\beqr
\label{subspace}
\n
&&\min\limits_{\{\mathcal{A}_j\}_{j=1}^r,\{\mathbf{f}_t\}_{t=1}^T}\sum_{t=1}^{T}\left\|\mathcal{X}_{t}-\sum_{j=1}^{r} f_{t,j} \mathcal{A}_{j}\right\|_F^{2}
=\min\limits_{\tilde{\mathbf{A}},\tilde{\mathbf{F}}}\|\mathbf{X}-\tilde{\mathbf{A}}\tilde{\mathbf{F}}^{\top}\|_{F}^2,\\
&&\min\limits_{\{\mathbf{A}_d\}_{d=1}^D,\{\mathcal{F}_t\}_{t=1}^T}\sum_{t=1}^{T}\left\|\mathcal{X}_{t}-\mathcal{F}_{t}\times_{1}\mathbf{A}_{1}\times_{2} \cdots \times_{D} \mathbf{A}_{D}\right\|_F^{2}
=\min\limits_{\mathbf{A}, \mathbf{F}}\|\mathbf{X}-\mathbf{A}\mathbf{F}^{\top}\|_{F}^2,
\eeqr
where $\mathbf{X}=\left(\vec(\mathcal{X}_1),\cdots,\vec(\mathcal{X}_T)\right)\in\mR^{p\times T}$,
$\tilde{\mathbf{A}}=\left(\vec(\mathcal{A}_1),\cdots,\vec(\mathcal{A}_r)\right)\in\mR^{p\times r}$, $\tilde{\mathbf{F}}=(\mathbf{f}_1,\cdots,\mathbf{f}_T)^{\top}\in\mR^{T\times r}$, $\mathbf{A}=\mathbf{A}_D\otimes\cdots\otimes \mathbf{A}_1\in\mR^{p\times k}$, $\mathbf{F}=\left(\vec(\mathcal{F}_1),\cdots,\vec(\mathcal{F}_T)\right)^{\top}\in\mR^{T\times k}$
{with $k = \Pi_{d=1} ^D k_d$, and $\|\cdot\|_F$ and $\vec(\cdot)$ are the Frobenius norm and the vectorization operator of a matrix or tensor, respectively.}

For the upper problem of expressions ({\ref{subspace}) (trivial case with simple eigenfaces $\mathcal{A}_j$), solving  $\min_{\tilde{\mathbf{A}},\tilde{\mathbf{F}}} \|\mathbf{X}-\tilde{\mathbf{A}}\tilde{\mathbf{F}}^{\top}\|_{F}^2$ with {the} corresponding vectorized tensor factor model $\vec(\mathcal{X}_t)=\left(\vec(\mathcal{A}_1),\cdots, \vec(\mathcal{A}_r)\right)\mathbf{f}_t+\vec(\mathcal{E}_t)$ for $t\in[T]$ is the same to {perform} the principal components method for the vector factor model (3) in the seminal work of \citet{Bai2003Econometrica-inferential} up to a normalized constant.
Therefore the trivial (vectorized) PCA estimation for the subspace learning analysis of the left side is to conduct {the truncated} eigen-decomposition on the sample variance-covariance matrix $\mathbf{X}\mathbf{X}^{\top}=\sum_{t=1}^{T}\vec(\mathcal{X}_t)\vec(\mathcal{X}_t)^{\top}$.

For the lower problem {of expressions (\ref{subspace}) (Tucker case with complicated rank-one eigenfaces)}, however, solving  $\min_{\mathbf{A},\mathbf{F}}\|\mathbf{X}-\mathbf{A}\mathbf{F}^{\top}\|_{F}^2$ {with the  corresponding vectorized model} $\vec(\mathcal{X}_t)=\left(\mathbf{A}_D\otimes\cdots\otimes \mathbf{A}_1\right)\vec(\mathcal{F}_t)+\vec(\mathcal{E}_t)$ by conducting trivial PCA can NOT recover the matrices {$\{\mathbf{A}_d\}_{d=1}^D$} from the estimator of $\mathbf{A}=\mathbf{A}_D\otimes\cdots\otimes \mathbf{A}_1$, and its direct minimization is NP-hard \citep{HillarLim2013ACM-most}.
Hence the PCA-type methods specifically for Tucker subspace learning should be re-investigated.
Note that the left side minimization can be transferred to the maximization of $\sum_{t=1}^{T}\left\|\mathcal{X}_{t}\times_{1}\mathbf{A}_{1}^{\top}\times_{2} \cdots \times_{D} \mathbf{A}_{D}^{\top}\right\|_F^{2}$ {(this sum of squared norm is named total tensor scatter in \citet{LuPlataniotisVenetsanopoulos2008IEEE-mpca})} or $\mbox{tr}\{\mathbf{A}_d^{\top}\sum_{t=1}^{T}\tilde{\mathcal{X}}_t^{(d)} \tilde{\mathcal{X}}_t^{(d)\top}\mathbf{A}_d\}$ \citep[equation (541)]{Petersen2012} with respect to $\{\mathbf{A}_d\}_{d=1}^D$, where $\tilde{\mathcal{X}}_t^{(d)}=\mathcal{X}_t^{(d)}\mathbf{A}_{[D]/\{d\}}$ {is the projection of $\mathcal{X}_t^{(d)}$ onto the column space of $\mathbf{A}_{[D]/\{d\}}$} \citep[Subsection 4.2]{KoldaBader2009SIAM-tensor}.
Then, a tensor PCA is doable by performing {truncated} eigen-decomposition on the \textit{mode-wise} sample variance-covariance matrix $\sum_{t=1}^{T}\tilde{\mathcal{X}}_t^{(d)}\tilde{\mathcal{X}}_t^{(d)\top}$ {over $[D]$ iteratively} given the estimation of other modes, which is a tractable approximate solution for the {original} optimization problem.
Based on different {choices of the estimation of $\mathbf{A}_{[D]/\{d\}}$, the mode-wise PCA has many variants, say, moPCA, PmoPCA, the estimation in \citet{BarigozziHeLi2022arXiv-statistical}, \citet{ChenLam2024AOS-rank}, and the estimation in the next section.}

\section{Iterative projected mode-wise PCA estimation}
\label{sec:IPmoPCA}

Note that, {in PmoPCA,} the loading matrices for data projection are neither updated in each projection (step) over $[D]$ nor iterated in equation (\ref{Y_td1}).
Involving updating and iteration will augment the signal-to-noise ratio and {accelerate} convergence rates of estimators;
refer to Proposition 6 and Corollary 1 in the \textit{Supplementary}.
This drives us to do updating in each projection step across modes iteratively, yielding the $(s+1)$st iteration in the projection-updated estimator of $\mathcal{Y}_{t,d}^{(d)}$
\beqr
\label{Y_td2}
\widehat{\mathcal{Y}}_{t,d}^{(d,s+1)}:=\frac{1}{p_{-d}}\mathcal{X}_{t}^{(d)}\widehat{\mathbf{A}}_{[D]/\{d\}}^{(s+1)},
\eeqr
where $\widehat{\mathbf{A}}_{[D]/\{d\}}^{(s+1)}=(\widehat{\mathbf{A}}_D^{(s)}\otimes\cdots\otimes
\widehat{\mathbf{A}}_{d+1}^{(s)}\otimes
\widehat{\mathbf{A}}_{d-1}^{(s+1)}\otimes\cdots\otimes
\widehat{\mathbf{A}}_{1}^{(s+1)})$ and $\widehat{\mathbf{A}}_{d^{\prime}}^{(s)}$ is the $s$th iteration estimator for $d^{\prime}\in[D]/\{d\}$.
Again the moPCA estimators $\{\widehat{\mathbf{A}}_{d^{\prime}}\}$ are good candidates for the initial estimators $\{\widehat{\mathbf{A}}_{d^{\prime}}^{(0)}\}$.
The induced mode-wise sample variance-covariance matrix is
\beqr
\label{Md_hat_s}
\widehat{\mathbf{M}}_{d}^{(s+1)}=\frac{1}{Tp_d}\sum_{t=1}^{T}
  \widehat{\mathcal{Y}}_{t,d}^{(d,s+1)} \widehat{\mathcal{Y}}_{t,d}^{(d,s+1)\top},~d\in[D].
\eeqr
By the method of principal components, the estimators of loading matrices, and the resulting latent factors and signal parts are obtained as
{\setlength\abovedisplayskip{3pt}
\setlength\belowdisplayskip{3pt}
\beqr
\label{IPmoPCA}\n
\widehat{\mathbf{A}}_d^{(s+1)}&=&\sqrt{p_d}~\mbox{eig}(\widehat{\mathbf{M}}_{d}^{(s+1)},k_d),~d\in[D],\\
\widehat{\mathcal{F}}_{t}^{(s+1)}&=&\frac{1}{p}\mathcal{X}_{t}\times_{1}\widehat{\mathbf{A}}_1^{(s+1)\top}
\times_{2} \cdots \times_{D} \widehat{\mathbf{A}}_{D}^{(s+1)\top},~t\in[T],\\\n
\widehat{\mathcal{S}}_{t}^{(s+1)}&=&\widehat{\mathcal{F}}_t^{(s+1)}\times_1 \widehat{\mathbf{A}}_{1}^{(s+1)}\times_2\cdots\times_D \widehat{\mathbf{A}}_D^{(s+1)},~t\in[T].
\eeqr}\noindent
Such an estimation procedure is named iterative projected mode-wise PCA (IPmoPCA) since it implements iteration and projection simultaneously.
Note that we still use the notations for moPCA in representing IPmoPCA estimators hereafter since the proof is under the moPCA-initial setting.
The IPmoPCA estimation is summarized in Algorithm \ref{algorithm}.

{
\begin{algorithm}
\caption{IPmoPCA}
\begin{algorithmic}[1]
{\STATE \textbf{Input}:
Tensor observations $\{\mathcal{X}_t\}_{t=1}^T$, factor numbers $\{k_d\}_{d=1}^D$, the error tolerance $\epsilon$, the maximum number of iterations $m$, the initial estimators  $\{\widehat{\mathbf{A}}_d^{(0)}\}_{d=1}^D$
\STATE \textbf{Output}: Estimation
$\{\widehat{\mathbf{A}}_d^{(s+1)}\}_{d=1}^D$,
$\{\widehat{\mathcal{F}}_t^{(s+1)}\}_{t=1}^T$,
$\{\widehat{\mathcal{S}}_t^{(s+1)}\}_{t=1}^T$
\FOR{$s$ in $0:m$}
\FOR{$d$ in $1:D$}
\STATE compute
$\widehat{\mathcal{Y}}_{t,d}^{(d,s+1)}$ in equation \eqref{Y_td2} for $t\in[T]$,
$\widehat{\mathbf{M}}_{d}^{(s+1)}$ in equation \eqref{Md_hat_s},    $\widehat{\mathbf{A}}_d^{(s+1)}$ in equation \eqref{IPmoPCA}
\ENDFOR
\STATE Break the for loop, if {\footnotesize$\max\limits_{d \in [D]} \left\{p_d^{-1}\left\|\widehat{\mathbf{A}}_d^{(s+1)}\widehat{\mathbf{A}}_d^{^{(s+1)}\top} -\widehat{\mathbf{A}}_d^{(s)}\widehat{\mathbf{A}}_d^{^{(s)}\top}\right\|\right\}\leq \epsilon$}
\ENDFOR
\FOR{$t$ in $1:T$}
\STATE compute $\widehat{\mathcal{F}}_{t}^{(s+1)}$ and $\widehat{\mathcal{S}}_{t}^{(s+1)}$ in equation \eqref{IPmoPCA}
\ENDFOR
\RETURN $\{\widehat{\mathbf{A}}_d^{(s+1)}\}_{d=1}^D$,
$\{\widehat{\mathcal{F}}_t^{(s+1)}\}_{t=1}^T$,
$\{\widehat{\mathcal{S}}_t^{(s+1)}\}_{t=1}^T$}
\end{algorithmic}
\label{algorithm}
\end{algorithm}}

\noindent
\textbf{Remark 1.} (Comparison with HOOI)
IPmoPCA estimation can be regarded as a special higher-order orthogonal iteration (HOOI) \citep{DeDeVandewalle2000SIAM-best} by taking the observations as a $(D+1)$-th order tensor with mode-$(d+1)$ being time points.
Model (\ref{Model}) can be rewritten as follows
{\setlength\abovedisplayskip{3pt}
\setlength\belowdisplayskip{3pt}
\beqrs
\mathcal{Z}=\mathcal{G}\times_1\mathbf{A}_1\times_2\cdots\times_D\mathbf{A}_D
+\mathcal{R},
\eeqrs}\noindent
where $\mathcal{Z}\in\mR^{p_1\times \cdots\times p_D\times T}$ is the $(D+1)$th-order tensor,
the signal part of which has been expressed as the Tucker decomposition formation \citep{Tucker1966Psychometrika-some, KoldaBader2009SIAM-tensor} with the $(D+1)$th loading matrix as the $T$-dim identity matrix.
Then $\mathcal{Z}_{:\cdots :t}=\mathcal{X}_t$, $\mathcal{G}_{:\cdots :t}=\mathcal{F}_t$, and $\mathcal{R}_{:\cdots :t}=\mathcal{E}_t$, for $t\in[T]$.
If one adopts the classical Tucker decomposition algorithm HOOI, then only the first $D$ loading matrices {need} to be calculated.
More specifically, for $d\in[D]$, when estimating $\mathbf{A}_d$, eigen-decomposition is conducted on
$\mathcal{Q}_d^{(d,s+1)}\mathcal{Q}_d^{(d,s+1)\top}
=\sum_{t=1}^T\widehat{\mathcal{Y}}_{t,d}^{(d,s+1)}\widehat{\mathcal{Y}}_{t,d}^{(d,s+1)\top},$
which has the same column space as $\widehat{\mathbf{M}}_d^{(s+1)}$ in equation (\ref{Md_hat_s}),
where
$\mathcal{Q}_d^{(d,s+1)}=\mathcal{Z}^{(d)}\left(\mathbf{I}_T\otimes\widehat{\mathbf{A}}_{[D]/\{d\}}^{(s+1)}\right)$.

{
\noindent
\textbf{Remark 2.} (Comparison with MPCA and MOP-UP)
IPmoPCA has the same algorithmic form as multilinear principal component analysis (MPCA) \citep{LuPlataniotisVenetsanopoulos2008IEEE-mpca} except the difference in the stopping rule
and up to some normalization constants.
MPCA is essentially a tensor object dimension reduction framework with the Tucker low-rank structure, where their mode-wise loading matrices $\{\mathbf{A}_d\}$ are directly obtained by maximizing the so-called total tensor scatter $\sum_{t=1}^{T}\left\|\mathcal{X}_{t}\times_{1}\mathbf{A}_{1}^{\top}\times_{2} \cdots \times_{D} \mathbf{A}_{D}^{\top}\right\|_F^{2}$ \citep[equation (4)]{LuPlataniotisVenetsanopoulos2008IEEE-mpca} (written in our notations).
This is exactly the transferred maximization problem of the lower minimization problem in our expressions \eqref{subspace}.
Then MPCA is equivalent to performing the mode-wise PCA on the iterated-projected estimator $\widehat{\mathcal{Y}}_{t,d}^{(d,s+1)}$ in equation \eqref{Y_td2} up to the constant $p_{-d}$, and the stopping rule of the surrounding for loop is based on the distance of total tensor scatters.
Similar to IPmoPCA and MPCA, a recent mode-wise principal subspace pursuit (MOP-UP) framework \citep{TangYuanZhang2023arXiv-mode}, though with a different additive mode-wise linear forms low-rank structure, is actually the mode-wise PCA on the iterated-projected estimator like   $\widehat{\mathcal{Y}}_{t,d}^{(d,s+1)}$ up to the constant $p_{-d}$, except that the projection matrices are replaced by $\widehat{\mathbf{A}}_{[D]/\{d\},\perp}^{(s+1)}$, i.e. the orthogonal complement of $\widehat{\mathbf{A}}_{[D]/\{d\}}^{(s+1)}$.
}

\noindent
\textbf{Remark 3.} (Factor numbers)
When we develop the mode-wise PCA estimations, we assume the factor numbers $\{k_d\}_{d=1}^D$ are fixed.
But when they are unknown, we need to determine the number of factors before the estimation.
Here, take the moPCA as an example, we use the general ratio-based estimators as follows
\beqr
\label{FactorNumber}
\widehat{k}_d=\underset{1 \leq j \leq k_{\max }}{\arg \max } \frac{\widehat{\lambda}_{j}(\widehat{\mathbf{M}}_{d})}{\widehat{\lambda}_{j+1}(\widehat{\mathbf{M}}_{d})},
\quad d\in[D],
\eeqr
where $\widehat{\lambda}_{1}(\widehat{\mathbf{M}}_{d}) \geq \widehat{\lambda}_{2}(\widehat{\mathbf{M}}_{d}) \geq \cdots \geq \widehat{\lambda}_{p_d}(\widehat{\mathbf{M}}_{d}) \geq 0$ are the ordered eigenvalues of $\widehat{\mathbf{M}}_{d}$ and $k_{\text{max}}$ is a preassigned upper bound.
Under some assumptions, the order of the top $k_d$ eigenvalues of $\widehat{\mathbf{M}}_d$ is higher than the last $p_d-k_d$ ones.
Thus the above estimator can provide a consistent estimator for $k_d$, which is proved in the \textit{Supplementary}.
Note that the similar ratio-based estimators obtained based on $\widetilde{\mathbf{M}}_d$ and $\widehat{\mathbf{M}}_{d}^{(s+1)}$  also have the consistency property.

\section{Theoretical results}
\label{sec:theoretical_results}

In this section, we develop the theoretical properties of the proposed methods.
In Subsection \ref{subsec:assumptions}, we develop the necessary assumptions.
In Subsection \ref{subsec:IPmoPCA_properties}, we first establish the convergence rates and the asymptotic distributions of
estimated loading matrices, and the convergence rates of the estimated latent factors and signal parts of PmoPCA in Theorems \ref{consistency_PmoPCA}-\ref{consistency_PmoPCA_S};
next, we establish the convergence rates of estimated loading matrices of IPmoPCA in Theorem \ref{consistency_IPmoPCA}.

\subsection{Assumptions}
\label{subsec:assumptions}

Our assumptions for inference of TuTFaM can be regarded as higher-order extensions of those used in the large-dimensional vector factor model \citep{Bai2003Econometrica-inferential} and in the high-dimensional matrix factor model \citep{ChenFan2023JASA-statistical, YuHeKongZhang2022JOE-projected}, respectively.

\noindent
\textbf{Assumption 1.} \textit{$\alpha$-mixing.} The vectorized factor process $\{\vec(\mathcal{F}_t)\}$ and noise process $\{\vec(\mathcal{E}_t)\}$ are $\alpha$-mixing.

A vector process $\{\boldsymbol{u}_t\}$ is said to be $\alpha$-mixing,
if $\sum_{h=1}^{\infty} \alpha(h)^{1-2 / \gamma}<\infty$ for some $\gamma > 2$, where the mixing coefficient
$\alpha(h)$ represents $\sup _{i} \sup _{A \in \mathcal{C}_{-\infty}^{i}, B \in \mathcal{C}_{i+h}^{\infty}}$ $|P(A \cap B)-P(A) P(B)|$
with $\mathcal{C}_{i}^{j}$ as the $\sigma$-field generated by
$\left\{\boldsymbol{u}_{t}: i \leq t \leq j\right\}$.
$\alpha$-mixing is more general than stationarity as a probabilistic tool characterizing variable dependence.
It states that the degree of dependence decreases as the distance between two given sets of random variables goes to infinity.
Assumption 1 is a regular condition for deriving asymptotic distribution.

\noindent
\textbf{Assumption 2.} \textit{Tensor factor.} For any $t\in[T]$, assume tensor factor $\mathcal{F}_{t}$ is of fixed dimension $k_1\times\cdots \times k_D$ with $\mE(\mathcal{F}_t)=0$ and $\mE\|\mathcal{F}_t\|_F^{4}\leq c < \infty$
for some constant $c$.
The mode-wise matricized latent tensor factor $\mathcal{F}_t^{(d)}$ satisfies
$$\frac{1}{T}\sum\limits_{t=1}^T\mathcal{F}_t^{(d)}\mathcal{F}_t^{(d)\top}\stackrel{p}{\rightarrow}\mathbf{\Sigma}_d, \hspace{3pt} d\in[D],$$
where $\mathbf{\Sigma}_d \in \mR^{k_d\times k_d}$
is a positive definite matrix and has distinct eigenvalues.

That the sample second-order moment of mode-$d$ matricization of $\mathcal{F}_t$ converges to some positive definite matrix $\mathbf{\Sigma}_d$ can be derived under the $\alpha$-mixing condition in Assumption 1.
Refer to \citet[Chapter 16]{AthreyaLahiri2006measure} and \citet[Appendix A.3]{FrancqZakoian2019garch} for more details.
Without loss of generality, we arrange the diagonal elements of $\mathbf{\Lambda}_d$ in descending order to guarantee unique eigen-decomposition and identifiable eigenvectors, where $\mathbf{\Lambda}_d$ is from the spectral decomposition,
$\mathbf{\Sigma}_d =  \mathbf{\Gamma}_d\mathbf{\Lambda}_d\mathbf{\Gamma}_d^{\top}$.
Assumption 2 is an extension of Assumption A in \citet{Bai2003Econometrica-inferential} and Assumption B in \citet{YuHeKongZhang2022JOE-projected} for higher-order tensor observations respectively.

\noindent
\textbf{Assumption 3.} \textit{Mode-wise loading matrix.} \\
1. For $d\in[D]$, there exists some positive constant $c_d$ such that $\|\mathbf{A}_d\|_{max} \leq c_d$, where $\|\cdot\|_{max}$ is the maximum absolute value of the entries of a matrix; \\
2. As $\min_d \left\{p_d\right\} \rightarrow \infty$, $\left\|p_{d}^{-1} \mathbf{A}_d^{\top}\mathbf{A}_d-\mathbf{I}_{k_{d}}\right\| \rightarrow 0$ for $d\in[D]$, where $\|\cdot\|$ is the matrix $L_2$-norm.

Assumption 3.1 assumes the boundedness of entries of mode-wise loading matrices;
Assumption 3.2 assumes the asymptotic normalization through orthogonalization and makes our model work within the strong factor regime.
They guarantee that the factors have nontrivial contribution to most of the entries of tensor observations.
Assumption 3 is standard in confining loading matrices, acting as the mode-wise counterpart of Assumption F1 (a) plus (c) in \citet{StockWatson2002JASA-forecasting}, and is consistent with similar assumptions in the literature \citep{Bai2003Econometrica-inferential, LamYaoBathia2011Biometrika-estimation, WangLiuChen2019JOE-factor, ChenFan2023JASA-statistical, YuHeKongZhang2022JOE-projected}.

\noindent
\textbf{Assumption 4.} \textit{Cross modes interrelation of noise.}
For any $i_d,~j_d\in[p_d]$, $d\in[D]$, and $t\in[T]$, there exists some positive constant $c$ such that, \\
1. $\mE(\mathcal{E}_t)=0$, $\mE(e_{t,i_1\cdots i_D})^8\leq c$;\\
2. $\sum_{i_1^{\prime}=1}^{p_1}\cdots\sum_{i_D^{\prime}=1}^{p_D}|\mE(e_{t,i_1\cdots i_D}e_{t,i_1^{\prime}\cdots i_D^{\prime}})|\leq c$;\\
3. $\sum_{s=1}^T\sum_{i_1^{\prime},j_1^{\prime}=1}^{p_1}     \cdots\sum_{i_D^{\prime},j_D^{\prime}=1}^{p_D} |\cov(e_{t,i_1\cdots i_D}e_{t,j_1\cdots j_D},   e_{s,i_1^{\prime}\cdots i_D^{\prime}}e_{s,j_1^{\prime}\cdots j_D^{\prime}})|\leq c$.

Assumption 4.1 assumes zero-mean and bounded eighth-order moment of each entry of the noise tensor.
It is consistent with Assumption C.1 in \citet{Bai2003Econometrica-inferential}, Assumption 2.2 in \citet{ChenFan2023JASA-statistical}, and Assumption D.1 in \citet{YuHeKongZhang2022JOE-projected}.
Recall that the entry of the $(2D)$th-order covariance tensor of noise is $\mE(e_{t,i_1\cdots i_D}e_{t,i_1^{\prime}\cdots i_D^{\prime}})$.
It describes the across-modes dependency, which is assumed to be weak in Assumption 4.2.
This reduces to what was stated in Assumption 4.2 in \citet{ChenFan2023JASA-statistical}, and Assumption D.2(2) in \citet{YuHeKongZhang2022JOE-projected}.
Assumption 4.3 gives bounds to restrict weak correlation across different modes of the $(4D)$th-order covariance tensor $\cov(\mathcal{E}_t\circ\mathcal{E}_t,\mathcal{E}_s\circ\mathcal{E}_s)$.
This reduces to the mode-2 tensor counterpart expressed in Assumption 4.3 in \citet{ChenFan2023JASA-statistical}, and Assumption D.3 in \citet{YuHeKongZhang2022JOE-projected}.
Overall, we allow weak dependence across modes and observations of the noise in Assumption 4, which is quite general incorporating independent tensor observations as special cases.

\noindent
\textbf{Assumption 5.} \textit{Weak dependence between factor and noise.}
There exists some positive constant $c$, such that, for $i_d,~j_d\in[p_d]$, and $d\in[D]$\\
1. $\mE\|T^{-1/2}\sum_{t=1}^T\mathcal{F}_{t}^{(d)}  \boldsymbol{u}^{\top}\mathcal{E}_{t}^{(d)}\boldsymbol{v}\|^2\leq c$, for any deterministic unit vectors $\boldsymbol{u} \in \mR^{p_d}$ and $\boldsymbol{v} \in \mR^{p_{-d}}$;\\
2. $\|\sum_{i_1^{\prime}=1}^{p_1}\cdots\sum_{i_D^{\prime}=1}^{p_D}
\mE(\boldsymbol{\xi}_{i_1\cdots i_D}\otimes \boldsymbol{\xi}_{i_1^{\prime}\cdots i_D^{\prime}})\|_{max}\leq c$, where
$\boldsymbol{\xi}_{i_1\cdots i_D}$ is defined as $\vec(T^{-1/2}\sum_{t=1}^T\mathcal{F}_te_{t,i_1\cdots i_D})$;\\
3. $\|\sum_{i_1^{\prime},j_1^{\prime}=1}^{p_1}\cdots\sum_{i_D^{\prime},j_D^{\prime}=1}^{p_D}   \cov(\boldsymbol{\xi}_{i_1\cdots i_D}\otimes\boldsymbol{\xi}_{j_1\cdots j_D},  \boldsymbol{\xi}_{i_1^{\prime}\cdots i_D^{\prime}}\otimes\boldsymbol{\xi}_{j_1^{\prime}\cdots j_D^{\prime}})\|_{max}\leq c.$

Assumption 5 assumes weak dependence between factors and idiosyncratic tensor errors.
Assumption 5.1 assumes the weak temporal correlation of the matricized process
$\{\mathcal{F}_{t}^{(d)}\boldsymbol{u}^{\top}\mathcal{E}_{t}^{(d)}\boldsymbol{v}\}$.
Assumptions 5.2 and 5.3 assume the weak correlation across modes for the higher-order moments between the factor tensor and noise tensor.
It is consistent with Assumption D in \citet{Bai2003Econometrica-inferential} and Assumption E in \citet{YuHeKongZhang2022JOE-projected}.
Note that \citet{ChenFan2023JASA-statistical} and \citet{ChenYangZhang2022JASA-factor} assume uncorrelation between the noise and factor terms directly.

\noindent
\textbf{Assumption 6.} \textit{Central limit theorems.}
Let $\mathcal{E}_{t,i\cdot}^{(d)}\in\mR^{p_d}$ be the $i$th row of $\mathcal{E}^{(d)}$ for $i\in[p_d]$,
\beqrs
  \frac{1}{\sqrt{Tp_{-d}}}\sum\limits_{t=1}^T \mathcal{F}_{t}^{(d)}\mathbf{A}_{[D]/\{d\}}^{\top}\mathcal{E}_{t,i\cdot}^{(d)}
   \stackrel{L}{\rightarrow}N_{k_d}(0,\mathbf{V}_{di}),~ d\in[D],
\eeqrs
where $\mathbf{V}_{di}$ denotes $\lim _{T, p_{1},\cdots, p_{D} \rightarrow \infty} (Tp_{-d})^{-1}\mE(\sum_{t=1}^T\sum_{s=1}^T \mathcal{F}_{t}^{(d)}\mathbf{A}_{[D]/\{d\}}^{\top}\mathcal{E}_{t,i\cdot}^{(d)}\mathcal{E}_{s,i\cdot}^{(d)\top}$ $\mathbf{A}_{[D]/\{d\}}\mathcal{F}_{s}^{(d)\top})$, which is a positive definite matrix with eigenvalues bounded away from 0 and infinity.

The process $\{\mathcal{F}_{t}^{(d)}\mathbf{A}_{[D]/\{d\}}^{\top}\mathcal{E}_{t,i\cdot}^{(d)}\}$ converges to Gaussian distribution by the central limit theorem for the $\alpha$-mixing process \citep{AthreyaLahiri2006measure, FrancqZakoian2019garch} under Assumptions 1-5.
Assumption 6 is the mode-wise extension of Assumption F.4 in \citet{Bai2003Econometrica-inferential}.

\subsection{Theoretical properties}
\label{subsec:IPmoPCA_properties}

Let $\widetilde{\mathbf{\Lambda}}_d$ be the diagonal matrix
with diagonal elements being the top $k_d$ eigenvalues of $\widetilde{\mathbf{M}}_d$, and denote the below asymptotic orthogonal transformation matrix,
$$\widetilde{\mathbf{H}}_d=\frac{1}{Tpp_{-d}}\sum\limits_{t=1}^T\mathcal{F}_{t}^{(d)}\mathbf{A}_{[D]/\{d\}}^{\top}\widehat{\mathbf{A}}_{[D]/\{d\}}\widehat{\mathbf{A}}_{[D]/\{d\}}^{\top}\mathbf{A}_{[D]/\{d\}}\mathcal{F}_{t}^{(d)^{\top}}\mathbf{A}_d^{\top}\widetilde{\mathbf{A}}_d\widetilde{\mathbf{\Lambda}}_d^{-1}.$$
We present the following convergence rate of the PmoPCA estimators $\widetilde{\mathbf{A}}_d$.
\bet
\label{consistency_PmoPCA}
Suppose that $T$ and $\{p_d\}_{d=1}^D$ tend to infinity, and $\{k_d\}_{d=1}^D$ are fixed.
If Assumptions 1-5 hold, then there exists an asymptotic orthogonal matrix $\widetilde{\mathbf{H}}_d$, such that
$$\frac{1}{p_d}\|\widetilde{\mathbf{A}}_d-\mathbf{A}_d\widetilde{\mathbf{H}}_d\|_{F}^2
=O_p\left\{\frac{1}{Tp_{-d}}+\frac{1}{p^2}
+\sum\limits_{d^{\prime} \neq d}\left(\frac{1}{Tp_{d^{\prime}}^2}+
\frac{1}{T^2p_{-d^{\prime}}^2}+\frac{1}{p_d^2p_{d^{\prime}}^4}\right)\right\}.$$
\eet
This result is the same as that in Theorem 3.1 for projected estimators in \citet{YuHeKongZhang2022JOE-projected} for $2$nd-order tensor observations.
Let $\widetilde{\mathbf{A}}_{d,i\cdot}\in\mR^{k_d}$ be the $i$th row of $\widetilde{\mathbf{A}}_{d}$ and we present the results on the row-wise asymptotic normality of PmoPCA estimators.

\bet
\label{clt_PmoPCA}
Suppose that $T$ and $\{p_d\}_{d=1}^D$ tend to infinity, $\{k_d\}_{d=1}^D$ are fixed, and Assumptions 1-6 hold.
For $d\in[D]$ and $i\in[p_d]$, \\
(i) if $Tp_{-d}=o\big\{p^2+\sum\limits_{d^{\prime} \neq d}\left(Tp_{d^{\prime}}^2
+T^2p_{-d^{\prime}}^2+p_d^2p_{d^{\prime}}^4\right)\big\}$, then
$$\sqrt{Tp_{-d}}(\widetilde{\mathbf{A}}_{d,i\cdot}-\widetilde{\mathbf{H}}_d^{\top}\mathbf{A}_{d,i\cdot})
\stackrel{d}{\rightarrow}N(0,\mathbf{\Lambda}_d^{-1}\mathbf{\Gamma}_d^{\top}\mathbf{V}_{di}\mathbf{\Gamma}_d\mathbf{\Lambda}_d^{-1});$$
(ii) if $p^2+\sum\limits_{d^{\prime} \neq d}\left(Tp_{d^{\prime}}^2
+T^2p_{-d^{\prime}}^2+p_d^2p_{d^{\prime}}^4\right)=O(Tp_{-d})$, then
{\setlength\abovedisplayskip{3pt}
\setlength\belowdisplayskip{3pt}
\beqrs\widetilde{\mathbf{A}}_{d,i\cdot}-\widetilde{\mathbf{H}}_d^{\top}\mathbf{A}_{d,i\cdot}=O_p\left\{\frac{1}{p}+
\sum\limits_{d^{\prime} \neq d}\left(\frac{1}{\sqrt{T}p_{d^{\prime}}} +\frac{1}{Tp_{-d^{\prime}}}
+ \frac{1}{p_dp_{d^{\prime}}^2}\right)\right\}.\eeqrs}
\eet
Theorem \ref{clt_PmoPCA} shows that $\widetilde{\mathbf{A}}_d$ is a good estimator of $\mathbf{A}_d\widetilde{\mathbf{H}}_d$ in the sense that, $\widetilde{\mathbf{A}}_{d,i\cdot}$ is asymptotic consistent without limit restriction on $T$ and $\{p_d\}_{d=1}^D$.
However, the asymptotic normality can not be achieved unless $D\leq 2$.
When $D=3$, for establishing asymptotic normality, the condition $p_2/p_3=o(1)$ and $p_3/p_2=o(1)$ need to hold simultaneously.

In the following theorem, we show the consistency of estimated tensor factors $\widetilde{\mathcal{F}}_{t}\in\mR^{k_1\times \cdots \times k_D}$ in terms of the Frobenius norm and signal parts $\widetilde{\mathcal{S}}_{t}\in\mR^{p_1\times \cdots \times p_D}$ element-wise.

\bet
\label{consistency_PmoPCA_S}
Suppose that $T$ and $\{p_d\}_{d=1}^D$ tend to infinity, and $\{k_d\}_{d=1}^D$ are fixed.
If Assumptions 1-5 hold, then for $i_d\in[p_d]$, $d\in[D]$ and $t\in[T]$, we have
{\small
\begin{align*}
&(a)~\left\|\widetilde{\mathcal{F}}_{t}
-\mathcal{F}_{t}\times_{1}\widetilde{\mathbf{H}}_{1}^{-1}\times_{2} \cdots \times_{D}\widetilde{\mathbf{H}}_{D}^{-1}\right\|_F\\
&=O_p\left\{\frac{1}{\sqrt{p}}
+\sum\limits_{d=1}^D\frac{1}{\sqrt{T}p_{-d}}
+\sum\limits_{d_1=1}^D\sum\limits_{d_2\neq d_1}^D
\left(\frac{1}{\sqrt{Tp_{d_1}}p_{d_2}}
+\frac{1}{\sqrt{Tp_{-d_1}}p_{d_2}}
+\frac{1}{p_{d_1}p_{d_2}^2}\right)\right\}.\\
&(b)~\left|\widetilde{s}_{t,i_1\cdots i_D}-s_{t,i_1\cdots i_D}\right|
=O_p\left\{\frac{1}{\sqrt{p}}
+\sum\limits_{d=1}^D\left(\frac{1}{\sqrt{T}p_d}+\frac{1}{\sqrt{Tp_{-d}}}\right)
+\sum\limits_{d_1=1}^D\sum\limits_{d_2\neq d_1}^D\frac{1}{p_{d_1}p_{d_2}^2}\right\}.
\end{align*}}
\eet
Note that, compared with the results of moPCA estimators (Proposition 4 in the Supplementary), the convergence rates of
$\widetilde{\mathcal{F}}_{t}$ and $\widetilde{s}_{t,i_1\cdots i_D}$ have been enhanced.

Recall that, the $(s+1)$st iteration loading matrix estimator $\widehat{\mathbf{A}}_d^{(s+1)}$ represented in equation (\ref{IPmoPCA}) is $\sqrt{p_d}$ times the top $k_d$ eigenvectors of $\widehat{\mathbf{M}}_{d}^{(s+1)}$ represented in equation (\ref{Md_hat_s}).
Let $\widehat{\mathbf{\Lambda}}_d^{(s+1)}$ be the diagonal matrix whose diagonal elements are the top $k_d$ eigenvalues of $\widehat{\mathbf{M}}_d^{(s+1)}$ in descending order.
The asymptotic orthogonal matrix constructed from the representation of $\widehat{\mathbf{A}}_d^{(s+1)}$ in the proof of Theorem \ref{consistency_IPmoPCA} is
{\small
\beqrs
\widehat{\mathbf{H}}_d^{(s+1)}&=&\frac{1}{Tpp_{-d}}\sum\limits_{t=1}^T\mathcal{F}_{t}^{(d)}\mathbf{A}_{[D]/\{d\}}^{\top}\widehat{\mathbf{A}}_{[D]/\{d\}}^{(s+1)}\widehat{\mathbf{A}}_{[D]/\{d\}}^{(s+1)\top}\mathbf{A}_{[D]/\{d\}}
\mathcal{F}_{t}^{(d)^{\top}}\mathbf{A}_d^{\top}\widehat{\mathbf{A}}_d^{(s+1)}\widehat{\mathbf{\Lambda}}_d^{(s+1)^{-1}}.
\eeqrs}\noindent
Let
$w_{d}^{(s)}=p_d^{-1}\|\widehat{\mathbf{A}}_d^{(s)}-\mathbf{A}_d\widehat{\mathbf{H}}_d^{(s)}\|_{F}^2$ for non-negative integers $s$.
Theorem \ref{consistency_IPmoPCA} below establishes the consistency of $\widehat{\mathbf{A}}_d^{(s+1)}$
in the Frobenius norm.

\bet
\label{consistency_IPmoPCA}
Suppose that $T$ and $\{p_d\}_{d=1}^D$ tend to infinity, and $\{k_d\}_{d=1}^D$ are fixed.
If Assumptions 1-5 hold, there exists an asymptotic orthogonal matrix $\widehat{\mathbf{H}}_d^{(s+1)}$, such that for integers $s\geq 0$, we have
\beqr
\label{theory:consistency_IPmoPCA}
w_{d}^{(s+1)}
=\frac{1}{p_d}\|\widehat{\mathbf{A}}_d^{(s+1)}-\mathbf{A}_d\widehat{\mathbf{H}}_d^{(s+1)}\|_{F}^2
=O_p\left\{\frac{1}{Tp_{-d}} + \gamma_d^{(s+1)}\right\},
\eeqr
where
{\scriptsize\beqrs\n
\gamma_d^{(1)}&=&\frac{1}{p^2}+ \sum\limits_{d^{\prime} \neq d}\frac{1}{T^2p_{-d^{\prime}}^2}
+ \left(\frac{1}{Tp_d}+\frac{1}{p_d^2}\right)
\left(\sum\limits_{d^{\prime} = d+1}^D w_{d^{\prime}}^{(0)2}
+ \sum\limits_{d^{\prime} = 1}^{d-1} w_{d^{\prime}}^{(1)2}\right)\\\n
&&+ \sum\limits_{d^{\prime} = d+1}^D \Bigg\{\frac{1}{Tp_{d^{\prime}}^2}
+ \left(\frac{1}{T^2p_{-d^{\prime}}}+\frac{1}{T p_{d^{\prime}}}\right){w_{d^{\prime}}^{(0)}}
\Bigg\}\\
&&+ \sum\limits_{d^{\prime} = 1}^{d-1} \Bigg\{
\left(\frac{1}{T^2p_{d^{\prime}}}+\frac{1}{Tp_{d^{\prime}}^2} + \frac{w_{d^{\prime}}^{(1)}}{T}\right)
\left(\sum\limits_{d^{\prime\prime} = d^{\prime}+1}^Dw_{d^{\prime\prime}}^{(0)2}
+\sum\limits_{d^{\prime\prime} = 1}^{d^{\prime}-1}w_{d^{\prime\prime}}^{(1)2}\right)
+ \frac{w_{d^{\prime}}^{(1)}}{Tp_{-d^{\prime}}^2}
\Bigg\},
\eeqrs}
and
{\scriptsize\beqrs
\gamma_d^{(s+1)}&=&\frac{1}{p^2}+ \sum\limits_{d^{\prime} \neq d}\frac{1}{T^2p_{-d^{\prime}}^2}
+ \left(\frac{1}{Tp_d}+\frac{1}{p_d^2}\right)
\left(\sum\limits_{d^{\prime} = d+1}^D w_{d^{\prime}}^{(s)2}
+ \sum\limits_{d^{\prime} = 1}^{d-1} w_{d^{\prime}}^{(s+1)2}\right)\\
&&+ \sum\limits_{d^{\prime} = d+1}^{D} \Bigg\{
\left(\frac{1}{T^2p_{d^{\prime}}}+\frac{1}{Tp_{d^{\prime}}^2} + \frac{w_{d^{\prime}}^{(s)}}{T}\right)
\left(\sum\limits_{d^{\prime\prime} = d^{\prime}+1}^D w_{d^{\prime\prime}}^{(s-1)2}
+\sum\limits_{d^{\prime\prime} = 1}^{d^{\prime}-1}w_{d^{\prime\prime}}^{(s)2}\right)
+ \frac{w_{d^{\prime}}^{(s)}}{Tp_{-d^{\prime}}^2}
\Bigg\}\\
&&+ \sum\limits_{d^{\prime} = 1}^{d-1} \Bigg\{
\left(\frac{1}{T^2p_{d^{\prime}}}+\frac{1}{Tp_{d^{\prime}}^2} + \frac{w_{d^{\prime}}^{(s+1)}}{T}\right)
\left(\sum\limits_{d^{\prime\prime} = d^{\prime}+1}^Dw_{d^{\prime\prime}}^{(s)2}
+\sum\limits_{d^{\prime\prime} = 1}^{d^{\prime}-1}w_{d^{\prime\prime}}^{(s+1)2}\right)
+ \frac{w_{d^{\prime}}^{(s+1)}}{Tp_{-d^{\prime}}^2}\Bigg\}.
\eeqrs}
\eet
In Theorem \ref{consistency_IPmoPCA}, one can see that convergence rates of IPmoPCA estimators for loading matrices of the $(s+1)$st iteration depend on that of former iterations, projection steps, and moPCA estimators.
Thus the convergence rate of IPmoPCA estimators may be enhanced compared to that of moPCA estimators (Proposition 2 in the Supplementary) since $Tp_{-d}$ may run faster than $p_d^2$ but slower than $1/\gamma_d^{(s+1)}$.
For tensor time series under the TuTFaM setting, \citet{HanChenYangZhang2020arXiv-tensor} presented iTOPUP and iTIPUP methods by adding iteration and projection on TOPUP and TIPUP methods in \citet{ChenYangZhang2022JASA-factor} respectively.
Under assumptions of strong factor and fixed factor numbers,
convergence rates of estimated loading matrices are all $O_p(1/Tp_{-d})$ for estimators of TIPUP, iTIPUP, and iTOPUP.

It is understandable that the above three estimators have faster convergence rates.
Under two stringent assumptions on noise $\mathcal{E}_t$, itself uncorrelated (white) across time and uncorrelated with factors $\mathcal{F}_t$ for all time lags, the expectation of their mode-wise sample auto-covariance matrix remains only the time-lag factor terms.
In contrast, under mild assumptions on noise $\mathcal{E}_t$,
itself weakly correlated and correlated with factors, our proposed IPmoPCA estimators integrate contemporaneous information from mode-wise sample variance-covariance matrices, whose expectation incorporates not only the factor terms but also the error terms and the interrelation terms between factor and noise.
The rich information of data beyond the factor terms,
despite complicating various orders of summands of $\gamma_d^{(1)}$
in the convergence rate expression, makes our estimation outperform auto-covariance based methods, and more effective in a wider application scope.
Refer to the numerical studies in assessing errors of estimated loading matrices, signal parts, and factor numbers.
Our strategy sacrifices the convergence rate but maintains more structural information so as to guarantee good estimation.
In addition, the IPmoPCA estimators may achieve the same convergence rate as the above three auto-covariance based estimators provided that the order of $1/Tp_{-d}$ is higher than $\gamma_d^{(s+1)}$ in equation (\ref{theory:consistency_IPmoPCA}).

\noindent
\textbf{Remark 4.} (Comparison of convergence rate)
Let IPmoPCA-$i$, $i\in \mathbb{N}$ be the IPmoPCA estimation with $i$ times iterations.
We compare convergence rates of estimators for $\mathbf{\mathbf{A}}_1$ by methods mentioned by far including IPmoPCA estimators with iteration up to thrice for tensors with equal mode size $q$, referred to Table \ref{Comparsion}.
The results indicate that the convergence rate of IPmoPCA-$3$ is comparable with that by auto-covariance based methods {and is faster than those of moPCA and PmoPCA} as the order of tensors increases.
Thus increasing the number of iterations is a way to acquire ideal theoretical rates.
\begin{table}[H]
\caption{The comparison of convergence rate when $p_1=\cdots=p_D=q$}
\label{Comparsion}
\begin{center}
\scalebox{0.9}{
\begin{tabular}{|ccccc|}
\hline
Methods  &  $D=2$ & $D=3$ & $D=4$ & $D=10$ \\
\hline
iTOPUP/iTIPUP        & $\frac{1}{Tq}$
       & $\frac{1}{Tq^2}$
       & $\frac{1}{Tq^{3}}$
       & $\frac{1}{Tq^{9}}$\\
moPCA         & $\frac{1}{Tq}+\frac{1}{q^2}$
       & $\frac{1}{q^2}$
       & $\frac{1}{q^2}$
       & $\frac{1}{q^2}$\\
PmoPCA        & $\frac{1}{Tq}+\frac{1}{q^{4}}$
       & $\frac{1}{Tq^2}+\frac{1}{q^{6}}$
       & $\frac{1}{Tq^2}+\frac{1}{q^{6}}$
       & $\frac{1}{Tq^2}+\frac{1}{q^{6}}$\\
IPmoPCA-$3$       & $\frac{1}{Tq}+\frac{1}{q^{4}}$
          & $\frac{1}{Tq^{2}}+\frac{1}{q^{6}}$
          & $\frac{1}{Tq^{3}}+\frac{1}{q^{6}}$
          & $\frac{1}{Tq^{9}}+\frac{1}{T^3q^{6}}+\frac{1}{T^4q^{5}}+\frac{1}{q^{20}}$\\
\hline
\end{tabular}}
\end{center}
\end{table}

\section{Simulation}
\label{sec:simulation}

In this section, we compare the performance of {five} methods including moPCA, PmoPCA, IPmoPCA, iTIPUP and {PROJ, where PROJ is proposed by \citet{ChenLam2024AOS-rank}, which is an iterative projection estimation based on a pre-averaging initial estimator, and the projection is on the direction aligning to the strongest estimated factor}.
The estimation errors of loading matrices, signal parts, and factor numbers estimators, as well as the computation time, are reported.
Note that, the moPCA estimators are used as projected matrices and initial estimators for PmoPCA and IPmoPCA, respectively.

\subsection{Simulation settings}
\label{subsec:setting}

The simulation data is generated by model (\ref{Model}) in the case of $D=3$ with $k_1=2$, $k_2=3$, and $k_3=4$.
The specific data generating procedure is as follows.
First, the loading matrix $\mathbf{A}_d$ is taken as $\sqrt{p_d}$ times the first $k_d$ left singular vectors of the SVD decomposition on the matrix with standard normal elements {for $d\in[3]$.}
Second, the tensor factor $\mathcal{F}_t \in \mR^{2\times 3 \times 4}$, $t\in [T]$ is generated from the following model
{\setlength\abovedisplayskip{3pt}
\setlength\belowdisplayskip{3pt}
\beqrs\mathcal{F}_{t} = \phi \mathcal{F}_{t-1}+\sqrt{1-\phi^2}\mathcal{V}_t,
\eeqrs}\noindent
where $\vec(\mathcal{V}_t)\sim N_k(\mathbf{0},\mathbf{I}_k)$, {$k=24$}, and
$\phi$ controls the temporal correlation of {tensor factors}.
Note that each element of $\mathcal{F}_{t}$ follows the $AR(1)$ model with unit variance.
At last, the tensor noise $\mathcal{E}_t\in \mR^{p_1 \times p_2 \times p_3}$, $t\in [T]$ is generated from the following model
{\setlength\abovedisplayskip{3pt}
\setlength\belowdisplayskip{3pt}
\beqrs
\mathcal{E}_{t} = \psi \mathcal{E}_{t-1}+\sqrt{1-\psi^2}\mathcal{U}_t,
\eeqrs}\noindent
where $\vec(\mathcal{U}_t)\sim N(0,\mathbf{\Delta}_D\otimes\cdots\otimes\mathbf{\Delta}_1)$, and the diagonal elements of $\mathbf{\Delta}_d$ are unit and the off diagonal elements are $1/p_{d}$.
Besides the parameter $\psi$ controls the temporal correlation of {tensor noises}, the adoption of the array-normal distribution of $\mathcal{U}_t$ \citep{Hoff2011BA-separable, FosdickHoff2014AOAS-separable} makes the tensor noise possessing intra-mode and inter-modes correlation as well.
We consider four different scenarios of $\phi$ and $\psi$ resulting in different type tensor observations.

\noindent
\textbf{Scenario I: (Uncorrelated)} Let $\phi=0$ and $\psi=0$, which means both $\mathcal{F}_t$ and $\mathcal{E}_t$, hence the tensor observations $\mathcal{X}_t$, have no temporal correlation.

\noindent
\textbf{Scenario II: (Correlated factors)} Let $\phi=0.6$ and $\psi=0$, which means the noise $\mathcal{E}_t$ has no temporal dependence and
the tensor observations $\mathcal{X}_t$ has temporal correlation due to the factor $\mathcal{F}_t$'s temporal correlation.

\noindent
\textbf{Scenario III: (Correlated noises)} Let $\phi=0$ and $\psi=0.8$, which means the factor $\mathcal{F}_t$ has no temporal dependence and
the tensor observations $\mathcal{X}_t$ has temporal correlation due to the noise $\mathcal{E}_t$'s temporal correlation.

\noindent
\textbf{Scenario IV: (Correlated factors and noises)} Let $\phi=0.6$ and $\psi=0.8$, which means both $\mathcal{F}_t$ and $\mathcal{E}_t$, hence the tensor observations $\mathcal{X}_t$, are temporal correlated.

For each scenario, five sizes $(20,20,20,20)$, $(50,20,20,20)$, $(50,50,50,50)$, $(100,50,50,50)$ and $(100,100,$ $100,100)$ are considered for $(T,p_1,p_2,p_3)$.
The number of replications is 100.
{Five methods, i.e., iTIPUP(1), moPCA, PmoPCA, IPmoPCA and PORJ}, are compared on the simulated data.
Note that, iTOPUP is excluded for its memory intractability and time consumption under high dimensionality, and iTIPUP(1) means the auto-covariance lag-$1$ case.

\subsection{Comparison of estimated loading matrices}

In this subsection, we display the results of the estimation error of estimated loading matrices.
Recall that the loading matrices are not identifiable, only the column spaces are uniquely determined.
Hence, the measurement focusing on the column space distance is used to evaluate the estimation errors, which are defined as follows
\beqrs
\mathcal{D}(\widehat{\mathbf{A}}_d,\mathbf{A}_d)
=p_d^{-1}\left\|\widehat{\mathbf{A}}_d
\widehat{\mathbf{A}}_d^{\top}-\mathbf{A}_d{\mathbf{A}}_d^{\top}\right\|,
\eeqrs
where $\widehat{\mathbf{A}}_d$ is moPCA estimator for mode-$d$.
The closer the column spaces of $\widehat{\mathbf{A}}_d$ and ${\mathbf{A}}_d$ get to each other, the closer $\mathcal{D}(\widehat{\mathbf{A}}_d,\mathbf{A}_d)$ goes to 0.
For PmoPCA estimator $\widetilde{\mathbf{A}}_d$ and IPmoPCA estimator $\widehat{\mathbf{A}}_d^{(s)}$, the measurement can be similarly defined.

We report the simulation results as the averaged column space distance over 100 replications.
Because the results for different modes are similar, only the ones for mode-$1$ are reported in Figure \ref{A1} for simplicity, and the ones for other modes are left in S4 of the \textit{Supplementary}.
From Figure \ref{A1}, we can conclude four points.
First, all {five} methods, i.e., iTIPUP(1), moPCA, PmoPCA, IPmoPCA, and {PROJ}, show consistency as the size increases in not only the sample size $T$ but also the mode sizes $p_d$ for $d\in[3]$.
Second, in Scenarios I and III, there exists some outliers for iTIPUP(1) because the temporal uncorrelatedness of the tensor factors ($\phi=0$) violates the basic assumptions of auto-covariance type methods.
{Third, in Scenarios I and III, PROJ performs well; while in Scenarios II and IV, its performance is not that satisfactory, it may be due to the information insufficiency of their single-direction projection.
At last, our methods (moPCA, PmoPCA, IPmoPCA) outperform iTIPUP(1) under all scenarios, and ourperform PROJ under Scenarios II and IV, which illustrates the robustness of our approach for a wider application scope.
In addition, moPCA, PmoPCA, and IPmoPCA have similar performance although IPmoPCA is slightly better sometimes.}

\begin{figure}[H]
  \centering
  \includegraphics[width=5.5in]{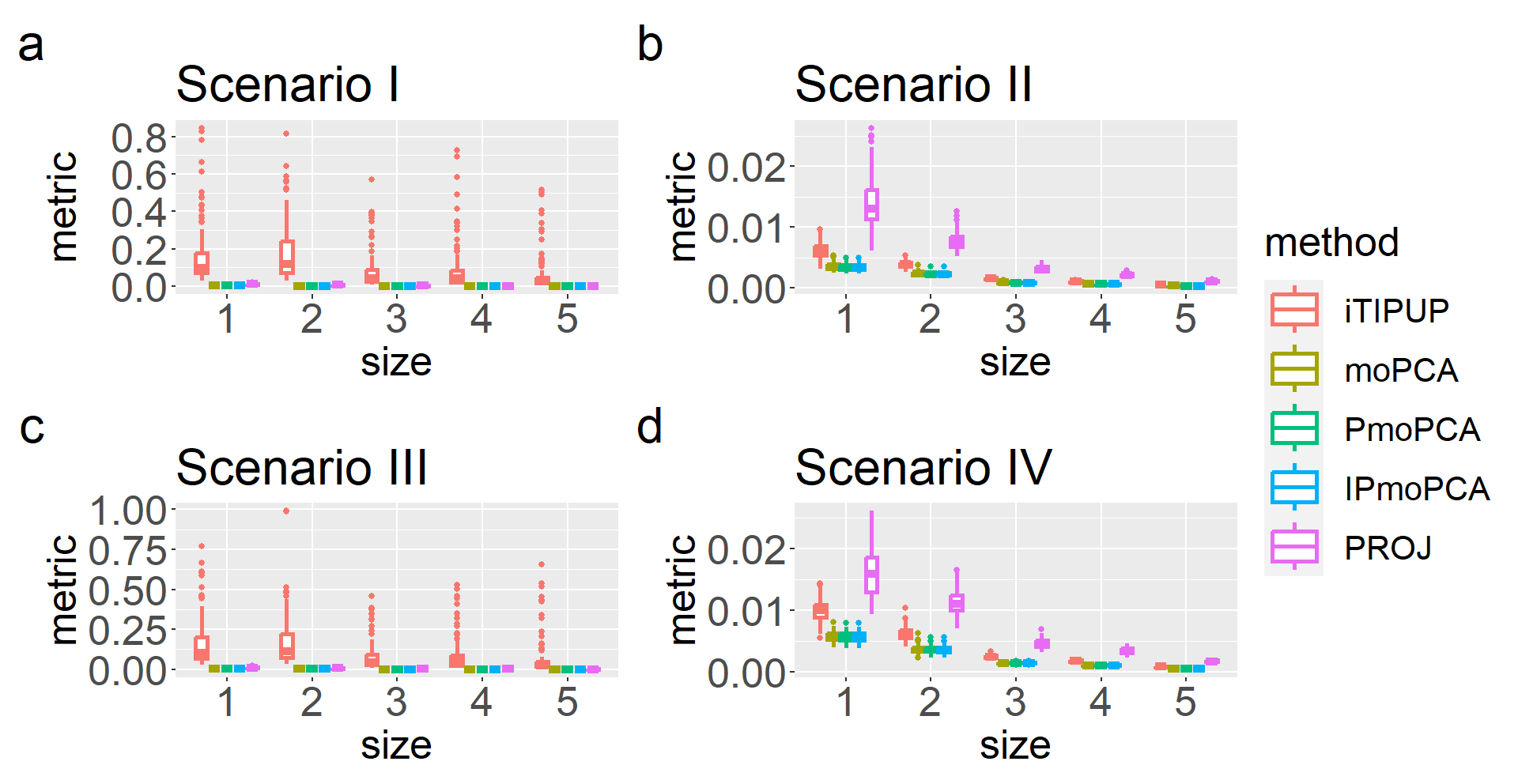}\\
  \caption{{The results for mode-$1$ loading matrix}}
  \label{A1}
\end{figure}

\begin{figure}[H]
  \centering
  \includegraphics[width=5.5in]{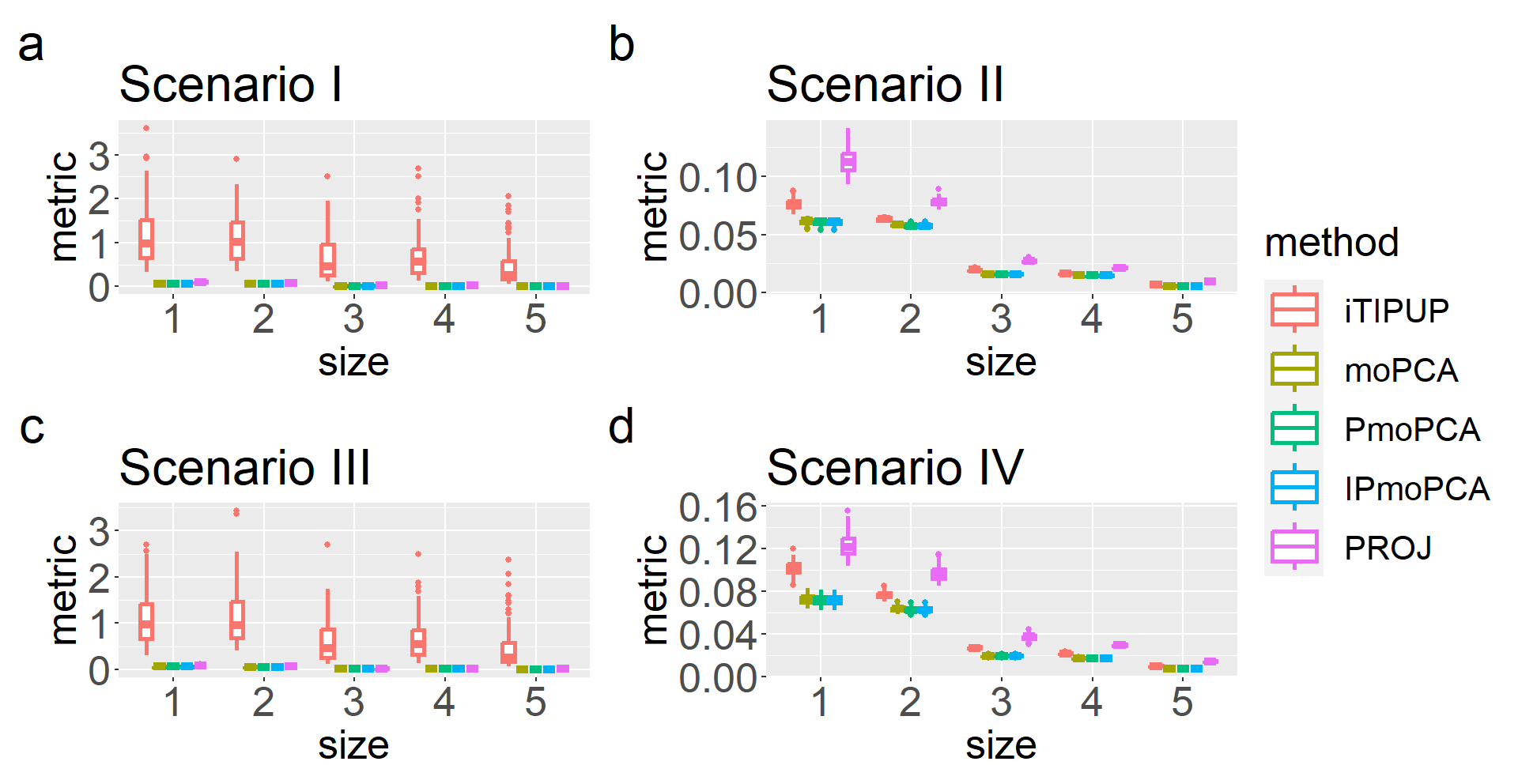}\\
  \caption{{The results for signal parts}}
  \label{X}
\end{figure}

\subsection{Comparison  of estimated signal parts}

In this subsection, we display the results of the estimation error of estimated signal parts, which is measured by the root mean square error (RMSE) defined below,
{\setlength\abovedisplayskip{3pt}
\setlength\belowdisplayskip{3pt}
\beqrs
\mbox{RMSE}_{\mathcal{S}}
=\sqrt{\frac{1}{Tp}\sum\limits_{t=1}^T\|\widehat{\mathcal{S}}_t-\mathcal{S}_t\|_F^2},
\eeqrs}\noindent
where $\widehat{\mathcal{S}}_t$ is moPCA estimator for the signal part.
For PmoPCA estimator $\widetilde{\mathcal{S}}_t$ and IPmoPCA estimator $\widehat{\mathcal{S}}_t^{(s)}$, the measurement can be similarly defined.

We report the simulation results as the averaged RMSE over 100 replications in Figure \ref{X}.
From Figure \ref{X}, we can see a similar trend to the estimated loading matrix in Figure \ref{A1}.
{All five methods show consistency as the size increases; there still exists some outliers for iTIPUP(1) in Scenarios I and III; the performance of PROJ is still not that satisfactory in Scenarios II and IV; our methods outperform iTIPUP(1) under all scenarios and outperform PROJ under Scenarios II and IV.}

\subsection{Comparison of estimated factor numbers}
\label{subsec:factor_number}

In this subsection, we display the results of the estimation error of estimated factor numbers, and compare the computation time.
{Note that the estimation of factor numbers with PROJ is not included considering its tedious  computation time.}

First, the estimation error of estimated factor numbers is measured by the accuracy defined as follows
{\setlength\abovedisplayskip{1pt}
\setlength\belowdisplayskip{1pt}
\beqrs
\mbox{acc}=\frac{1}{D}\sum\limits_{d=1}^D\mathbf{I}(\widehat{k}_d=k_d)*100\%,
\eeqrs}\noindent
where $\widehat{k}_d$ is defined in equation (\ref{FactorNumber}).
Then the average values of $\mbox{acc}$ among 100 replications are reported as the final results in Table \ref{FacNum}, from which we can see that our approach can estimate the factor numbers correctly under all scenarios, but iTIPUP(1) can not identify the right number of factors when the latent tensor factor is temporal uncorrelated (Scenarios I and III).
The result is similar to that of factor loadings and signal parts, which shows that our approach is robust in a wider application scope again.

\begin{table}
\caption{The result for estimated factor numbers}
\label{FacNum}
\begin{center}
\scalebox{0.9}{
\begin{tabular}{|c|ccccc|ccccc|}
\hline
& \multicolumn{5}{c|}{Scenario I} & \multicolumn{5}{c|}{Scenario II}\\
\hline
methods & size 1 & size 2 & size 3 & size 4 & size 5 & size 1 & size 2 & size 3 & size 4 & size 5\\
\hline
$\mbox{acc}_{iTIPUP}$ & 82.0 & 78.0 & 88.0 & 88.3 & 91.3 & 100& 100& 100& 100& 100\\
$\mbox{acc}_{moPCA}$ & 100 & 100& 100& 100& 100& 100& 100& 100& 100& 100\\
\hline
& \multicolumn{5}{c|}{Scenario III} & \multicolumn{5}{c|}{Scenario IV}\\
\hline
methods & size 1 & size 2 & size 3 & size 4 & size 5 & size 1 & size 2 & size 3 & size 4 & size 5\\
\hline
$\mbox{acc}_{iTIPUP}$ & 81.0 & 79.7 & 89.3 & 89.0 & 90.0 & 100& 100& 100& 100& 100\\
$\mbox{acc}_{moPCA}$ & 100 & 100& 100& 100& 100& 100& 100& 100& 100& 100\\
\hline
\end{tabular}}
\end{center}
\end{table}

Second, the averaged computation time over 100 replications of the five methods are reported in Figure \ref{TIME}, from which we can see that our methods have faster computational speed, especially for large sizes.
The computation time of PmoPCA estimation starts from using moPCA as projected matrices in representation (\ref{Y_td1}), and that of IPmoPCA estimation starts from using moPCA as initial estimators in representation (\ref{Y_td2}).

\begin{figure}[H]
  \centering
  \includegraphics[width=4.5in]{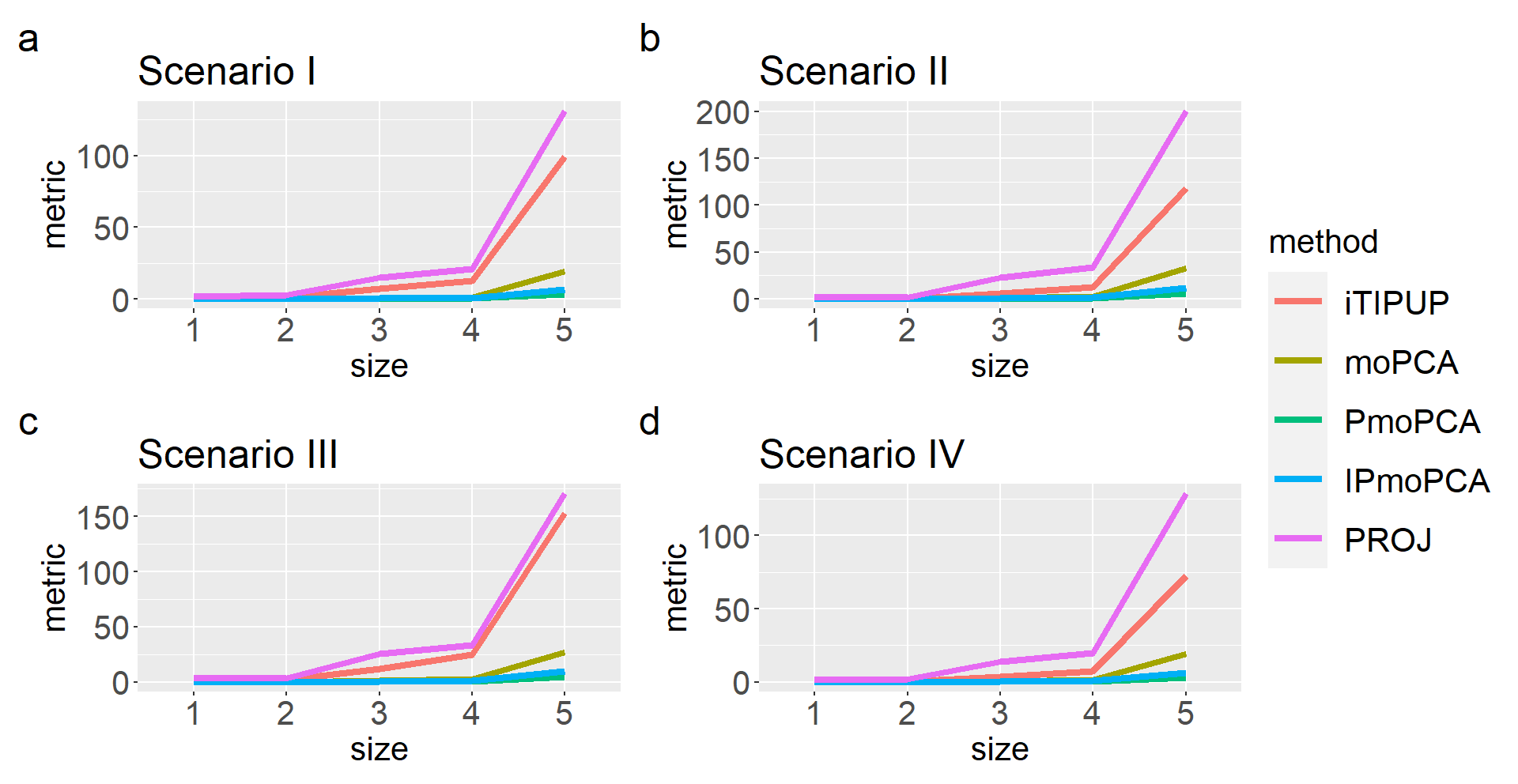}\\
  \caption{{The results for running time}}
  \label{TIME}
\end{figure}

\section{Real data analysis}

We demonstrate the proposed approaches on two real datasets.
The first one is the Moving MNIST data \citep{SrivastavaMansimovSalakhutdinov2015ICML-unsupervised}
(\url{http://www.cs.toronto.edu/~nitish/unsupervised_video/}), which has serially uncorrelated tensor observations.
The second one is the import-export transport networks data, which is analyzed in \citet{ChenYangZhang2022JASA-factor}, and has temporally correlated tensor observations.
The goal of the analysis is from two aspects, one is about the explanation of the estimated loading matrix as well as the corresponding mode-wise clustering result, and the other is about the performance of tensor reconstruction.
{Five} methods, i.e., iTIPUP, moPCA, PmoPCA, IPmoPCA, and {PROJ}, are adopted in accordance with simulation studies.
Note that, the moPCA estimators are used as projected matrices and initial estimators for PmoPCA and IPmoPCA respectively.

\subsection{Moving MNIST data{: serial independent}}

The Moving MNIST data contains 10,000 {serial independent video sequences.
Each video contains two numbers rotating in a counterclockwise direction to switch relative positions across 20 frames; refer to the frames of the first video sample in Figure \ref{example}.
Each frame has a spatial resolution of $64\times 64$ pixels, then} a video can be formulated as a $3$rd-order tensor $\mathcal{X}_t\in\mR^{20\times 64\times 64}$ for $t\in[10000]$, where the $1$st mode is the frame mode, {and} the $2$nd and the $3$rd modes are the pixel row and column {coordinate} modes, {respectively}.
{The data is centralized to satisfy our assumptions, and $({k}_1,{k}_2,{k}_3)=(2,3,4)$ is used as the size of the latent factors.}

\begin{figure}[H]
  \centering
  \includegraphics[width=4in]{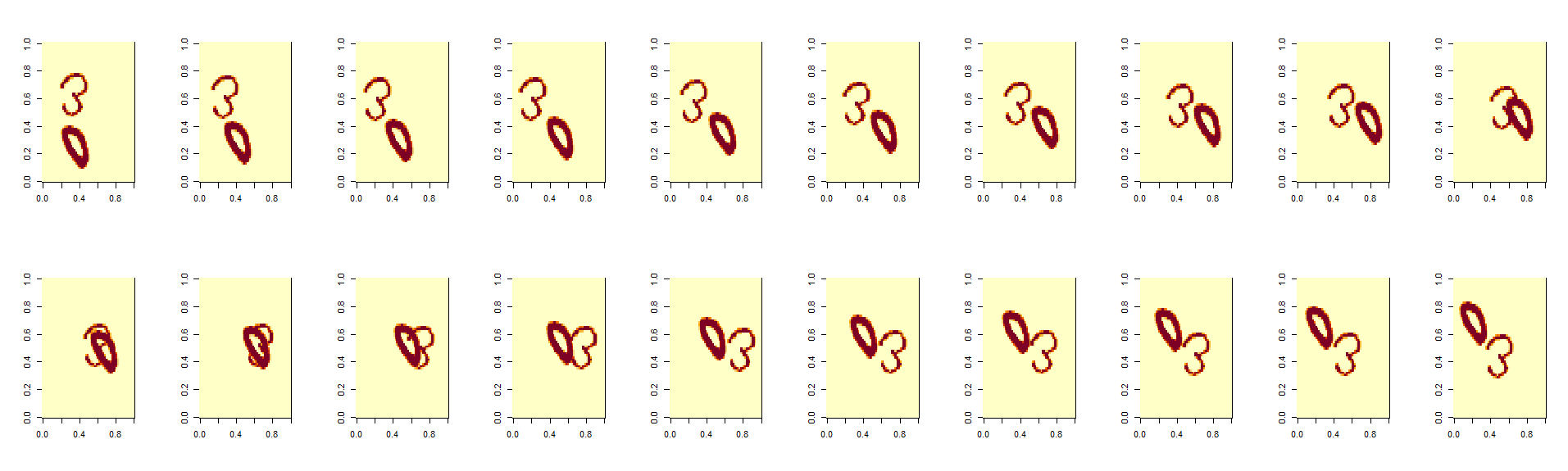}
  \caption{{The 20 frames of video 1 are represented from left to right, top to bottom sequentially}}
  \label{example}
\end{figure}

\noindent\textbf{\textit{Clustering of frames}}
We first show the mode-wise clustering results based on the estimated loading matrices.
Note that, for the sake of interpretability, we only consider the clustering of frames by conducting hierarchical clustering on the estimated mode-$1$ loading matrix.
{The hierarchical clustering plots of the five methods are displayed in Figure \ref{heatmap}.}
From Figure \ref{heatmap}, we can see that,  {our methods (moPCA, PmoPCA, IPmoPCA) can clearly separate the two groups of frames before and after relative position interchange, while iTIPUP(1) and PROJ can not.}

\noindent\textbf{\textit{Tensor reconstruction}}
We evaluate the performance of the tensor reconstruction  numerically by the following reconstruction error (RE)
\beqrs
RE=\frac{\|\hat{\mathcal{X}}-\mathcal{X}\|_F}
{\|\mathcal{X}\|_F},
\eeqrs
where $\mathcal{X}\in\mR^{20\times 64\times 64\times 10000}$ is the $4$th-order tensor, mode-$4$ slices of which are the tensor observations.
The results of intra-sample RE are reported in the first row of Table \ref{MM}.
We can see that, the proposed three approaches are competitive and outperform iTIPUP(1) and {PROJ} methods in the RE merit.
Furthermore, we also report the computation time of the above methods in the second row of Table \ref{MM}.
The computation time of PmoPCA estimation starts from using moPCA as projected matrices, and that of IPmoPCA estimation starts from using moPCA as initial estimators.
We can see that the proposed three approaches are much faster than iTIPUP(1) and {PROJ}.

\begin{table}[H]
\caption{{Comparison of reconstruction errors for Moving MNIST data}}
\vspace{-0.35cm}
\label{MM}
\begin{center}
\scalebox{0.9}{
\begin{tabular}{c|ccccc}
\hline
& iTIPUP & moPCA & PmoPCA & IPmoPCA & PROJ\\
\hline
RE & 0.9516 & 0.9166 &0.9146 & 0.9146 & 0.9247\\
TIME & 587.83s & 97.28s & 27.35s & 73.53s& 1804.03s\\
\hline
\end{tabular}}
\end{center}
\end{table}

\begin{figure}[H]
\centering
\subfloat[iTIPUP]
{\includegraphics[height=1.3in,width=1.8in]{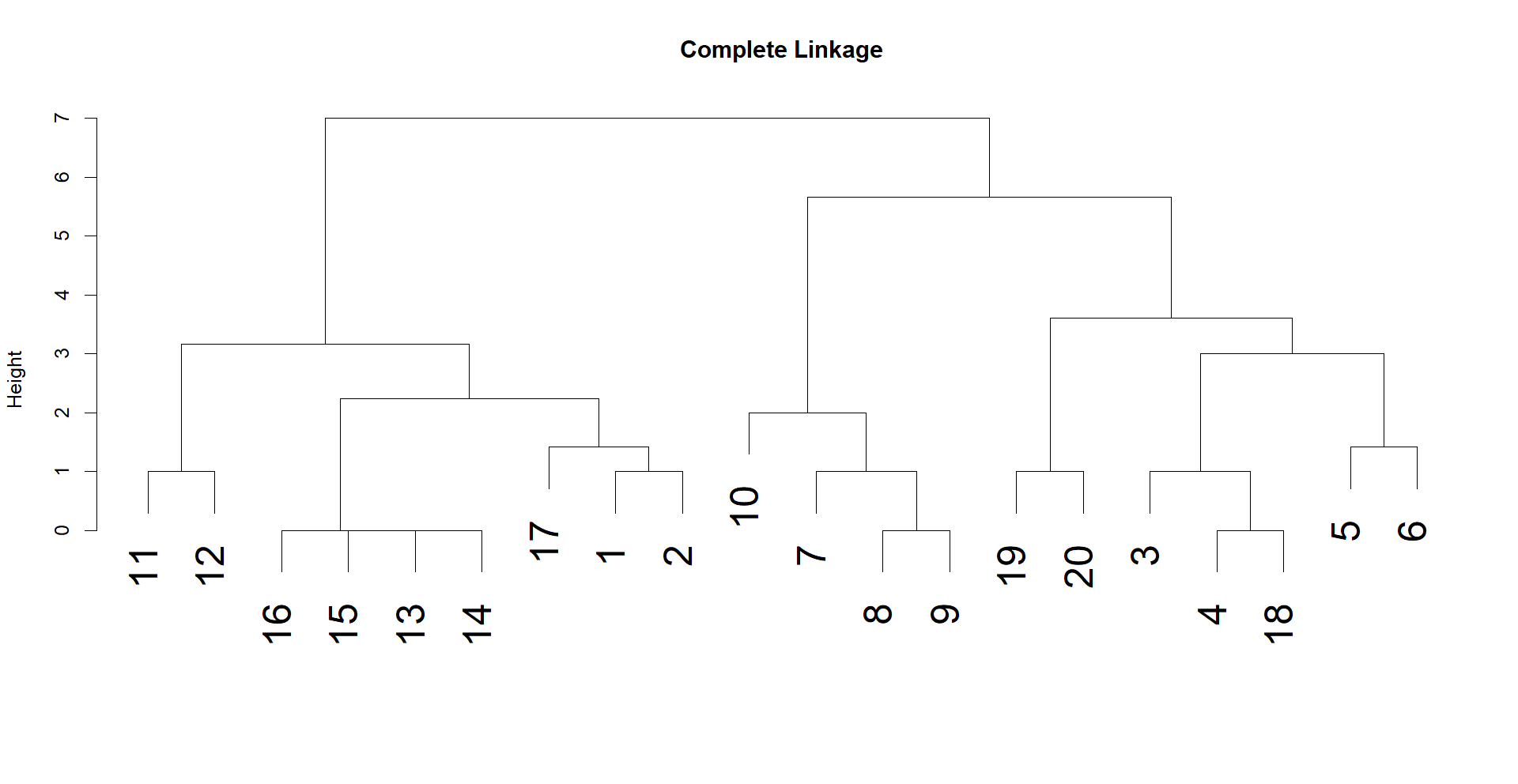}}
\vspace{-0.1cm}
\subfloat[moPCA]
{\includegraphics[height=1.3in,width=1.8in]{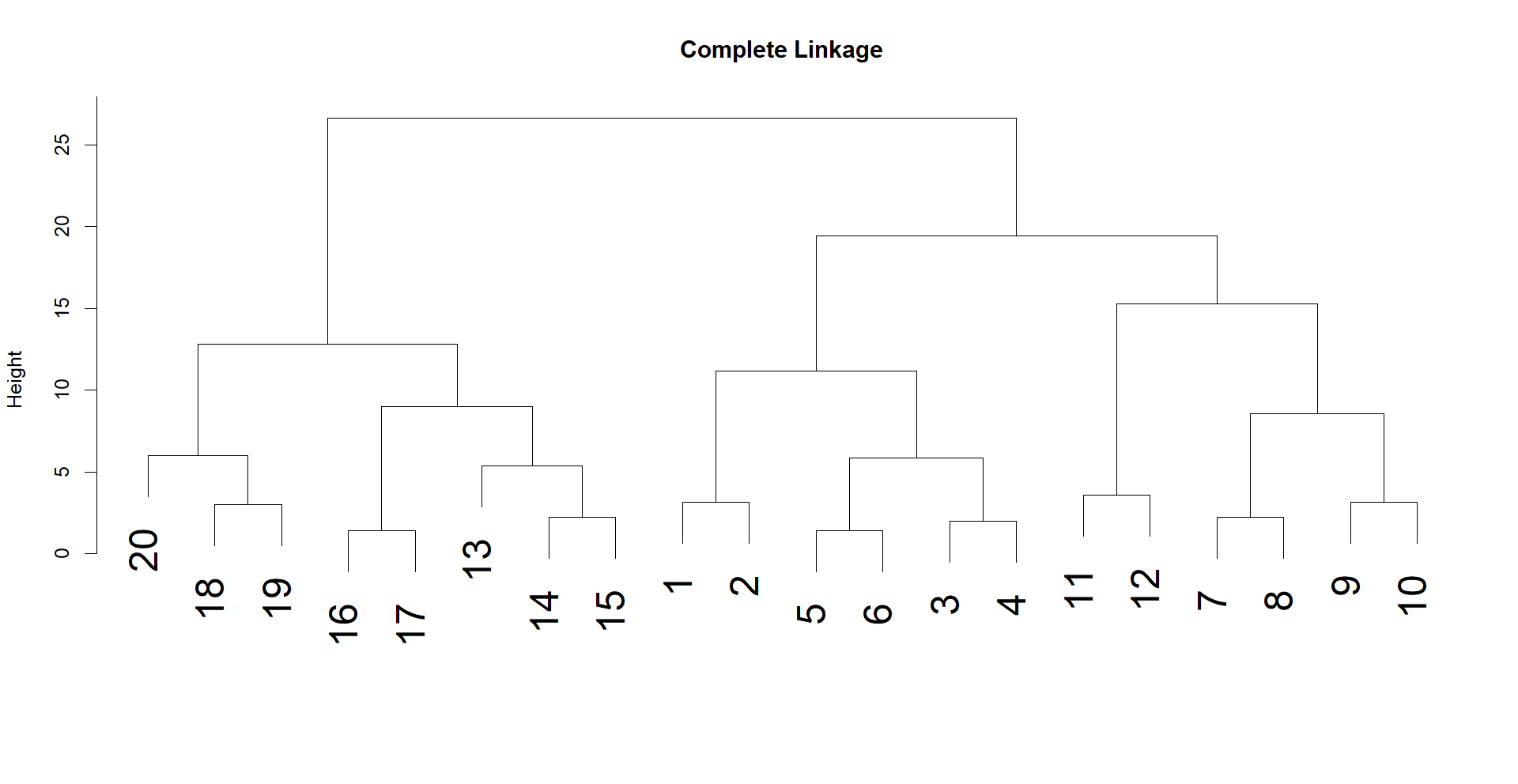}}
\vspace{-0.1cm}
\subfloat[PmoPCA]
{\includegraphics[height=1.3in,width=1.8in]{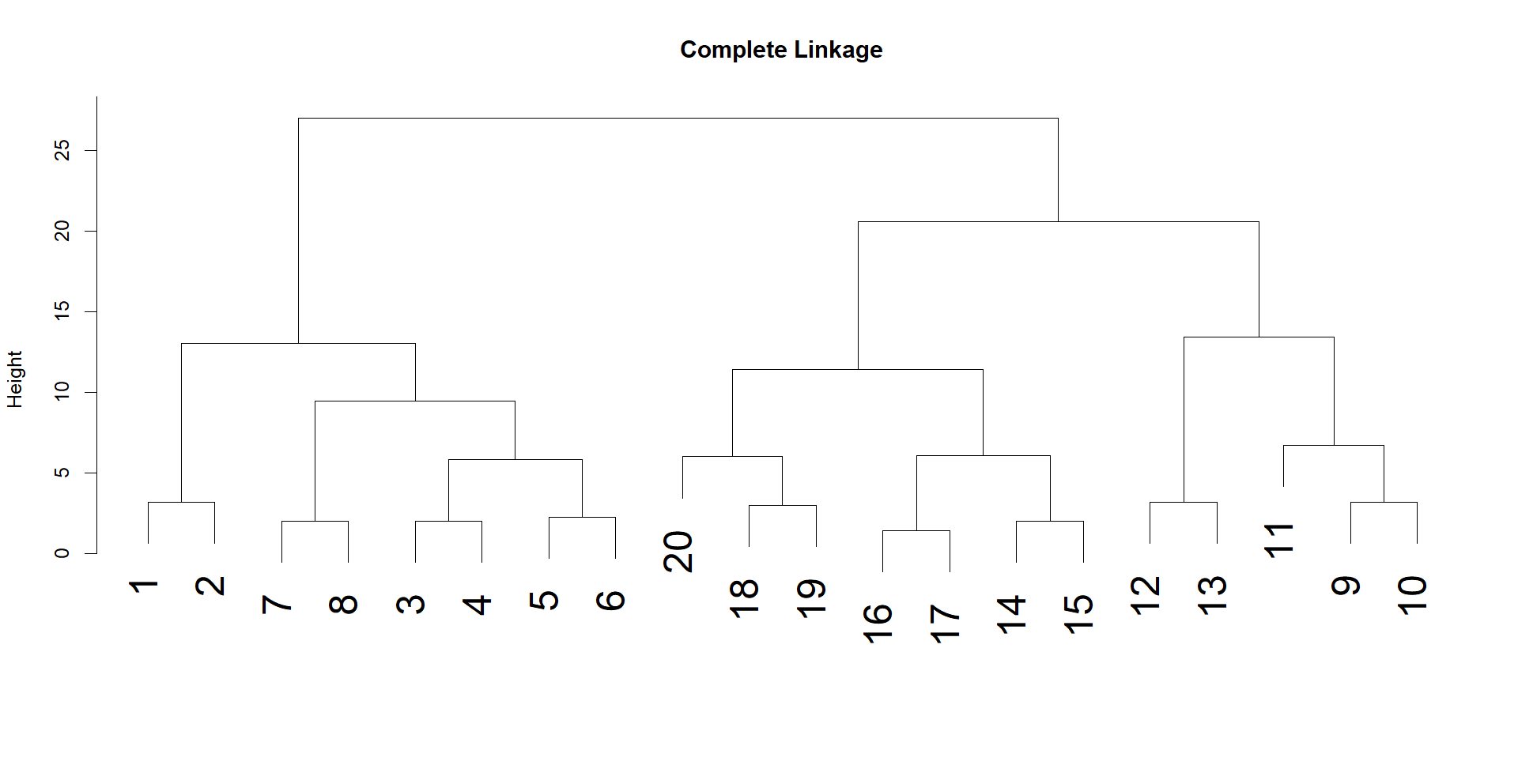}}
\vspace{-0.1cm}
\subfloat[IPmoPCA]
{\includegraphics[height=1.3in,width=1.8in]{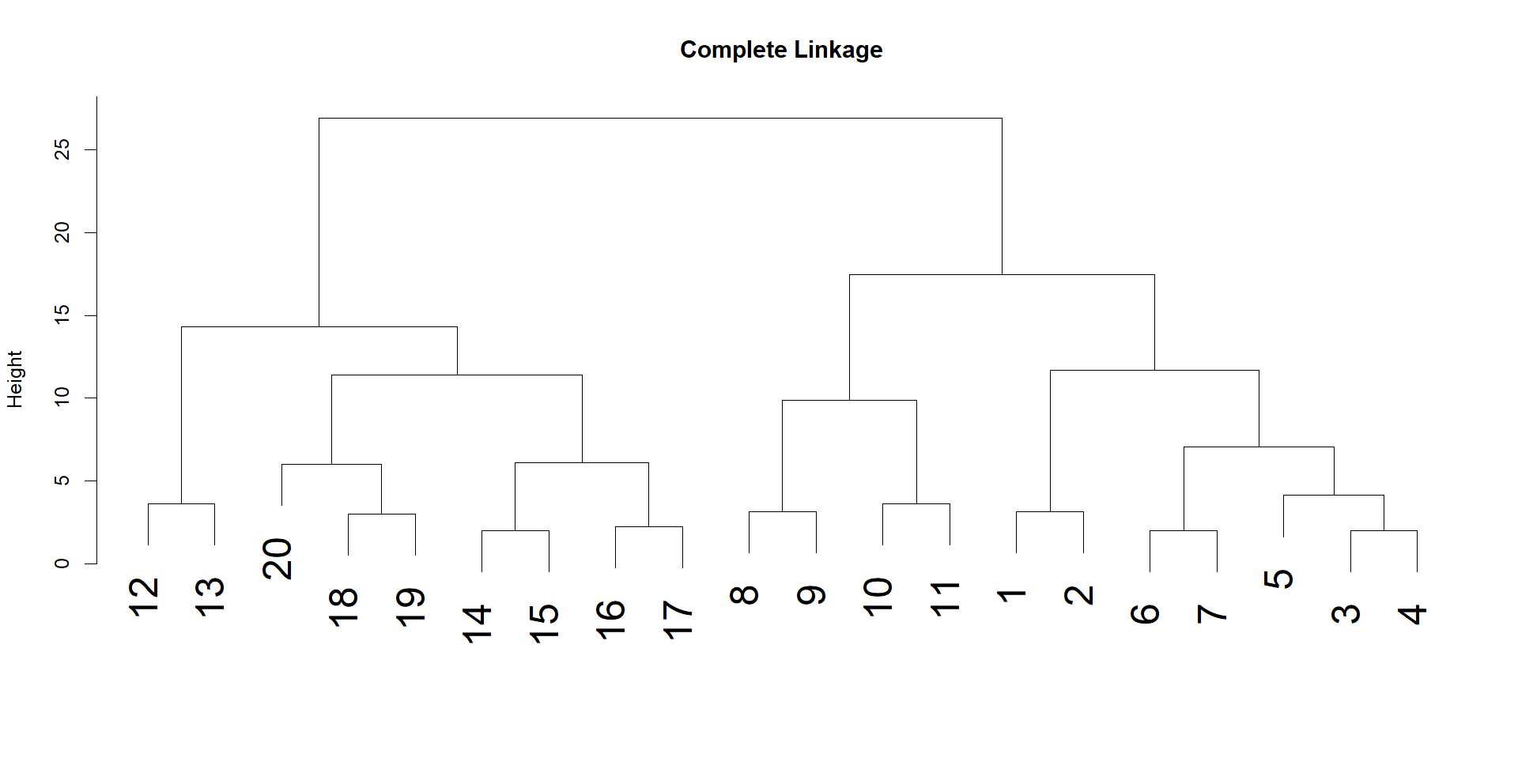}}
\vspace{-0.1cm}
\subfloat[PROJ]
{\includegraphics[height=1.3in,width=1.8in]{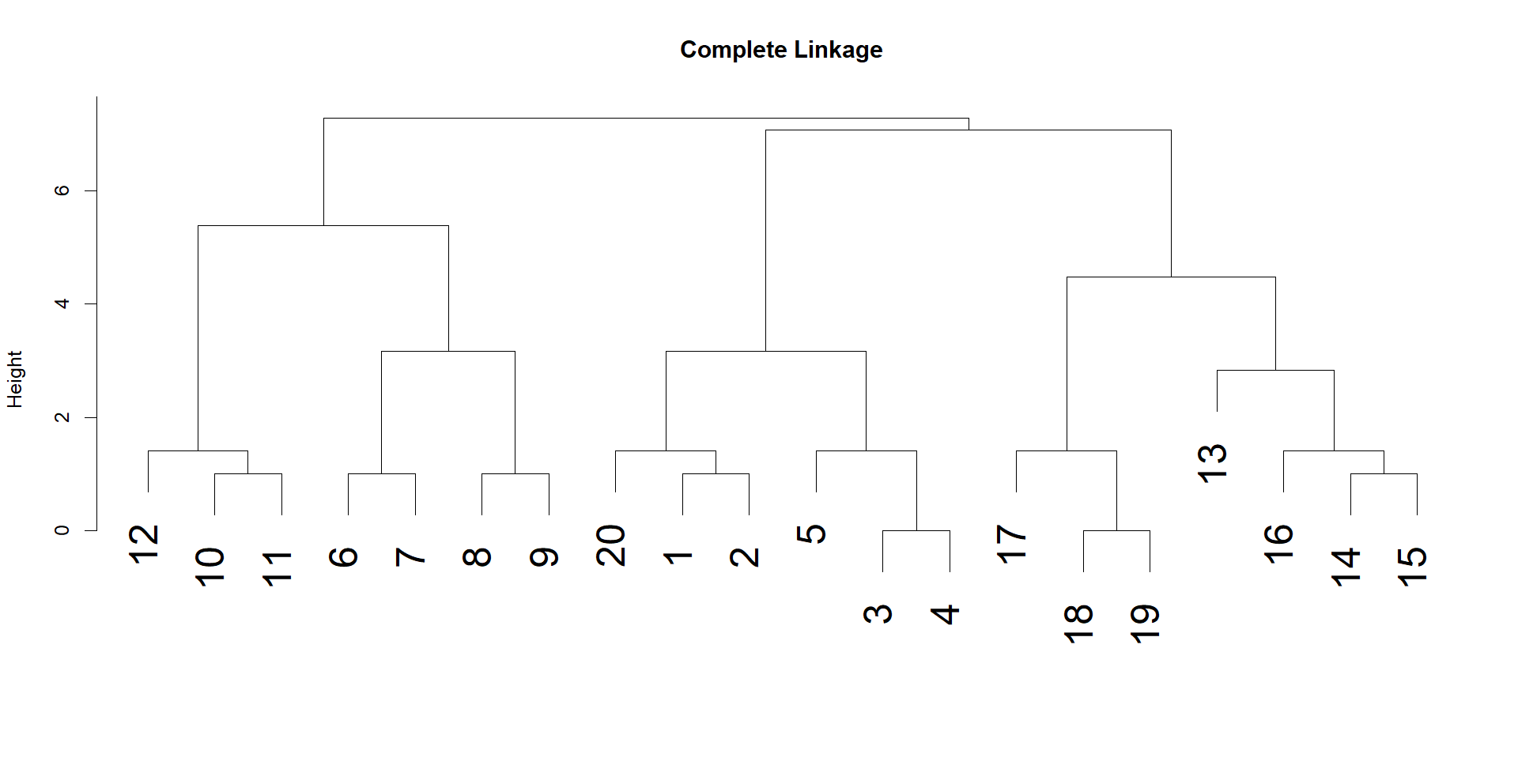}}
\caption{{The results for estimated mode-$1$ loading matrix}}
\label{heatmap}
\end{figure}

\subsection{Import-export transport network data{: serial dependent}}
The import-export transport networks data is shared by authors of \citet{ChenYangZhang2022JASA-factor}.
It contains the monthly export volume among $22$ countries on $15$ product categories from January 2010 to December 2016 ($84$ months).
The detailed list of countries and products is left in S5 of the \textit{Supplementary}.
It is preprocessed by conducting the three-month moving average to reduce the effect of incidental transactions of large trades or unusual shipping delays.
The resulting $3$rd-order tensor time series are denoted as $\mathcal{X}_t\in\mR^{22\times 22\times 15}$, $t\in[82]$,
where the three modes are export countries, import countries, and product categories, respectively.
We use factor numbers $({k}_1,{k}_2,{k}_3)=(4,4,6)$ following \citet{ChenYangZhang2022JASA-factor}, then {six} methods, i.e., iTIPUP(1), iTIPUP(2), moPCA, PmoPCA, IPmoPCA and {PROJ} are used to fit the whole dataset.
To improve interpretability, after applying the varimax rotation, the rotated loadings are multiplied by 30 and truncated to integers.

\begin{table}[H]
\caption{Estimation of loading matrix $\mathbf{A}_1$}
\label{real_A1}
\begin{center}
\renewcommand\arraystretch{0.9}
\resizebox{\columnwidth}{!}{
\begin{tabular}{|cc|cccccccccccccccccccccc|}
\hline
method & factor & BE & BU & CA & DK & FI & FR & DE & GR & HU & IS & IR & IT & MX & NO & PO & PT & ES & SE & CH & TR & US & UK\\
\hline
         &1&-1&1&\textbf{-29}&0&0& 0&          1&0&1&0&-1&1&          1&0&1&0& 0&0&0&0&          0&-1\\
iTIPUP(1)&2& 1&1&           0&1&1& 1&         -1&1&1&1& 0&1&          1&1&1&1& 0&1&1&1&\textbf{30}& 1\\
         &3& 8&1&          -1&2&1&10&\textbf{24}&1&3&1& 6&8&         -1&1&4&1& 6&3&5&2&          1& 7\\
         &4&-2&0&           1&0&0& 1&          3&0&1&0&-2&1&\textbf{30}&0&0&0&-1&1&0&0&          1& 1\\
\hline
         &1&-1&0&\textbf{-29}& 0&0& 0&           1&0& 1&0&-1& 1&          1&0& 1&0& 0& 0& 0& 0&          0&-1\\
iTIPUP(2)&2& 1&1&           0& 1&1& 1&          -1&1& 1&1& 0& 1&          1&1& 1&1& 0& 1& 1& 1&\textbf{30}& 1\\
         &3&-7&0&           2&-1&0&-9&\textbf{-23}&0&-2&0&-5&-7&          2&0&-3&0&-5&-2&-4&-1&          0&-6\\
         &4&-2&0&           1& 0&0& 1&           3&0& 1&0&-2& 1&\textbf{30}&0& 0&0&-1& 1& 0& 0&          1& 1\\
\hline
     &1& 3&1&          -1&1&1& 1&         -4&1&1&1& 3&0&\textbf{-29}& 1&1&1&2&0& 1&1&          0& 0\\
moPCA&2& 0&1&          -1&1&1& 1&         -1&1&1&0& 0&1&           1& 1&1&1&0&1& 1&1&\textbf{30}& 1\\
     &3& 8&1&           0&2&1&10&\textbf{24}&1&2&1& 6&8&          -2& 1&4&1&6&3& 5&2&          2& 7\\
     &4&-1&0&\textbf{-29}&0&0& 0&          2&0&1&0&-1&1&           2&-1&0&0&0&0&-1&0&          0&-1\\
\hline
      &1&-1&0&\textbf{-29}&0&0& 0&          2&0&1&0&-1&1&          1&0&1&0& 0&0&0&0&          0&-1\\
PmoPCA&2& 0&1&           0&1&1& 1&         -1&1&1&0& 0&1&          1&1&1&1& 0&1&1&1&\textbf{30}& 1\\
      &3& 8&1&           0&2&1&10&\textbf{24}&1&2&1& 6&8&         -2&1&4&1& 6&3&5&2&          1& 7\\
      &4&-2&0&           1&0&0& 0&          5&0&0&0&-2&1&\textbf{30}&0&0&0&-1&1&0&0&          1& 1\\
\hline
       &1&-1&1&\textbf{-29}& 0&0& 0&           0&0& 1&0&-1& 1&           1&0& 1&0& 0& 0& 0& 0&          0&-1\\
IPmoPCA&2& 1&1&           0& 1&1& 1&          -1&1& 1&1& 0& 1&           1&1& 1&1& 0& 1& 1& 1&\textbf{30}& 1\\
       &3&-7&0&           2&-1&0&-9&\textbf{-23}&0&-2&0&-5&-7&           2&0&-3&0&-5&-2&-4&-1&          0&-6\\
       &4& 3&1&           0& 1&1& 0&          -2&1& 0&1& 3& 0&\textbf{-29}&1& 1&1& 2& 0& 1& 1&          0& 0\\
\hline
    &1&-1&1&\textbf{-29}& 0& 0& 0&           2&1& 1&1& 0& 1&          -1&           2&0&0& 0& 0& 1& 0&-1&  0\\
{PROJ}&2& 0&1&          -1& 1& 0& 2&           4&0& 1&0& 0& 3& \textbf{30}&           3&1&0& 1& 1& 1& 1& 1&  2\\
    &3&-2&1&          -1& 1& 1&-1&\textbf{-25}&0& 0&0& 0&-1&           4&           3&0&1& 1& 0&-2& 0& 5&\textbf{-14}\\
    &4& 0&0&           0& 0&-1& 3&           8&0& 0&0&-1& 0&           2&\textbf{-22}&1&1&-4& 0& 3& 0& 4&\textbf{-17}\\
\hline
\end{tabular}}
\end{center}
\end{table}

\noindent\textbf{\textit{Export countries}}
For the export countries, results for the {six} methods are shown in Table \ref{real_A1}.
{We can see that the results of iTIPUP(1), iTIPUP(2), PmoPCA, and IPmoPCA are very close}, where Canada, the United States, and Mexico load heavily on the export factors 1, 2, and 4 respectively, and most of the European countries mainly load on the factor 3, especially Germany.
{For} the result of the moPCA, the order of factors is changed, where Canada and Mexico load heavily on export factors 4 and 1 respectively.
{For the result of PROJ, Canada and Mexico load heavily on the export factors 1 and 2 respectively, Germany and Norway mainly load on factors 3 and 4, respectively, and the United Kingdom mainly loads on both factors 3 and 4.}
These results are coincident with the heatmap and hierarchical clustering results based on the loading matrices of the {six} methods in S5 of the \textit{Supplementary}.

\begin{table}[H]
\caption{Estimation of loading matrix $\mathbf{A}_2$}
\label{real_A2}
\begin{center}
\renewcommand\arraystretch{0.9}
\resizebox{\columnwidth}{!}{
\begin{tabular}{|cc|cccccccccccccccccccccc|}
\hline
method & factor & BE & BU & CA & DK & FI & FR & DE & GR & HU & IS & IR & IT & MX & NO & PO & PT & ES & SE & CH & TR & US & UK\\
\hline
         & 1& 1&0&0&0&0&1&1&1&0&0&1&1&1&0&1&1&0&0&1&0&\textbf{-29}&0\\
iTIPUP(1)& 2&-2&1&\textbf{-29}&0&1&-2&-5&0&1&0&-2&0&2&0&1&1&0&1&-2&0&1&-1\\
         & 3&-8&0&6&-2&-1&\textbf{-16}&\textbf{-10}&0&-3&0&0&-9&0&-1&-4&-1&-7&-3&-6&-3&0&\textbf{-12}\\
         & 4& 2&1&-1&1&0&1&0&1&0&0&2&1&\textbf{-29}&1&1&1&1&0&3&1&0&-4\\
\hline
         &1& 1&0&0&0&0&1&1&1&0&0&1&1&1&0&1&1&0&0&1&0&\textbf{-29}&0\\
iTIPUP(2)&2&-2&1&\textbf{-29}&0&1&-2&-5&0&1&0&-2&0&2&0&1&1&0&1&-2&0&1&-1\\
         &3&-8&0&6&-2&-1&\textbf{-16}&\textbf{-10}&0&-3&0&0&-9&0&-1&-5&-1&-7&-3&-6&-3&0&\textbf{-12}\\
         &4& 2&0&-1&1&0&1&0&1&0&0&2&1&\textbf{-29}&1&1&1&1&0&3&1&0&-3\\
\hline
&1&0&0&0&1&0&1&1&1&0&1&0&1&1&0&1&1&0&1&0&0&\textbf{-29}&1\\
moPCA& 2&1&0&\textbf{24}&0&0&1&-1&0&0&1&1&0&\textbf{19}&1&0&0&0&0&0&1&0&3\\
&3&-8&0&2&-2&-1&\textbf{-18}&3&0&-3&0&0&-9&1&-1&-5&-1&-5&-4&-7&-3&0&\textbf{-14}\\
& 4&3&1&2&1&1&0&\textbf{30}&1&-1&1&2&2&1&1&0&1&4&1&-1&1&1&2\\
\hline
& 1&1&0&0&0&0&1&1&1&0&0&1&1&1&0&1&1&0&0&1&0&\textbf{-29}&0\\
PmoPCA& 2&3&0&\textbf{30}&1&0&3&5&1&0&1&3&1&-1&1&0&0&1&0&3&1&0&2\\
& 3&-8&0&6&-2&-1&\textbf{-16}&\textbf{-9}&0&-3&0&0&-9&0&-1&-5&-1&-6&-4&-6&-3&0&\textbf{-12}\\
& 4&2&1&-1&1&0&2&-1&1&0&0&2&2&\textbf{-29}&1&1&1&1&0&4&1&0&-4\\
\hline
&1&1&0&0&0&0&1&1&1&0&0&1&1&1&0&1&1&0&0&1&0&\textbf{-29}&0\\
IPmoPCA&2&-2&1&\textbf{-29}&0&1&-2&-5&0&1&0&-2&0&2&0&1&1&0&1&-1&0&1&-1\\
& 3&9&1&-5&3&2&\textbf{17}&\textbf{11}&1&4&1&2&10&1&2&5&2&8&4&7&4&1&\textbf{13}\\
& 4&2&1&-1&1&0&1&-1&1&0&0&2&1&\textbf{-29}&1&1&1&1&0&3&1&0&-3\\
\hline
    &1& 1&0&           1& 0& 1&            1&  0& 1&  0&0& 0& 0& 0& 1& 1& 1&  1&  1& 0& 0&\textbf{-29}&          1\\
{PROJ}&2& 0&0&\textbf{-28}& 1& 0&           -4&  1& 0&  1&0&-1&-4& 2& 0& 0& 0&  1&  0& 2& 1&           0&          8\\
    &3&-1&0&          -8& 1& 0&            7& -3& 1& -2&1&-2& 3& 1&-1&-1& 1&  0& -1&-5& 0&           0&\textbf{-26}\\
    &4&-6&0&           3& 0& 0& \textbf{-22}& -7&-1&  1&1& 0&-2&-4& 0& 0&-2&\textbf{-14}& -2&-5&-2&           0&          -3\\
\hline
\end{tabular}}
\end{center}
\end{table}

\noindent\textbf{\textit{Import countries}}
For the import countries, results for the {six} methods are shown in Table \ref{real_A2}.
The results of the moPCA show that the United States, Mexico and Canada, and Germany load heavily on export factors 1, 2, and 4 respectively, and most of the European countries mainly load on factor 3, especially for France and the United Kingdom.
{For} the results of iTIPUP(1), iTIPUP(2), PmoPCA, and IPmoPCA, the United States, Canada, and Mexico load heavily on export factors 1, 2, and 4 respectively, and most of the European countries mainly load on factor 3, especially for France, the United Kingdom, and Germany.
{For the results of PROJ, the United States, Canada, and the United Kingdom load heavily on export factors 1, 2, and 3 respectively, and most of the European countries mainly load on factor 4, especially France and Spain.}
These results are coincident with the heatmap and hierarchical clustering results based on the loading matrices of the {six} methods in S5 of the \textit{Supplementary}.

\begin{table}[H]
\caption{Estimation of loading matrix $\mathbf{A}_3$}
\label{real_A3}
\begin{center}
\scalebox{0.7}{\renewcommand\arraystretch{0.8}
\begin{tabular}{|cc|ccccccccccccccc|}
\hline
method & factor & $C_1$ & $C_2$ & $C_3$ & $C_4$ & $C_5$ & $C_6$ & $C_7$ & $C_8$ & $C_9$ & $C_{10}$
& $C_{11}$ & $C_{12}$ & $C_{13}$ & $C_{14}$ & $C_{15}$\\
\hline
& 1&1&-2&1&1&1&-1&0&3&-1&0&-1&0&\textbf{-29}&0&-4\\
& 2&0&0&1&\textbf{-29}&1&2&1&-1&1&1&-2&0&1&0&0\\
iTIPUP(1)&3&0&-2&0&0&\textbf{29}&1&1&0&1&1&4&3&0&0&7\\
&4&0&1&-4&1&1&3&0&-2&0&0&-2&-1&1&\textbf{-29}&-2\\
& 5&6&7&6&1&1&\textbf{21}&1&10&2&0&-5&\textbf{16}&1&1&-3\\
& 6&2&-2&3&0&5&-3&1&-5&8&3&10&4&4&2&\textbf{-25}\\
\hline
& 1&1&-2&1&1&1&-1&0&3&-1&0&-1&0&\textbf{-29}&0&-4\\
& 2&0&0&1&\textbf{-29}&1&2&1&-1&1&1&-2&0&1&0&0\\
iTIPUP(2)&3&0&-2&0&0&\textbf{29}&1&1&0&1&1&4&3&0&0&7\\
& 4&0&1&-4&1&1&3&0&-2&0&0&-2&-1&1&\textbf{-29}&-2\\
&5&6&7&6&1&1&\textbf{21}&1&10&2&0&-5&\textbf{16}&1&1&-3\\
&6&2&-1&3&0&5&-3&1&-5&8&3&10&4&4&2&\textbf{-25}\\
\hline
&1&1&-1&0&0&2&1&0&3&-1&0&1&1&\textbf{-29}&0&-6\\
& 2&0&0&1&\textbf{-29}&1&1&1&-2&1&1&1&0&1&1&-2\\
moPCA&3&1&2&0&1&\textbf{-28}&0&0&1&1&0&1&0&1&1&-7\\
& 4&0&0&2&0&0&0&0&3&-1&0&0&2&0&\textbf{30}&3\\
&5&6&6&7&1&3&\textbf{17}&2&8&7&2&4&\textbf{19}&4&0&-9\\
& 6&0&0&2&0&0&-2&1&1&1&1&\textbf{30}&0&0&0&5\\
\hline
&1&1&-2&0&1&1&-1&1&3&-1&1&-2&0&\textbf{-29}&1&-3\\
& 2&0&0&1&\textbf{-29}&1&2&1&-1&1&1&-2&0&1&1&0\\
PmoPCA& 3&0&-2&0&0&\textbf{29}&1&1&0&1&1&4&2&0&0&8\\
& 4&5&7&7&1&2&\textbf{20}&0&10&2&0&-4&\textbf{17}&1&4&-4\\
& 5&2&0&2&0&0&-3&1&1&1&1&1&0&0&\textbf{30}&6\\
&6&-2&4&-1&1&-4&5&-1&6&-7&-2&-9&-2&-2&-3&\textbf{26}\\
\hline
&1&1&-2&1&1&1&-1&0&3&-1&0&-1&0&\textbf{-29}&0&-4\\
& 2&0&0&1&\textbf{-29}&1&2&1&-1&1&1&-2&0&1&0&0\\
IPmoPCA& 3&1&3&1&1&\textbf{-28}&0&0&1&0&0&-3&-2&1&1&-6\\
& 4&1&0&5&0&0&-2&1&3&1&1&3&2&0&\textbf{30}&3\\
& 5&6&7&6&1&1&\textbf{21}&1&10&2&0&-5&\textbf{16}&1&1&-3\\
&6&2&-2&3&1&5&-3&1&-5&7&3&10&4&4&2&\textbf{-25}\\
\hline
&1&0&1&1&\textbf{30}&0&1&1&1&1&1&1&1&0&1&-1\\
&2&1&-1&0&0&-1&1&0&2&1&1&2&1&-3&\textbf{-29}&0\\
{PROJ}&3&6&8&-2&1&\textbf{21}&8&1&4&1&0&1&\textbf{16}&8&-1&4\\
&4&-5&5&1&1&0&0&1&-6&0&0&\textbf{28}&-1&7&0&-1\\
&5&1&\textbf{14}&-1&0&5&1&2&1&-1&0&5&-2&\textbf{-24}&3&5\\
&6&3&4&9&1&-5&4&1&6&1&1&1&-6&6&0&\textbf{26}\\
\hline
\end{tabular}}
\end{center}
\end{table}

\noindent\textbf{\textit{Product categories}}
As for the product categories,
results for the {six} methods are shown in Table \ref{real_A3}.
The results of the moPCA show that factors 1, 2, 3, 4, and 6 are the Machinery $\&$ Electrical factor, Mineral Products factor, Chemicals $\&$ Allied Industries factor, Transportation factor, and Stone $\&$ Glass factor, respectively.
Factor 5 has heavy loadings on Plastics $\&$ Rubbers and Metals.
The results of iTIPUP(1), iTIPUP(2), and IPmoPCA are similar, where factors 1, 2, 3, 4, and 6 are the Machinery $\&$ Electrical factor, Mineral Products factor, Chemicals $\&$ Allied Industries factor,
Transportation factor, and Miscellaneous factor, respectively.
Factor 5 has heavy loadings on Plastics $\&$ Rubbers and Metals.
{For} PmoPCA, factor 4 is the mixed factor of Plastics $\&$ Rubbers and Metals, and factor 5 is the Transportation factor.
For PROJ, factors 1, 2, 4, and 6 are the Mineral Products factor, Transportation factor, Stone $\&$ Glass factor, and Miscellaneous factor, respectively.
Factors 3 and 5 are both mixed factors of two product categories.
These results can be illustrated by the heatmap and hierarchical clustering results based on the loading matrices in S5 of the \textit{Supplementary}.

\begin{table}[H]
\caption{RE values for the Import-Export Transport dataset}
\label{IET}
\begin{center}
\scalebox{0.9}{
\begin{tabular}{c|cccccc}
\hline
& iTIPUP(1) & iTIPUP(2) & moPCA & PmoPCA & IPmoPCA & {PROJ}\\
\hline
(2,2,2)   & 0.566542&0.566548&0.575356&0.567267&0.566519&0.737432\\
(4,2,2)   & 0.492059&0.492065&0.499865&0.493276&0.491989&0.717259\\
(5,5,5)   & 0.283764&0.283767&0.294388&0.284301&0.283761&0.609906\\
(10,10,5) & 0.162273&0.162268&0.165199&0.162877&0.162268&0.363559\\
time      &   12.53s& 14.90s &  1.54s & 1.73s  & 10.73  & 83.05s\\
\hline
\end{tabular}}
\end{center}
\end{table}

\noindent\textbf{\textit{Tensor reconstruction}}
The results of averaged RE values over 10 folds cross-validation for aforementioned {six} methods under various factor number settings are in Table \ref{IET}.
We can see that our proposed three methods are doable for highly correlated tensor time series even though we designed our methodology for weakly dependent and independent tensor observations.

\section{Conclusion}
\label{sec:conclusion}

In this article, we bond the estimation of low-rank structure under {TuTFaM} with that of the well-developed high-dimensional vector factor models through tensor matricization; refer to S6.2 of the \textit{Supplementary}.
Under such a framework, we develop two sets of tensor PCA variants to estimate the components in the low-rank signal under {TuTFaM}.
Particularly, IPmoPCA which is based on both iteration and projection, is {a} refinement of PmoPCA that involves projection only, in terms of enhancement in convergence rate and signal-to-noise ratio; refer to \textit{Supplementary} for proof.
{Note that though our angle of view is on the decomposition of data objects directly, there is another line of work focusing on the estimation of covariance tensors with low-rank and/or sparsity assumptions \citep{FanLiLiao2021ARFE-recent,
FanLiaoMincheva2013JRSSB-large, 
FanLiaoLiu2016EJ-overview, 
TsiligkaridisHeroZhou2013IEEETSP-convergence, 
Zhou2014AOS-gemini, 
GreenewaldZhouHeroIII2019JRSSB-tensor}; refer to S6.1 of the \textit{Supplementary} for moPCA of TuTFaM.}

Our intention of contribution is to show that, the PCA or spectral method for any order tensor observations follows the same vein of the mode-wise thought, though tensor objects have larger sizes and more complicated structures with the growing number of modes.
This article is essentially the extension of the following two works.
On the one hand, it follows the line of {CVC-based PCA estimation methods} in \citet{ChenFan2023JASA-statistical} and extends the estimation of factor models from the $2$nd-order tensor (matrix) to the higher-order tensor.
On the other hand, it extends the work of \citet{ChenYangZhang2022JASA-factor} in terms of application scope, i.e., from the tensor time series to the more general uncorrelated or weakly correlated tensor observations.
{In addition,
our work is under the assumption of homogeneous factor strength.
As one reviewer pointed out,
recently there has been arising research interest in factor analysis under
the practical heterogeneous factor strength situation \citep{BaiNg2023JOE-approximate, Freyaldenhoven2022JOE-factor, UematsuYamagata2022JBES-estimation, ChenLam2024AOS-rank, ZhangPanYaoZhou2023}, where there coexist strong and weak factors owing to varying column strength measures.
Note that the sporadic such works evidenced their superiority in estimating factor numbers, not to mention the model flexibility and practicability.
This motivates us to extend the proposed IPmoPCA approach under the heterogeneous factor case in the near future.}

\bibliographystyle{chicago}
\bibliography{TuTFaM}

\end{document}


\date{}
\title{Supplementary material for ``Tucker tensor factor models: matricization and mode-wise PCA estimation"}
\maketitle

The supplementary material is organized as follows.
First, the properties of moPCA estimators and estimated factor numbers are provided with their proofs in \textbf{S1}.
Second, the theoretical proofs of PmoPCA and IPmoPCA are provided in \textbf{S2}.
Third, necessary lemmas are provided in \textbf{S3}.
Fourth, the additional results for simulation are provided in \textbf{S4}.
Fifth, the extra results of import-export transport network data are provided in \textbf{S5}.
Sixth, for moPCA estimators, the illustration from the variance-covariance learning perspective, and the streamline from vectors to tensors are provided in \textbf{S6}.
At last, the symbol table and the running environments of codes are in \textbf{S7} and \textbf{S8}, respectively. 

Note that, although we assume $\left\|p_{d}^{-1} \mathbf{A}_d^{\top}\mathbf{A}_d-\mathbf{I}_{k_{d}}\right\| \rightarrow 0$ as $\min_{d\in[D]} \left\{p_d\right\} \rightarrow \infty$ in Assumption 3, we use the equations $p_{d}^{-1} \mathbf{A}_d^{\top}\mathbf{A}_d=\mathbf{I}_{k_{d}}$ directly to simplify the derivation, which will not affect the order of results because the remainder induced by the approximation are small order of the main term.
Besides, as $\{k_{d}\}_{d=1}^D$ are fixed, hence we may assume $k_{d}=1$ during the procedure of proofs for simplifying notations, which has no impact on the order of results.
\par

\setcounter{section}{0}
\setcounter{equation}{0}
\def\theequation{S\arabic{section}.\arabic{equation}}
\def\thesection{S\arabic{section}}

\fontsize{12}{14pt plus.8pt minus .6pt}\selectfont

\section{Properties of moPCA Estimators}
\label{subsec:moPCA_properties}

We first introduce the following proposition about the order of eigenvalues of $\widehat{\mathbf{M}}_d$.

\bep
Suppose that $T$ and $\{p_d\}_{d=1}^D$ tend to infinity, and $\{k_d\}_{d=1}^D$ are fixed. If Assumptions 1-5 hold, then for $d\in[D]$, the order of eigenvalues of $\widehat{\mathbf{M}}_d$ are
{\small\beqr
\lambda_j(\widehat{\mathbf{M}}_d)=\left\{\begin{array}{l}
\vspace{0.5cm}
\lambda_j({\mathbf{\Sigma}}_d)+o_p(1), \quad j\leq k_d;\\
O_p(\frac{1}{\sqrt{p_d}}+\frac{1}{\sqrt{Tp_{-d}}}), \quad j>k_d.
\end{array}\right.
\eeqr}
\eep

\textit{Proof}: Without loss of generality, we only prove the result of $\widehat{\mathbf{M}}_1$, and it holds that
{\small\beqrs
\widehat{\mathbf{M}}_1&=&\frac{1}{Tp}\sum\limits_{t=1}^T \mathcal{X}_{t}^{(1)}\mathcal{X}_{t}^{(1)^{\top}}\\
&=&\frac{1}{Tp}\sum\limits_{t=1}^T \mathbf{A}_1\mathcal{F}_{t}^{(1)}\mathbf{A}_{[D]/\{1\}}^{\top}\mathbf{A}_{[D]/\{1\}}\mathcal{F}_{t}^{(1)^{\top}}\mathbf{A}_1^{\top}
+\frac{1}{Tp}\sum\limits_{t=1}^T \mathbf{A}_1\mathcal{F}_{t}^{(1)}\mathbf{A}_{[D]/\{1\}}^{\top}\mathcal{E}_{t}^{(1)^{\top}}\\
&&+\frac{1}{Tp}\sum\limits_{t=1}^T \mathcal{E}_{t}^{(1)}\mathbf{A}_{[D]/\{1\}}
\mathcal{F}_{t}^{(1)^{\top}}\mathbf{A}_1^{\top}
+\frac{1}{Tp}\sum\limits_{t=1}^T \mathcal{E}_{t}^{(1)}\mathcal{E}_{t}^{(1)^{\top}}\\
&=&\frac{1}{Tp_1}\sum\limits_{t=1}^T \mathbf{A}_1\mathcal{F}_{t}^{(1)}\mathcal{F}_{t}^{(1)^{\top}}\mathbf{A}_1^{\top}
+\frac{1}{Tp}\sum\limits_{t=1}^T \mathbf{A}_1\mathcal{F}_{t}^{(1)}\mathbf{A}_{[D]/\{1\}}^{\top}\mathcal{E}_{t}^{(1)^{\top}}\\
&&+\frac{1}{Tp}\sum\limits_{t=1}^T \mathcal{E}_{t}^{(1)}\mathbf{A}_{[D]/\{1\}}
\mathcal{F}_{t}^{(1)^{\top}}\mathbf{A}_1^{\top}
+\frac{1}{Tp}\sum\limits_{t=1}^T \mathcal{E}_{t}^{(1)}\mathcal{E}_{t}^{(1)^{\top}}.
\eeqrs}

For the first term, the $k_1$ largest eigenvalues of $(Tp_1)^{-1}\sum_{t=1}^T \mathbf{A}_1\mathcal{F}_{t}^{(1)}\mathcal{F}_{t}^{(1)^{\top}}\mathbf{A}_1^{\top}$ are same as those of
$T^{-1}\sum_{t=1}^T \mathcal{F}_{t}^{(1)}\mathcal{F}_{t}^{(1)^{\top}}$,
while the other $p_1-k_1$ eigenvalues are zero.
Moreover, $T^{-1}\sum_{t=1}^T \mathcal{F}_{t}^{(1)}\mathcal{F}_{t}^{(1)^{\top}}
\stackrel{p}{\rightarrow}{\mathbf{\Sigma}}_1$, and it is then implied by Weyl's inequality that
{\small\beqrs
\lambda_j\left(\frac{1}{Tp_1}\sum\limits_{t=1}^T \mathbf{A}_1\mathcal{F}_{t}^{(1)}\mathcal{F}_{t}^{(1)^{\top}}\mathbf{A}_1^{\top}\right)
=\left\{\begin{array}{l}
\vspace{0.5cm}
\lambda_j({\mathbf{\Sigma}}_1)+o_p(1), \quad j\leq k_1;\\
0, \quad j>k_1.
\end{array}\right.
\eeqrs}
For the second term, by Lemma 1, we have
{\small\beqrs
\left\|\frac{1}{Tp}\sum\limits_{t=1}^T \mathbf{A}_1\mathcal{F}_{t}^{(1)}\mathbf{A}_{[D]/\{1\}}^{\top}\mathcal{E}_{t}^{(1)^{\top}}\right\|_{F}
\leq\frac{\sqrt{p_1}}{Tp}\left\|\sum\limits_{t=1}^T \mathcal{F}_{t}^{(1)}\mathbf{A}_{[D]/\{1\}}^{\top}\mathcal{E}_{t}^{(1)^{\top}}\right\|_F=O_p\left(\frac{1}{\sqrt{Tp_{-1}}}\right).
\eeqrs}
For the third term, similar to the second term, it can be verified that
{\small\beqrs
\left\|\frac{1}{Tp}\sum\limits_{t=1}^T \mathcal{E}_{t}^{(1)}\mathbf{A}_{[D]/\{1\}}
\mathcal{F}_{t}^{(1)^{\top}}\mathbf{A}_1^{\top}\right\|_{F}=O_p\left(\frac{1}{\sqrt{Tp_{-1}}}\right).
\eeqrs}
For the fourth term, it is implied by Lemma 2 that
{\small\beqrs
\left\|\frac{1}{Tp}\sum\limits_{t=1}^T \mathcal{E}_{t}^{(1)}\mathcal{E}_{t}^{(1)^{\top}}\right\|_F
=O_p\left(\frac{1}{\sqrt{Tp_{-1}}}+\frac{1}{\sqrt{p_1}}\right).
\eeqrs}
We hence accomplish the proof by Weyl's inequality. $\square$\\

Note that asymptotic normalization in Assumption 3 for mode-wise loading matrices guarantees uniqueness of the column space of $\mathbf{A}_d$ up to an asymptotical orthogonal matrix $\widehat{\mathbf{H}}_d\in\mR^{k_d\times k_d}$ in the sense that $\widehat{\mathbf{H}}_d^{\top}\widehat{\mathbf{H}}_d$ converges to a ${k_d}$-dim identity matrix in probability. 
In our methodology, we construct the mode-wise orthogonal matrix below
\beqr
\widehat{\mathbf{H}}_d=\frac{1}{Tp_d}\sum\limits_{t=1}^T \mathcal{F}_{t}^{(d)}\mathcal{F}_{t}^{(d)^{\top}}
\mathbf{A}_d^{\top}\widehat{\mathbf{A}}_d\widehat{\mathbf{\Lambda}}_d^{-1},
\eeqr
\noindent
where $\widehat{\mathbf{\Lambda}}_d$ is a diagonal matrix with diagonal elements being the top $k_d$ eigenvalues of $\widehat{\mathbf{M}}_d$ in the descending order.
Then the asymptotic consistency of $\widehat{\mathbf{A}}_d$ is assured by the bound in the measure of Frobenius norm in Proposition \ref{consistency_pca} as follows.
\bep
\label{consistency_pca}
Suppose that $T$ and $\{p_d\}_{d=1}^D$ tend to infinity, and $\{k_d\}_{d=1}^D$ are fixed.
If Assumptions 1-5 hold, then there exists an asymptotic orthogonal matrix $\widehat{\mathbf{H}}_d$, such that along modes, we have
$$
\frac{1}{p_d}\|\widehat{\mathbf{A}}_d-\mathbf{A}_d\widehat{\mathbf{H}}_d\|_{F}^2=O_p(\frac{1}{Tp_{-d}}+\frac{1}{p_d^2}),
\hspace{3pt} d\in[D].
$$
\eep
The mode-wise estimator $\widehat{\mathbf{A}}_d$ is developed  when we take the fiber vector factor model $
\mathbf{X}^{(d)}=\mathbf{A}_d\mathbf{Z}^{(d)}+\mathbf{E}^{(d)}$ as the matricized working model for the TuTFaM mode-wisely.
Thus its convergence rate above is the same as that in Bai (2003) \cite{Bai2003Econometrica-inferential} if regarding $Tp_{-d}$ as the sample size, and reduces to Theorem 3.3 in Yu et al. (2022) \cite{YuHeKongZhang2022JOE-projected} when $D=2$.
Our mode-wise estimator has a faster convergence rate compared to the convergence rate $O_p(1/T)$ of the mode-wise type TOPUP method in Chen et al. (2022) \cite{ChenYangZhang2022JASA-factor} under strong factor and fixed factor numbers settings when $p_d^2$ runs faster than $T$.

\textit{Proof}: Without loss of generality, we only prove the result for $\widehat{\mathbf{A}}_1$.
Note that, for the moPCA estimator, $\widehat{\mathbf{A}}_1$ is set to be the top $k_{1}$ eigenvectors of the sample variance-covariance matrix $\widehat{\mathbf{M}}_{1}$ up to a scalar $\sqrt{p_{_1}}$.
Specifically, assume that the spectral eigen-decompostion of $\widehat{\mathbf{M}}_{1}$ has the form of $\sum_{j=1}^{p_1} \lambda_{j} \boldsymbol{\xi}_{j} \boldsymbol{\xi}_{j}^{T}$ with distinct eigenvalues $\{\lambda_j\}_{j=1}^{p_1}$ in descending order, and then $\widehat{\mathbf{A}}_1=\sqrt{p_1}(\boldsymbol{\xi}_{1},\cdots,\boldsymbol{\xi}_{k_1})$.
Hence, by the definition of eigenvectors and eigenvalues, it holds that 
$$\widehat{\mathbf{M}}_1\widehat{\mathbf{A}}_1
=\sqrt{p_1}(\widehat{\mathbf{M}}_{1}\boldsymbol{\xi}_{1},\cdots,\widehat{\mathbf{M}}_{1}\boldsymbol{\xi}_{k_1})
=\sqrt{p_1}(\lambda_1\boldsymbol{\xi}_{1},\cdots,\lambda_{k_1}\boldsymbol{\xi}_{k_1})
=\widehat{\mathbf{A}}_1\widehat{\mathbf{\Lambda}}_1,$$
where $\widehat{\mathbf{\Lambda}}_1=\mbox{diag}(\lambda_1,\cdots,\lambda_{k_1})$.
As a result,
{\small\beqr\n\widehat{\mathbf{A}}_1
&=&\widehat{\mathbf{M}}_1\widehat{\mathbf{A}}_1\widehat{\mathbf{\Lambda}}_1^{-1}\\\n
&=& \mathbf{A}_1 \left\{
\frac{1}{Tp_1}\sum\limits_{t=1}^T \mathcal{F}_{t}^{(1)}\mathcal{F}_{t}^{(1)^{\top}}\mathbf{A}_1^{\top}\widehat{\mathbf{A}}_1\widehat{\mathbf{\Lambda}}_1^{-1}\right\}
+\frac{1}{Tp}\sum\limits_{t=1}^T \mathbf{A}_1\mathcal{F}_{t}^{(1)}\mathbf{A}_{[D]/\{1\}}^{\top}\mathcal{E}_{t}^{(1)^{\top}}\widehat{\mathbf{A}}_1\widehat{\mathbf{\Lambda}}_1^{-1}\\
&&+\frac{1}{Tp}\sum\limits_{t=1}^T \mathcal{E}_{t}^{(1)}\mathbf{A}_{[D]/\{1\}}
\mathcal{F}_{t}^{(1)^{\top}}\mathbf{A}_1^{\top}\widehat{\mathbf{A}}_1\widehat{\mathbf{\Lambda}}_1^{-1}
+\frac{1}{Tp}\sum\limits_{t=1}^T \mathcal{E}_{t}^{(1)}\mathcal{E}_{t}^{(1)^{\top}}\widehat{\mathbf{A}}_1\widehat{\mathbf{\Lambda}}_1^{-1}.
\eeqr}
Denote $\widehat{\mathbf{H}}_1=(Tp_1)^{-1}\sum\limits_{t=1}^T \mathcal{F}_{t}^{(1)}\mathcal{F}_{t}^{(1)^{\top}}\mathbf{A}_1^{\top}\widehat{\mathbf{A}}_1\widehat{\mathbf{\Lambda}}_1^{-1}$, then
{\small\beqrs
\widehat{\mathbf{A}}_1-\mathbf{A}_1\widehat{\mathbf{H}}_1&=&\frac{1}{Tp}\sum\limits_{t=1}^T \mathbf{A}_1\mathcal{F}_{t}^{(1)}\mathbf{A}_{[D]/\{1\}}^{\top}\mathcal{E}_{t}^{(1)^{\top}}\widehat{\mathbf{A}}_1\widehat{\mathbf{\Lambda}}_1^{-1}\\
&&+\frac{1}{Tp}\sum\limits_{t=1}^T \mathcal{E}_{t}^{(1)}\mathbf{A}_{[D]/\{1\}}
\mathcal{F}_{t}^{(1)^{\top}}\mathbf{A}_1^{\top}\widehat{\mathbf{A}}_1\widehat{\mathbf{\Lambda}}_1^{-1}
+\frac{1}{Tp}\sum\limits_{t=1}^T \mathcal{E}_{t}^{(1)}\mathcal{E}_{t}^{(1)^{\top}}\widehat{\mathbf{A}}_1\widehat{\mathbf{\Lambda}}_1^{-1}.
\eeqrs}
\noindent
According to Proposition 1, 
$p_{-d}^{-1}\|\widehat{\mathbf{A}}_d-\mathbf{A}_d\widehat{\mathbf{H}}_d\|_{F}^2=O_p((Tp_{-d})^{-1}+p_d^{-1})$, 
$\|\widehat{\mathbf{\Lambda}}_1\|_F^2=O_p(1)$ and $\|\widehat{\mathbf{H}}_1\|_F^2=O_p(1)$,
then for the first term,
{\small\beqrs
&&\left\|\frac{1}{Tp}\sum\limits_{t=1}^T \mathbf{A}_1\mathcal{F}_{t}^{(1)}\mathbf{A}_{[D]/\{1\}}^{\top}\mathcal{E}_{t}^{(1)^{\top}}\widehat{\mathbf{A}}_1\widehat{\mathbf{\Lambda}}_1^{-1}\right\|_{F}^2\\
&\leq& \frac{p_1}{T^2p^2}\left\|\sum\limits_{t=1}^T\mathcal{F}_{t}^{(1)}\mathbf{A}_{[D]/\{1\}}^{\top}\mathcal{E}_{t}^{(1)^{\top}}\mathbf{A}_1\right\|_{F}^2
+\frac{p_1}{T^2p^2}\left\|\sum\limits_{t=1}^T\mathcal{F}_{t}^{(1)}\mathbf{A}_{[D]/\{1\}}^{\top}\mathcal{E}_{t}^{(1)^{\top}}\right\|_{F}^2\left\|\widehat{\mathbf{A}}_1-\mathbf{A}_1\widehat{\mathbf{H}}_1\right\|_{F}^2\\
&=&O_p(\frac{p_1}{Tp})+O_p(\frac{p_1}{Tp})\left\|\widehat{\mathbf{A}}_1-\mathbf{A}_1\widehat{\mathbf{H}}_1\right\|_{F}^2,
\eeqrs}
where the last equation is according to Lemmas 1 and 3.

For the second term,
{\small\beqrs
\left\|\frac{1}{Tp}\sum\limits_{t=1}^T \mathcal{E}_{t}^{(1)}\mathbf{A}_{[D]/\{1\}}
\mathcal{F}_{t}^{(1)^{\top}}\mathbf{A}_1^{\top}\widehat{\mathbf{A}}_1\widehat{\mathbf{\Lambda}}_1^{-1}\right\|_{F}^2
\leq\frac{p_1^2}{T^2p^2}\left\|\sum\limits_{t=1}^T \mathcal{F}_{t}^{(1)}
\mathbf{A}_{[D]/\{1\}}^{\top}\mathcal{E}_{t}^{(1)^{\top}}\right\|_F^2
=O_p(\frac{p_1^2}{Tp}),
\eeqrs}
where the last equation is according to Lemma 1.

For the third term,
{\small\beqrs
\left\|\frac{1}{Tp}\sum\limits_{t=1}^T \mathcal{E}_{t}^{(1)}\mathcal{E}_{t}^{(1)^{\top}}\widehat{\mathbf{A}}_1\widehat{\mathbf{\Lambda}}_1^{-1}\right\|_F^2
&\leq&\left\|\frac{1}{Tp}\sum\limits_{t=1}^T \mathcal{E}_{t}^{(1)}\mathcal{E}_{t}^{(1)^{\top}}\mathbf{A}_1\right\|_F^2
+\left\|\frac{1}{Tp}\sum\limits_{t=1}^T \mathcal{E}_{t}^{(1)}\mathcal{E}_{t}^{(1)^{\top}}\right\|_F^2
\left\|\widehat{\mathbf{A}}_1-\mathbf{A}_1\widehat{\mathbf{H}}_1\right\|_{F}^2\\
&=&O_p(\frac{p_1}{Tp}+\frac{1}{p_1})+O_p(\frac{p_1}{Tp}+\frac{1}{p_1})
\left\|\widehat{\mathbf{A}}_1-\mathbf{A}_1\widehat{\mathbf{H}}_1\right\|_{F}^2,
\eeqrs}
where the last equation is according to Lemmas 2 and 4.
As a result, {\small\beqrs
\frac{1}{p_1}\left\|\widehat{\mathbf{A}}_1-\mathbf{A}_1\widehat{\mathbf{H}}_1\right\|_{F}^2=O_p(\frac{p_1}{Tp}+\frac{1}{p_1^2}).
\eeqrs}

Next, we prove that $\widehat{\mathbf{H}}_1$ is an asymptotic orthogonal matrix.
Because $\|{p_1^{-1}}\mathbf{A}_1^{\top}(\widehat{\mathbf{A}}_1-\mathbf{A}_1\widehat{\mathbf{H}}_1)\|_F^2
=\|{p_1^{-1}}\mathbf{A}_1^{\top}\widehat{\mathbf{A}}_1-\widehat{\mathbf{H}}_1\|_F^2=o_p(1)$
and $\|{p_1^{-1}}\widehat{\mathbf{A}}_1^{\top}(\widehat{\mathbf{A}}_1-\mathbf{A}_1\widehat{\mathbf{H}}_1)\|_F^2
=\|\mathbf{I}_{k_1}-{p_1^{-1}}\widehat{\mathbf{A}}_1^{\top}\mathbf{A}_1\widehat{\mathbf{H}}_1\|_F^2=o_p(1)$,
thus $\widehat{\mathbf{H}}_1^{\top}\widehat{\mathbf{H}}_1 =\mathbf{I}_{k_1}+o_p(\mathbf{1}_{k_1\times k_1})$. $\square$

The formulation of orthogonal matrix $\widehat{\mathbf{H}}_d$ originates from the representation of $\widehat{\mathbf{A}}_d$ in the proof of Proposition \ref{consistency_pca}, while is impacted by the cross products of loading matrices, which plays an important role in tackling $\widehat{\mathbf{M}}_d$ in Proposition 1.
In the literature of high-dimensional factor models, from tensor perspectives, $\mathbf{H}$ in Bai (2003) \cite{Bai2003Econometrica-inferential} and $\mathbf{H}_R$, $\mathbf{H}_C$ in Chen and Fan (2023) \cite{ChenFan2023JASA-statistical} are counterparts of $\widehat{\mathbf{H}}_d$ with respect to the $1$st-order and the $2$nd-order tensors, respectively.
They allow the cross product of the loading matrix to be positive definite approximately, whereas our cross products of the mode-wise loading matrices are required to be asymptotic orthogonal, sort of restricted though for the concern of simplifying the theoretical derivation procedure considering the complicated data structure, and yet not influencing the convergence rate, where both assumptions can ensure the strong factor condition.
This leads to the invertible transformation in Bai (2003) \cite{Bai2003Econometrica-inferential} and Chen and Fan (2023) \cite{ChenFan2023JASA-statistical}, and our orthogonal transformation in line with Stock and Watson (2002) \cite{StockWatson2002JASA-forecasting} and Yu et al. (2022)  \cite{YuHeKongZhang2022JOE-projected} for determining the unique column space of mode-wise loading matrices.

Next, we present the theory on the asymptotic normality of moPCA estimator $\widehat{\mathbf{A}}_d\in\mR^{p_d\times k_d}$ of the mode-wise loading matrices.
Note that $k_d$ is known and fixed, and thus we derive the asymptotic distribution of $\widehat{\mathbf{A}}_{d,i\cdot}\in\mR^{k_d}$, the $i$th row of $\widehat{\mathbf{A}}_d$.
\bep
\label{clt_pca}
Suppose that $T$ and $\{p_d\}_{d=1}^D$ tend to infinity, $\{k_d\}_{d=1}^D$ are fixed, and Assumptions 1-6 hold. For $d\in[D]$ and $i\in[p_d]$,\\\noindent
i) if $Tp_{-d}=o(p_d^2)$, then $\sqrt{Tp_{-d}}(\widehat{\mathbf{A}}_{d,i\cdot}-\widehat{\mathbf{H}}_1^{\top}\mathbf{A}_{d,i\cdot})
\stackrel{d}{\rightarrow}N(0,\mathbf{\Lambda}_d^{-1}\mathbf{\Gamma}_d^{\top}\mathbf{V}_{di}\mathbf{\Gamma}_d\mathbf{\Lambda}_d^{-1})$;\\\noindent
ii) if $p_d^2=O(Tp_{-d})$, then  $\widehat{\mathbf{A}}_{d,i\cdot}-\widehat{\mathbf{H}}_d^{\top}\mathbf{A}_{d,i\cdot}=O_p(\frac{1}{p_d})\mathbf{1}_{k_d},$
where $\mathbf{1}_{k_d}$ is a $k_d$-dim vector with elements 1.
\eep

Proposition \ref{clt_pca} shows that $\widehat{\mathbf{A}}_d$ is a good estimator of $\mathbf{A}_d\widehat{\mathbf{H}}_d$ in the sense that, $\widehat{\mathbf{A}}_{d,i\cdot}$ is asymptotically consistent without limit restriction on $T$ and $\{p_d\}_{d=1}^D$, and it can reach asymptotic normality when $p_d$ is sufficiently large.
Proposition \ref{clt_pca} is a higher-order extension of Theorem 3.4 in Yu et al. (2022) \cite{YuHeKongZhang2022JOE-projected}, and is consistent with Theorem 1 in Bai (2003) \cite{Bai2003Econometrica-inferential} and Theorem 2 in Chen and Fan (2023) \cite{ChenFan2023JASA-statistical}, despite distinct in the form of asymptotic covariance matrices owning to impact of assumptions on cross products of loading matrices.
Note that, for higher-order tensor factor models, the condition $Tp_{-d}=o(p_d^2)$ is hard to be satisfied on every mode except some specific modes with much larger size than others.
Take the $3$rd-order tensor for instance.
The restriction $Tp_{-d}=o(p_d^2)$ on all three modes implies $T^3$ must be $o(1)$, a paradox of the fact that $T$ tends to infinity.

\textit{Proof}: Without loss of generality, we only prove the result of $\widehat{\mathbf{A}}_1$.
According to the equation of $\widehat{\mathbf{A}}_1-\mathbf{A}_1\widehat{\mathbf{H}}_1$, we have
{\small\beqrs
\widehat{\mathbf{A}}_{1,i\cdot}-\widehat{\mathbf{H}}_1^{\top}\mathbf{A}_{1,i\cdot}
&=&\frac{1}{Tp}\sum\limits_{t=1}^T
\widehat{\mathbf{\Lambda}}_1^{-1}\widehat{\mathbf{A}}_1^{\top}\mathcal{E}_{t}^{(1)}
\mathbf{A}_{[D]/\{1\}}\mathcal{F}_{t}^{(1)^{\top}}\mathbf{A}_{1,i\cdot}\\
&&+\frac{1}{Tp}\sum\limits_{t=1}^T
\widehat{\mathbf{\Lambda}}_1^{-1}\widehat{\mathbf{A}}_1^{\top}\mathbf{A}_1\mathcal{F}_{t}^{(1)}
\mathbf{A}_{[D]/\{1\}}^{\top}\mathcal{E}_{t,i\cdot}^{(1)}
+\frac{1}{Tp}\sum\limits_{t=1}^T \widehat{\mathbf{\Lambda}}_1^{-1}\widehat{\mathbf{A}}_1^{\top}\mathcal{E}_{t}^{(1)}\mathcal{E}_{t,i\cdot}^{(1)}
\eeqrs}
For the first term,
{\small\beqrs
&&\left\|\frac{1}{Tp}\sum\limits_{t=1}^T
\widehat{\mathbf{\Lambda}}_1^{-1}\widehat{\mathbf{A}}_1^{\top}\mathcal{E}_{t}^{(1)}
\mathbf{A}_{[D]/\{1\}}\mathcal{F}_{t}^{(1)^{\top}}\right\|_F
\leq\left\|\frac{1}{Tp}\sum\limits_{t=1}^T
\mathbf{A}_1^{\top}\mathcal{E}_{t}^{(1)}
\mathbf{A}_{[D]/\{1\}}\mathcal{F}_{t}^{(1)^{\top}}\right\|_F\\
&&~~~+\left\|\frac{1}{Tp}\sum\limits_{t=1}^T\mathcal{E}_{t}^{(1)}
\mathbf{A}_{[D]/\{1\}}\mathcal{F}_{t}^{(1)^{\top}}\right\|_F \left\|\widehat{\mathbf{A}}_1-\mathbf{A}_1\widehat{\mathbf{H}}_1\right\|_F
=O_p(\frac{p_1}{Tp})+O_p(\frac{1}{\sqrt{Tp}}),
\eeqrs}
where the last equation is derived according to Lemmas 1, 3, and Proposition 2.

For the second term,
{\small\beqrs
\left\|\frac{1}{Tp}\sum\limits_{t=1}^T
\widehat{\mathbf{\Lambda}}_1^{-1}\widehat{\mathbf{A}}_1^{\top}\mathbf{A}_1\mathcal{F}_{t}^{(1)}
\mathbf{A}_{[D]/\{1\}}^{\top}\mathcal{E}_{t,i\cdot}^{(1)}\right\|_F
\leq\frac{p_1}{Tp}\left\|\sum\limits_{t=1}^T \mathcal{F}_{t}^{(1)}\mathbf{A}_{[D]/\{1\}}^{\top}\mathcal{E}_{t,i\cdot}^{(1)}\right\|_F =O_p(\sqrt{\frac{p_1}{Tp}}),
\eeqrs}
where the last equation is derived according to Lemma 5.

For the third term,
{\small\beqrs
&&\left\| \frac{1}{Tp}\sum\limits_{t=1}^T \widehat{\mathbf{\Lambda}}_1^{-1}\widehat{\mathbf{A}}_1^{\top}
\mathcal{E}_{t}^{(1)}\mathcal{E}_{t,i\cdot}^{(1)}\right\|_F\\
&\leq&\left\|\frac{1}{Tp}\sum\limits_{t=1}^T \mathcal{E}_{t,i\cdot}^{(1)^{\top}}\mathcal{E}_{t}^{(1)^{\top}}\mathbf{A}_1\right\|_F
+\left\|\frac{1}{Tp}\sum\limits_{t=1}^T \mathcal{E}_{t,i\cdot}^{(1)^{\top}}\mathcal{E}_{t}^{(1)^{\top}}\right\|_F
\left\|\widehat{\mathbf{A}}_1-\mathbf{A}_1\widehat{\mathbf{H}}_1\right\|_F\\
&=&O_p(\frac{p_1}{Tp}+\frac{1}{\sqrt{Tp}}+\frac{1}{p_1}),
\eeqrs}
where the last equation is derived according to Lemmas 6, 7, and Proposition 2.

If $Tp_{-1}=o(p_1^2)$, then
$$\widehat{\mathbf{A}}_{1,i\cdot}-\widehat{\mathbf{H}}_1^{\top}\mathbf{A}_{1,i\cdot}=
\frac{1}{Tp}\sum\limits_{t=1}^T\widehat{\mathbf{\Lambda}}_1^{-1}\widehat{\mathbf{A}}_1^{\top}\mathbf{A}_1\mathcal{F}_{t}^{(1)}
\mathbf{A}_{[D]/\{1\}}^{\top}\mathcal{E}_{t,i\cdot}^{(1)}+o_p(\frac{1}{\sqrt{Tp_{-1}}}).$$
According to Assumption 2, Propositions 1 and 2, we have
{\footnotesize$$\widehat{\mathbf{H}}_1=\frac{1}{Tp_1}\sum\limits_{t=1}^T \mathcal{F}_{t}^{(1)}\mathcal{F}_{t}^{(1)^{\top}}\mathbf{A}_1^{\top}\widehat{\mathbf{A}}_1\widehat{\mathbf{\Lambda}}_1^{-1}
={\mathbf{\Sigma}}_1\widehat{\mathbf{H}}_1\mathbf{\Lambda}_1^{-1}+o_p(\mathbf{1}_{k_1\times k_1})
=\mathbf{\Gamma}_1\mathbf{\Lambda}_1\mathbf{\Gamma}_1^{\top}\widehat{\mathbf{H}}_1\mathbf{\Lambda}_1^{-1}+o_p(\mathbf{1}_{k_1\times k_1}),$$}
hence $\mathbf{\Gamma}_1^{\top}\widehat{\mathbf{H}}_1\mathbf{\Lambda}_1=\mathbf{\Lambda}_1\mathbf{\Gamma}^{\top}\widehat{\mathbf{H}}_1+o_p(\mathbf{1}_{k_1\times k_1})$.
Because $\mathbf{\Lambda}_1$ is diagonal matrix, $\mathbf{\Gamma}_1^{\top}\widehat{\mathbf{H}}_1$ is diagonal as well.
Note that $(\mathbf{\Gamma}_1^{\top}\widehat{\mathbf{H}}_1)^{\top}\mathbf{\Gamma}_1^{\top}\widehat{\mathbf{H}}_1=\mathbf{I}_{k_1}+o_p(\mathbf{1}_{k_1\times k_1})$.
Taking $\mathbf{\Gamma}_1^{\top}\widehat{\mathbf{H}}_1=\mathbf{I}_{k_1}+o_p(\mathbf{1}_{k_1\times k_1})$, then
$p_1^{-1}\widehat{\mathbf{A}}_1^{\top}\mathbf{A}_1=\widehat{\mathbf{H}}_1^{\top}+o_p(\mathbf{1}_{k_1\times k_1})
=\mathbf{\Gamma}_1^{\top}+o_p(\mathbf{1}_{k_1\times k_1})$.
As a result,
{\small$$\sqrt{Tp_{-1}}(\widehat{\mathbf{A}}_{1,i\cdot}-\widehat{\mathbf{H}}_1^{\top}\mathbf{A}_{1,i\cdot})
\stackrel{d}{\rightarrow}N(0,\mathbf{\Lambda}_1^{-1}\mathbf{\Gamma}_1^{\top}\mathbf{V}_{1i}\mathbf{\Gamma}_1\mathbf{\Lambda}_1^{-1}).$$}
If $p_1^2=O(Tp_{-1})$, then
{\small$$\widehat{\mathbf{A}}_{1,i\cdot}-\widehat{\mathbf{H}}_1^{\top}\mathbf{A}_{1,i\cdot}=O_p(\frac{1}{p_1}).~\square$$} 

In the following proposition, we show the consistency of estimated tensor factors $\widehat{\mathcal{F}}_{t}\in\mR^{k_1\times \cdots \times k_D}$ in terms of the Forbenius norm and signal parts $\widehat{\mathcal{S}}_{t}\in\mR^{p_1\times \cdots \times p_D}$ element-wisely.
\bep
\label{PCA_X} 
Suppose that $T$ and $\{p_d\}_{d=1}^D$ tend to infinity, and $\{k_d\}_{d=1}^D$ are fixed.
If Assumptions 1-5 hold, then for any $i_d\in[p_d]$, $d\in[D]$ and $t\in[T]$, we have\\\noindent
(a) $\left\|\widehat{\mathcal{F}}_{t}
-\mathcal{F}_{t}\times_{1}\widehat{\mathbf{H}}_{1}^{-1}\times_{2} \cdots \times_{D}\widehat{\mathbf{H}}_{D}^{-1}\right\|_F
=O_p(\frac{1}{\sqrt{p}}+\sum\limits_{d=1}^D\frac{1}{p_{d}})$;\\\noindent
(b) $\left|\widehat{s}_{t,i_1\cdots i_D}-s_{t,i_1\cdots i_D}\right|
=O_p\left\{\frac{1}{\sqrt{p}}+\sum\limits_{d=1}^D\left(\frac{1}{p_{d}}+\frac{1}{\sqrt{Tp_{-d}}}\right)\right\}.$
\eep
Note that, the convergence rate is the same as that in Theorems 3 and 4 of Chen and Fan (2023) \cite{ChenFan2023JASA-statistical} for the matrix-variate factor model. 

\textit{Proof}: (a) The factor score in terms of moPCA estimator is
{\small\beqrs
\widehat{\mathcal{F}}_{t}
&=&\frac{1}{p}\mathcal{X}_{t}\times_{1}\widehat{\mathbf{A}}_1^{\top}\times_{2} \cdots \times_{D} \widehat{\mathbf{A}}_{D}^{\top}\\
&=&\frac{1}{p}\mathcal{F}_{t}\times_{1}\widehat{\mathbf{A}}_1^{\top}\mathbf{A}_{1}
\times_{2} \cdots \times_{D} \widehat{\mathbf{A}}_{D}^{\top}\mathbf{A}_{D}
+\frac{1}{p}\mathcal{E}_{t}\times_{1}\widehat{\mathbf{A}}_1^{\top}\times_{2} \cdots \times_{D} \widehat{\mathbf{A}}_{D}^{\top}\\
&=&\frac{1}{p}\mathcal{F}_{t}\times_{1}\widehat{\mathbf{A}}_1^{\top}
(\mathbf{A}_{1}-\widehat{\mathbf{A}}_{1}\widehat{\mathbf{H}}_{1}^{-1}+\widehat{\mathbf{A}}_{1}\widehat{\mathbf{H}}_{1}^{-1})
\times_{2} \cdots \times_{D}
\widehat{\mathbf{A}}_{D}^{\top}(\mathbf{A}_{D}-\widehat{\mathbf{A}}_{D}\widehat{\mathbf{H}}_{D}^{-1}+\widehat{\mathbf{A}}_{D}\widehat{\mathbf{H}}_{D}^{-1})\\
&&+\frac{1}{p}\mathcal{E}_{t}\times_{1}(\widehat{\mathbf{A}}_{1}-\mathbf{A}_{1}\widehat{\mathbf{H}}_{1}+\mathbf{A}_{1}\widehat{\mathbf{H}}_{1})^{\top}
\times_{2} \cdots \times_{D}(\widehat{\mathbf{A}}_{D}-\mathbf{A}_{D}\widehat{\mathbf{H}}_{D}+\mathbf{A}_{D}\widehat{\mathbf{H}}_{D})^{\top}.
\eeqrs}
Some algebra shows that
{\small\beqrs
&&\widehat{\mathcal{F}}_{t}
-\mathcal{F}_{t}\times_{1}\widehat{\mathbf{H}}_{1}^{-1}\times_{2} \cdots \times_{D}\widehat{\mathbf{H}}_{D}^{-1}\\
&=&\frac{1}{p}\mathcal{E}_{t}\times_{1}\widehat{\mathbf{H}}_{1}^{\top}\mathbf{A}_1^{\top}\times_{2} \cdots \times_{D}\widehat{\mathbf{H}}_{D}^{\top}\mathbf{A}_D^{\top}\\
&&+\frac{1}{p}\sum\limits_{d=1}^D\mathcal{E}_{t}\times_{1}\widehat{\mathbf{H}}_{1}^{\top}\mathbf{A}_1^{\top}\times_{2} \cdots \times_{d}(\widehat{\mathbf{A}}_d^{\top}-\mathbf{A}_d\widehat{\mathbf{H}}_d)\times_{d+1}\cdots\times_{D}\widehat{\mathbf{H}}_{D}^{\top}\mathbf{A}_D^{\top}
+\cdots\\
&&+\sum\limits_{d=1}^D\frac{1}{p_d}\mathcal{F}_{t}\times_{1}\widehat{\mathbf{H}}_{1}^{-1}\times_{2} \cdots \times_{d}\widehat{\mathbf{A}}_d^{\top}(\mathbf{A}_d-\widehat{\mathbf{A}}_d\widehat{\mathbf{H}}_d^{-1})\times_{d+1}\cdots\times_{D}\widehat{\mathbf{H}}_{D}^{-1}
+\cdots,
\eeqrs}
where the omitted terms are small order of the presented terms.
For the first term,
{\small\beqrs
\frac{1}{p}\left\|\mathcal{E}_{t}\times_{1}\widehat{\mathbf{H}}_{1}^{\top}\mathbf{A}_1^{\top}\times_{2} \cdots \times_{D}\widehat{\mathbf{H}}_{D}^{\top}\mathbf{A}_D^{\top} \right\|_F
&=&\frac{1}{p} \left\|(\mathbf{A}_D\widehat{\mathbf{H}}_{D}\otimes\cdots\otimes \mathbf{A}_1\widehat{\mathbf{H}}_1)^{\top}\vec(\mathcal{E}_t)\right\|_F\\
&\leq&\frac{1}{p}\left\| \mathbf{A}_{[D]/\{1\}}^{\top}\vec(\mathcal{E}_t) \right\|_F\\
&=&\frac{1}{p}\sqrt{\sum\limits_{i_1=1}^{p_1}\cdots\sum\limits_{i_D=1}^{p_D}(a_{1,i_1}\cdots a_{D,i_D}e_{t,i_1\cdots i_D})^2}\\&=&O_p(\frac{1}{\sqrt{p}}).
\eeqrs}
For the $d$th part in the second term,
{\small\beqrs
&&\frac{1}{p}\left\|\mathcal{E}_{t}\times_{1}\widehat{\mathbf{H}}_{1}^{\top}\mathbf{A}_1^{\top}\times_{2} \cdots \times_{d}(\widehat{\mathbf{A}}_d-\mathbf{A}_d\widehat{\mathbf{H}}_d)^{\top}\times_{d+1}\cdots\times_{D}\widehat{\mathbf{H}}_{D}^{\top}\mathbf{A}_D^{\top}\right\|_F\\
&\leq&\frac{1}{p}\left\|\left\{\mathbf{A}_D\otimes\cdots\otimes(\widehat{\mathbf{A}}_d-\mathbf{A}_{d}\widehat{\mathbf{H}}_d)\otimes\cdots\otimes \mathbf{A}_1\right\}^{\top}\vec(\mathcal{E}_t)\right\|_F\\
&\leq& \frac{1}{p}\|\widehat{\mathbf{A}}_d-\mathbf{A}_{d}\widehat{\mathbf{H}}_d\|_F
\left\| \mathcal{E}_t^{(d)}\mathbf{A}_{[D]/\{d\}}^{\top} \right\|_F\\
&\leq&\frac{1}{p}\sqrt{\sum_{i_1,i_1^{\prime}=1}^{p_1}\cdots\sum_{i_d=1}^{p_d}\cdots\sum_{i_D,i_D^{\prime}=1}^{p_D}
e_{t,i_1\cdots i_d\cdots i_D}e_{t,i_1^{\prime}\cdots i_d\cdots i_D^{\prime}}}\|\widehat{\mathbf{A}}_d-\mathbf{A}_{d}\widehat{\mathbf{H}}_d\|_F\\
&=&O_p(\frac{1}{\sqrt{T}p_{-d}}+\frac{1}{\sqrt{pp_d}}).
\eeqrs}

For the $d$th part in the third term, according to Lemma 8,
{\small\beqrs
&&\left\| \frac{1}{p_d}\mathcal{F}_{t}\times_{1}\widehat{\mathbf{H}}_{1}^{-1}\times_{2} \cdots \times_{d}\widehat{\mathbf{A}}_d^{\top}(\mathbf{A}_d-\widehat{\mathbf{A}}_d\widehat{\mathbf{H}}_d^{-1})\times_{d+1}\cdots\times_{D}\widehat{\mathbf{H}}_{D}^{-1} \right\|_F\\
&\leq&\left\| \mathcal{F}_{t}\right\|_F\frac{1}{p_d} \left\| \widehat{\mathbf{A}}_d^{\top}(\mathbf{A}_d-\widehat{\mathbf{A}}_d\widehat{\mathbf{H}}_d^{-1})\right\|_F=O_p(\frac{1}{\sqrt{Tp}}+\frac{1}{Tp_{-d}}+\frac{1}{p_d}).
\eeqrs}
As a result, 
{\small$$\left\|\widehat{\mathcal{F}}_{t}
-\mathcal{F}_{t}\times_{1}\widehat{\mathbf{H}}_{1}^{-1}\times_{2} \cdots \times_{D}\widehat{\mathbf{H}}_{D}^{-1}\right\|_F=O_p(\frac{1}{\sqrt{p}}+\sum\limits_{d=1}^D\frac{1}{p_{d}}).$$}

(b) Denote the $(i_1,\cdots,i_D)$th elements of the true signal part as $s_{t,i_1\cdots i_D}$ and
the estimated signal part as $\widehat{s}_{t,i_1\cdots i_D}$, then
{\small\beqrs
&&\widehat{s}_{t,i_1\cdots i_D}-s_{t,i_1\cdots i_D}\\
&=&\widehat{\mathcal{F}}_t\times_1 \widehat{\mathbf{A}}_{1,i_1\cdot}^{\top}\times_2\cdots\times_D \widehat{\mathbf{A}}_{D,i_D\cdot}^{\top}
-\mathcal{F}_t\times_1 \mathbf{A}_{1,i_1\cdot}^{\top}\times_2\cdots\times_D \mathbf{A}_{D,i_D\cdot}^{\top}\\
&=&\widehat{\mathcal{F}}_t\times_1 (\widehat{\mathbf{A}}_{1,i_1\cdot}-\widehat{\mathbf{H}}_1^{\top}\mathbf{A}_{1,i_1\cdot}+\widehat{\mathbf{H}}_1^{\top}\mathbf{A}_{1,i_1\cdot})^{\top}\times_2\cdots\times_D (\widehat{\mathbf{A}}_{D,i_D\cdot}-\widehat{\mathbf{H}}_D^{\top}\mathbf{A}_{D,i_D\cdot}+\widehat{\mathbf{H}}_D^{\top}\mathbf{A}_{D,i_D\cdot})^{\top}\\
&&-\mathcal{F}_t\times_1 \mathbf{A}_{1,i_1\cdot}^{\top}\times_2\cdots\times_D \mathbf{A}_{D,i_D\cdot}^{\top}\\
&=&(\widehat{\mathcal{F}}_t\times_1\widehat{\mathbf{H}}_1\times_2\cdots\times_D\widehat{\mathbf{H}}_D-\mathcal{F}_t)\times_1 \mathbf{A}_{1,i_1\cdot}^{\top}\times_2\cdots\times_D \mathbf{A}_{D,i_D\cdot}^{\top}\\
&&+\sum\limits_{d=1}^D\widehat{\mathcal{F}}_t\times_1 \mathbf{A}_{1,i_1\cdot}^{\top}\widehat{\mathbf{H}}_1
\times_2\cdots\times_d(\widehat{\mathbf{A}}_{d,i_d\cdot}-\widehat{\mathbf{H}}_d^{\top}\mathbf{A}_{d,i_d\cdot})^{\top}\times_{d+1}\cdots
\times_D\widehat{\mathbf{A}}_{D,i_D\cdot}^{\top}\widehat{\mathbf{H}}_D+\cdots,
\eeqrs}
\noindent 
where the omitted terms are small order of the existing term.
According to Propositions 3 and 4 (a), we have
{\small$$\left|\widehat{s}_{t,i_1\cdots i_D}-s_{t,i_1\cdots i_D}\right|
=O_p\left\{\frac{1}{\sqrt{p}}+\sum\limits_{d=1}^D\left(\frac{1}{p_{d}}+\frac{1}{\sqrt{Tp_{-d}}}\right)\right\}.~\square$$}

Propositions \ref{consistency_pca}-\ref{PCA_X} present asymptotic properties when mode-wise factor numbers are known;
otherwise, the ratio-type estimators in equation (3.4) of the main manuscript for the unknown factor numbers can replace instead, guaranteed by the consistency property below.
\bep
\label{consistency_factornumber}
Suppose that $T$ and $\{p_d\}_{d=1}^D$ tend to infinity, and $k_{max}$ is not smaller than $\max\{k_d\}_{d=1}^D$.
If Assumptions 1-5 hold, then the estimators of factor numbers defined in equation (3.4) satisfy
\beqr
P(\widehat{k}_d=k_d)\rightarrow 1,~d\in[D].
\eeqr
\eep
Proposition \ref{consistency_factornumber} illustrates that if $k_d$ is bounded by $k_{max}$ for $d\in[D]$,
the factor numbers can be estimated consistently.

\textit{Proof}: For $\widehat{\mathbf{M}}_d$ ($d\in[D]$), according to Proposition 1,
only the first $k_d$ leading eigenvalues are $O_p(1)$ and
the $p_d-k_d$ eigenvalues are 
$O_p\{p_d^{-1/2}+({Tp_{-d}})^{-1/2}\}$.
Hence, the ratio-type estimator can find out the true factor number $k_d$. $\square$

\section{Theoretical Proofs of IPmoPCA and PmoPCA Estimators}
\setcounter{equation}{0}

To prove the theorems of PmoPCA estimator and IPmoPCA estimator, the following proposition and corollary about the order of eigenvalues of $\widehat{\mathbf{M}}_d^{(s+1)}$ and $\widetilde{\mathbf{M}}_d$ are needed.

\bep
Suppose that $T$ and $\{p_d\}_{d=1}^D$ tend to infinity, and $\{k_d\}_{d=1}^D$ are fixed. If Assumptions 1-5 hold, then the order of eigenvalues of $\widehat{\mathbf{M}}_d^{(s+1)}$ ($d\in[D]$) are\vspace{0.3cm}\\
(a) for $j\leq k_d$, $\lambda_j(\widehat{\mathbf{M}}_d^{(s+1)})=\lambda_j({\mathbf{\Sigma}}_d)+o_p(1)$;\vspace{0.3cm}\\
(b) for $j>k_d$ and $s=0$
{\small\beqrs
\lambda_j(\widehat{\mathbf{M}}_d^{(1)})&=&
O_p\Bigg[\frac{1}{\sqrt{Tp_{-d}}} + \frac{1}{p_{-d}}+ \sum\limits_{d^{\prime} \neq d}\frac{1}{Tp_{-d^{\prime}}}
+ \sum\limits_{d^{\prime} = d+1}^D w_{d^{\prime}}^{(0)}
+ \sum\limits_{d^{\prime} = 1}^{d-1} w_{d^{\prime}}^{(1)}\\
&&+ \sum\limits_{d^{\prime} = d+1}^D \Bigg\{\frac{1}{\sqrt{T}p_{d^{\prime}}}
+ \left(\frac{1}{T\sqrt{p_{-d^{\prime}}}}+\frac{1}{\sqrt{T p_{d^{\prime}}}}\right)\sqrt{w_{d^{\prime}}^{(0)}}
\Bigg\}\\
&&+ \sum\limits_{d^{\prime} = 1}^{d-1} \Bigg\{
\left(\frac{1}{T\sqrt{p_{d^{\prime}}}}+\frac{1}{\sqrt{T}p_{d^{\prime}}} + \sqrt{\frac{w_{d^{\prime}}^{(1)}}{T}}\right)
\left(\sum\limits_{d^{\prime\prime} = d^{\prime}+1}^Dw_{d^{\prime\prime}}^{(0)}
+\sum\limits_{d^{\prime\prime} = 1}^{d^{\prime}-1}w_{d^{\prime\prime}}^{(1)}\right)
+ \frac{\sqrt{w_{d^{\prime}}^{(1)}}}{\sqrt{T}p_{-d^{\prime}}}
\Bigg\}
\Bigg];
\eeqrs}
(c) for $j>k_d$ and $s\geq 1$
{\small\begin{align*}
&~\lambda_j(\widehat{\mathbf{M}}_d^{(s+1)})=
O_p\Bigg[\frac{1}{\sqrt{Tp_{-d}}} + \frac{1}{p_{-d}}+ \sum\limits_{d^{\prime} \neq d}\frac{1}{Tp_{-d^{\prime}}}
+ \sum\limits_{d^{\prime} = d+1}^D w_{d^{\prime}}^{(s)}
+ \sum\limits_{d^{\prime} = 1}^{d-1} w_{d^{\prime}}^{(s+1)}\\
&~~~~~~~~+ \sum\limits_{d^{\prime} = d+1}^D \Bigg\{
\left(\frac{1}{T\sqrt{p_{d^{\prime}}}}+\frac{1}{\sqrt{T}p_{d^{\prime}}} + \sqrt{\frac{w_{d^{\prime}}^{(s)}}{T}}\right)
\left(\sum\limits_{d^{\prime\prime} = d^{\prime}+1}^Dw_{d^{\prime\prime}}^{(s-1)}
+\sum\limits_{d^{\prime\prime} = 1}^{d^{\prime}-1}w_{d^{\prime\prime}}^{(s)}\right)
+ \frac{\sqrt{w_{d^{\prime}}^{(s)}}}{\sqrt{T}p_{-d^{\prime}}}
\Bigg\}\\
&~~~~~~~~+ \sum\limits_{d^{\prime} = 1}^{d-1} \Bigg\{
\left(\frac{1}{T\sqrt{p_{d^{\prime}}}}+\frac{1}{\sqrt{T}p_{d^{\prime}}} + \sqrt{\frac{w_{d^{\prime}}^{(s+1)}}{T}}\right)
\left(\sum\limits_{d^{\prime\prime} = d^{\prime}+1}^Dw_{d^{\prime\prime}}^{(s)}
+\sum\limits_{d^{\prime\prime} = 1}^{d^{\prime}-1}w_{d^{\prime\prime}}^{(s+1)}\right)
+ \frac{\sqrt{w_{d^{\prime}}^{(s+1)}}}{\sqrt{T}p_{-d^{\prime}}}
\Bigg\}
\Bigg].
\end{align*}}
\eep

\textit{Proof}: For simplicity, we only prove the case of $s\geq 1$. Note that
\beqrs
\widehat{\mathbf{M}}_d^{(s+1)}&=&\frac{1}{Tp_d}\sum\limits_{t=1}^T \widehat{\mathcal{Y}}_{t,d}^{(d,s+1)}\widehat{\mathcal{Y}}_{t,d}^{(d,s+1)\top}\\
&=&\frac{1}{Tpp_{-d}}\sum\limits_{t=1}^T\mathbf{A}_d \mathcal{F}_{t}^{(d)}\mathbf{A}_{[D]/\{d\}}^{\top}\widehat{\mathbf{A}}_{[D]/\{d\}}^{(s+1)}\widehat{\mathbf{A}}_{[D]/\{d\}}^{(s+1)\top}\mathbf{A}_{[D]/\{d\}}\mathcal{F}_{t}^{(d)\top}\mathbf{A}_d^{\top}\\
&&+\frac{1}{Tpp_{-d}}\sum\limits_{t=1}^T \mathbf{A}_d\mathcal{F}_{t}^{(d)}\mathbf{A}_{[D]/\{d\}}^{\top}\widehat{\mathbf{A}}_{[D]/\{d\}}^{(s+1)}\widehat{\mathbf{A}}_{[D]/\{d\}}^{(s+1)\top}\mathcal{E}_{t}^{(d)\top}\\
&&+\frac{1}{Tpp_{-d}}\sum\limits_{t=1}^T \mathcal{E}_{t}^{(d)}\widehat{\mathbf{A}}_{[D]/\{d\}}^{(s+1)}\widehat{\mathbf{A}}_{[D]/\{d\}}^{(s+1)\top}\mathbf{A}_{[D]/\{d\}}\mathcal{F}_{t}^{(d)\top}\mathbf{A}_d^{\top}\\
&&+\frac{1}{Tpp_{-d}}\sum\limits_{t=1}^T \mathcal{E}_{t}^{(d)}\widehat{\mathbf{A}}_{[D]/\{d\}}^{(s+1)}\widehat{\mathbf{A}}_{[D]/\{d\}}^{(s+1)\top}\mathcal{E}_{t}^{(d)\top}.
\eeqrs
For the first term, it approximates
$(Tp_{d})^{-1}\sum\limits_{t=1}^T \mathbf{A}_d\mathcal{F}_{t}^{(d)}\mathcal{F}_{t}^{(d)\top}\mathbf{A}_d^{\top}$. 
Hence its $k_1$ largest eigenvalues are same as those of $T^{-1}\sum\limits_{t=1}^T\mathcal{F}_{t}^{(d)}\mathcal{F}_{t}^{(d)\top}$, while the other $p_d-k_d$ eigenvalues are zero.
Moreover, $T^{-1}\sum\limits_{t=1}^T \mathcal{F}_{t}^{(d)}\mathcal{F}_{t}^{(d)^{\top}}
\stackrel{p}{\rightarrow}{\mathbf{\Sigma}}_d$, and it is then implied by Weyl’s inequality that
\beqrs
\lambda_j\left(\frac{1}{Tp_d}\sum\limits_{t=1}^T \mathbf{A}_d\mathcal{F}_{t}^{(d)}\mathcal{F}_{t}^{(d)^{\top}}\mathbf{A}_d^{\top}\right)
=\left\{\begin{array}{l}
\vspace{0.5cm}
\lambda_j({\mathbf{\Sigma}}_d)+o_p(1), \quad j\leq k_d;\\
0, \quad j>k_d.
\end{array}\right.
\eeqrs
For the second term, which has the same order as the third one due to the symmetry, by Lemma 11 (c), we have
{\scriptsize\begin{align*}
&\left\|\frac{1}{Tp}\sum\limits_{t=1}^T \mathbf{A}_d\mathcal{F}_{t}^{(d)}
\widehat{\mathbf{A}}_{[D]/\{d\}}^{(s+1){\top}}
\mathcal{E}_{t}^{(d)\top}\right\|_F
\leq\frac{\sqrt{p_d}}{Tp}\left\|\sum\limits_{t=1}^T\mathcal{E}_{t}^{(d)}
\widehat{\mathbf{A}}_{[D]/\{d\}}^{(s+1)}\mathcal{F}_{t}^{(d)\top}\right\|_F\\
&=O_p\Bigg[\frac{1}{\sqrt{Tp_{-d}}} + \sum\limits_{d^{\prime} \neq d}\frac{1}{Tp_{-d^{\prime}}}
+ \sum\limits_{d^{\prime} = d+1}^D \Bigg\{
\left(\frac{1}{T\sqrt{p_{d^{\prime}}}}+\frac{1}{\sqrt{T}p_{d^{\prime}}} + \sqrt{\frac{w_{d^{\prime}}^{(s)}}{T}}\right)
\left(\sum\limits_{d^{\prime\prime} = d^{\prime}+1}^Dw_{d^{\prime\prime}}^{(s-1)}
+\sum\limits_{d^{\prime\prime} = 1}^{d^{\prime}-1}w_{d^{\prime\prime}}^{(s)}\right)
+ \frac{\sqrt{w_{d^{\prime}}^{(s)}}}{\sqrt{T}p_{-d^{\prime}}}
\Bigg\}\\
&+ \sum\limits_{d^{\prime} = 1}^{d-1} \Bigg\{
\left(\frac{1}{T\sqrt{p_{d^{\prime}}}}+\frac{1}{\sqrt{T}p_{d^{\prime}}} + \sqrt{\frac{w_{d^{\prime}}^{(s+1)}}{T}}\right)
\left(\sum\limits_{d^{\prime\prime} = d^{\prime}+1}^Dw_{d^{\prime\prime}}^{(s)}
+\sum\limits_{d^{\prime\prime} = 1}^{d^{\prime}-1}w_{d^{\prime\prime}}^{(s+1)}\right)
+ \frac{\sqrt{w_{d^{\prime}}^{(s+1)}}}{\sqrt{T}p_{-d^{\prime}}}
\Bigg\}
\Bigg].
\end{align*}}
For the fourth term, it is implied by Lemma 9 that
\beqrs
\left\|\frac{1}{Tpp_{-d}}\sum\limits_{t=1}^T \mathcal{E}_{t}^{(d)}
\widehat{\mathbf{A}}_{[D]/\{d\}}^{(s+1)}\widehat{\mathbf{A}}_{[D]/\{d\}}^{(s+1)\top}\mathcal{E}_{t}^{(d)\top}\right\|_F=O_p\left(\frac{1}{p_{-d}}
+ \sum\limits_{d^{\prime} = d+1}^D w_{d^{\prime}}^{(s)}
+ \sum\limits_{d^{\prime} = 1}^{d-1} w_{d^{\prime}}^{(s+1)}\right).
\eeqrs
We hence accomplish the proof by Weyl's inequality.
Similarly, the result can be proved for the case of $s=0$. $\square$

\bec
Suppose that $T$ and $\{p_d\}_{d=1}^D$ tend to infinity, and $\{k_d\}_{d=1}^D$ are fixed. If Assumptions 1-5 hold, then the order of eigenvalues of $\widetilde{\mathbf{M}}_d$ ($d\in[D]$) are
{\small\beqr
\lambda_j(\widetilde{\mathbf{M}}_d)=\left\{\begin{array}{l}
\vspace{0.5cm}
\lambda_j({\mathbf{\Sigma}}_d)+o_p(1), \quad j\leq k_d;\\
O_p\left\{\frac{1}{\sqrt{Tp_{-d}}} + \frac{1}{p_{-d}}
+ \sum\limits_{d^{\prime} \neq d}\left(\frac{1}{\sqrt{T}p_{d^{\prime}}}
+ \frac{1}{p_{d^{\prime}}^2} + \frac{1}{Tp_{-d^{\prime}}}\right)\right\}, \quad j>k_d.
\end{array}\right.
\eeqr}
\eec

\textit{Proof}:
According to the proof of Proposition 6, for $j > k_d$, we have
{\scriptsize\beqrs
\lambda_j(\widetilde{\mathbf{M}}_d)&=&
O_p\Bigg[\frac{1}{\sqrt{Tp_{-d}}} + \frac{1}{p_{-d}}
+ \sum\limits_{d^{\prime} \neq d}\Bigg\{
\frac{1}{Tp_{-d^{\prime}}}+ w_{d^{\prime}}^{(0)}\frac{1}{\sqrt{T}p_{d^{\prime}}}
+ \left(\frac{1}{T\sqrt{p_{-d^{\prime}}}}+\frac{1}{\sqrt{T p_{d^{\prime}}}}\right)\sqrt{w_{d^{\prime}}^{(0)}}
\Bigg\}
\Bigg].
\eeqrs}
Taking $w_d^{(0)}$ into the above equation results in the corollary. $\square$

\subsection*{Proof of Theorem 4.4}

Note that IPmoPCA estimation for loading matrices reduces to PmoPCA estimation when iteration is only once and projected loading matrices do not update. 
So we first prove the result of Theorem 4.4, and Theorem 4.1 can be proved accordingly.\\

\textit{Proof}: For simplicity, we only prove the case of $s\geq 1$. Note that the IPmoPCA estimator  $\widehat{\mathbf{A}}_d^{(s+1)}$ is set to be the top
$k_d$ eigenvectors of the sample variance-covariance
$\widehat{\mathbf{M}}_d^{(s+1)}$ up to a scalar $\sqrt{p_{_d}}$.
Similar to the proof of Proposition 2, we have $\widehat{\mathbf{M}}_d^{(s+1)}\widehat{\mathbf{A}}_d^{(s+1)}=\widehat{\mathbf{A}}_d^{(s+1)}\widehat{\mathbf{\Lambda}}_d^{(s+1)}$, where $\widehat{\mathbf{\Lambda}}_d^{(s+1)}$ is the diagonal matrix with elements as the $k_d$ largest eigenvalues of $\widehat{\mathbf{M}}_d^{(s+1)}$.
As a result, 
{\small\begin{align*}
&\widehat{\mathbf{A}}_d^{(s+1)}=\widehat{\mathbf{M}}_d^{(s+1)}\widehat{\mathbf{A}}_d^{(s+1)}\widehat{\mathbf{\Lambda}}_d^{(s+1)^{-1}}\\
&=\mathbf{A}_d\left\{\frac{1}{Tpp_{-d}}\sum\limits_{t=1}^T \mathcal{F}_{t}^{(d)}\mathbf{A}_{[D]/\{d\}}^{\top}\widehat{\mathbf{A}}_{[D]/\{d\}}^{(s+1)}\widehat{\mathbf{A}}_{[D]/\{d\}}^{(s+1)\top}\mathbf{A}_{[D]/\{d\}}\mathcal{F}_{t}^{(d)\top}\mathbf{A}_d^{\top}\widehat{\mathbf{A}}_d^{(s+1)}\widehat{\mathbf{\Lambda}}_d^{(s+1)^{-1}}\right\}\\
&~+\frac{1}{Tpp_{-d}}\sum\limits_{t=1}^T \mathbf{A}_d\mathcal{F}_{t}^{(d)}\mathbf{A}_{[D]/\{d\}}^{\top}\widehat{\mathbf{A}}_{[D]/\{d\}}^{(s+1)}\widehat{\mathbf{A}}_{[D]/\{d\}}^{(s+1)\top}\mathcal{E}_{t}^{(d)\top}\widehat{\mathbf{A}}_d^{(s+1)}\widehat{\mathbf{\Lambda}}_d^{(s+1)^{-1}}\\
&~+\frac{1}{Tpp_{-d}}\sum\limits_{t=1}^T \mathcal{E}_{t}^{(d)}\widehat{\mathbf{A}}_{[D]/\{d\}}^{(s+1)}\widehat{\mathbf{A}}_{[D]/\{d\}}^{(s+1)\top}\mathbf{A}_{[D]/\{d\}}\mathcal{F}_{t}^{(d)\top}\mathbf{A}_d^{\top}\widehat{\mathbf{A}}_d^{(s+1)}\widehat{\mathbf{\Lambda}}_d^{(s+1)^{-1}}\\
&~+\frac{1}{Tpp_{-d}}\sum\limits_{t=1}^T \mathcal{E}_{t}^{(d)}\widehat{\mathbf{A}}_{[D]/\{d\}}^{(s+1)}\widehat{\mathbf{A}}_{[D]/\{d\}}^{(s+1)\top}\mathcal{E}_{t}^{(d)\top}\widehat{\mathbf{A}}_d^{(s+1)}\widehat{\mathbf{\Lambda}}_d^{(s+1)^{-1}}.
\end{align*}}
Denote 
{\small
$$\widehat{\mathbf{H}}_d^{(s+1)}=(Tpp_{-d})^{-1}\sum\limits_{t=1}^T\mathcal{F}_{t}^{(d)}\mathbf{A}_{[D]/\{d\}}^{\top}\widehat{\mathbf{A}}_{[D]/\{d\}}^{(s+1)}\widehat{\mathbf{A}}_{[D]/\{d\}}^{(s+1)\top}\mathbf{A}_{[D]/\{d\}}\mathcal{F}_{t}^{(d)^{\top}}\mathbf{A}_d^{\top}\widehat{\mathbf{A}}_d^{(s+1)}\widehat{\mathbf{\Lambda}}_d^{(s+1)^{-1}},$$}
then
\beqrs
&&\widehat{\mathbf{A}}_d^{(s+1)}-\mathbf{A}_d\widehat{\mathbf{H}}_d^{(s+1)}\\
&=&\frac{1}{Tpp_{-d}}\sum\limits_{t=1}^T \mathbf{A}_d\mathcal{F}_{t}^{(d)}\mathbf{A}_{[D]/\{d\}}^{\top}\widehat{\mathbf{A}}_{[D]/\{d\}}^{(s+1)}\widehat{\mathbf{A}}_{[D]/\{d\}}^{(s+1)\top}\mathcal{E}_{t}^{(d)\top}\widehat{\mathbf{A}}_d^{(s+1)}\widehat{\mathbf{\Lambda}}_d^{(s+1)^{-1}}\\
&&~+\frac{1}{Tpp_{-d}}\sum\limits_{t=1}^T \mathcal{E}_{t}^{(d)}\widehat{\mathbf{A}}_{[D]/\{d\}}^{(s+1)}\widehat{\mathbf{A}}_{[D]/\{d\}}^{(s+1)\top}\mathbf{A}_{[D]/\{d\}}\mathcal{F}_{t}^{(d)\top}\mathbf{A}_d^{\top}\widehat{\mathbf{A}}_d^{(s+1)}\widehat{\mathbf{\Lambda}}_d^{(s+1)^{-1}}\\
&&~+\frac{1}{Tpp_{-d}}\sum\limits_{t=1}^T \mathcal{E}_{t}^{(d)}\widehat{\mathbf{A}}_{[D]/\{d\}}^{(s+1)}\widehat{\mathbf{A}}_{[D]/\{d\}}^{(s+1)\top}\mathcal{E}_{t}^{(d)\top}\widehat{\mathbf{A}}_d^{(s+1)}\widehat{\mathbf{\Lambda}}_d^{(s+1)^{-1}}\\
&&\approx\frac{1}{Tp}\sum\limits_{t=1}^T \mathbf{A}_d\mathcal{F}_{t}^{(d)}\widehat{\mathbf{A}}_{[D]/\{d\}}^{(s+1)\top}\mathcal{E}_{t}^{(d)\top}\widehat{\mathbf{A}}_d^{(s+1)}\widehat{\mathbf{\Lambda}}_d^{(s+1)^{-1}}\\
&&~+\frac{1}{Tp}\sum\limits_{t=1}^T \mathcal{E}_{t}^{(d)}\widehat{\mathbf{A}}_{[D]/\{d\}}^{(s+1)}\mathcal{F}_{t}^{(d)\top}\mathbf{A}_d^{\top}\widehat{\mathbf{A}}_d^{(s+1)}\widehat{\mathbf{\Lambda}}_d^{(s+1)^{-1}}\\
&&~+\frac{1}{Tpp_{-d}}\sum\limits_{t=1}^T \mathcal{E}_{t}^{(d)}\widehat{\mathbf{A}}_{[D]/\{d\}}^{(s+1)}\widehat{\mathbf{A}}_{[D]/\{d\}}^{(s+1)\top}\mathcal{E}_{t}^{(d)\top}\widehat{\mathbf{A}}_d^{(s+1)}\widehat{\mathbf{\Lambda}}_d^{(s+1)^{-1}},
\eeqrs
where $\approx$ means that the order is the same after approximation.
Similar to the proof of Proposition 2, according to Lemmas 10 and 11, the result can be proved. $\square$

\subsection*{Proof of Theorem 4.1}

\textit{Proof}: Note that
\beqrs\small
&&\frac{1}{p_d}\|\widetilde{\mathbf{A}}_d-\mathbf{A}_d\widetilde{\mathbf{H}}_d\|_{F}^2=
O_p\Bigg[\frac{1}{Tp_{-d}} + \frac{1}{p^2}+ \sum\limits_{d^{\prime} \neq d}\frac{1}{T^2p_{-d^{\prime}}^2}
+ \left(\frac{1}{Tp_d}+\frac{1}{p_d^2}\right)
\left(\sum\limits_{d^{\prime} \neq d} w_{d^{\prime}}^{(0)2}\right)\\
&&~~~~~~~~~~~~~~~~~~+ \sum\limits_{d^{\prime}\neq d} \Bigg\{\frac{1}{Tp_{d^{\prime}}^2}
+ \left(\frac{1}{T^2p_{-d^{\prime}}}+\frac{1}{T p_{d^{\prime}}}\right){w_{d^{\prime}}^{(0)}}
\Bigg\}
\Bigg],
\eeqrs
by taking $w_{d}^{(0)}$ into the above equation, the result can be proved. $\square$

\subsection*{Proof of Theorem 4.2}

\textit{Proof}: Without loss of generality, we only prove the result of $\widetilde{\mathbf{A}}_1$.
According to the equation of $\widetilde{\mathbf{A}}_1-\mathbf{A}_1\widetilde{\mathbf{H}}_1$, we have
\beqrs
\widetilde{\mathbf{A}}_{1,i\cdot}-\widetilde{\mathbf{H}}_1^{\top}\mathbf{A}_{1,i\cdot}
&=&\frac{1}{Tpp_{-1}}\sum\limits_{t=1}^T
\widetilde{\mathbf{\Lambda}}_1^{-1}\widetilde{\mathbf{A}}_1^{\top}\mathcal{E}_{t}^{(1)}
\widehat{\mathbf{A}}_{[D]/\{1\}}\widehat{\mathbf{A}}_{[D]/\{1\}}^{\top}{\mathbf{A}}_{[D]/\{1\}}
\mathcal{F}_{t}^{(1)^{\top}}\mathbf{A}_{1,i\cdot}\\
&&+\frac{1}{Tpp_{-1}}\sum\limits_{t=1}^T
\widetilde{\mathbf{\Lambda}}_1^{-1}\widetilde{\mathbf{A}}_1^{\top}\mathbf{A}_1\mathcal{F}_{t}^{(1)}
{\mathbf{A}}_{[D]/\{1\}}^{\top}\widehat{\mathbf{A}}_{[D]/\{1\}}\widehat{\mathbf{A}}_{[D]/\{1\}}^{\top}
\mathcal{E}_{t,i\cdot}^{(1)}\\
&&+\frac{1}{Tpp_{-1}}\sum\limits_{t=1}^T \widetilde{\mathbf{\Lambda}}_1^{-1}\widetilde{\mathbf{A}}_1^{\top}\mathcal{E}_{t}^{(1)}
\widehat{\mathbf{A}}_{[D]/\{1\}}\widehat{\mathbf{A}}_{[D]/\{1\}}^{\top}\mathcal{E}_{t,i\cdot}^{(1)}
\eeqrs
For the first term,
\beqrs
&&\|\frac{1}{Tpp_{-1}}\sum\limits_{t=1}^T
\widetilde{\mathbf{\Lambda}}_1^{-1}\widetilde{\mathbf{A}}_1^{\top}\mathcal{E}_{t}^{(1)}
\widehat{\mathbf{A}}_{[D]/\{1\}}\widehat{\mathbf{A}}_{[D]/\{1\}}^{\top}{\mathbf{A}}_{[D]/\{1\}}\mathcal{F}_{t}^{(1)^{\top}}\|_F\\
&\leq&\|\frac{1}{Tp}\sum\limits_{t=1}^T
{A}_1^{\top}\mathcal{E}_{t}^{(1)}
\widehat{\mathbf{A}}_{[D]/\{1\}}\mathcal{F}_{t}^{(1)^{\top}}\|_F+\|\frac{1}{Tp}\sum\limits_{t=1}^T\mathcal{E}_{t}^{(1)}
\widehat{\mathbf{A}}_{[D]/\{1\}}\mathcal{F}_{t}^{(1)^{\top}}\|_F
\|\widetilde{\mathbf{A}}_1-\mathbf{A}_1\widetilde{\mathbf{H}}_1\|_F\\
&=&O_p\left\{\frac{p_1}{Tp}+\frac{1}{\sqrt{Tp}}+
\sum\limits_{d=2}^D\left(\frac{p_d}{Tp}
+\frac{1}{\sqrt{Tp_1}p_{d}}+\frac{1}{T\sqrt{p_{-1}}p_{d}}\right)
+\sum\limits_{d_1\neq 1}\sum\limits_{d_2\neq 1,d_1}\frac{1}{Tp_{d_1}p_{d_2}}\right\},
\eeqrs
where the last equation is derived according to Lemmas 11 (a),  11 (d), and Theorem 4.1 in the main text.

For the second term,
\beqrs
&&\| \frac{1}{Tpp_{-1}}\sum\limits_{t=1}^T
\widetilde{\mathbf{\Lambda}}_1^{-1}\widetilde{\mathbf{A}}_1^{\top}\mathbf{A}_1\mathcal{F}_{t}^{(1)}
{\mathbf{A}}_{[D]/\{1\}}^{\top}\widehat{\mathbf{A}}_{[D]/\{1\}}\widehat{\mathbf{A}}_{[D]/\{1\}}^{\top}\mathcal{E}_{t,i\cdot}^{(1)} \|_F\\
&\leq&\frac{1}{Tp_{-1}}\| \sum\limits_{t=1}^T \mathcal{F}_{t}^{(1)}\widehat{\mathbf{A}}_{[D]/\{1\}}^{\top}\mathcal{E}_{t,i\cdot}^{(1)}\|_F
=O_p\left\{\frac{1}{\sqrt{Tp_{-1}}}+\sum\limits_{d=2}^D\left(
\frac{1}{\sqrt{T}p_d}+\frac{1}{Tp_{-d}}\right)\right\}.
\eeqrs
where the last equation is derived according to Lemma 12.

For the third term,
\beqrs
&&\| \frac{1}{Tpp_{-1}}\sum\limits_{t=1}^T \widetilde{\mathbf{\Lambda}}_1^{-1}\widetilde{\mathbf{A}}_1^{\top}\mathcal{E}_{t}^{(1)}
\widehat{\mathbf{A}}_{[D]/\{1\}}\widehat{\mathbf{A}}_{[D]/\{1\}}^{\top}
\mathcal{E}_{t,i\cdot}^{(1)} \|_F\\
&\leq& \| \frac{1}{Tpp_{-1}}\sum\limits_{t=1}^T {A}_1^{\top}\mathcal{E}_{t}^{(1)}
\widehat{\mathbf{A}}_{[D]/\{1\}}\widehat{\mathbf{A}}_{[D]/\{1\}}^{\top}\mathcal{E}_{t,i\cdot}^{(1)} \|_F\\
&&+\| \frac{1}{Tpp_{-1}}\sum\limits_{t=1}^T\mathcal{E}_{t}^{(1)}
\widehat{\mathbf{A}}_{[D]/\{1\}}\widehat{\mathbf{A}}_{[D]/\{1\}}^{\top}\mathcal{E}_{t,i\cdot}^{(1)} \|_F
\|\widetilde{\mathbf{A}}_1-\mathbf{A}_1\widetilde{\mathbf{H}}_1\|_F\\
&=&O_p(\frac{1}{p}+\sum\limits_{d=2}^D\frac{1}{p_1p_d^2})
+o_p\left\{\frac{1}{\sqrt{Tp_{-1}}}+\frac{1}{p}+\sum\limits_{d=2}^D
\left(\frac{1}{\sqrt{T}p_d^2}+\frac{1}{Tp_{-d}}+\frac{1}{p_1p_{d}^2}
\right)\right\}
\eeqrs
where the last equation is derived according to Lemmas 13, 14, and Theorem 4.1 in the main text.

If $Tp_{-1}=o\left\{p^2+\sum\limits_{d=2}^D\left(Tp_d^2+T^2p_{-d}^2+p_1^2p_d^4\right)\right\}$, then $$\widetilde{\mathbf{A}}_{1,i\cdot}-\widetilde{\mathbf{H}}_1^{\top}\mathbf{A}_{1,i\cdot}=
\frac{1}{Tpp_{-1}}\sum\limits_{t=1}^T
\widetilde{\mathbf{\Lambda}}_1^{-1}\widetilde{\mathbf{A}}_1^{\top}\mathbf{A}_1\mathcal{F}_{t}^{(1)}
{\mathbf{A}}_{[D]/\{1\}}^{\top}\widehat{\mathbf{A}}_{[D]/\{1\}}\widehat{\mathbf{A}}_{[D]/\{1\}}^{\top}
\mathcal{E}_{t,i\cdot}^{(1)}+o_p(\frac{1}{\sqrt{Tp_{-1}}}).$$
According to Assumption 2, Propositions 2 and 6, Theorem 4.2, we have
\beqrs
\widetilde{\mathbf{H}}_1&=&\frac{1}{Tpp_{-1}}\sum\limits_{t=1}^T \mathcal{F}_{t}^{(1)}
{\mathbf{A}}_{[D]/\{1\}}^{\top}\widehat{\mathbf{A}}_{[D]/\{1\}}\widehat{\mathbf{A}}_{[D]/\{1\}}^{\top}{\mathbf{A}}_{[D]/\{1\}}
\mathcal{F}_{t}^{(1)^{\top}}\mathbf{A}_1^{\top}\widetilde{\mathbf{A}}_1\widetilde{\mathbf{\Lambda}}_1^{-1}\\
&=&{\mathbf{\Sigma}}_1\widetilde{\mathbf{H}}_1\mathbf{\Lambda}_1^{-1}+o_p(\mathbf{1}),
\eeqrs
hence $\widetilde{\mathbf{H}}_1=\mathbf{\Gamma}_1\mathbf{\Lambda}_1\mathbf{\Gamma}_1^{\top}\widetilde{\mathbf{H}}_1\mathbf{\Lambda}_1^{-1}+o_p(\mathbf{1})$
and $\mathbf{\Gamma}_1^{\top}\widetilde{\mathbf{H}}_1\mathbf{\Lambda}_1=\mathbf{\Lambda}_1\mathbf{\Gamma}^{\top}\widetilde{\mathbf{H}}_1+o_p(\mathbf{1})$.
Because $\mathbf{\Lambda}_1$ is diagonal matrix, $\mathbf{\Gamma}_1\widetilde{\mathbf{H}}_1$ is diagonal as well.
Note that $(\mathbf{\Gamma}_1\widetilde{\mathbf{H}}_1)^{\top}\mathbf{\Gamma}_1\widetilde{\mathbf{H}}_1=\mathbf{I}_{k_1}+o_p(\mathbf{1})$.
Taking $\mathbf{\Gamma}_1\widetilde{\mathbf{H}}_1=\mathbf{I}_{k_1}+o_p(\mathbf{1})$, then
$p_1^{-1}\widetilde{\mathbf{A}}_1^{\top}\mathbf{A}_1=\widetilde{\mathbf{H}}_1^{\top}+o_p(\mathbf{1})=\mathbf{\Gamma}_1^{\top}+o_p(\mathbf{1})$.
As a result,
$$\sqrt{Tp_{-1}}(\widetilde{\mathbf{A}}_{1,i\cdot}-\widetilde{\mathbf{H}}_1^{\top}\mathbf{A}_{1,i\cdot})
\stackrel{d}{\rightarrow}N(0,\mathbf{\Lambda}_1^{-1}\mathbf{\Gamma}_1^{\top}\mathbf{V}_{1i}\mathbf{\Gamma}_1\mathbf{\Lambda}_1^{-1}).$$
Note that, the result is only true for $D=2$.
When $D\geq3$, we need to make $p_{-1}=o(p_d^{2})$ ($d=2,\cdots,D$),
which will result in contradiction.
For example, for $D=3$, we need $p_2=o(p_3)$ and $p_3=o(p_2)$ simultaneously.

If $p^2+\sum\limits_{d=2}^D\left(Tp_d^2+T^2p_{-d}^2+p_1^2p_d^4\right)=O(Tp_{-1})$, then $$\widetilde{\mathbf{A}}_{1,i\cdot}-\widetilde{\mathbf{H}}_1^{\top}\mathbf{A}_{1,i\cdot}
=O_p\left\{\frac{1}{p}+\sum\limits_{d=2}^D\left(
\frac{1}{\sqrt{T}p_d}+\frac{1}{Tp_{-d}}+\frac{1}{p_1p_d^2}\right)\right\}.~\square$$

\subsection*{Proof of Theorem 4.3}

\textit{Proof}: (a) The factor score in terms of PmoPCA estimator is
\beqrs
\widetilde{\mathcal{F}}_{t}
&=&\frac{1}{p}\mathcal{X}_{t}\times_{1}\widetilde{\mathbf{A}}_1^{\top}\times_{2} \cdots \times_{D} \widetilde{\mathbf{A}}_{D}^{\top}\\
&=&\frac{1}{p}\mathcal{F}_{t}\times_{1}\widetilde{\mathbf{A}}_1^{\top}\mathbf{A}_{1}
\times_{2} \cdots \times_{D} \widetilde{\mathbf{A}}_{D}^{\top}\mathbf{A}_{D}
+\frac{1}{p}\mathcal{E}_{t}\times_{1}\widetilde{\mathbf{A}}_1^{\top}\times_{2} \cdots \times_{D} \widetilde{\mathbf{A}}_{D}^{\top}\\
&=&\frac{1}{p}\mathcal{F}_{t}\times_{1}\widetilde{\mathbf{A}}_1^{\top}
(\mathbf{A}_{1}-\widetilde{\mathbf{A}}_{1}\widetilde{\mathbf{H}}_{1}^{-1}+\widetilde{\mathbf{A}}_{1}\widetilde{\mathbf{H}}_{1}^{-1})
\times_{2} \cdots \times_{D}
\widetilde{\mathbf{A}}_{D}^{\top}(\mathbf{A}_{D}-\widetilde{\mathbf{A}}_{D}\widetilde{\mathbf{H}}_{D}^{-1}+\widetilde{\mathbf{A}}_{D}\widetilde{\mathbf{H}}_{D}^{-1})\\
&&+\frac{1}{p}\mathcal{E}_{t}\times_{1}(\widetilde{\mathbf{A}}_{1}-\mathbf{A}_{1}\widetilde{\mathbf{H}}_{1}+\mathbf{A}_{1}\widetilde{\mathbf{H}}_{1})^{\top}
\times_{2} \cdots \times_{D}(\widetilde{\mathbf{A}}_{D}-\mathbf{A}_{D}\widetilde{\mathbf{H}}_{D}+\mathbf{A}_{D}\widetilde{\mathbf{H}}_{D})^{\top}.
\eeqrs
Some algebra shows that
\beqrs
&&\widetilde{\mathcal{F}}_{t}
-\mathcal{F}_{t}\times_{1}\widetilde{\mathbf{H}}_{1}^{-1}\times_{2} \cdots \times_{D}\widetilde{\mathbf{H}}_{D}^{-1}\\
&=&\frac{1}{p}\mathcal{E}_{t}\times_{1}\widetilde{\mathbf{H}}_{1}^{\top}\mathbf{A}_1^{\top}\times_{2} \cdots \times_{D}\widetilde{\mathbf{H}}_{D}^{\top}\mathbf{A}_D^{\top}\\
&&+\frac{1}{p}\sum\limits_{d=1}^D\mathcal{E}_{t}\times_{1}\widetilde{\mathbf{H}}_{1}^{\top}\mathbf{A}_1^{\top}\times_{2} \cdots \times_{d}(\widetilde{\mathbf{A}}_d^{\top}-\mathbf{A}_d\widetilde{\mathbf{H}}_d)\times_{d+1}\cdots\times_{D}\widetilde{\mathbf{H}}_{D}^{\top}\mathbf{A}_D^{\top}
+\cdots\\
&&+\sum\limits_{d=1}^D\frac{1}{p_d}\mathcal{F}_{t}\times_{1}\widetilde{\mathbf{H}}_{1}^{-1}\times_{2} \cdots \times_{d}\widetilde{\mathbf{A}}_d^{\top}(\mathbf{A}_d-\widetilde{\mathbf{A}}_d\widetilde{\mathbf{H}}_d^{-1})\times_{d+1}\cdots\times_{D}\widetilde{\mathbf{H}}_{D}^{-1}
+\cdots,
\eeqrs
where the omitted terms are small order of the presented terms.

For the first term,
\beqrs
&&\frac{1}{p}\left\|\mathcal{E}_{t}\times_{1}\widetilde{\mathbf{H}}_{1}^{\top}\mathbf{A}_1^{\top}\times_{2} \cdots \times_{D}\widetilde{\mathbf{H}}_{D}^{\top}\mathbf{A}_D^{\top} \right\|_F
=\frac{1}{p} \left\|(\mathbf{A}_D\widetilde{\mathbf{H}}_{D}\otimes\cdots\otimes \mathbf{A}_1\widetilde{\mathbf{H}}_1)^{\top}\vec(\mathcal{E}_t)\right\|_F\\
&&~~~\leq\frac{1}{p}\left\| {\mathbf{A}}_{[D]/\{1\}}^{\top}\vec(\mathcal{E}_t) \right\|_F
=\frac{1}{p}\sqrt{\sum\limits_{i_1=1}^{p_1}\cdots\sum\limits_{i_D=1}^{p_D}(a_{1,i_1}\cdots a_{D,i_D}e_{t,i_1\cdots i_D})^2}
=O_p(\frac{1}{\sqrt{p}}).
\eeqrs

For the $d$th part in the second term,
\beqrs
&&\frac{1}{p}\left\|\mathcal{E}_{t}\times_{1}\widetilde{\mathbf{H}}_{1}^{\top}\mathbf{A}_1^{\top}\times_{2} \cdots \times_{d}(\widetilde{\mathbf{A}}_d-\mathbf{A}_d\widetilde{\mathbf{H}}_d)^{\top}\times_{d+1}\cdots\times_{D}\widetilde{\mathbf{H}}_{D}^{\top}\mathbf{A}_D^{\top}\right\|_F\\
&\leq&\frac{1}{p}\left\|\left\{\mathbf{A}_D\otimes\cdots\otimes(\widetilde{\mathbf{A}}_d-\mathbf{A}_{d}\widetilde{\mathbf{H}}_d)\otimes\cdots\otimes \mathbf{A}_1\right\}^{\top}\vec(\mathcal{E}_t)\right\|_F\\
&\leq& \frac{1}{p}\|\widetilde{\mathbf{A}}_d-\mathbf{A}_{d}\widetilde{\mathbf{H}}_d\|_F
\left\| \mathcal{E}_t^{(d)}{\mathbf{A}}_{[D]/\{1\}}^{\top} \right\|_F\\
&\leq&\frac{1}{p}\sqrt{\sum_{i_1,i_1^{\prime}=1}^{p_1}\cdots\sum_{i_d=1}^{p_d}\cdots\sum_{i_D,i_D^{\prime}=1}^{p_D}
e_{t,i_1\cdots i_d\cdots i_D}e_{t,i_1^{\prime}\cdots i_d\cdots i_D^{\prime}}}\|\widetilde{\mathbf{A}}_d-\mathbf{A}_{d}\widetilde{\mathbf{H}}_d\|_F\\
&=&O_p\left\{\frac{1}{\sqrt{T}p_{-d}}+\frac{1}{p\sqrt{p_{-d}}}
+\sum\limits_{d^{\prime}\neq d}\left(\frac{1}{\sqrt{Tp_{-d}}p_{d^{\prime}}}
+\frac{p_{d^{\prime}}}{Tp\sqrt{p_{-d}}}+\frac{1}{\sqrt{pp_{d}}p_{d^{\prime}}^2}\right)\right\}.
\eeqrs

For the $d$th part in the third term, by Lemma 15, we have
{\small\beqrs
&&\left\| \frac{1}{p_d}\mathcal{F}_{t}\times_{1}\widetilde{\mathbf{H}}_{1}^{-1}\times_{2} \cdots \times_{d}\widetilde{\mathbf{A}}_d^{\top}(\mathbf{A}_d-\widetilde{\mathbf{A}}_d\widetilde{\mathbf{H}}_d^{-1})\times_{d+1}\cdots\times_{D}\widetilde{\mathbf{H}}_{D}^{-1} \right\|_F\\
&=&O_p\left\{\frac{1}{\sqrt{Tp}}+\frac{1}{p}+\sum\limits_{d_1=1}^D\frac{p_{d_1}}{Tp}+
\sum\limits_{d^{\prime}\neq d}\left(\frac{1}{\sqrt{Tp_d}p_{d^{\prime}}}
+\frac{1}{T\sqrt{p_{-d}}p_{d^{\prime}}}+\frac{1}{p_dp_{d^{\prime}}^2}
\right)
+\sum\limits_{d_1\neq d}\sum\limits_{d_2\neq d}\frac{1}{Tp_{d_1}p_{d_2}}\right\}.
\eeqrs}
As a result, 
\beqrs
&&\left\|\widetilde{\mathcal{F}}_{t}
-\mathcal{F}_{t}\times_{1}\widetilde{\mathbf{H}}_{1}^{-1}\times_{2} \cdots \times_{D}\widetilde{\mathbf{H}}_{D}^{-1}\right\|_F\\
&=&O_p\left\{\frac{1}{\sqrt{p}}
+\sum\limits_{d=1}^D\frac{1}{\sqrt{T}p_{-d}}
+\sum\limits_{d_1=1}^D\sum\limits_{d_2\neq d_1}^D
\left(\frac{1}{\sqrt{Tp_{d_1}}p_{d_2}}
+\frac{1}{\sqrt{Tp_{-d_1}}p_{d_2}}
+\frac{1}{p_{d_1}p_{d_2}^2}\right)\right\}.
\eeqrs

(b) Denote the $(i_1,\cdots,i_D)$ elements of the true signal part as $s_{t,i_1\cdots i_D}$ and
the estimated signal part as $\widehat{s}_{t,i_1\cdots i_D}$, then
{\small\beqrs
&&\widetilde{s}_{t,i_1\cdots i_D}-s_{t,i_1\cdots i_D}\\
&=&\widetilde{\mathcal{F}}_t\times_1 \widetilde{\mathbf{A}}_{1,i_1\cdot}^{\top}\times_2\cdots\times_D \widetilde{\mathbf{A}}_{D,i_D\cdot}^{\top}
-\mathcal{F}_t\times_1 \mathbf{A}_{1,i_1\cdot}^{\top}\times_2\cdots\times_D \mathbf{A}_{D,i_D\cdot}^{\top}\\
&=&\widetilde{\mathcal{F}}_t\times_1 (\widetilde{\mathbf{A}}_{1,i_1\cdot}-\widetilde{\mathbf{H}}_1^{\top}\mathbf{A}_{1,i_1\cdot}+\widetilde{\mathbf{H}}_1^{\top}\mathbf{A}_{1,i_1\cdot})^{\top}\times_2\cdots\times_D (\widetilde{\mathbf{A}}_{D,i_D\cdot}-\widetilde{\mathbf{H}}_D^{\top}\mathbf{A}_{D,i_D\cdot}+\widetilde{\mathbf{H}}_D^{\top}\mathbf{A}_{D,i_D\cdot})^{\top}\\
&&-\mathcal{F}_t\times_1 \mathbf{A}_{1,i_1\cdot}^{\top}\times_2\cdots\times_D \mathbf{A}_{D,i_D\cdot}^{\top}\\
&=&(\widetilde{\mathcal{F}}_t\times_1\widetilde{\mathbf{H}}_1\times_2\cdots\times_D\widetilde{\mathbf{H}}_D-\mathcal{F}_t)\times_1 \mathbf{A}_{1,i_1\cdot}^{\top}\times_2\cdots\times_D \mathbf{A}_{D,i_D\cdot}^{\top}\\
&&+\sum\limits_{d=1}^D\widetilde{\mathcal{F}}_t\times_1 \mathbf{A}_{1,i_1\cdot}^{\top}\widetilde{\mathbf{H}}_1
\times_2\cdots\times_d(\widetilde{\mathbf{A}}_{d,i_d\cdot}-\widetilde{\mathbf{H}}_d^{\top}\mathbf{A}_{d,i_d\cdot})^{\top}\times_{d+1}\cdots
\times_D\widetilde{\mathbf{A}}_{D,i_D\cdot}^{\top}\widetilde{\mathbf{H}}_D+\cdots,
\eeqrs}
where the omitted terms are small order of the existing term.
According to Theorem 4.2 and Theorem 4.3 (a), we have
$$\left|\widetilde{s}_{t,i_1\cdots i_D}-s_{t,i_1\cdots i_D}\right|
=O_p\left\{\frac{1}{\sqrt{p}}
+\sum\limits_{d=1}^D\left(\frac{1}{\sqrt{T}p_d}+\frac{1}{\sqrt{Tp_{-d}}}\right)
+\sum\limits_{d_1=1}^D\sum\limits_{d_2\neq d_1}^D\frac{1}{p_{d_1}p_{d_2}^2}\right\}.~\square$$

\section{Technique Lemmas}

\bel
Suppose that $T$ and $\{p_d\}_{d=1}^D$ tend to infinity, and $\{k_d\}_{d=1}^D$ are fixed. 
If Assumptions 1-5 hold, then for $d\in[D]$, we have
$$\left\|\sum\limits_{t=1}^T\mathcal{F}_{t}^{(d)}{\mathbf{A}}_{[D]/\{d\}}^{\top}\mathcal{E}_{t}^{(d)\top}\right\|_F^2=O_p(Tp).$$
\eel

\textit{Proof}: For simplicity, we only consider the proof when $d=1$ and fix $k_1=\cdots=k_D=1$.
\begin{align*}
\mE \left\| \sum\limits_{t=1}^T\mathcal{F}_{t}^{(1)}{\mathbf{A}}_{[D]/\{1\}}^{\top}\mathcal{E}_{t}^{(1)^{\top}} \right\|_F^2
&=\sum\limits_{i_1=1}^{p_1}\mE\left(\sum\limits_{t=1}^T\sum\limits_{i_2=1}^{p_2}\cdots\sum\limits_{i_D=1}^{p_D}
F_t e_{t,i_1\cdots i_D}a_{2,i_2}\cdots a_{D,i_D}\right)^2\\
&=T\sum\limits_{i_1=1}^{p_1}\mE(\sum\limits_{i_2=1}^{p_2}\cdots\sum\limits_{i_D=1}^{p_D}
\boldsymbol\xi_{i_1\cdots i_D}a_{2,i_2}\cdots a_{D,i_D})^2\\
&=T\sum\limits_{i_1=1}^{p_1}\sum\limits_{i_2,i_2^{\prime}=1}^{p_2}
\cdots\sum\limits_{i_D,i_D^{\prime}=1}^{p_D}\mE(\boldsymbol\xi_{i_1\cdots i_D}\boldsymbol\xi_{i_1^{\prime}\cdots i_D^{\prime}})
a_{2,i_2}a_{2,i_2^{\prime}}\cdots a_{D,i_D}a_{D,i_D^{\prime}}\\
&=O(Tp),
\end{align*}
where $a_{d,i_d}$ represents the $i_d$th element of vector $\mathbf{A}_d$, and the last equation is derived according to Assumption 5.2.

\bel
Suppose that $T$ and $\{p_d\}_{d=1}^D$ tend to infinity, and $\{k_d\}_{d=1}^D$ are fixed. If Assumptions 1-5 hold, then for $d\in[D]$, we have
$$\left\|\sum\limits_{t=1}^T\mathcal{E}_{t}^{(d)}\mathcal{E}_{t}^{(d)\top}\right\|_F^2=O_p(Tpp_d+\frac{T^2p^2}{p_d}).$$
\eel

\textit{Proof}: For simplicity, we only consider the proof when $d=1$ and fix $k_1=\cdots=k_D=1$.
\beqrs
\mE \left\| \sum\limits_{t=1}^T\mathcal{E}_{t}^{(1)}\mathcal{E}_{t}^{(1)^{\top}} \right\|_F^2
&=&\mE\left\| \sum\limits_{t=1}^T\mathcal{E}_{t}^{(1)}\mathcal{E}_{t}^{(1)^{\top}}
-\sum\limits_{t=1}^T\mE\left(\mathcal{E}_{t}^{(1)}\mathcal{E}_{t}^{(1)^{\top}}\right)
+\sum\limits_{t=1}^T\mE\left(\mathcal{E}_{t}^{(1)}\mathcal{E}_{t}^{(1)^{\top}}\right)\right\|_F^2\\
&\leq& \mE\left\| \sum\limits_{t=1}^T\mathcal{E}_{t}^{(1)}\mathcal{E}_{t}^{(1)^{\top}}
-\sum\limits_{t=1}^T\mE\left(\mathcal{E}_{t}^{(1)}\mathcal{E}_{t}^{(1)^{\top}}\right)\right\|_F^2
+\left\|\sum\limits_{t=1}^T\mE\left(\mathcal{E}_{t}^{(1)}\mathcal{E}_{t}^{(1)^{\top}}\right)\right\|_F^2.
\eeqrs

For the first term,
\beqrs
&&\mE\left\| \sum\limits_{t=1}^T\mathcal{E}_{t}^{(1)}\mathcal{E}_{t}^{(1)^{\top}}
-\sum\limits_{t=1}^T\mE\left(\mathcal{E}_{t}^{(1)}\mathcal{E}_{t}^{(1)^{\top}}\right)\right\|_F^2\\
&=&\sum\limits_{i_1,i_1^{\prime}=1}^{p_1}
\mE\left[\sum\limits_{t=1}^T\sum\limits_{i_2=1}^{p_2}\cdots\sum\limits_{i_D=1}^{p_D}
\left\{e_{t,i_1i_2\cdots i_D}e_{t,i_1^{\prime}i_2\cdots i_D}-
\mE\left(e_{t,i_1i_2\cdots i_D}e_{t,i_1^{\prime}i_2\cdots i_D}\right)\right\}\right]^2\\
&=&\sum\limits_{s,t=1}^T\sum\limits_{i_1,i_1^{\prime}=1}^{p_1}\sum\limits_{i_2,i_2^{\prime}=1}^{p_2}
\cdots\sum\limits_{i_D,i_D^{\prime}=1}^{p_D}
\cov\left(e_{t,i_1i_2\cdots i_D}e_{t,i_1^{\prime}i_2\cdots i_D},
e_{s,i_1i_2^{\prime}\cdots i_D^{\prime}}e_{s,i_1^{\prime}i_2^{\prime}\cdots i_D^{\prime}}\right)\\
&=&O(T p p_1),
\eeqrs
where the last equation is derived according to Assumption 4.3.

For the second term,
\begin{align*}
\left\|\sum\limits_{t=1}^T\mE\left(\mathcal{E}_{t}^{(1)}\mathcal{E}_{t}^{(1)^{\top}}\right)\right\|_F^2
&=\sum\limits_{i_1,i_1^{\prime}}^{p_1}
\left\{\sum\limits_{t=1}^T\sum\limits_{i_2=1}^{p_2}\cdots\sum\limits_{i_D=1}^{p_D}
\mE(e_{t,i_1i_2\cdots i_D}e_{t,i_1^{\prime}i_2\cdots i_D})\right\}^2\\
&=\sum\limits_{s,t=1}^T\sum\limits_{i_1,i_1^{\prime}=1}^{p_1}\sum\limits_{i_2,i_2^{\prime}=1}^{p_2}
\cdots\sum\limits_{i_D,i_D^{\prime}=1}^{p_D}
\mE(e_{t,i_1i_2\cdots i_D}e_{t,i_1^{\prime}i_2\cdots i_D})
\mE(e_{s,i_1i_2^{\prime}\cdots i_D^{\prime}}e_{s,i_1^{\prime}i_2^{\prime}\cdots i_D^{\prime}})\\
&=O(\frac{T^2p^2}{p_1}),
\end{align*}
where the last equation is derived according to Assumption 4.2.

\bel
Suppose that $T$ and $\{p_d\}_{d=1}^D$ tend to infinity, and $\{k_d\}_{d=1}^D$ are fixed. If Assumptions 1-5 hold, then for $d\in[D]$, we have
$$\left\|\sum\limits_{t=1}^T\mathcal{F}_{t}^{(d)}{\mathbf{A}}_{[D]/\{d\}}^{\top}\mathcal{E}_{t}^{(d)\top}\mathbf{A}_D\right\|_F^2=O_p(Tp).$$
\eel

\textit{Proof}: For simplicity, we only consider the proof when $d=1$ and fix $k_1=\cdots=k_D=1$.
\beqrs
&&\mE \left\| \sum\limits_{t=1}^T\mathcal{F}_{t}^{(1)}
{\mathbf{A}}_{[D]/\{1\}}^{\top}\mathcal{E}_{t}^{(1)^{\top}}\mathbf{A}_D \right\|_F^2\\
&=&\mE\left(\sum\limits_{t=1}^T\sum\limits_{i_1=1}^{p_1}\sum\limits_{i_2=1}^{p_2}\cdots\sum\limits_{i_D=1}^{p_D}
F_t e_{t,i_1\cdots i_D}a_{1,i_1}a_{2,i_2}\cdots a_{D,i_D}\right)^2\\
&=&T\mE(\sum\limits_{i_1=1}^{p_1}\sum\limits_{i_2=1}^{p_2}\cdots\sum\limits_{i_D=1}^{p_D}
\boldsymbol\xi_{i_1\cdots i_D}a_{1,i_1}a_{2,i_2}\cdots a_{D,i_D})^2\\
&=&T\sum\limits_{i_1,i_1^{\prime}=1}^{p_1}\sum\limits_{i_2,i_2^{\prime}=1}^{p_2}
\cdots\sum\limits_{i_D,i_D^{\prime}=1}^{p_D}
\mE(\boldsymbol\xi_{i_1\cdots i_D}\boldsymbol\xi_{i_1^{\prime}\cdots i_D^{\prime}})a_{1,i_1}a_{1,i_1^{\prime}}
a_{2,i_2}a_{2,i_2^{\prime}}\cdots a_{D,i_D}a_{D,i_D^{\prime}}\\
&=&O(Tp),
\eeqrs
where the last equation is derived according to Assumption 5.2.

\bel
Suppose that $T$ and $\{p_d\}_{d=1}^D$ tend to infinity, and $\{k_d\}_{d=1}^D$ are fixed. If Assumptions 1-5 hold, then for $d\in[D]$, we have
$$\left\|\sum\limits_{t=1}^T\mathcal{E}_{t}^{(d)}\mathcal{E}_{t}^{(d)\top}\mathbf{A}_d\right\|_F^2=O_p(Tpp_d+\frac{T^2p^2}{p_d}).$$
\eel

\textit{Proof}: For simplicity, we only consider the proof when $d=1$ and fix $k_1=\cdots=k_D=1$.
\beqrs
&&\mE \left\| \sum\limits_{t=1}^T\mathcal{E}_{t}^{(1)}\mathcal{E}_{t}^{(1)^{\top}}\mathbf{A}_1 \right\|_F^2\\
&=&\mE\left\| \sum\limits_{t=1}^T\mathcal{E}_{t}^{(1)}\mathcal{E}_{t}^{(1)^{\top}}\mathbf{A}_1
-\sum\limits_{t=1}^T\mE\left(\mathcal{E}_{t}^{(1)}\mathcal{E}_{t}^{(1)^{\top}}\mathbf{A}_1\right)
+\sum\limits_{t=1}^T\mE\left(\mathcal{E}_{t}^{(1)}\mathcal{E}_{t}^{(1)^{\top}}\mathbf{A}_1\right)\right\|_F^2\\
&\leq& \mE\left\| \sum\limits_{t=1}^T\mathcal{E}_{t}^{(1)}\mathcal{E}_{t}^{(1)^{\top}}\mathbf{A}_1
-\sum\limits_{t=1}^T\mE\left(\mathcal{E}_{t}^{(1)}\mathcal{E}_{t}^{(1)^{\top}}\mathbf{A}_1\right)\right\|_F^2
+\left\|\sum\limits_{t=1}^T\mE\left(\mathcal{E}_{t}^{(1)}\mathcal{E}_{t}^{(1)^{\top}}\mathbf{A}_1\right)\right\|_F^2.
\eeqrs

For the first term,
\beqrs
&&\mE\left\| \sum\limits_{t=1}^T\mathcal{E}_{t}^{(1)}\mathcal{E}_{t}^{(1)^{\top}}\mathbf{A}_1
-\sum\limits_{t=1}^T\mE\left(\mathcal{E}_{t}^{(1)}\mathcal{E}_{t}^{(1)^{\top}}\mathbf{A}_1\right)\right\|_F^2\\
&=&\sum\limits_{i=1}^{p_1}
\mE\left[\sum\limits_{t=1}^T\sum\limits_{i_1=1}^{p_1}\sum\limits_{i_2=1}^{p_2}\cdots\sum\limits_{i_D=1}^{p_D}
a_{1,i_1}\left\{e_{t,i_1i_2\cdots i_D}e_{t,ii_2\cdots i_D}-
\mE\left(e_{t,i_1i_2\cdots i_D}e_{t,ii_2\cdots i_D}\right)\right\}\right]^2\\
&=&\sum\limits_{s,t=1}^T\sum\limits_{i,i_1,i_1^{\prime}=1}^{p_1}\sum\limits_{i_2,i_2^{\prime}=1}^{p_2}
\cdots\sum\limits_{i_D,i_D^{\prime}=1}^{p_D}a_{1,i_1}a_{1,i_1^{\prime}}
\cov\left(e_{t,i_1i_2\cdots i_D}e_{t,ii_2\cdots i_D},
e_{s,i_1^{\prime}i_2^{\prime}\cdots i_D^{\prime}}e_{s,ii_2^{\prime}\cdots i_D^{\prime}}\right)\\
&=&O(T p p_1),
\eeqrs
where the last equation is derived according to Assumption 4.3.

For the second term,
\begin{align*}
\left\|\sum\limits_{t=1}^T\mE\left(\mathcal{E}_{t}^{(1)}\mathcal{E}_{t}^{(1)^{\top}}\mathbf{A}_1\right)\right\|_F^2
&=\sum\limits_{i=1}^{p_1}
\left\{\sum\limits_{t=1}^T\sum\limits_{i_1=1}^{p_1}\sum\limits_{i_2=1}^{p_2}\cdots\sum\limits_{i_D=1}^{p_D}
a_{1,i_1}\mE(e_{t,i_1i_2\cdots i_D}e_{t,ii_2\cdots i_D})\right\}^2\\
&=\sum\limits_{s,t=1}^T\sum\limits_{i,i_1,i_1^{\prime}=1}^{p_1}\sum\limits_{i_2,i_2^{\prime}=1}^{p_2}
\cdots\sum\limits_{i_D,i_D^{\prime}=1}^{p_D}
\mE(e_{t,i_1i_2\cdots i_D}e_{t,ii_2\cdots i_D})
\mE(e_{s,i_1^{\prime}i_2^{\prime}\cdots i_D^{\prime}}e_{s,ii_2^{\prime}\cdots i_D^{\prime}})\\
&=O(\frac{T^2p^2}{p_1}),
\end{align*}
where the last equation is derived according to Assumption 4.2.

\bel
Suppose that $T$ and $\{p_d\}_{d=1}^D$ tend to infinity, and $\{k_d\}_{d=1}^D$ are fixed. If Assumptions 1-5 hold, then for $d\in[D]$, $i\in[p_d]$, we have
$$\left\|\sum\limits_{t=1}^T\mathcal{F}_{t}^{(d)}{\mathbf{A}}_{[D]/\{d\}}^{\top}\mathcal{E}_{t,i\cdot}^{(d)}\right\|_F^2=O_p(\frac{Tp}{p_d}).$$
\eel

\textit{Proof}: For simplicity, we only consider the proof when $d=1$ and fix $k_1=\cdots=k_D=1$.
\begin{align*}
\mE \left\| \sum\limits_{t=1}^T\mathcal{F}_{t}^{(1)}
{\mathbf{A}}_{[D]/\{1\}}^{\top}\mathcal{E}_{t,i\cdot}^{(1)} \right\|_F^2
&=\mE\left(\sum\limits_{t=1}^T\sum\limits_{i_2=1}^{p_2}\cdots\sum\limits_{i_D=1}^{p_D}
F_t e_{t,ii_2\cdots i_D}a_{2,i_2}\cdots a_{D,i_D}\right)^2\\
&=T\mE(\sum\limits_{i_2=1}^{p_2}\cdots\sum\limits_{i_D=1}^{p_D}
\boldsymbol\xi_{ii_2\cdots i_D}a_{2,i_2}\cdots a_{D,i_D})^2\\
&=T\sum\limits_{i_2,i_2^{\prime}=1}^{p_2}\cdots\sum\limits_{i_D,i_D^{\prime}=1}^{p_D}
\mE(\boldsymbol\xi_{ii_2\cdots i_D}\boldsymbol\xi_{ii_2^{\prime}\cdots i_D^{\prime}})
a_{2,i_2}a_{2,i_2^{\prime}}\cdots a_{D,i_D}a_{D,i_D^{\prime}}\\
&=O(\frac{Tp}{p_1}),
\end{align*}
where the last equation is derived according to Assumption 5.2.

\bel
Suppose that $T$ and $\{p_d\}_{d=1}^D$ tend to infinity, and $\{k_d\}_{d=1}^D$ are fixed. If Assumptions 1-5 hold, then for $d\in[D]$, $i\in[p_d]$, we have
$$\left\|\sum\limits_{t=1}^T\mathcal{E}_{t}^{(d)}\mathcal{E}_{t,i\cdot}^{(d)}\right\|_F^2=O_p(Tp+\frac{T^2p^2}{p_d^2}).$$
\eel

\textit{Proof}: For simplicity, we only consider the proof when $d=1$ and fix $k_1=\cdots=k_D=1$.
\beqrs
\mE \left\| \sum\limits_{t=1}^T\mathcal{E}_{t}^{(1)}\mathcal{E}_{t,i\cdot}^{(1)} \right\|_F^2
&=&\mE\left\| \sum\limits_{t=1}^T\mathcal{E}_{t}^{(1)}\mathcal{E}_{t,i\cdot}^{(1)}
-\sum\limits_{t=1}^T\mE\left(\mathcal{E}_{t}^{(1)}\mathcal{E}_{t,i\cdot}^{(1)}\right)
+\sum\limits_{t=1}^T\mE\left(\mathcal{E}_{t}^{(1)}\mathcal{E}_{t,i\cdot}^{(1)}\right)\right\|_F^2\\
&\leq& \mE\left\| \sum\limits_{t=1}^T\mathcal{E}_{t}^{(1)}\mathcal{E}_{t,i\cdot}^{(1)}
-\sum\limits_{t=1}^T\mE\left(\mathcal{E}_{t}^{(1)}\mathcal{E}_{t,i\cdot}^{(1)}\right)\right\|_F^2
+\left\|\sum\limits_{t=1}^T\mE\left(\mathcal{E}_{t}^{(1)}\mathcal{E}_{t,i\cdot}^{(1)}\right)\right\|_F^2.
\eeqrs

For the first term,
\beqrs
&&\mE\left\| \sum\limits_{t=1}^T\mathcal{E}_{t}^{(1)}\mathcal{E}_{t,i\cdot}^{(1)}
-\sum\limits_{t=1}^T\mE\left(\mathcal{E}_{t}^{(1)}\mathcal{E}_{t,i\cdot}^{(1)}\right)\right\|_F^2\\
&=&\sum\limits_{i_1=1}^{p_1}\mE\left[\sum\limits_{t=1}^T\sum\limits_{i_2=1}^{p_2}\cdots\sum\limits_{i_D=1}^{p_D}
\left\{e_{t,i_1i_2\cdots i_D}e_{t,ii_2\cdots i_D}-
\mE\left(e_{t,i_1i_2\cdots i_D}e_{t,ii_2\cdots i_D}\right)\right\}\right]^2\\
&=&\sum\limits_{s,t=1}^T\sum\limits_{i_1=1}^{p_1}\sum\limits_{i_2,i_2^{\prime}=1}^{p_2}
\cdots\sum\limits_{i_D,i_D^{\prime}=1}^{p_D}
\cov\left(e_{t,i_1i_2\cdots i_D}e_{t,ii_2\cdots i_D},
e_{s,i_1i_2^{\prime}\cdots i_D^{\prime}}e_{s,ii_2^{\prime}\cdots i_D^{\prime}}\right)\\
&=&O(T p),
\eeqrs
where the last equation is derived according to Assumption 4.3.

For the second term,
\begin{align*}
\left\|\sum\limits_{t=1}^T\mE\left(\mathcal{E}_{t}^{(1)}\mathcal{E}_{t,i\cdot}^{(1)}\right)\right\|_F^2
&=\sum\limits_{i_1=1}^{p_1}
\left\{\sum\limits_{t=1}^T\sum\limits_{i_2=1}^{p_2}\cdots\sum\limits_{i_D=1}^{p_D}
\mE(e_{t,i_1i_2\cdots i_D}e_{t,ii_2\cdots i_D})\right\}^2\\
&=\sum\limits_{s,t=1}^T\sum\limits_{i_1=1}^{p_1}\sum\limits_{i_2,i_2^{\prime}=1}^{p_2}
\cdots\sum\limits_{i_D,i_D^{\prime}=1}^{p_D}
\mE(e_{t,i_1i_2\cdots i_D}e_{t,ii_2\cdots i_D})
\mE(e_{s,i_1i_2^{\prime}\cdots i_D^{\prime}}e_{s,ii_2^{\prime}\cdots i_D^{\prime}})\\
&=O(\frac{T^2p^2}{p_1^2}),
\end{align*}
where the last equation is derived according to Assumption 4.2.

\bel
Suppose that $T$ and $\{p_d\}_{d=1}^D$ tend to infinity, and $\{k_d\}_{d=1}^D$ are fixed. If Assumptions 1-5 hold, then for $d\in[D]$, $i\in[p_d]$, we have
$$\left\|\sum\limits_{t=1}^T\mathcal{E}_{t,i\cdot}^{(d)\top}\mathcal{E}_{t}^{(d)\top}\mathbf{A}_d\right\|_F^2
=O_p(Tp+\frac{T^2p^2}{p_d^2}).$$
\eel

\textit{Proof}: For simplicity, we only consider the proof when $d=1$ and fix $k_1=\cdots=k_D=1$.
\beqrs
&&\mE \left\| \sum\limits_{t=1}^T\mathcal{E}_{t,i\cdot}^{(1)^{\top}}\mathcal{E}_{t}^{(1)^{\top}}\mathbf{A}_1 \right\|_F^2\\
&=&\mE\left\| \sum\limits_{t=1}^T\mathcal{E}_{t,i\cdot}^{(1)^{\top}}\mathcal{E}_{t}^{(1)^{\top}}\mathbf{A}_1
-\sum\limits_{t=1}^T\mE\left(\mathcal{E}_{t,i\cdot}^{(1)^{\top}}\mathcal{E}_{t}^{(1)^{\top}}\mathbf{A}_1\right)
+\sum\limits_{t=1}^T\mE\left(\mathcal{E}_{t,i\cdot}^{(1)^{\top}}\mathcal{E}_{t}^{(1)^{\top}}\mathbf{A}_1\right)\right\|_F^2\\
&\leq& \mE\left\| \sum\limits_{t=1}^T\mathcal{E}_{t,i\cdot}^{(1)^{\top}}\mathcal{E}_{t}^{(1)^{\top}}\mathbf{A}_1
-\sum\limits_{t=1}^T\mE\left(\mathcal{E}_{t,i\cdot}^{(1)^{\top}}\mathcal{E}_{t}^{(1)^{\top}}\mathbf{A}_1\right)\right\|_F^2
+\left\|\sum\limits_{t=1}^T\mE\left(\mathcal{E}_{t,i\cdot}^{(1)^{\top}}\mathcal{E}_{t}^{(1)^{\top}}\mathbf{A}_1\right)\right\|_F^2.
\eeqrs

For the first term,
\beqrs
&&\mE\left\| \sum\limits_{t=1}^T\mathcal{E}_{t,i\cdot}^{(1)^{\top}}\mathcal{E}_{t}^{(1)^{\top}}\mathbf{A}_1
-\sum\limits_{t=1}^T\mE\left(\mathcal{E}_{t,i\cdot}^{(1)^{\top}}\mathcal{E}_{t}^{(1)^{\top}}\mathbf{A}_1\right)\right\|_F^2\\
&=&\mE\left[\sum\limits_{t=1}^T\sum\limits_{i_1=1}^{p_1}\sum\limits_{i_2=1}^{p_2}\cdots\sum\limits_{i_D=1}^{p_D}
a_{1,i_1}\left\{e_{t,i_1i_2\cdots i_D}e_{t,ii_2\cdots i_D}-
\mE\left(e_{t,i_1i_2\cdots i_D}e_{t,ii_2\cdots i_D}\right)\right\}\right]^2\\
&=&\sum\limits_{s,t=1}^T\sum\limits_{i_1,i_1^{\prime}=1}^{p_1}\sum\limits_{i_2,i_2^{\prime}=1}^{p_2}
\cdots\sum\limits_{i_D,i_D^{\prime}=1}^{p_D}a_{1,i_1}a_{1,i_1^{\prime}}
\cov\left(e_{t,i_1i_2\cdots i_D}e_{t,ii_2\cdots i_D},
e_{s,i_1^{\prime}i_2^{\prime}\cdots i_D^{\prime}}e_{s,ii_2^{\prime}\cdots i_D^{\prime}}\right)\\
&=&O(T p),
\eeqrs
where the last equation is derived according to Assumption 4.3.

For the second term,
\beqrs
&&\left\|\sum\limits_{t=1}^T\mE\left(\mathcal{E}_{t,i\cdot}^{(1)^{\top}}\mathcal{E}_{t}^{(1)^{\top}}\mathbf{A}_1\right)\right\|_F^2\\
&=&\left\{\sum\limits_{t=1}^T\sum\limits_{i_1=1}^{p_1}\sum\limits_{i_2=1}^{p_2}\cdots\sum\limits_{i_D=1}^{p_D}
a_{1,i_1}\mE(e_{t,i_1i_2\cdots i_D}e_{t,ii_2\cdots i_D})\right\}^2\\
&=&\sum\limits_{s,t=1}^T\sum\limits_{i_1,i_1^{\prime}=1}^{p_1}\sum\limits_{i_2,i_2^{\prime}=1}^{p_2}
\cdots\sum\limits_{i_D,i_D^{\prime}=1}^{p_D}a_{1,i_1}a_{1,i_1^{\prime}}
\mE(e_{t,i_1i_2\cdots i_D}e_{t,ii_2\cdots i_D})
\mE(e_{s,i_1^{\prime}i_2^{\prime}\cdots i_D^{\prime}}e_{s,ii_2^{\prime}\cdots i_D^{\prime}})\\
&=&O(\frac{T^2p^2}{p_1^2}),
\eeqrs
where the last equation is derived according to Assumption 4.2.

\bel
Suppose that $T$ and $\{p_d\}_{d=1}^D$ tend to infinity, and $\{k_d\}_{d=1}^D$ are fixed. If Assumptions 1-5 hold, then for $d\in[D]$, we have
$$\frac{1}{p_d}\left\| \mathbf{A}_d^{\top}(\widehat{\mathbf{A}}_d-\mathbf{A}_d\widehat{\mathbf{H}}_d)\right\|_F
=O_p(\frac{1}{\sqrt{Tp}}+\frac{1}{Tp_{-d}}+\frac{1}{p_d}).$$
\eel

\textit{Proof}: Note that
\beqrs
\frac{1}{p_d}\mathbf{A}_d^{\top}(\widehat{\mathbf{A}}_d-\mathbf{A}_d\widehat{\mathbf{H}}_d)
&=&\frac{1}{Tp}\sum\limits_{t=1}^T \mathcal{F}_t^{(d)}
{\mathbf{A}}_{[D]/\{d\}}^{\top}
\mathcal{E}_t^{(d)\top}\widehat{\mathbf{A}}_d\widehat{\mathbf{\Lambda}}_d^{-1}\\
&&+\frac{1}{Tpp_d}\sum\limits_{t=1}^T \mathbf{A}_d^{\top}\mathcal{E}_t^{(d)}
{\mathbf{A}}_{[D]/\{d\}}
\mathcal{F}_t^{(d)\top}\mathbf{A}_d^{\top}\widehat{\mathbf{A}}_d\widehat{\mathbf{\Lambda}}_d^{-1}\\
&&+\frac{1}{Tpp_d}\sum\limits_{t=1}^T \mathbf{A}_d^{\top}\mathcal{E}_t^{(d)}\mathcal{E}_t^{(d)\top}\widehat{\mathbf{A}}_d\widehat{\mathbf{\Lambda}}_d^{-1}.
\eeqrs

For the first term,
{\small\beqrs
&&\left\| \frac{1}{Tp}\sum\limits_{t=1}^T \mathcal{F}_t^{(d)}
{\mathbf{A}}_{[D]/\{d\}}^{\top}
\mathcal{E}_t^{(d)\top}\widehat{\mathbf{A}}_d\widehat{\mathbf{\Lambda}}_d^{-1} \right\|_F\\
&\leq&\left\| \frac{1}{Tp}\sum\limits_{t=1}^T \mathcal{F}_t^{(d)}
{\mathbf{A}}_{[D]/\{d\}}^{\top}
\mathcal{E}_t^{(d)\top}\mathbf{A}_d \right\|_F+\left\| \frac{1}{Tp}\sum\limits_{t=1}^T \mathcal{F}_t^{(d)}
{\mathbf{A}}_{[D]/\{d\}}^{\top}
\mathcal{E}_t^{(d)\top}\right\|_F\left\|\widehat{\mathbf{A}}_d-\mathbf{A}_d\widehat{\mathbf{H}}_d\right\|_F\\
&=&O_p(\frac{1}{\sqrt{Tp}}+\frac{1}{Tp_{-d}}),
\eeqrs}
where the last equation is derived according to Lemmas 1, 3, and Proposition 2.

For the second term,
\beqrs
&&\left\| \frac{1}{Tpp_d}\sum\limits_{t=1}^T \mathbf{A}_d^{\top}\mathcal{E}_t^{(d)}
{\mathbf{A}}_{[D]/\{d\}}
\mathcal{F}_t^{(d)\top}\mathbf{A}_d^{\top}\widehat{\mathbf{A}}_d\widehat{\mathbf{\Lambda}}_d^{-1} \right\|_F\\
&\leq&\left\| \frac{1}{Tpp_d}\sum\limits_{t=1}^T \mathbf{A}_d^{\top}\mathcal{E}_t^{(d)}
{\mathbf{A}}_{[D]/\{d\}}
\mathcal{F}_t^{(d)\top}\right\|_F \|\mathbf{A}_d^{\top}\widehat{\mathbf{A}}_d\|_F=O_p(\frac{1}{\sqrt{Tp}}),
\eeqrs
where the last equation is derived according to Lemma 3.

For the third term,
\beqrs
&&\left\| \frac{1}{Tpp_d}\sum\limits_{t=1}^T \mathbf{A}_d^{\top}\mathcal{E}_t^{(d)}\mathcal{E}_t^{(d)\top}\widehat{\mathbf{A}}_d\widehat{\mathbf{\Lambda}}_d^{-1} \right\|_F\\
&\leq& \left\| \frac{1}{Tpp_d}\sum\limits_{t=1}^T \mathbf{A}_d^{\top}\mathcal{E}_t^{(d)}\mathcal{E}_t^{(d)\top}\mathbf{A}_d\right\|_F
+\left\| \frac{1}{Tpp_d}\sum\limits_{t=1}^T \mathbf{A}_d^{\top}\mathcal{E}_t^{(d)}\mathcal{E}_t^{(d)\top}\right\|_F
\left\|\widehat{\mathbf{A}}_d-\mathbf{A}_d\widehat{\mathbf{H}}_d\right\|_F\\
&=&O_p(\frac{1}{p_d}+\frac{\sqrt{p_d}}{Tp}+\frac{1}{\sqrt{Tpp_d}}),
\eeqrs
where the last equation is derived according to Lemma 4, Proposition 2 and
{\small\beqrs
\left\|\sum\limits_{t=1}^T \mathbf{A}_d^{\top}\mathcal{E}_t^{(d)}\mathcal{E}_t^{(d)\top}\mathbf{A}_d\right\|_F^2
=\sum\limits_{t=1}^T\sum\limits_{i_1=1}^{p_1}\cdots\sum\limits_{i_d,i_d^{\prime}=1}^{p_d}\cdots\sum\limits_{i_D=1}^{p_D}
a_{d,i_d}a_{d,i_d^{\prime}}e_{t,i_1\cdots i_d\cdots i_D}e_{t,i_1\cdots i_d^{\prime}\cdots i_D}
=O_p(Tp).
\eeqrs}

\bel
Suppose that $T$ and $\{p_d\}_{d=1}^D$ tend to infinity, and $\{k_d\}_{d=1}^D$ are fixed. If Assumptions 1-5 hold, then for $d\in[D]$, we have
{\beqrs
\sum\limits_{t=1}^T \left\|\mathcal{E}_{t}^{(d)}
\widehat{\mathbf{A}}_{[D]/\{d\}}^{(s+1)}\right\|_F^2
=O_p\left\{T p + \frac{Tp^2}{p_d}
\left(\sum\limits_{d^{\prime}= d+1}^Dw_{d^{\prime}}^{(s)}
+\sum\limits_{d^{\prime}= 1}^{d-1}w_{d^{\prime}}^{(s+1)}\right)\right\},
\eeqrs}
and
{\beqrs
\sum\limits_{t=1}^T\left\|\mathcal{E}_{t}^{(d)}\widehat{\mathbf{A}}_{[D]/\{d\}}\right\|_F^2
=O_p\left(Tp+\sum\limits_{d^{\prime}\neq d}^D \frac{Tp^2}{p_d}w_{d^{\prime}}^{(0)}\right).
\eeqrs}
\eel

\textit{Proof}: Note that

{\scriptsize\beqrs
&&\sum\limits_{t=1}^T\left\|\mathcal{E}_{t}^{(d)}
\widehat{\mathbf{A}}_{[D]/\{d\}}^{(s+1)}\right\|_F^2\\
&\leq& \sum\limits_{t=1}^T\left\|\mathcal{E}_{t}^{(d)}{\mathbf{A}}_{[D]/\{d\}}\right\|_F^2
+\sum_{d^{\prime}=d+1}^D\sum\limits_{t=1}^T\left\|\mathcal{E}_{t}^{(d)}
\left\{\mathbf{A}_D \otimes\cdots\otimes \left(\widehat{\mathbf{A}}_{d^{\prime}}^{(s)}-A_{d^{\prime}}\widehat{\mathbf{H}}_{d^{\prime}}^{(s)}\right)
\otimes\cdots\otimes \mathbf{A}_{d+1} \otimes \mathbf{A}_{d-1}\otimes\cdots\otimes \mathbf{A}_1\right\}\right\|_F^2\\
&&+\sum_{d^{\prime}=1}^{d-1}\sum\limits_{t=1}^T\left\|\mathcal{E}_{t}^{(d)}
\left\{\mathbf{A}_D \otimes\cdots\otimes \mathbf{A}_{d+1} \otimes \mathbf{A}_{d-1}\otimes\cdots\otimes \left(\widehat{\mathbf{A}}_{d^{\prime}}^{(s+1)}-A_{d^{\prime}}\widehat{\mathbf{H}}_{d^{\prime}}^{(s+1)}\right)
\otimes\cdots\otimes \mathbf{A}_1\right\}\right\|_F^2+\cdots,
\eeqrs}

\noindent
where the omitted terms are small order of the existing terms.
Without loss of generality, assume $k_1=\cdots=k_D=1$ in the following proof.

For the first term,
{\scriptsize\beqrs
&&\sum\limits_{t=1}^T\left\|\mathcal{E}_{t}^{(d)}
{\mathbf{A}}_{[D]/\{d\}}\right\|_F^2\\
&=&\sum\limits_{t=1}^T\sum\limits_{i_d=1}^{p_d}\left(\sum\limits_{i_1=1}^{p_1}\cdots
\sum\limits_{i_{d-1}=1}^{p_{d-1}}\sum\limits_{i_{d+1}=1}^{p_{d+1}}\cdots\sum\limits_{i_D=1}^{p_D}
e_{t,i_1\cdots i_{d-1}i_di_{d+1}\cdots i_D}a_{1,i_1}\cdots a_{d-1,i_{d-1}} a_{d+1,i_{d+1}}\cdots a_{D,i_D}\right)^2\\
&\leq& \sum\limits_{t=1}^T\sum\limits_{i_1,i_1^{\prime}=1}^{p_1}
\cdots\sum\limits_{i_{d-1},i_{d-1}^{\prime}=1}^{p_{d-1}}\sum\limits_{i_d=1}^{p_d}
\sum\limits_{i_{d+1},i_{d+1}^{\prime}=1}^{p_{d+1}}\cdots\sum\limits_{i_D,i_D^{\prime}=1}^{p_D}
e_{t,i_1\cdots i_{d-1}i_di_{d+1}\cdots i_D}e_{t,i_1^{\prime}\cdots i_{d-1}^{\prime}i_di_{d+1}^{\prime}\cdots i_D^{\prime}}\\
&=&O_p(Tp).
\eeqrs}
For the $d^{\prime}$th part of the second term,
{\scriptsize\beqrs
&&\sum\limits_{t=1}^T\left\|\mathcal{E}_{t}^{(d)}
\left\{\mathbf{A}_D \otimes\cdots\otimes \left(\widehat{\mathbf{A}}_{d^{\prime}}^{(s)}-A_{d^{\prime}}\widehat{\mathbf{H}}_{d^{\prime}}^{(s)}\right)
\otimes\cdots\otimes \mathbf{A}_{d+1} \otimes \mathbf{A}_{d-1}\otimes\cdots\otimes \mathbf{A}_1\right\}\right\|_F^2\\
&\leq& \sum\limits_{t=1}^T\left\|\mathcal{E}_t^{(d)}\right\|_F^2\left\|\mathbf{A}_D\right\|_F^2
\cdots\left\|A_{d^{\prime}+1}\right\|_F^2\left\|A_{d^{\prime}-1}\right\|_F^2
\cdots\left\|\mathbf{A}_{d+1}\right\|_F^2\left\|\mathbf{A}_{d-1}\right\|_F^2\cdots\left\|\mathbf{A}_1\right\|_F^2
\left\|\widehat{\mathbf{A}}_{d^{\prime}}^{(s)}-A_{d^{\prime}}\widehat{\mathbf{H}}_{d^{\prime}}^{(s)}\right\|_F^2\\
&=&O_p\left(\frac{Tp^2}{p_d}w_{d^{\prime}}^{(s)}\right).
\eeqrs}
Similarly, for the $d^{\prime}$th part of the third term,
{\scriptsize$$\sum\limits_{t=1}^T\left\|\mathcal{E}_{t}^{(d)}
\left\{\mathbf{A}_D \otimes\cdots\otimes \mathbf{A}_{d+1} \otimes \mathbf{A}_{d-1}\otimes\cdots\otimes \left(\widehat{\mathbf{A}}_{d^{\prime}}^{(s+1)}-A_{d^{\prime}}\widehat{\mathbf{H}}_{d^{\prime}}^{(s+1)}\right)
\otimes\cdots\otimes \mathbf{A}_1\right\}\right\|_F^2=O_p\left(\frac{Tp^2}{p_d}w_{d^{\prime}}^{(s+1)}\right).$$}
As a result, 
{\scriptsize\beqrs
\sum\limits_{t=1}^T \left\|\mathcal{E}_{t}^{(d)}
\widehat{\mathbf{A}}_{[D]/\{d\}}^{(s+1)}\right\|_F^2
=O_p\left\{T p + \frac{Tp^2}{p_d}
\left(\sum\limits_{d^{\prime}= d+1}^Dw_{d^{\prime}}^{(s)}
+\sum\limits_{d^{\prime}= 1}^{d-1}w_{d^{\prime}}^{(s+1)}\right)\right\}.
\eeqrs}
A special case is adopting $\widehat{\mathbf{A}}_d^{(0)}$ for $d\in[D]$, then
{\scriptsize\beqrs
\sum\limits_{t=1}^T\left\|\mathcal{E}_{t}^{(d)}\widehat{\mathbf{A}}_{[D]/\{d\}}\right\|_F^2
=O_p\left(Tp+\sum\limits_{d^{\prime}\neq d}^D \frac{Tp^2}{p_d}w_{d^{\prime}}^{(0)}\right).
\eeqrs}

\bel
Suppose that $T$ and $\{p_d\}_{d=1}^D$ tend to infinity, and $\{k_d\}_{d=1}^D$ are fixed. If Assumptions 1-5 hold, then for $d\in[D]$, we have
{\scriptsize\beqrs
\left\|\sum\limits_{t=1}^T\mathcal{E}_{t}^{(d)}
\widehat{\mathbf{A}}_{[D]/\{d\}}^{(s+1)}\widehat{\mathbf{A}}_{[D]/\{d\}}^{(s+1)^{\top}}\mathcal{E}_{t}^{(d)\top}\mathbf{A}_d\right\|_F^2
=O_p\Bigg\{\frac{Tp^3}{p_d}+\frac{T^2p^2}{p_d}
+\left(\frac{Tp^4}{p_d^2}+\frac{T^2p^4}{p_d^3}\right)
\left(\sum\limits_{d^{\prime}= d+1}^Dw_{d^{\prime}}^{(s)^2}
+\sum\limits_{d^{\prime}= 1}^{d-1}w_{d^{\prime}}^{(s+1)^2}\right)\Bigg\},
\eeqrs}
and
{\scriptsize\beqrs
\left\|\sum\limits_{t=1}^T\mathcal{E}_{t}^{(d)}
\widehat{\mathbf{A}}_{[D]/\{d\}}\widehat{\mathbf{A}}_{[D]/\{d\}}^{{\top}}\mathcal{E}_{t}^{(d)\top}\mathbf{A}_d\right\|_F^2
=O_p\left\{\frac{Tp^3}{p_d}+\frac{T^2p^2}{p_d}
+\left(\frac{Tp^4}{p_d^2}+\frac{T^2p^4}{p_d^3}\right)
\left(\sum\limits_{d^{\prime}\neq d}^Dw_{d^{\prime}}^{(0)^2}\right)\right\}.
\eeqrs}
\eel

\textit{Proof}: For simplicity, we fix $k_1=\cdots=k_D=1$.
Note that

{\scriptsize\begin{align*}
&\left\|\sum\limits_{t=1}^T\mathcal{E}_{t}^{(d)}
\widehat{\mathbf{A}}_{[D]/\{d\}}^{(s+1)}\widehat{\mathbf{A}}_{[D]/\{d\}}^{(s+1)^{\top}}\mathcal{E}_{t}^{(d)\top}\mathbf{A}_d\right\|_F^2\leq \left\|\sum\limits_{t=1}^T\mathcal{E}_{t}^{(d)}
{\mathbf{A}}_{[D]/\{d\}}{\mathbf{A}}_{[D]/\{d\}}^{{\top}}\mathcal{E}_{t}^{(d)\top}\mathbf{A}_d\right\|_F^2\\
&~+\sum_{d^{\prime}=d+1}^D\Bigg\|\sum\limits_{t=1}^T\mathcal{E}_{t}^{(d)}
\Big\{\mathbf{A}_D\mathbf{A}_D^{\top}\otimes\cdots\otimes
\left(\widehat{\mathbf{A}}_{d^{\prime}}^{(s)}-A_{d^{\prime}}\widehat{\mathbf{H}}_{d^{\prime}}^{(s)}\right)
\left(\widehat{\mathbf{A}}_{d^{\prime}}^{(s)}-A_{d^{\prime}}\widehat{\mathbf{H}}_{d^{\prime}}^{(s)}\right)^{\top}
\otimes\cdots\\
&~~~~~~~~~~~~~~~~~~~~~~~~\otimes\mathbf{A}_{d+1}\mathbf{A}_{d+1}^{\top}\otimes \mathbf{A}_{d-1}\mathbf{A}_{d-1}^{\top}\otimes\cdots\otimes \mathbf{A}_{1}\mathbf{A}_{1}^{\top}\Big\}\mathcal{E}_{t}^{(d)\top}\mathbf{A}_d\Bigg\|_F^2\\
&~+\sum_{d^{\prime}=1}^{d-1}\Bigg\|\sum\limits_{t=1}^T\mathcal{E}_{t}^{(d)}
\Big\{\mathbf{A}_D\mathbf{A}_D^{\top}\otimes\cdots\otimes \mathbf{A}_{d+1}\mathbf{A}_{d+1}^{\top}\otimes \mathbf{A}_{d-1}\mathbf{A}_{d-1}^{\top}\otimes\cdots\\
&~~~~~~~~~~~~~~~~~~~~~~~~~~\otimes
\left(\widehat{\mathbf{A}}_{d^{\prime}}^{(s+1)}-A_{d^{\prime}}\widehat{\mathbf{H}}_{d^{\prime}}^{(s+1)}\right)
\left(\widehat{\mathbf{A}}_{d^{\prime}}^{(s+1)}-A_{d^{\prime}}\widehat{\mathbf{H}}_{d^{\prime}}^{(s+1)}\right)^{\top}
\otimes\cdots\otimes \mathbf{A}_{1}\mathbf{A}_{1}^{\top}\Big\}\mathcal{E}_{t}^{(d)\top}\mathbf{A}_d\Bigg\|_F^2\\
&~+\cdots,
\end{align*}}

where the omitted terms are small order of the existing terms.
For the first term,
{\scriptsize\beqrs
\left\|\sum\limits_{t=1}^T\mathcal{E}_{t}^{(d)}
{\mathbf{A}}_{[D]/\{d\}}{\mathbf{A}}_{[D]/\{d\}}^{{\top}}\mathcal{E}_{t}^{(d)\top}\mathbf{A}_d\right\|_F^2
&\leq& \left\|\mathbf{A}_D\right\|_F^2\cdots\left\|\mathbf{A}_{d+1}\right\|_F^2\left\|\mathbf{A}_{d-1}\right\|_F^2
\cdots\left\|\mathbf{A}_1\right\|_F^2
\left\|\sum\limits_{t=1}^T\mathcal{E}_{t}^{(d)}
\mathbf{A}_{[D]/\{d\}}^{\top}\mathcal{E}_{t}^{(d)\top}\mathbf{A}_d\right\|_F^2\\
&\leq&p_{-d}\sum\limits_{i_1=1}^{p_1}\cdots\sum\limits_{i_D=1}^{p_D}
\left(\sum\limits_{t=1}^T\sum\limits_{i_1^{\prime}=1}^{p_1}\cdots\sum\limits_{i_D^{\prime}=1}^{p_D}
e_{t,i_1\cdots i_D}e_{t,i_1^{\prime}\cdots i_D^{\prime}}a_{1,i_1^{\prime}}\cdots a_{D,i_D^{\prime}}\right)^2\\
&=&O_p(\frac{Tp^3}{p_d}+\frac{T^2p^2}{p_d}),
\eeqrs}
where
{\scriptsize
\beqrs
&&\mE\left(\sum\limits_{t=1}^T\sum\limits_{i_1^{\prime}=1}^{p_1}\cdots\sum\limits_{i_D^{\prime}=1}^{p_D}
e_{t,i_1\cdots i_D}e_{t,i_1^{\prime}\cdots i_D^{\prime}}a_{1,i_1^{\prime}}\cdots a_{D,i_D^{\prime}}\right)^2\\
&\leq& \sum\limits_{s,t=1}^T\sum\limits_{i_1^{\prime}i_1^{\prime\prime}=1}^{p_1}
\cdots\sum\limits_{i_D^{\prime},i_D^{\prime\prime}=1}^{p_D}
\cov\left(e_{t,i_1\cdots i_D}e_{t,i_1^{\prime}\cdots i_D^{\prime}},
e_{s,i_1\cdots i_D}e_{s,i_1^{\prime\prime}\cdots i_D^{\prime\prime}}\right)
+\left\{\sum\limits_{t=1}^T\sum\limits_{i_1^{\prime}=1}^{p_1}\cdots\sum\limits_{i_D^{\prime}=1}^{p_D}
\mE\left(e_{t,i_1\cdots i_D}e_{t,i_1^{\prime}\cdots i_D^{\prime}}\right)\right\}^2\\
&=&O(Tp+T^2).
\eeqrs}

For the $d^{\prime}$th part in the second term,
{\scriptsize\beqrs
&&\Bigg\|\sum\limits_{t=1}^T\mathcal{E}_{t}^{(d)}
\Big\{\mathbf{A}_D\mathbf{A}_D^{\top}\otimes\cdots\otimes
\left(\widehat{\mathbf{A}}_{d^{\prime}}^{(s)}-A_{d^{\prime}}\widehat{\mathbf{H}}_{d^{\prime}}^{(s)}\right)
\left(\widehat{\mathbf{A}}_{d^{\prime}}^{(s)}-A_{d^{\prime}}\widehat{\mathbf{H}}_{d^{\prime}}^{(s)}\right)^{\top}
\otimes\cdots\\
&&~~~~~~~~~~~~~~~~~~~~~\otimes \mathbf{A}_{d+1}\mathbf{A}_{d+1}^{\top}\otimes \mathbf{A}_{d-1}\mathbf{A}_{d-1}^{\top}\otimes\cdots\otimes \mathbf{A}_{1}\mathbf{A}_{1}^{\top}\Big\}\mathcal{E}_{t}^{(d)\top}\mathbf{A}_d\Bigg\|_F^2\\
&\leq& \left\|\mathbf{A}_D\right\|_F^2\cdots
\left\|\widehat{\mathbf{A}}_{d^{\prime}}^{(s)}-A_{d^{\prime}}\widehat{\mathbf{H}}_{d^{\prime}}^{(s)}\right\|_F^2
\cdots\left\|\mathbf{A}_{d+1}\right\|_F^2\left\|\mathbf{A}_{d-1}\right\|_F^2\cdots\left\|\mathbf{A}_1\right\|_F^2\\
&&~~~~~~~~~~~~~~~~~~~\cdot
\left\|\sum\limits_{t=1}^T\mathcal{E}_{t}^{(d)}\left\{{A}_D\otimes\cdots\otimes
\left(\widehat{\mathbf{A}}_{d^{\prime}}^{(s)}-A_{d^{\prime}}\widehat{\mathbf{H}}_{d^{\prime}}^{(s)}\right)\otimes\cdots\otimes
\mathbf{A}_{d+1}\otimes \mathbf{A}_{d-1}\otimes\cdots\otimes{A}_1\right\}\mathbf{A}_d^{\top}\mathcal{E}_{t}^{(d)}\right\|_F^2\\
&\leq& \left\|\mathbf{A}_D\right\|_F^2\cdots
\left\|\widehat{\mathbf{A}}_{d^{\prime}}^{(s)}-A_{d^{\prime}}\widehat{\mathbf{H}}_{d^{\prime}}^{(s)}\right\|_F^2
\cdots\left\|\mathbf{A}_{d+1}\right\|_F^2\left\|\mathbf{A}_{d-1}\right\|_F^2\cdots\left\|\mathbf{A}_1\right\|_F^2
\sum\limits_{i_1=1}^{p_1}\cdots\sum\limits_{i_D=1}^{p_D}\\
&&~~~\cdot\left\|\sum\limits_{t=1}^T
\left\{{A}_D\otimes\cdots\otimes
\left(\widehat{\mathbf{A}}_{d^{\prime}}^{(s)}-A_{d^{\prime}}\widehat{\mathbf{H}}_{d^{\prime}}^{(s)}\right)\otimes\cdots\otimes
\mathbf{A}_{d+1}\otimes \mathbf{A}_{d-1}\otimes\cdots\otimes{A}_1\right\}^{\top}
\mathcal{E}_{t,i_d\cdot}^{(d)}\mathbf{A}_d^{\top}{e}_{t,i_1\cdots i_{d-1}\cdot i_{d+1}\cdots i_D}\right\|_F^2\\
&\leq& \left\|\mathbf{A}_D\right\|_F^4\cdots
\left\|\widehat{\mathbf{A}}_{d^{\prime}}^{(s)}-A_{d^{\prime}}\widehat{\mathbf{H}}_{d^{\prime}}^{(s)}\right\|_F^4
\cdots\left\|\mathbf{A}_{d+1}\right\|_F^4\left\|\mathbf{A}_{d-1}\right\|_F^4\cdots\left\|\mathbf{A}_1\right\|_F^4
\sum\limits_{i_1,i_1^{\prime}=1}^{p_1}\cdots\sum\limits_{i_{d-1},i_{d-1}^{\prime}=1}^{p_{d-1}}
\sum\limits_{i_d=1}^{p_d}\sum\limits_{i_{d+1},i_{d+1}^{\prime}=1}^{p_{d+1}}\cdots\sum\limits_{i_D,i_D^{\prime}=1}^{p_D}\\
&&~~~~~~~~~~~~~\cdot\left(\sum\limits_{t=1}^T\sum\limits_{i_d^{\prime}=1}^{p_d}a_{d,i_{d}^{\prime}}
e_{t,i_1\cdots i_{d-1}i_d^{\prime}i_{d+1}\cdots i_D}
e_{t,i_{1}^{\prime}\cdots i_{d-1}^{\prime}i_d i_{d+1}^{\prime}\cdots i_D^{\prime}}\right)^2\\
&=&O_p\left\{\left(\frac{Tp^4}{p_d^2}+\frac{T^2p^4}{p_d^3}\right)w_{d^{\prime}}^{(s)^2}\right\},
\eeqrs}
where
{\tiny\beqrs
&&\sum\limits_{i_1,i_1^{\prime}=1}^{p_1}\cdots\sum\limits_{i_{d-1},i_{d-1}^{\prime}=1}^{p_{d-1}}
\sum\limits_{i_d=1}^{p_d}\sum\limits_{i_{d+1},i_{d+1}^{\prime}=1}^{p_{d+1}}
\cdots\sum\limits_{i_D,i_D^{\prime}=1}^{p_D}
\mE\left(\sum\limits_{t=1}^T\sum\limits_{i_d^{\prime}=1}^{p_d}a_{d,i_{d}^{\prime}}
e_{t,i_1\cdots i_{d-1}i_d^{\prime}i_{d+1}\cdots i_D}
e_{t,i_{1}^{\prime}\cdots i_{d-1}^{\prime}i_d i_{d+1}^{\prime}\cdots i_D^{\prime}}\right)^2\\
&\leq&\sum\limits_{i_1,i_1^{\prime}=1}^{p_1}\cdots\sum\limits_{i_{d-1},i_{d-1}^{\prime}=1}^{p_{d-1}}
\sum\limits_{i_d=1}^{p_d}\sum\limits_{i_{d+1},i_{d+1}^{\prime}=1}^{p_{d+1}}
\cdots\sum\limits_{i_D,i_D^{\prime}=1}^{p_D}
\left\{\sum\limits_{t=1}^T\sum\limits_{i_d^{\prime}=1}^{p_d}\mE\left(
e_{t,i_1\cdots i_{d-1}i_d^{\prime}i_{d+1}\cdots i_D}
e_{t,i_{1}^{\prime}\cdots i_{d-1}^{\prime}i_d i_{d+1}^{\prime}\cdots i_D^{\prime}}\right)\right\}^2\\
&&+\sum\limits_{s,t}^T\sum\limits_{i_1,i_1^{\prime}}^{p_1}\cdots\sum\limits_{i_{d-1},i_{d-1}^{\prime}}^{p_{d-1}}
\sum\limits_{i_d,i_d^{\prime},i_d^{\prime\prime}}^{p_d}\sum\limits_{i_{d+1},i_{d+1}^{\prime}}^{p_{d+1}}
\cdots\sum\limits_{i_D,i_D^{\prime}}^{p_D}\\
&&~~~~~~~~~~\cdot\cov\left(e_{t,i_1\cdots i_{d-1}i_d^{\prime}i_{d+1}\cdots i_D}
e_{t,i_{1}^{\prime}\cdots i_{d-1}^{\prime}i_d i_{d+1}^{\prime}\cdots i_D^{\prime}},
e_{s,i_1\cdots i_{d-1}i_d^{\prime\prime}i_{d+1}\cdots i_D}
e_{s,i_{1}^{\prime}\cdots i_{d-1}^{\prime}i_d i_{d+1}^{\prime}\cdots i_D^{\prime}}\right)\\
&=&O\left(\frac{T^2p^2}{p_d}+Tp^2\right).
\eeqrs}

Similarly, for the $d^{\prime}$th part of the third term,
{\scriptsize\beqrs
&&\Bigg\|\sum\limits_{t=1}^T\mathcal{E}_{t}^{(d)}
\Big\{\mathbf{A}_D\mathbf{A}_D^{\top}\otimes\cdots\otimes \mathbf{A}_{d+1}\mathbf{A}_{d+1}^{\top}\otimes \mathbf{A}_{d-1}\mathbf{A}_{d-1}^{\top}\otimes\cdots\\
&&~~~~~~~~~~~~~~~~~\otimes
\left(\widehat{\mathbf{A}}_{d^{\prime}}^{(s+1)}-A_{d^{\prime}}\widehat{\mathbf{H}}_{d^{\prime}}^{(s+1)}\right)
\left(\widehat{\mathbf{A}}_{d^{\prime}}^{(s+1)}-A_{d^{\prime}}\widehat{\mathbf{H}}_{d^{\prime}}^{(s+1)}\right)^{\top}
\otimes\cdots\otimes \mathbf{A}_{1}\mathbf{A}_{1}^{\top}\Big\}\mathcal{E}_{t}^{(d)\top}\mathbf{A}_d\Bigg\|_F^2\\
&=&O_p\left\{\left(\frac{Tp^4}{p_d^2}+\frac{T^2p^4}{p_d^3}\right)w_{d^{\prime}}^{(s+1)^2}\right\}.
\eeqrs}
As a result, 
{\scriptsize\beqrs
\left\|\sum\limits_{t=1}^T\mathcal{E}_{t}^{(d)}
\widehat{\mathbf{A}}_{[D]/\{d\}}^{(s+1)}\widehat{\mathbf{A}}_{[D]/\{d\}}^{(s+1)^{\top}}\mathcal{E}_{t}^{(d)\top}\mathbf{A}_d\right\|_F^2=O_p\Bigg\{\frac{Tp^3}{p_d}+\frac{T^2p^2}{p_d}
+\left(\frac{Tp^4}{p_d^2}+\frac{T^2p^4}{p_d^3}\right)
\left(\sum\limits_{d^{\prime}= d+1}^Dw_{d^{\prime}}^{(s)^2}
+\sum\limits_{d^{\prime}= 1}^{d-1}w_{d^{\prime}}^{(s+1)^2}\right)\Bigg\},
\eeqrs}
A special case is adopting $\widehat{\mathbf{A}}_d^{(0)}$ for $d\in[D]$, then
{\scriptsize\beqrs
\left\|\sum\limits_{t=1}^T\mathcal{E}_{t}^{(d)}
\widehat{\mathbf{A}}_{[D]/\{d\}}\widehat{\mathbf{A}}_{[D]/\{d\}}^{{\top}}\mathcal{E}_{t}^{(d)\top}\mathbf{A}_d\right\|_F^2=O_p\left\{\frac{Tp^3}{p_d}+\frac{T^2p^2}{p_d}
+\left(\frac{Tp^4}{p_d^2}+\frac{T^2p^4}{p_d^3}\right)
\left(\sum\limits_{d^{\prime}\neq d}^Dw_{d^{\prime}}^{(0)^2}\right)\right\}.
\eeqrs}

\bel
Suppose that $T$ and $\{p_d\}_{d=1}^D$ tend to infinity, and $\{k_d\}_{d=1}^D$ are fixed. If Assumptions 1-5 hold, then for $d\in[D]$, we have

{\scriptsize
\begin{align*}
&(a)~~\left\|\sum\limits_{t=1}^T\mathcal{E}_{t}^{(d)}\widehat{\mathbf{A}}_{[D]/\{d\}}\mathcal{F}_{t}^{(d)\top}\right\|_F^2
=O_p\Bigg[Tp + \sum\limits_{d^{\prime} \neq d}
\Bigg\{\frac{p_{d^{\prime}}^2}{p_d}+\frac{Tp^2}{p_d p_{d^{\prime}}^2}
+ \left(\frac{pp_{d^{\prime}}}{p_d}+\frac{Tp^2}{p_dp_{d^{\prime}}}\right)w_{d^{\prime}}^{(0)}\Bigg\}\Bigg].\\
&(b)~~\left\|\sum\limits_{t=1}^T\mathcal{E}_{t}^{(d)}
\widehat{\mathbf{A}}_{[D]/\{d\}}^{(1)}\mathcal{F}_{t}^{(d)\top}\right\|_F^2=O_p\Bigg[Tp + \sum\limits_{d^{\prime} \neq d}\frac{p_{d^{\prime}}^2}{p_d}
+ \sum\limits_{d^{\prime} = d+1}^D
\Bigg\{\frac{Tp^2}{p_d p_{d^{\prime}}^2}
+ \left(\frac{pp_{d^{\prime}}}{p_d}+\frac{Tp^2}{p_dp_{d^{\prime}}}\right)w_{d^{\prime}}^{(0)}\Bigg\}\\
&~~~~~~
+ \sum\limits_{d^{\prime} = 1}^{d-1} \Bigg\{
\left(\frac{p^2}{p_dp_{d^{\prime}}}+\frac{Tp^2}{p_dp_{d^{\prime}}^2} + \frac{Tp^2}{p_d}w_{d^{\prime}}^{(1)}\right)
\left(\sum\limits_{d^{\prime\prime} = d^{\prime}+1}^Dw_{d^{\prime\prime}}^{(0)^2}
+\sum\limits_{d^{\prime\prime} = 1}^{d^{\prime}-1}w_{d^{\prime\prime}}^{(1)^2}\right)
+ \frac{Tp_{d^{\prime}}^2}{p_d}w_{d^{\prime}}^{(1)}\Bigg\}\Bigg].\\
&(c)~~\left\|\sum\limits_{t=1}^T\mathcal{E}_{t}^{(d)}
\widehat{\mathbf{A}}_{[D]/\{d\}}^{(s+1)}\mathcal{F}_{t}^{(d)\top}\right\|_F^2\\
&~~~~~=O_p\Bigg[Tp + \sum\limits_{d^{\prime} \neq d}\frac{p_{d^{\prime}}^2}{p_d}
+ \sum\limits_{d^{\prime} = d+1}^D
\Bigg\{
\left(\frac{p^2}{p_dp_{d^{\prime}}}+\frac{Tp^2}{p_dp_{d^{\prime}}^2} + \frac{Tp^2}{p_d}w_{d^{\prime}}^{(s)}\right)
\left(\sum\limits_{d^{\prime\prime} = d^{\prime}+1}^Dw_{d^{\prime\prime}}^{(s-1)^2}
+\sum\limits_{d^{\prime\prime} = 1}^{d^{\prime}-1}w_{d^{\prime\prime}}^{(s)^2}\right)
+ \frac{Tp_{d^{\prime}}^2}{p_d}w_{d^{\prime}}^{(s)}\Bigg\}\\
&~~~~~~+ \sum\limits_{d^{\prime} = 1}^{d-1} \Bigg\{
\left(\frac{p^2}{p_dp_{d^{\prime}}}+\frac{Tp^2}{p_dp_{d^{\prime}}^2} + \frac{Tp^2}{p_d}w_{d^{\prime}}^{(s+1)}\right)
\left(\sum\limits_{d^{\prime\prime} = d^{\prime}+1}^Dw_{d^{\prime\prime}}^{(s)^2}
+\sum\limits_{d^{\prime\prime} = 1}^{d^{\prime}-1}w_{d^{\prime\prime}}^{(s+1)^2}\right)
+ \frac{Tp_{d^{\prime}}^2}{p_d}w_{d^{\prime}}^{(s+1)}\Bigg\}\Bigg].\\
&(d)~~\left\|\sum\limits_{t=1}^T\mathbf{A}_d^{\top}\mathcal{E}_{t}^{(d)}
\widehat{\mathbf{A}}_{[D]/\{d\}}\mathcal{F}_{t}^{(d)\top}\right\|_F^2
=O_p\Bigg[Tp + \sum\limits_{d^{\prime} \neq d}\Bigg\{p_{d^{\prime}}^2+\frac{Tp^2}{p_d p_{d^{\prime}}^2}
+ \left(\frac{pp_{d^{\prime}}}{p_d}+\frac{Tp^2}{p_dp_{d^{\prime}}}\right)w_{d^{\prime}}^{(0)}\Bigg\}\Bigg].\\
&(e)~~\left\|\sum\limits_{t=1}^T\mathbf{A}_d^{\top}\mathcal{E}_{t}^{(d)}
\widehat{\mathbf{A}}_{[D]/\{d\}}^{(1)}\mathcal{F}_{t}^{(d)\top}\right\|_F^2=O_p\Bigg[Tp + \sum\limits_{d^{\prime} \neq d}p_{d^{\prime}}^2
+ \sum\limits_{d^{\prime} = d+1}^D
\Bigg\{\frac{Tp^2}{p_d p_{d^{\prime}}^2}
+ \left(\frac{pp_{d^{\prime}}}{p_d}+\frac{Tp^2}{p_dp_{d^{\prime}}}\right)w_{d^{\prime}}^{(0)}\Bigg\}\\
&~~~~~~+ \sum\limits_{d^{\prime} = 1}^{d-1} \Bigg\{
\left(\frac{p^2}{p_dp_{d^{\prime}}}+\frac{Tp^2}{p_dp_{d^{\prime}}^2} + \frac{Tp^2}{p_d}w_{d^{\prime}}^{(1)}\right)
\left(\sum\limits_{d^{\prime\prime} = d^{\prime}+1}^Dw_{d^{\prime\prime}}^{(0)^2}
+\sum\limits_{d^{\prime\prime} = 1}^{d^{\prime}-1}w_{d^{\prime\prime}}^{(1)^2}\right)
+ \frac{Tp_{d^{\prime}}^2}{p_d}w_{d^{\prime}}^{(1)}\Bigg\}\Bigg].\\
&(f)~~\left\|\sum\limits_{t=1}^T\mathbf{A}_d^{\top}\mathcal{E}_{t}^{(d)}
\widehat{\mathbf{A}}_{[D]/\{d\}}^{(s+1)}\mathcal{F}_{t}^{(d)\top}\right\|_F^2\\
&~~~~~=O_p\Bigg[Tp + \sum\limits_{d^{\prime} \neq d}p_{d^{\prime}}^2
+ \sum\limits_{d^{\prime} = d+1}^D
\Bigg\{
\left(\frac{p^2}{p_dp_{d^{\prime}}}+\frac{Tp^2}{p_dp_{d^{\prime}}^2} + \frac{Tp^2}{p_d}w_{d^{\prime}}^{(s)}\right)
\left(\sum\limits_{d^{\prime\prime} = d^{\prime}+1}^Dw_{d^{\prime\prime}}^{(s-1)^2}
+\sum\limits_{d^{\prime\prime} = 1}^{d^{\prime}-1}w_{d^{\prime\prime}}^{(s)^2}\right)
+ \frac{Tp_{d^{\prime}}^2}{p_d}w_{d^{\prime}}^{(s)}\Bigg\}\\
&~~~~~~+ \sum\limits_{d^{\prime} = 1}^{d-1} \Bigg\{
\left(\frac{p^2}{p_dp_{d^{\prime}}}+\frac{Tp^2}{p_dp_{d^{\prime}}^2} + \frac{Tp^2}{p_d}w_{d^{\prime}}^{(s+1)}\right)
\left(\sum\limits_{d^{\prime\prime} = d^{\prime}+1}^Dw_{d^{\prime\prime}}^{(s)^2}
+\sum\limits_{d^{\prime\prime} = 1}^{d^{\prime}-1}w_{d^{\prime\prime}}^{(s+1)^2}\right)
+ \frac{Tp_{d^{\prime}}^2}{p_d}w_{d^{\prime}}^{(s+1)}\Bigg\}\Bigg].
\end{align*}}

\eel

\textit{Proof}: The mathematical induction is adopted to derive the results.
For simplicity, we only prove the case of $s=0$, i.e., (b) and (e),
and assume $k_1=\cdots=k_D=1$ if necessary.

First, when $d=1$,

{\scriptsize
\beqrs
&&\left\|\sum\limits_{t=1}^T\mathcal{E}_{t}^{(1)}\widehat{\mathbf{A}}_{[D]/\{1\}}
\mathcal{F}_{t}^{(1)^{\top}}\right\|_F^2\\
&\leq& \left\|\sum\limits_{t=1}^T\mathcal{E}_{t}^{(1)}{\mathbf{A}}_{[D]/\{1\}}
\mathcal{F}_{t}^{(1)^{\top}}\right\|_F^2
+\sum_{d=2}^D\left\|\sum\limits_{t=1}^T\mathcal{E}_{t}^{(1)}\left\{{A}_D\otimes\cdots
\otimes\left(\widehat{\mathbf{A}}_d^{(0)}-\mathbf{A}_d\widehat{\mathbf{H}}_d^{(0)}\right)\otimes\cdots\otimes{A}_2\right\}
\mathcal{F}_{t}^{(1)^{\top}}\right\|_F^2+\cdots,
\eeqrs}

where the omitted terms are small order of the existing terms. Note that
{\scriptsize\beqrs
&&\widehat{\mathbf{A}}_d^{(0)}-\mathbf{A}_d\widehat{\mathbf{H}}_d^{(0)}\\
&=&\frac{1}{Tp}\sum\limits_{t=1}^T \mathbf{A}_d\mathcal{F}_{t}^{(d)}{\mathbf{A}}_{[D]/\{1\}}^{\top}
\mathcal{E}_{t}^{(d)\top}\widehat{\mathbf{A}}_d^{(0)}\widehat{\mathbf{\Lambda}}_d^{(0)^{-1}}+\frac{1}{Tp}\sum\limits_{t=1}^T \mathcal{E}_{t}^{(d)}
{\mathbf{A}}_{[D]/\{1\}}
\mathcal{F}_{t}^{(d)\top}\mathbf{A}_d^{\top}\widehat{\mathbf{A}}_d^{(0)}\widehat{\mathbf{\Lambda}}_d^{(0)^{-1}}+\frac{1}{Tp}\sum\limits_{t=1}^T \mathcal{E}_{t}^{(d)}\mathcal{E}_{t}^{(d)\top}\widehat{\mathbf{A}}_d^{(0)}\widehat{\mathbf{\Lambda}}_d^{(0)^{-1}}\\
&=:& T_{d,1}^{(0)}+T_{d,2}^{(0)}+T_{d,3}^{(0)}.
\eeqrs}
Then
{\scriptsize\beqrs
&&\left\|\sum\limits_{t=1}^T\mathcal{E}_{t}^{(1)}
\left\{\mathbf{A}_D\otimes\cdots\otimes \mathbf{A}_{d+1}\otimes \left(\widehat{\mathbf{A}}_{d}^{(0)}-{A}_{d}\widehat{\mathbf{H}}_d^{(0)}\right)
\otimes \mathbf{A}_{d-1}\otimes\cdots\otimes \mathbf{A}_2\right\}\mathcal{F}_{t}^{(1)^{\top}}\right\|_F^2\\
&\leq&\|\sum\limits_{t=1}^T\mathcal{E}_t^{(1)}
\left(\mathbf{A}_D\otimes\cdots\otimes T_{d,1}^{(0)}\otimes\cdots\otimes \mathbf{A}_2\right)\mathcal{F}_{t}^{(1)^{\top}}\|_F^2\\
&&+\|\sum\limits_{t=1}^T\mathcal{E}_t^{(1)}
\left(\mathbf{A}_D\otimes\cdots\otimes T_{d,2}^{(0)}\otimes\cdots\otimes \mathbf{A}_2\right)\mathcal{F}_{t}^{(1)^{\top}}\|_F^2\\
&&+\|\sum\limits_{t=1}^T\mathcal{E}_t^{(1)}
\left(\mathbf{A}_D\otimes\cdots\otimes T_{d,3}^{(0)}\otimes\cdots\otimes \mathbf{A}_2\right)\mathcal{F}_{t}^{(1)^{\top}}\|_F^2.
\eeqrs}
For the first term,
{\scriptsize\beqrs
&&\|\sum\limits_{t=1}^T\mathcal{E}_t^{(1)}
\left(\mathbf{A}_D\otimes\cdots\otimes T_{d,1}^{(0)}\otimes\cdots\otimes \mathbf{A}_2\right)\mathcal{F}_{t}^{(1)^{\top}}\|_F^2\\
&\leq& \frac{1}{T^2p^2}
\left\|\sum\limits_{t=1}^T\mathcal{E}_t^{(1)}{\mathbf{A}}_{[D]/\{1\}}\mathcal{F}_{t}^{(1)^{\top}}\right\|_F^2
\cdot\left\|\sum\limits_{t=1}^T \mathcal{F}_{t}^{(d)}{\mathbf{A}}_{[D]/\{d\}}^{\top}\mathcal{E}_{t}^{(d)\top}\widehat{\mathbf{A}}_d^{(0)}\right\|_F^2\\
&\leq& \frac{1}{Tp}
\left\|\sum\limits_{t=1}^T \mathcal{F}_{t}^{(d)}{\mathbf{A}}_{[D]/\{d\}}^{\top}\mathcal{E}_{t}^{(d)\top}\mathbf{A}_D\right\|_F^2+\frac{p_dw_d^{(0)}}{Tp}\left\|\sum\limits_{t=1}^T \mathcal{F}_{t}^{(d)}{\mathbf{A}}_{[D]/\{d\}}^{\top}\mathcal{E}_{t}^{(d)\top}\right\|_F^2\\
&=& O_p(1+p_dw_d^{(0)}),
\eeqrs}

For the second term,
{\scriptsize
\beqrs
&&\left\|\sum\limits_{t=1}^T\mathcal{E}_t^{(1)}
\left(\mathbf{A}_D\otimes\cdots\otimes T_{d,2}^{(0)}\otimes\cdots\otimes \mathbf{A}_2\right)\mathcal{F}_{t}^{(1)^{\top}}\right\|_F^2\\
&\leq&
\frac{p_d^2}{T^2p^2}\Bigg\|\sum\limits_{t=1}^T\mathcal{E}_t^{(1)}[\mathbf{A}_D\otimes\cdots\otimes
\{\sum\limits_{s=1}^T \mathcal{E}_{s}^{(d)}
{\mathbf{A}}_{[D]/\{d\}}
\mathcal{F}_{s}^{(d)\top}\}
\otimes\cdots\otimes \mathbf{A}_2]\mathcal{F}_{t}^{(1)^{\top}}\Bigg\|_F^2\\
&\leq& \frac{1}{T^2p_{-d}^2}\sum\limits_{i=1}^{p_1}
\Bigg(\sum\limits_{s,t}^T\sum\limits_{i_1}^{p_1}\sum\limits_{i_2,i_2^{\prime}}^{p_2}\cdots
\sum\limits_{i_d}^{p_d}\cdots\sum\limits_{i_D,i_D^{\prime}}^{p_D}
e_{t,ii_2\cdots i_d\cdots i_D}e_{s,i_1i_2^{\prime}\cdots i_d\cdots i_D^{\prime}}
F_tF_s\\
&&a_{1,i_1}a_{2,i_2}a_{2,i_2^{\prime}}\cdots a_{d+1,i_{d+1}}a_{d+1,i_{d+1}^{\prime}}
a_{d-1,i_{d-1}}a_{d-1,i_{d-1}^{\prime}}\cdots a_{D,i_{D}}a_{D,i_{D}^{\prime}}\Bigg)^2\\
&\leq&\frac{1}{p_{-d}^2}\sum\limits_{i=1}^{p_1}
\Bigg(\sum\limits_{i_1}^{p_1}\sum\limits_{i_2,i_2^{\prime}}^{p_2}\cdots
\sum\limits_{i_d}^{p_d}\cdots\sum\limits_{i_D,i_D^{\prime}}^{p_D}
\boldsymbol\xi_{ii_2\cdots i_d\cdots i_D}\boldsymbol\xi_{i_1i_2^{\prime}\cdots i_d\cdots i_D^{\prime}}\Bigg)^2,
\eeqrs}
where
{\scriptsize\beqrs
&&\sum\limits_{i=1}^{p_1}
\mE\Bigg(\sum\limits_{i_1}^{p_1}\sum\limits_{i_2,i_2^{\prime}}^{p_2}\cdots
\sum\limits_{i_d}^{p_d}\cdots\sum\limits_{i_D,i_D^{\prime}}^{p_D}
\boldsymbol\xi_{ii_2\cdots i_d\cdots i_D}\boldsymbol\xi_{i_1i_2^{\prime}\cdots i_d\cdots i_D^{\prime}}\Bigg)^2\\
&=&\sum\limits_{i=1}^{p_1}\sum\limits_{i_1,i_1^{\prime}}^{p_1}
\sum\limits_{i_2,i_2^{\prime},i_2^{\prime\prime},i_2^{\prime\prime\prime}}^{p_2}\cdots
\sum\limits_{i_d,i_d^{\prime}}^{p_d}\cdots
\sum\limits_{i_D,i_D^{\prime},i_D^{\prime\prime},i_D^{\prime\prime\prime}}^{p_D}
\cov(\boldsymbol\xi_{ii_2\cdots i_d\cdots i_D}\boldsymbol\xi_{i_1i_2^{\prime}\cdots i_d\cdots i_D^{\prime}},
\boldsymbol\xi_{ii_2^{\prime\prime}\cdots i_d^{\prime}\cdots i_D^{\prime\prime}}
\boldsymbol\xi_{i_1^{\prime}i_2^{\prime\prime\prime}\cdots i_d^{\prime}\cdots i_D^{\prime\prime\prime}})\\
&&+\sum\limits_{i=1}^{p_1}
\left\{\sum\limits_{i_1}^{p_1}\sum\limits_{i_2,i_2^{\prime}}^{p_2}\cdots
\sum\limits_{i_d}^{p_d}\cdots\sum\limits_{i_D,i_D^{\prime}}^{p_D}
\mE\left(\boldsymbol\xi_{ii_2\cdots i_d\cdots i_D}\boldsymbol\xi_{i_1i_2^{\prime}\cdots i_d\cdots i_D^{\prime}}\right)\right\}^2\\
&=&O(pp_{-d}+pp_{-1}),
\eeqrs}
then
{\scriptsize\beqrs
\left\|\sum\limits_{t=1}^T\mathcal{E}_t^{(1)}
\left(\mathbf{A}_D\otimes\cdots\otimes T_{d,2}^{(0)}\otimes\cdots\otimes \mathbf{A}_2\right)\mathcal{F}_{t}^{(1)^{\top}}\right\|_F^2
=O_p(p_d+\frac{p_d^2}{p_1}).
\eeqrs}

For the third term,
{\scriptsize\beqrs
&&\|\sum\limits_{t=1}^T\mathcal{E}_t^{(1)}
\left(\mathbf{A}_D\otimes\cdots\otimes T_{d,3}^{(0)}\otimes\cdots\otimes \mathbf{A}_2\right)\mathcal{F}_{t}^{(1)^{\top}}\|_F^2\\
&\leq&\frac{1}{T^2p^2}\left\|\sum\limits_{t=1}^T\mathcal{E}^{(1)}_t\mathcal{F}_{t}^{(1)^{\top}}\right\|_F^2
\cdot\left\|\sum\limits_{t=1}^T\mathcal{E}^{(d)}_t\mathcal{E}^{(d)\top}_t\widehat{\mathbf{A}}_d^{(0)}\widehat{\mathbf{\Lambda}}_d^{-1}\right\|_F^2
\left\|\mathbf{A}_D\right\|_F^2\cdots\left\|\mathbf{A}_{d+1}\right\|_F^2\left\|\mathbf{A}_{d-1}\right\|_F^2\cdots\left\|\mathbf{A}_2\right\|_F^2\\
&\leq&\frac{1}{T^2pp_1p_d}\left\{\sum\limits_{i_1,\cdots,i_D}\left(\sum\limits_{t=1}^TF_te_{t,i_1\cdots i_D}\right)^2\right\}\left\{
\left\|\sum\limits_{t=1}^T\mathcal{E}^{(d)}_t\mathcal{E}^{(d)\top}_t{A}_d\right\|_F^2
+\left\|\sum\limits_{t=1}^T\mathcal{E}^{(d)}_t\mathcal{E}^{(d)\top}_t\right\|_F^2
\cdot\left\|\widehat{\mathbf{A}}_d^{(0)}-\mathbf{A}_D\widehat{\mathbf{H}}_d^{(0)}\right\|_F^2\right\}\\
&=&\frac{1}{T^2pp_1p_d}\left\{\sum\limits_{i_1,\cdots,i_D}\left(\sqrt{T}\boldsymbol\xi_{i_1\cdots i_D}\right)^2\right\}\left\{
\left\|\sum\limits_{t=1}^T\mathcal{E}^{(d)}_t\mathcal{E}^{(d)\top}_t{A}_d\right\|_F^2
+\left\|\sum\limits_{t=1}^T\mathcal{E}^{(d)}_t\mathcal{E}^{(d)\top}_t\right\|_F^2
\cdot\left\|\widehat{\mathbf{A}}_d^{(0)}-\mathbf{A}_D\widehat{\mathbf{H}}_d^{(0)}\right\|_F^2\right\}\\
&=&O_p\left\{\left(p_{-1}+\frac{Tp_{-d}^2}{p_1}\right)(1+p_dw_d^{(0)})\right\}.
\eeqrs}

As a result,
{\scriptsize\beqrs
\left\|\sum\limits_{t=1}^T\mathcal{E}_{t}^{(1)}\widehat{\mathbf{A}}_{[D]/\{1\}}
\mathcal{F}_{t}^{(1)^{\top}}\right\|_F^2=O_p\left[Tp+\sum\limits_{d\neq 1}\frac{p_d^2}{p_1}
+\sum\limits_{d=2}^D\left\{\frac{Tp^2}{p_1p_d^2}+\left(\frac{pp_d}{p_1}+\frac{Tp^2}{p_1p_d}\right)w_d^{(0)}\right\}\right],
\eeqrs}
and because
{\tiny\beqrs
\|\sum\limits_{t=1}^T\mathcal{E}_t^{(1)}
\left(\mathbf{A}_D\otimes\cdots\otimes T_{d,1}^{(0)}\otimes\cdots\otimes \mathbf{A}_2\right)\mathcal{F}_{t}^{(1)^{\top}}\|_F^2
\asymp
\|\sum\limits_{t=1}^T\mathbf{A}_1^{\top}\mathcal{E}_t^{(1)}
\left(\mathbf{A}_D\otimes\cdots\otimes T_{d,1}^{(0)}\otimes\cdots\otimes \mathbf{A}_2\right)\mathcal{F}_{t}^{(1)^{\top}}\|_F^2,
\eeqrs}

{\scriptsize\beqrs
\|\sum\limits_{t=1}^T\mathcal{E}_t^{(1)}
\left(\mathbf{A}_D\otimes\cdots\otimes T_{d,3}^{(0)}\otimes\cdots\otimes \mathbf{A}_2\right)\mathcal{F}_{t}^{(1)^{\top}}\|_F^2
\asymp
\|\sum\limits_{t=1}^T\mathbf{A}_1^{\top}\mathcal{E}_t^{(1)}
\left(\mathbf{A}_D\otimes\cdots\otimes T_{d,3}^{(0)}\otimes\cdots\otimes \mathbf{A}_2\right)\mathcal{F}_{t}^{(1)^{\top}}\|_F^2
\eeqrs}
and
{\scriptsize
\beqrs
&&\left\|\sum\limits_{t=1}^T\mathbf{A}_1^{\top}\mathcal{E}_t^{(1)}
\left(\mathbf{A}_D\otimes\cdots\otimes T_{d,2}^{(0)}\otimes\cdots\otimes \mathbf{A}_2\right)\mathcal{F}_{t}^{(1)^{\top}}\right\|_F^2\\
&\leq&
\frac{p_d^2}{T^2p^2}\Bigg\|\sum\limits_{t=1}^T
\mathbf{A}_1^{\top}\mathcal{E}_t^{(1)}[\mathbf{A}_D\otimes\cdots\otimes
\{\sum\limits_{s=1}^T \mathcal{E}_{s}^{(d)}
{\mathbf{A}}_{[D]/\{d\}}
\mathcal{F}_{s}^{(d)\top}\}
\otimes\cdots\otimes \mathbf{A}_2]\mathcal{F}_{t}^{(1)^{\top}}\Bigg\|_F^2\\
&\leq& \frac{1}{T^2p_{-d}^2}
\Bigg(\sum\limits_{s,t}^T\sum\limits_{i_1,i_1^{\prime}}^{p_1}\sum\limits_{i_2,i_2^{\prime}}^{p_2}\cdots
\sum\limits_{i_d}^{p_d}\cdots\sum\limits_{i_D,i_D^{\prime}}^{p_D}
e_{t,i_1i_2\cdots i_d\cdots i_D}e_{s,i_1^{\prime}i_2^{\prime}\cdots i_d\cdots i_D^{\prime}}F_tF_s\\
&&a_{1,i_1}a_{1,i_1^{\prime}}a_{2,i_2}a_{2,i_2^{\prime}}\cdots a_{d+1,i_{d+1}}a_{d+1,i_{d+1}^{\prime}}
a_{d-1,i_{d-1}}a_{d-1,i_{d-1}^{\prime}}\cdots a_{D,i_{D}}a_{D,i_{D}^{\prime}}\Bigg)^2\\
&\leq&\frac{1}{p_{-d}^2}
\Bigg(\sum\limits_{i_1,i_1^{\prime}}^{p_1}\sum\limits_{i_2,i_2^{\prime}}^{p_2}\cdots
\sum\limits_{i_d}^{p_d}\cdots\sum\limits_{i_D,i_D^{\prime}}^{p_D}
\boldsymbol\xi_{i_1i_2\cdots i_d\cdots i_D}\boldsymbol\xi_{i_1^{\prime}i_2^{\prime}\cdots i_d\cdots i_D^{\prime}}\Bigg)^2,
\eeqrs}
where
{\scriptsize\beqrs
&&\mE\Bigg(\sum\limits_{i_1,i_1^{\prime}}^{p_1}\sum\limits_{i_2,i_2^{\prime}}^{p_2}\cdots
\sum\limits_{i_d}^{p_d}\cdots\sum\limits_{i_D,i_D^{\prime}}^{p_D}
\boldsymbol\xi_{i_1i_2\cdots i_d\cdots i_D}\boldsymbol\xi_{i_1^{\prime}i_2^{\prime}\cdots i_d\cdots i_D^{\prime}}\Bigg)^2\\
&=&\sum\limits_{i_1,i_1^{\prime},i_1^{\prime\prime},i_1^{\prime\prime\prime}}^{p_1}
\sum\limits_{i_2,i_2^{\prime},i_2^{\prime\prime},i_2^{\prime\prime\prime}}^{p_2}\cdots
\sum\limits_{i_d,i_d^{\prime}}^{p_d}\cdots
\sum\limits_{i_D,i_D^{\prime},i_D^{\prime\prime},i_D^{\prime\prime\prime}}^{p_D}
\cov(\boldsymbol\xi_{i_1i_2\cdots i_d\cdots i_D}\boldsymbol\xi_{i_1^{\prime}i_2^{\prime}\cdots i_d\cdots i_D^{\prime}},
\boldsymbol\xi_{i_1^{\prime\prime}i_2^{\prime\prime}\cdots i_d^{\prime}\cdots i_D^{\prime\prime}}
\boldsymbol\xi_{i_1^{\prime\prime\prime}i_2^{\prime\prime\prime}\cdots i_d^{\prime}\cdots i_D^{\prime\prime\prime}})\\
&&+\left\{\sum\limits_{i_1,i_1^{\prime}}^{p_1}\sum\limits_{i_2,i_2^{\prime}}^{p_2}\cdots
\sum\limits_{i_d}^{p_d}\cdots\sum\limits_{i_D,i_D^{\prime}}^{p_D}
\mE\left(\boldsymbol\xi_{i_1i_2\cdots i_d\cdots i_D}\boldsymbol\xi_{i_1^{\prime}i_2^{\prime}\cdots i_d\cdots i_D^{\prime}}\right)\right\}^2\\
&=&O(p^2),
\eeqrs}
then
{\scriptsize\beqrs
\left\|\sum\limits_{t=1}^T\mathbf{A}_1^{\top}\mathcal{E}_t^{(1)}
\left(\mathbf{A}_D\otimes\cdots\otimes T_{d,2}^{(0)}\otimes\cdots\otimes \mathbf{A}_2\right)\mathcal{F}_{t}^{(1)^{\top}}\right\|_F^2
=O_p(p_d^2).
\eeqrs}
Hence
{\scriptsize\beqrs
\left\|\sum\limits_{t=1}^T\mathbf{A}_1^{\top}\mathcal{E}_{t}^{(1)}\widehat{\mathbf{A}}_{[D]/\{1\}}
\mathcal{F}_{t}^{(1)^{\top}}\right\|_F^2=O_p\left[Tp+\sum\limits_{d\neq 1}p_d^2
+\sum\limits_{d=2}^D\left\{\frac{Tp^2}{p_1p_d^2}+\left(\frac{pp_d}{p_1}+\frac{Tp^2}{p_1p_d}\right)w_d^{(0)}\right\}\right].
\eeqrs}

Second, when $d=2$, similar to the case of $d=1$, we have
{\scriptsize
\beqrs
&&\left\|\sum\limits_{t=1}^T\mathcal{E}_{t}^{(2)}
\widehat{\mathbf{A}}_{[D]/\{2\}}^{(1)}
\mathcal{F}_{t}^{(2)\top}\right\|_F^2\\
&\leq& \left\|\sum\limits_{t=1}^T\mathcal{E}_{t}^{(2)}{\mathbf{A}}_{[D]/\{2\}}
\mathcal{F}_{t}^{(1)^{\top}}\right\|_F^2+\sum_{d=3}^D\left\|\sum\limits_{t=1}^T\mathcal{E}_{t}^{(2)}
\left\{{A}_D\otimes\cdots\otimes\left(\widehat{\mathbf{A}}_d^{(0)}-\mathbf{A}_d\widehat{\mathbf{H}}_d^{(0)}\right)
\otimes\cdots\otimes{A}_3\otimes{A}_1\right\}\mathcal{F}_{t}^{(2)\top}\right\|_F^2\\
&&+\left\|\sum\limits_{t=1}^T\mathcal{E}_{t}^{(2)}\left\{{A}_D\otimes\cdots\otimes A_3\otimes
\left(\widehat{\mathbf{A}}_1^{(1)}-\mathbf{A}_1\widehat{\mathbf{H}}_1^{(1)}\right)\right\}\mathcal{F}_{t}^{(2)\top}\right\|_F^2+\cdots\\
&\leq&O_p\left[Tp+\frac{p_1^2}{p_2}+\sum\limits_{d=3}^D
\left\{\frac{p_d^2}{p_2}+\frac{Tp^2}{p_2p_d^2}+\left(\frac{pp_d}{p_2}+\frac{Tp^2}{p_2p_d}\right)w_d^{(0)}\right\}\right]\\
&&+\frac{1}{Tp}\left\|\sum\limits_{t=1}^T\widehat{\mathbf{A}}_1^{(1)^{\top}}\mathcal{E}_{t}^{(1)}\widehat{\mathbf{A}}_{[D]/\{1\}}
\mathcal{F}_{t}^{(1)^{\top}}\right\|_F^2+\frac{p_1}{Tp^2p_2}\left\|\sum\limits_{t=1}^T\mathcal{E}_{t}^{(1)}
\widehat{\mathbf{A}}_{[D]/\{1\}}\widehat{\mathbf{A}}_{[D]/\{1\}}^{\top}\mathcal{E}_{t}^{(1)^{\top}}\widehat{\mathbf{A}}_1^{(1)}\right\|_F^2\\
&=&O_p\Bigg[Tp+\sum\limits_{d\neq 2}\frac{p_d^2}{p_2}+
\sum\limits_{d=3}^D\left\{\frac{Tp^2}{p_2p_d^2}+\left(\frac{pp_d}{p_2}+\frac{Tp^2}{p_2p_d}\right)w_d^{(0)}\right\}
+\left(\frac{p^2}{p_1p_2}+\frac{Tp^2}{p_1^2p_2} + \frac{Tp^2}{p_2}w_{1}^{(1)}\right)
\sum\limits_{d = 2}^Dw_{d}^{(0)^2}+\frac{Tp_{1}^2}{p_2}w_{1}^{(1)}\Bigg],
\eeqrs}
and
{\scriptsize
\beqrs
&&\left\|\sum\limits_{t=1}^T\mathbf{A}_2^{\top}\mathcal{E}_{t}^{(2)}
\widehat{\mathbf{A}}_{[D]/\{2\}}^{(1)}
\mathcal{F}_{t}^{(2)\top}\right\|_F^2\\
&=&O_p\Bigg[Tp+\sum\limits_{d\neq 2}p_d^2+
\sum\limits_{d=3}^D\left\{\frac{Tp^2}{p_2p_d^2}+\left(\frac{pp_d}{p_2}+\frac{Tp^2}{p_2p_d}\right)w_d^{(0)}\right\}
+\left(\frac{p^2}{p_1p_2}+\frac{Tp^2}{p_1^2p_2} + \frac{Tp^2}{p_2}w_{1}^{(1)}\right)
\sum\limits_{d = 2}^Dw_{d}^{(0)^2}+\frac{Tp_{1}^2}{p_2}w_{1}^{(1)}\Bigg].
\eeqrs}

Third, assume for $d-1$, we have
{\scriptsize \beqrs
&&\left\|\sum\limits_{t=1}^T\mathcal{E}_{t}^{(d-1)}
\widehat{\mathbf{A}}_{[D]/\{d-1\}}^{(1)}\mathcal{F}_{t}^{(d-1)\top}\right\|_F^2\\
&=&O_p\Bigg[Tp + \sum\limits_{d^{\prime} \neq d-1}\frac{p_{d^{\prime}}^2}{p_{d-1}}
+ \sum\limits_{d^{\prime} = d}^D
\Bigg\{\frac{Tp^2}{p_{d-1} p_{d^{\prime}}^2}
+ \left(\frac{pp_{d^{\prime}}}{p_{d-1}}+\frac{Tp^2}{p_{d-1}p_{d^{\prime}}}\right)w_{d^{\prime}}^{(0)}\Bigg\}\\
&&+ \sum\limits_{d^{\prime} = 1}^{d-2} \Bigg\{
\left(\frac{p^2}{p_{d-1}p_{d^{\prime}}}+\frac{Tp^2}{p_{d-1}p_{d^{\prime}}^2} + \frac{Tp^2}{p_{d-1}}w_{d^{\prime}}^{(1)}\right)
\left(\sum\limits_{d^{\prime\prime} = d^{\prime}+1}^Dw_{d^{\prime\prime}}^{(0)^2}
+\sum\limits_{d^{\prime\prime} = 1}^{d^{\prime}-1}w_{d^{\prime\prime}}^{(1)^2}\right)
+ \frac{Tp_{d^{\prime}}^2}{p_{d-1}}w_{d^{\prime}}^{(1)}\Bigg\}\Bigg],
\eeqrs}
and
{\scriptsize \beqrs
&&\left\|\sum\limits_{t=1}^T\mathbf{A}_{d-1}\mathcal{E}_{t}^{(d-1)}
\widehat{\mathbf{A}}_{[D]/\{d-1\}}^{(1)}\mathcal{F}_{t}^{(d-1)\top}\right\|_F^2\\
&=&O_p\Bigg[Tp + \sum\limits_{d^{\prime} \neq d-1}p_{d^{\prime}}^2
+ \sum\limits_{d^{\prime} = d}^D
\Bigg\{\frac{Tp^2}{p_{d-1} p_{d^{\prime}}^2}
+ \left(\frac{pp_{d^{\prime}}}{p_{d-1}}+\frac{Tp^2}{p_{d-1}p_{d^{\prime}}}\right)w_{d^{\prime}}^{(0)}\Bigg\}\\
&&+ \sum\limits_{d^{\prime} = 1}^{d-2} \Bigg\{
\left(\frac{p^2}{p_{d-1}p_{d^{\prime}}}+\frac{Tp^2}{p_{d-1}p_{d^{\prime}}^2} + \frac{Tp^2}{p_{d-1}}w_{d^{\prime}}^{(1)}\right)
\left(\sum\limits_{d^{\prime\prime} = d^{\prime}+1}^Dw_{d^{\prime\prime}}^{(0)^2}
+\sum\limits_{d^{\prime\prime} = 1}^{d^{\prime}-1}w_{d^{\prime\prime}}^{(1)^2}\right)
+ \frac{Tp_{d^{\prime}}^2}{p_{d-1}}w_{d^{\prime}}^{(1)}\Bigg\}\Bigg],
\eeqrs}

For the general $d$,
{\scriptsize\beqrs
&&\left\|\sum\limits_{t=1}^T\mathcal{E}_{t}^{(d)}
\widehat{\mathbf{A}}_{[D]/\{d\}}^{(1)}\mathcal{F}_t^{(d)^{\top}}\right\|_F^2\\
&\leq& \left\|\sum\limits_{t=1}^T\mathcal{E}_{t}^{(d)}
{\mathbf{A}}_{[D]/\{d\}}\mathcal{F}_t^{(d)^{\top}}\right\|_F^2\\
&&+\sum_{d^{\prime}=d+1}^D\left\|\sum\limits_{t=1}^T\mathcal{E}_{t}^{(d)}
\left\{\mathbf{A}_D \otimes\cdots\otimes \left(\widehat{\mathbf{A}}_{d^{\prime}}^{(0)}-A_{d^{\prime}}\widehat{\mathbf{H}}_{d^{\prime}}^{(0)}\right)
\otimes\cdots\otimes \mathbf{A}_{d+1} \otimes \mathbf{A}_{d-1}\otimes\cdots\otimes \mathbf{A}_1\right\}\mathcal{F}_t^{(d)^{\top}}\right\|_F^2\\
&&+\sum_{d^{\prime}=1}^{d-1}\left\|\sum\limits_{t=1}^T\mathcal{E}_{t}^{(d)}
\left\{\mathbf{A}_D \otimes\cdots\otimes \mathbf{A}_{d+1} \otimes \mathbf{A}_{d-1}\otimes\cdots\otimes \left(\widehat{\mathbf{A}}_{d^{\prime}}^{(1)}-A_{d^{\prime}}\widehat{\mathbf{H}}_{d^{\prime}}^{(1)}\right)
\otimes\cdots\otimes \mathbf{A}_1\right\}\mathcal{F}_t^{(d)^{\top}}\right\|_F^2\\
&=&
O_p\Bigg[Tp + \sum\limits_{d^{\prime} = d+1}^D
\Bigg\{\frac{p_{d^{\prime}}^2}{p_d}+\frac{Tp^2}{p_d p_{d^{\prime}}^2}
+ \left(\frac{pp_{d^{\prime}}}{p_d}+\frac{Tp^2}{p_dp_{d^{\prime}}}\right)w_{d^{\prime}}^{(0)}\Bigg\}\Bigg]\\
&&+ \sum\limits_{d^{\prime} = 1}^{d-1} \Bigg\{
\frac{1}{Tp}\left\|\sum\limits_{t=1}^T\widehat{\mathbf{A}}_{d^{\prime}}^{(1)^{\top}}\mathcal{E}_{t}^{(d^{\prime})}
\widehat{\mathbf{A}}_{[D]/\{d^{\prime}\}}^{(1)}\mathcal{F}_{t}^{(d^{\prime})\top}\right\|_F^2+\frac{p_{d^{\prime}}^2}{p_d}+\frac{p_d^{\prime}}{Tp^2p_d}\left\|\sum\limits_{t=1}^T\mathcal{E}_{t}^{(d^{\prime})}
\widehat{\mathbf{A}}_{[D]/\{d^{\prime}\}}^{(1)}\widehat{\mathbf{A}}_{[D]/\{d^{\prime}\}}^{(1)^{\top}}\mathcal{E}_{t}^{(d^{\prime})^{\top}}\widehat{\mathbf{A}}_{d^{\prime}}^{(1)}\right\|_F^2
\Bigg\}\\
&=&O_p\Bigg[Tp + \sum\limits_{d^{\prime} \neq d}\frac{p_{d^{\prime}}^2}{p_d}
+ \sum\limits_{d^{\prime} = d+1}^D
\Bigg\{\frac{Tp^2}{p_d p_{d^{\prime}}^2}
+ \left(\frac{pp_{d^{\prime}}}{p_d}+\frac{Tp^2}{p_dp_{d^{\prime}}}\right)w_{d^{\prime}}^{(0)}\Bigg\}\\
&&+ \sum\limits_{d^{\prime} = 1}^{d-1} \Bigg\{
\left(\frac{p^2}{p_dp_{d^{\prime}}}+\frac{Tp^2}{p_dp_{d^{\prime}}^2} + \frac{Tp^2}{p_d}w_{d^{\prime}}^{(1)}\right)
\left(\sum\limits_{d^{\prime\prime} = d^{\prime}+1}^Dw_{d^{\prime\prime}}^{(0)^2}
+\sum\limits_{d^{\prime\prime} = 1}^{d^{\prime}-1}w_{d^{\prime\prime}}^{(1)^2}\right)
+ \frac{Tp_{d^{\prime}}^2}{p_d}w_{d^{\prime}}^{(1)}\Bigg\}\Bigg],
\eeqrs}
and
{\scriptsize\beqrs
\left\|\sum\limits_{t=1}^T\mathbf{A}_d^{\top}\mathcal{E}_{t}^{(d)}
\widehat{\mathbf{A}}_{[D]/\{d\}}^{(1)}\mathcal{F}_{t}^{(d)\top}\right\|_F^2
&=&O_p\Bigg[Tp + \sum\limits_{d^{\prime} \neq d}p_{d^{\prime}}^2
+ \sum\limits_{d^{\prime} = d+1}^D
\Bigg\{\frac{Tp^2}{p_d p_{d^{\prime}}^2}
+ \left(\frac{pp_{d^{\prime}}}{p_d}+\frac{Tp^2}{p_dp_{d^{\prime}}}\right)w_{d^{\prime}}^{(0)}\Bigg\}\\
&&+ \sum\limits_{d^{\prime} = 1}^{d-1} \Bigg\{
\left(\frac{p^2}{p_dp_{d^{\prime}}}+\frac{Tp^2}{p_dp_{d^{\prime}}^2} + \frac{Tp^2}{p_d}w_{d^{\prime}}^{(1)}\right)
\left(\sum\limits_{d^{\prime\prime} = d^{\prime}+1}^Dw_{d^{\prime\prime}}^{(0)^2}
+\sum\limits_{d^{\prime\prime} = 1}^{d^{\prime}-1}w_{d^{\prime\prime}}^{(1)^2}\right)
+ \frac{Tp_{d^{\prime}}^2}{p_d}w_{d^{\prime}}^{(1)}\Bigg\}\Bigg].
\eeqrs}

\bel
Suppose that $T$ and $\{p_d\}_{d=1}^D$ tend to infinity, and $\{k_d\}_{d=1}^D$ are fixed. If Assumptions 1-5 hold, then we have
$$\left\|\sum\limits_{t=1}^T\mathcal{E}_{t,i\cdot}^{(d)\top}\widehat{\mathbf{A}}_{[D]/\{d\}}\mathcal{F}_{t}^{(d)\top}\right\|_F^2=O_p\left\{Tp_{-d}+
\sum\limits_{d^{\prime}\neq d}\left(\frac{Tp^2}{p_d^2p_{d^{\prime}}^2}
+\frac{p_{d^{\prime}}^2}{p_d^2}\right)\right\}.$$
\eel

\textit{Proof}: The proof is similar to Lemma 11.

\bel
Suppose that $T$ and $\{p_d\}_{d=1}^D$ tend to infinity, and $\{k_d\}_{d=1}^D$ are fixed. If Assumptions 1-5 hold, then we have
{\footnotesize\beqrs
\left\|\sum\limits_{t=1}^T\mathcal{E}_{t,i\cdot}^{(d)\top}\widehat{\mathbf{A}}_{[D]/\{d\}}\widehat{\mathbf{A}}_{[D]/\{d\}}^{\top}\mathcal{E}_{t}^{(d)\top}\mathbf{A}_d\right\|_F^2=O_p\left\{\frac{Tp^3}{p_d^2}+\frac{T^2p^2}{p_d^2}+
\sum\limits_{d^{\prime}\neq d}\left(\frac{p^2p_{d^{\prime}}^2}{Tp_d^3}
+\frac{Tp^4}{p_d^3p_{d^{\prime}}^4}
+\frac{p^2p_{d^{\prime}}^2}{p_d^4}
+\frac{T^2p^4}{p_d^4p_{d^{\prime}}^4}
\right)\right\}.
\eeqrs}
\eel

\textit{Proof}: The proof is similar to Lemma 10.

\bel
Suppose that $T$ and $\{p_d\}_{d=1}^D$ tend to infinity, and $\{k_d\}_{d=1}^D$ are fixed. If Assumptions 1-5 hold, then we have
\beqrs
\left\|\sum\limits_{t=1}^T\mathcal{E}_{t,i\cdot}^{(d)\top}\widehat{\mathbf{A}}_{[D]/\{d\}}\widehat{\mathbf{A}}_{[D]/\{d\}}^{\top}\mathcal{E}_{t}^{(d)\top}\right\|_F=O_p\left\{\frac{Tp}{\sqrt{p_d}}+\sum\limits_{d^{\prime}\neq d}^D\left(
\frac{pp_d^{\prime}}{\sqrt{p_d}p_d}+\frac{Tp^2}{\sqrt{p_d}p_dp_{d^{\prime}}^2}\right)\right\}.
\eeqrs
\eel

\textit{Proof}: For simplicity, we only consider the proof when $d=1$ and fix $k_1=\cdots=k_D=1$.
Note that
\beqrs
\left\|\sum\limits_{t=1}^T\mathcal{E}_{t,i\cdot}^{(1)^{\top}}\widehat{\mathbf{A}}_{[D]/\{1\}}\widehat{\mathbf{A}}_{[D]/\{1\}}^{\top}\mathcal{E}_{t}^{(1)^{\top}}\right\|_F
&\leq&\sqrt{\sum\limits_{t=1}^T\left\|\mathcal{E}_{t}^{(1)}\widehat{\mathbf{A}}_{[D]/\{1\}}\right\|_F^2}
\sqrt{\sum\limits_{t=1}^T\left\|\widehat{\mathbf{A}}_{[D]/\{1\}}^{\top}\mathcal{E}_{t,i\cdot}^{(1)}\right\|_F^2}\\
&=&O_p\left\{\frac{Tp}{\sqrt{p_1}}+\sum\limits_{d=2}^D\left(
\frac{pp_d}{\sqrt{p_1}p_1}+\frac{Tp^2}{\sqrt{p_1}p_1p_d^2}\right)\right\}.
\eeqrs

\bel
Suppose that $T$ and $\{p_d\}_{d=1}^D$ tend to infinity, and $\{k_d\}_{d=1}^D$ are fixed. If Assumptions 1-5 hold, then for $d\in[D]$, we have
{\small\beqrs
&&\frac{1}{p_d}\left\| \mathbf{A}_d^{\top}(\widetilde{\mathbf{A}}_d-\mathbf{A}_d\widetilde{\mathbf{H}}_d)\right\|_F\\
&=&O_p\left\{\frac{1}{\sqrt{Tp}}+\frac{1}{p}+\sum\limits_{d_1=1}^D\frac{p_{d_1}}{Tp}
+\sum\limits_{d^{\prime}\neq d}\left(\frac{1}{\sqrt{Tp_d}p_{d^{\prime}}}
+\frac{1}{T\sqrt{p_{-d}}p_{d^{\prime}}}+\frac{1}{p_dp_{d^{\prime}}^2}
\right)
+\sum\limits_{d_1\neq d}\sum\limits_{d_2\neq d} \frac{1}{Tp_{d_1}p_{d_2}}\right\}.
\eeqrs}
\eel

\textit{Proof}: Note that
{\footnotesize\beqrs
&&\frac{1}{p_d}\|\mathbf{A}_d^{\top}(\widetilde{\mathbf{A}}_d-\mathbf{A}_d\widetilde{\mathbf{H}}_d)\|_F
\leq\frac{1}{Tpp_{-d}}\Bigg\|\sum\limits_{t=1}^T\mathcal{F}_t^{(d)}
\mathbf{A}_{[D]/\{d\}}^{\top}\widehat{\mathbf{A}}_{[D]/\{d\}}\widehat{\mathbf{A}}_{[D]/\{d\}}^{\top}
\mathcal{E}_t^{(d)\top}\widetilde{\mathbf{A}}_d\widetilde{\mathbf{\Lambda}}_d^{-1}\Bigg\|_F\\
&&~+\frac{1}{Tp^2}\Bigg\|\sum\limits_{t=1}^T\mathbf{A}_d^{\top}\mathcal{E}_t^{(d)}
\widehat{\mathbf{A}}_{[D]/\{d\}}\widehat{\mathbf{A}}_{[D]/\{d\}}^{\top}\mathbf{A}_{[D]/\{d\}}\mathcal{F}_t^{(d)\top}\mathbf{A}_d^{\top}\widetilde{\mathbf{A}}_d\widetilde{\mathbf{\Lambda}}_d^{-1}\Bigg\|_F\\
&&~+\frac{1}{Tp^2}\left\|\sum\limits_{t=1}^T\mathbf{A}_d^{\top}\mathcal{E}_t^{(d)}
\widehat{\mathbf{A}}_{[D]/\{d\}}\widehat{\mathbf{A}}_{[D]/\{d\}}^{\top}
\mathcal{E}_t^{(d)\top}\widetilde{\mathbf{A}}_d\widetilde{\mathbf{\Lambda}}_d^{-1}\right\|_F\\
&&~\leq \frac{1}{Tp}\left\|\sum\limits_{t=1}^T\mathcal{F}_t^{(d)}
\widehat{\mathbf{A}}_{[D]/\{d\}}^{\top}
\mathcal{E}_t^{(d)\top}\mathbf{A}_d\right\|_F+\frac{1}{Tp}\left\|\sum\limits_{t=1}^T\mathcal{F}_t^{(d)}
\widehat{\mathbf{A}}_{[D]/\{d\}}^{\top}
\mathcal{E}_t^{(d)\top}\right\|_F\cdot\left\|\widetilde{\mathbf{A}}_d-\mathbf{A}_d\widetilde{\mathbf{H}}_d\right\|_F\\
&&~+\frac{1}{Tp}\Bigg\|\sum\limits_{t=1}^T\mathbf{A}_d^{\top}\mathcal{E}_t^{(d)}
\widehat{\mathbf{A}}_{[D]/\{d\}}\mathcal{F}_t^{(d)\top}\Bigg\|_F
+\frac{1}{Tp^2}\left\|\sum\limits_{t=1}^T\mathbf{A}_d^{\top}\mathcal{E}_t^{(d)}\widehat{\mathbf{A}}_{[D]/\{d\}}
\widehat{\mathbf{A}}_{[D]/\{d\}}^{\top}
\mathcal{E}_t^{(d)\top}\mathbf{A}_d\right\|_F\\
&&~+\frac{1}{Tp^2}\left\|\sum\limits_{t=1}^T\mathbf{A}_d^{\top}\mathcal{E}_t^{(d)}
\widehat{\mathbf{A}}_{[D]/\{d\}}\widehat{\mathbf{A}}_{[D]/\{d\}}^{\top}
\mathcal{E}_t^{(d)\top}\right\|_F\cdot\left\|\widetilde{\mathbf{A}}_d-\mathbf{A}_d\widetilde{\mathbf{H}}_d\right\|_F\\
&&~=O_p\left\{\frac{1}{\sqrt{Tp}}+\frac{1}{p}+
\sum\limits_{d^{\prime}\neq d}\left(\frac{1}{\sqrt{Tp_d}p_{d^{\prime}}}
+\frac{1}{T\sqrt{p_{-d}}p_{d^{\prime}}}+\frac{1}{p_dp_{d^{\prime}}^2}
\right)
+\sum\limits_{d_1=1}^D\frac{p_{d_1}}{Tp}
+\sum\limits_{d_1\neq d}\sum\limits_{d_2\neq d}\frac{1}{Tp_{d_1}p_{d_2}}\right\}.
\eeqrs

\section{Additional Simulation Results}

The results of the estimated mode-2 and mode-3 loading matrices are shown in Figures \ref{A2} and \ref{A3}, and have similar explanations to the results of mode-1 in the manuscript.

\begin{figure}[H]
  \centering
  \includegraphics[width=5in]{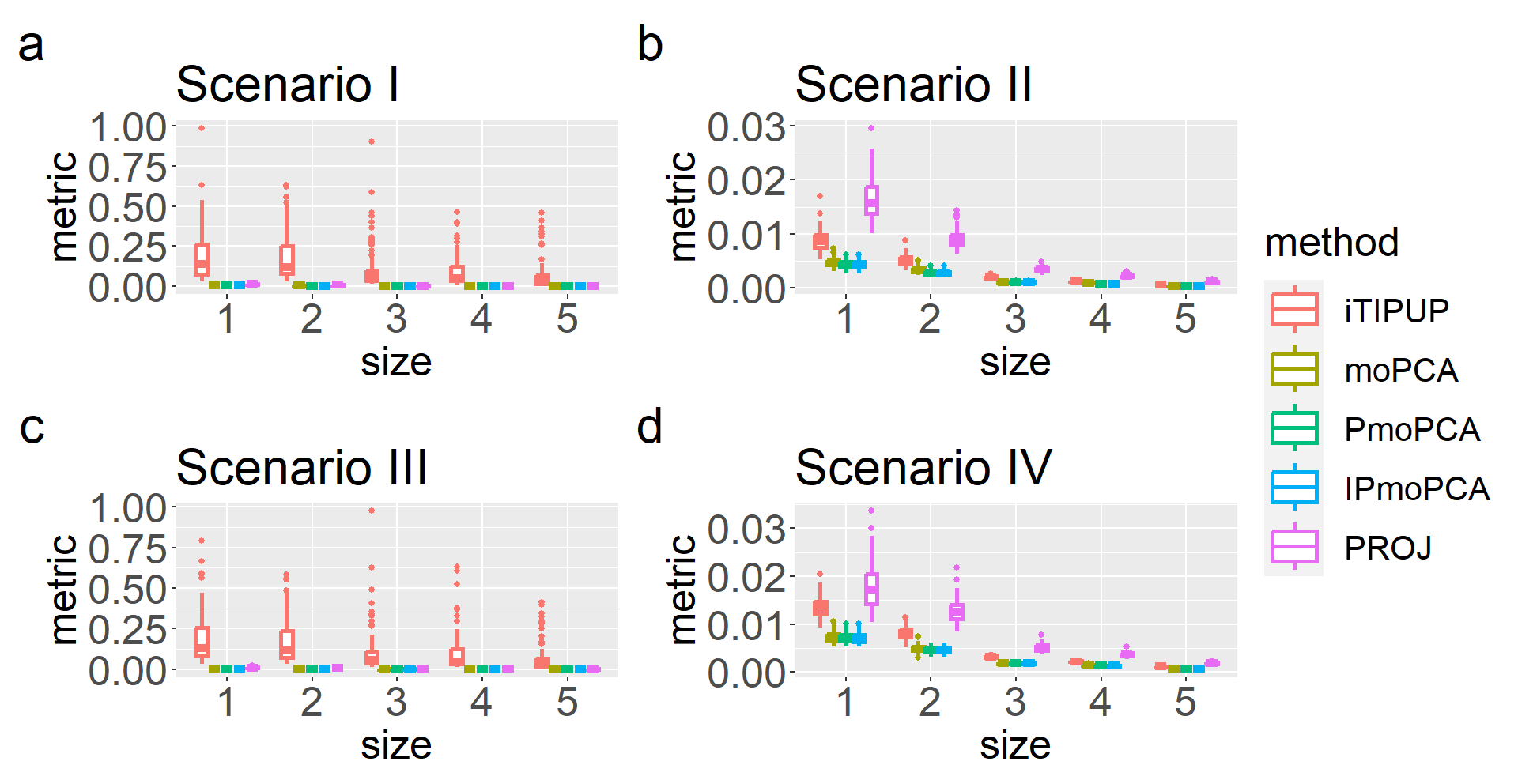}
  \caption{The results for estimated mode-$2$ loading matrix.}
  \label{A2}
\end{figure}

\begin{figure}[H]
  \centering
  \includegraphics[width=5in]{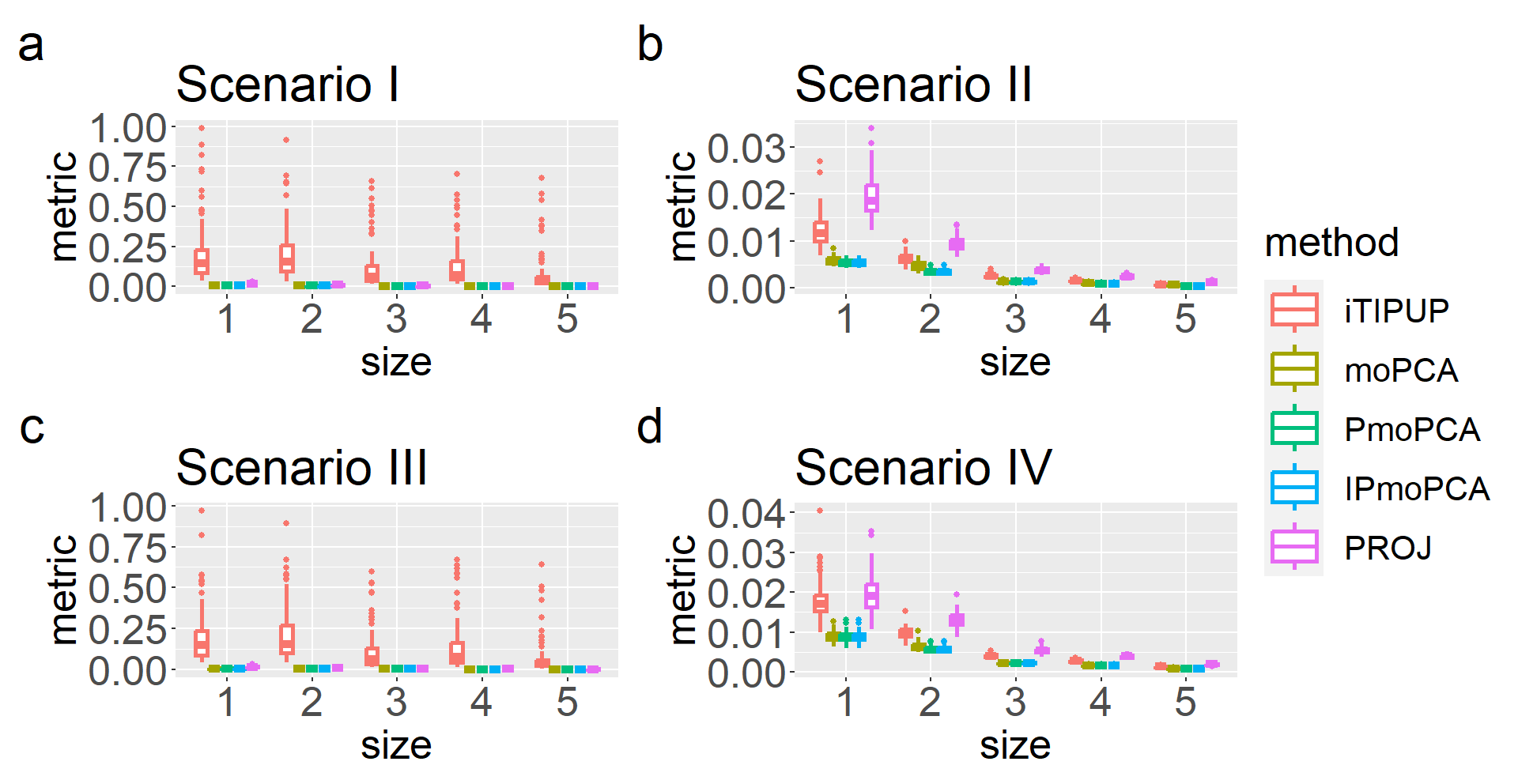}\\
  \caption{The results for estimated mode-$3$ loading matrix.}
  \label{A3}
\end{figure}

\section{Additional Results of Import-Export Data}

\subsection*{Data Description}

\begin{table}[H]
\caption{Countries and Product Categories}
\label{Notation}
\begin{center}
\scalebox{0.8}
{\renewcommand\arraystretch{0.5}
\begin{tabular}{cc|cc}
\hline
Country & Abbreviation & Product Category & Label\\
\hline
Belgium        & BE & Animal $\&$ Animal Products (HS code 01-05) & C1\\
Bulgaria       & BU & Vegetable Products (06-15) & C2\\
Canada         & CA & Foodstuffs (16-24) & C3\\
Denmark        & DK & Mineral Products (25-27) & C4\\
Finland        & FI & Chemicals $\&$ Allied Industries(28-38) & C5\\
France         & FR & Plastics $\&$ Rubbers (39-40) & C6\\
Germany        & DE & Raw Hides, Skins, Leather $\&$ Furs (41-43) & C7\\
Greece         & GR & Wood $\&$ WoodProducts (44-49) & C8\\
Hungary        & HU & Textiles (50-63) & C9\\
Iceland        & IS & Footwear $\&$ Headgear (64-67) & C10\\
Ireland        & IR & Stone $\&$ Glass (68-71) & C11\\
Italy          & IT & Metals(72-83) & C12\\
Mexico         & MX & Machinery $\&$ Electrical (84-85) & C13\\
Norway         & NO & Transportation (86-89) & C14\\
Poland         & PO & Miscellaneous (90-97) & C15\\
Portugal       & PT &&\\
Spain          & ES &&\\
Sweden         & SE &&\\
Switzerland    & CH &&\\
Turkey         & TR &&\\
United States  & US &&\\
United Kingdom & UK&&\\
\hline
\end{tabular}}
\end{center}
\end{table}

\subsubsection*{Hierarchical Clustering Results of Export Countries}

\begin{figure}[H]
\centering
\subfigure[Heatmap for iTIPUP(1).]
{\includegraphics[width=2.5in,height=2in]{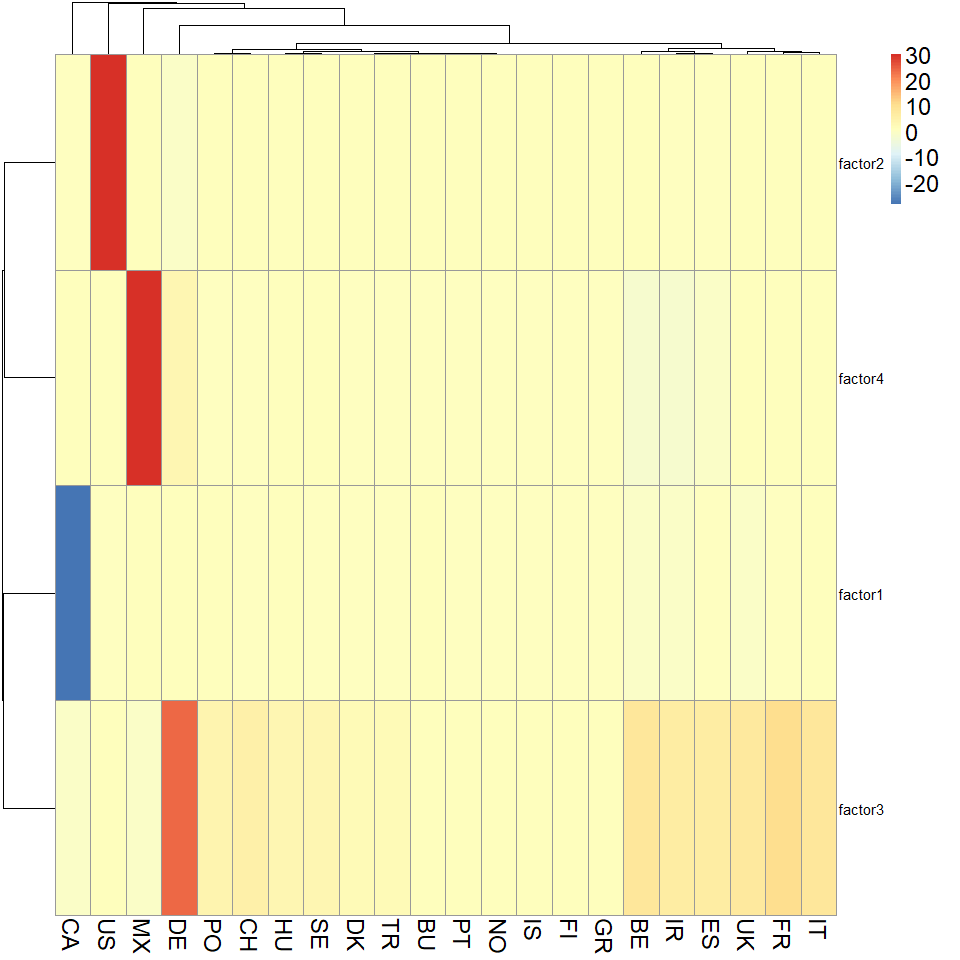}}
\quad
\subfigure[Hierarchical cluster for iTIPUP(1).]
{\includegraphics[width=2.5in,height=2in]{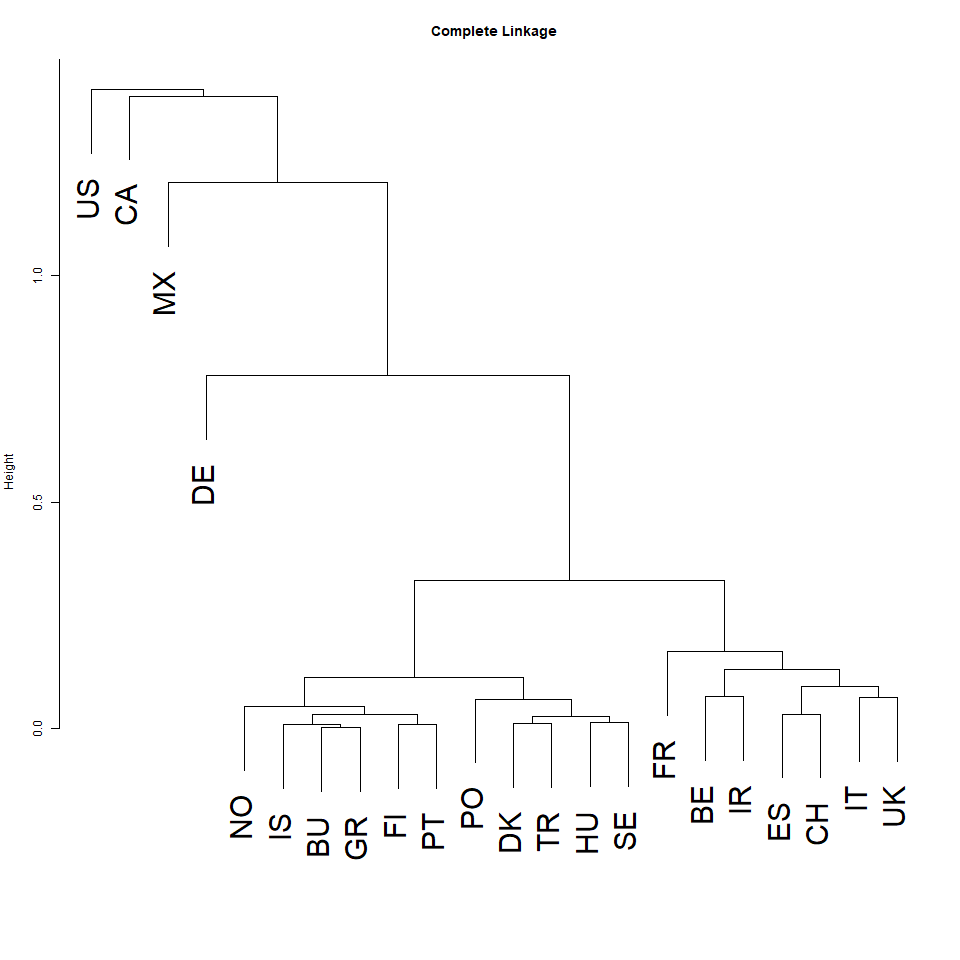}}
  \caption{The results for iTIPUP(1)}
  \label{heatmap11}
\end{figure}

\begin{figure}[H]
\centering
\subfigure[Heatmap for iTIPUP(2).]
{\includegraphics[width=2.5in,height=2in]{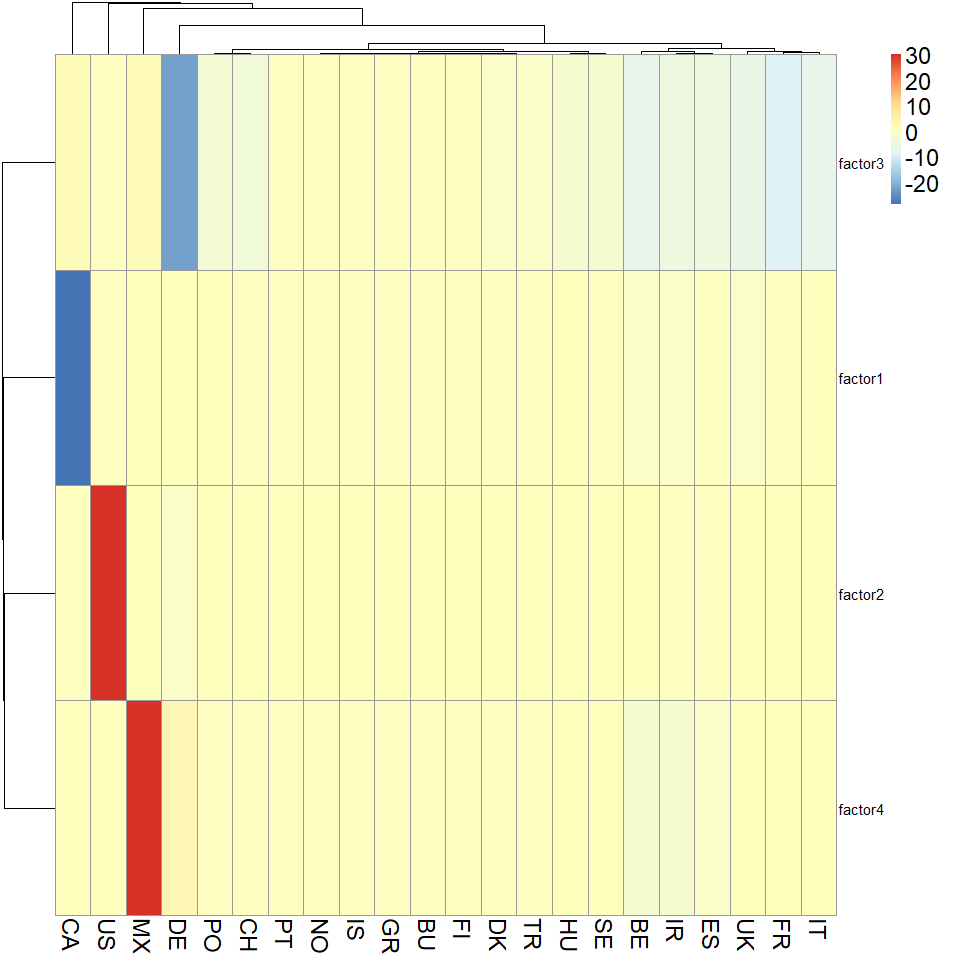}}
\quad
\subfigure[Hierarchical cluster for iTIPUP(2).]
{\includegraphics[width=2.5in,height=2in]{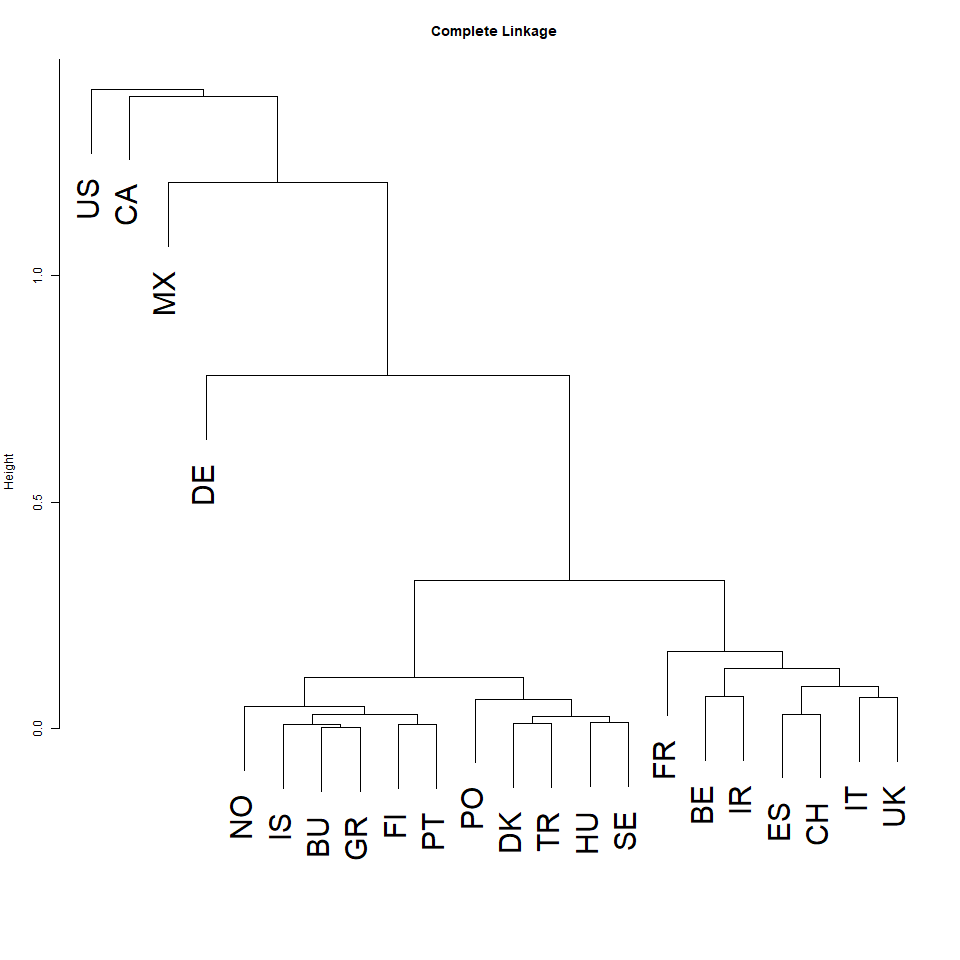}}
  \caption{The results for iTIPUP(2)}
  \label{heatmap12}
\end{figure}

\begin{figure}[H]
\centering
\subfigure[Heatmap for moPCA.]
{\includegraphics[width=2.5in,height=2in]{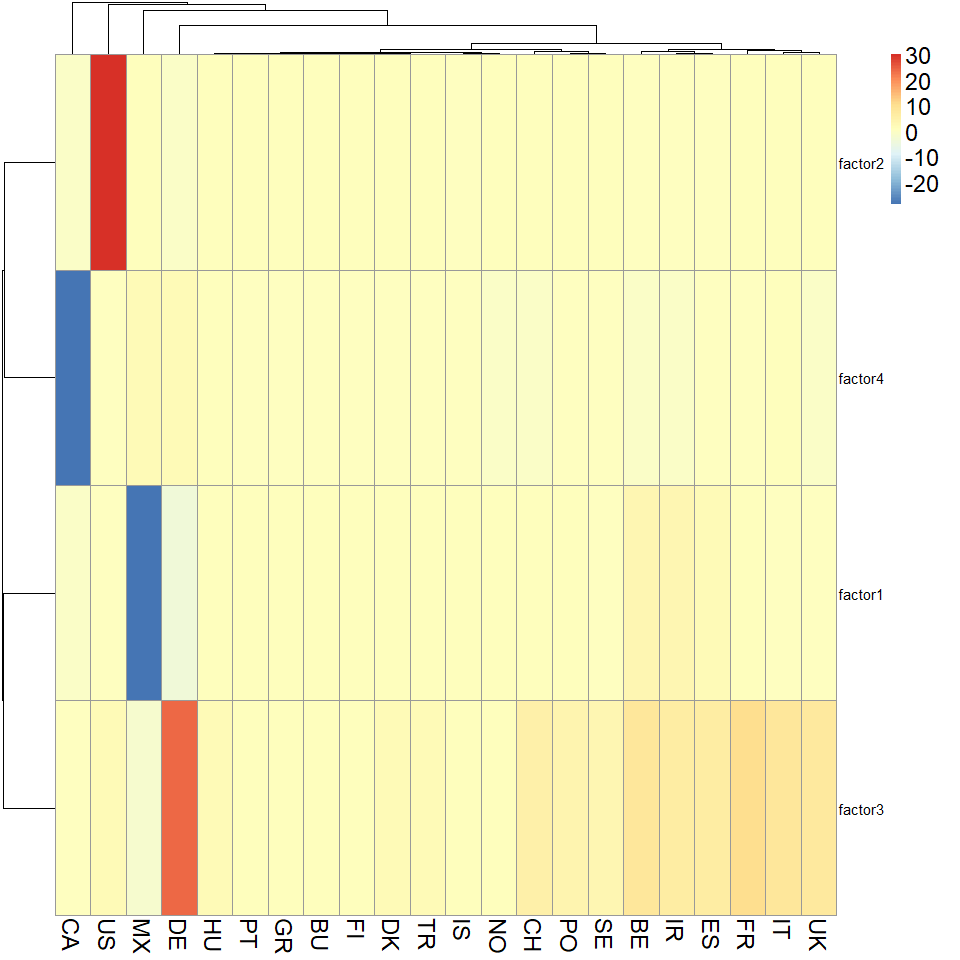}}
\quad
\subfigure[Hierarchical cluster for moPCA.]
{\includegraphics[width=2.5in,height=2in]{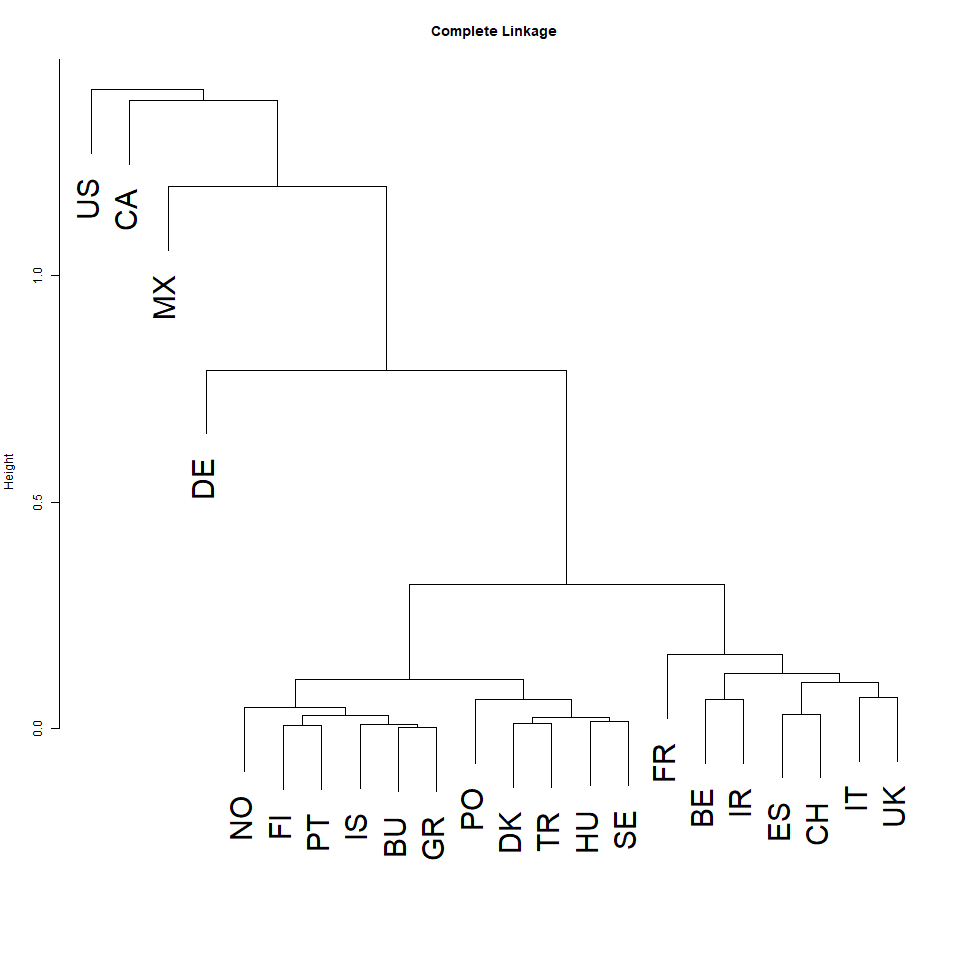}}
  \caption{The results for moPCA}
  \label{heatmap13}
\end{figure}

\begin{figure}[H]
\centering
\subfigure[Heatmap for PmoPCA.]
{\includegraphics[width=2.5in,height=2in]{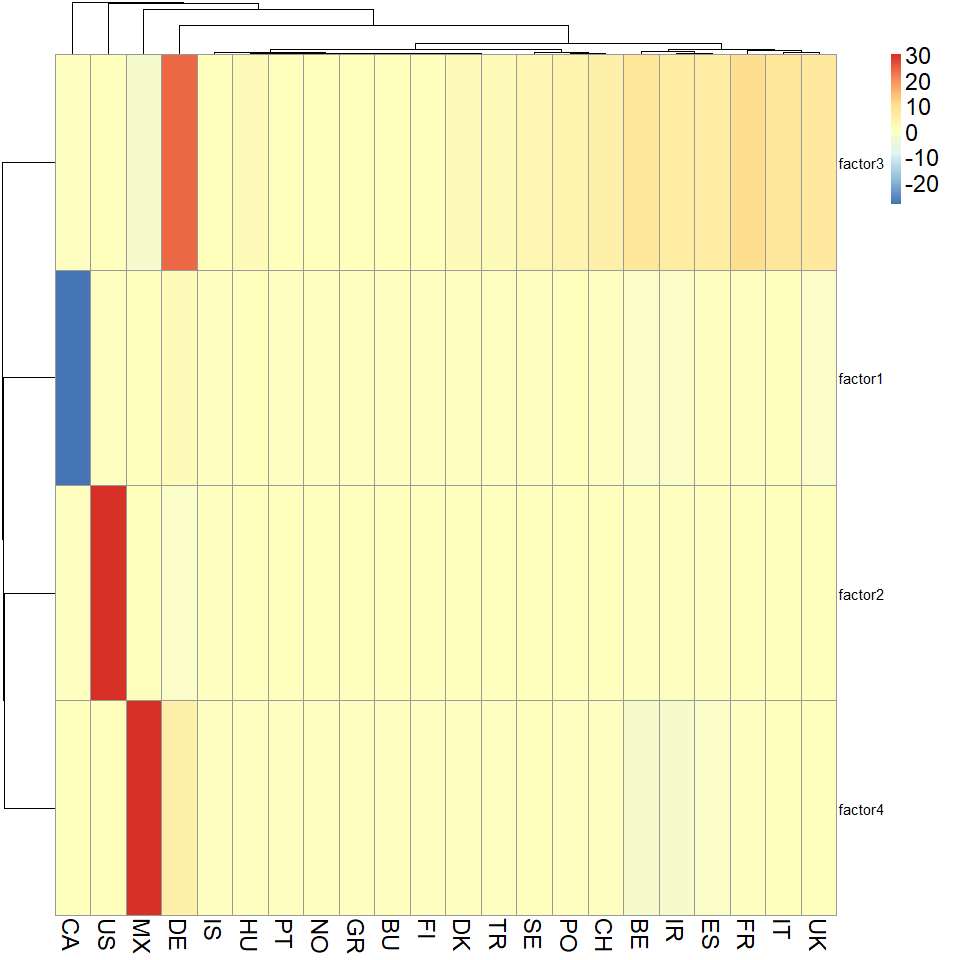}}
\quad
\subfigure[Hierarchical cluster for PmoPCA.]
{\includegraphics[width=2.5in,height=2in]{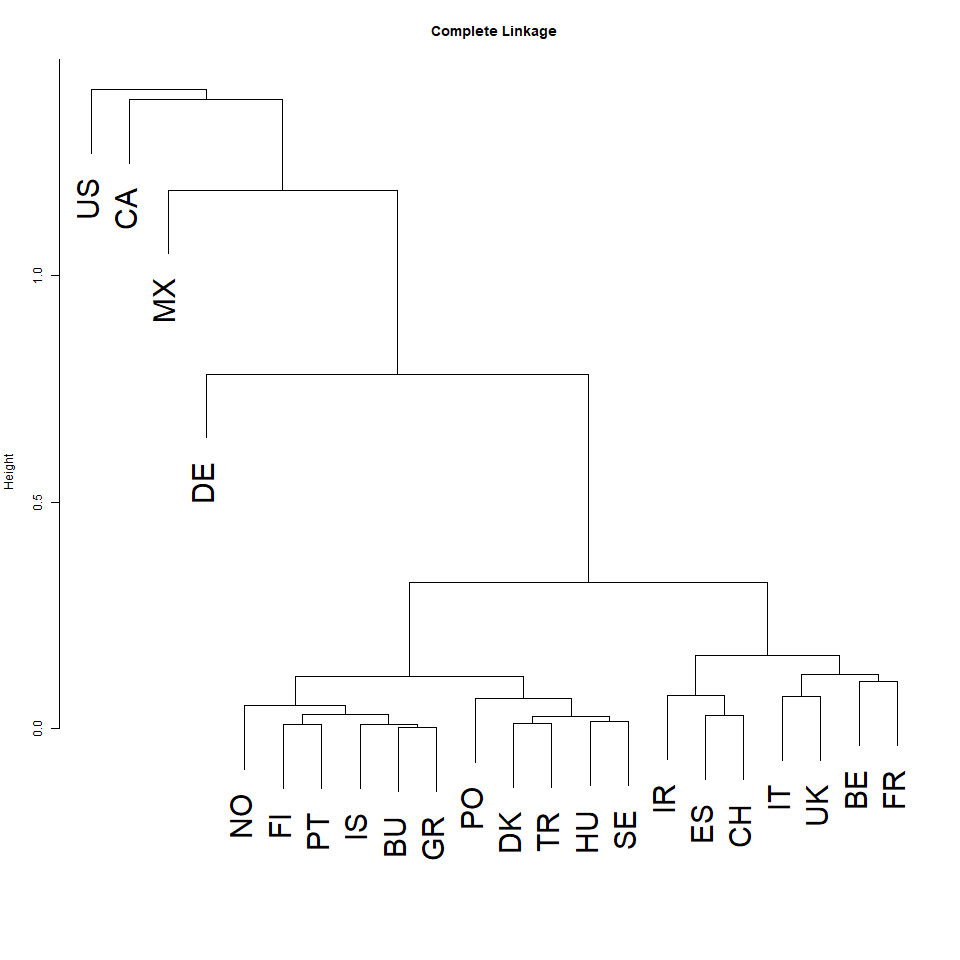}}
  \caption{The results for PmoPCA}
  \label{heatmap14}
\end{figure}

\begin{figure}[H]
\centering
\subfigure[Heatmap for IPmoPCA.]
{\includegraphics[width=2.5in,height=2in]{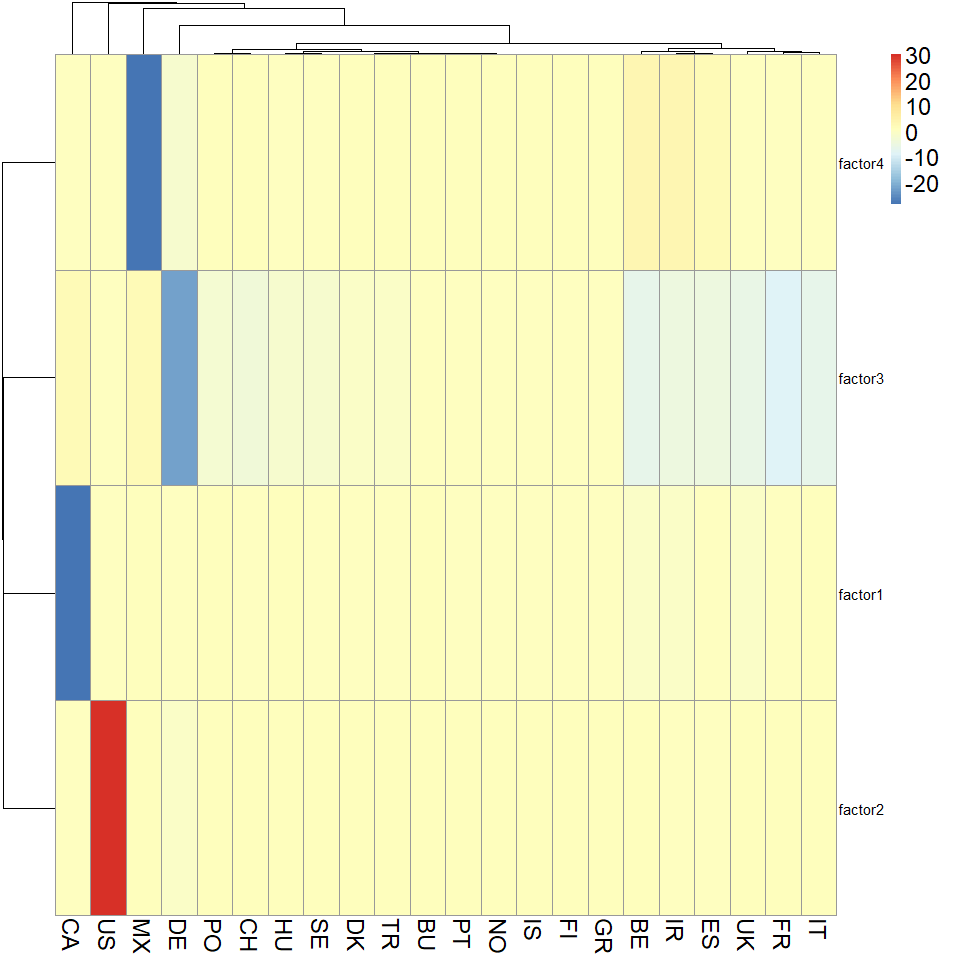}}
\quad
\subfigure[Hierarchical cluster for IPmoPCA.]
{\includegraphics[width=2.5in,height=2in]{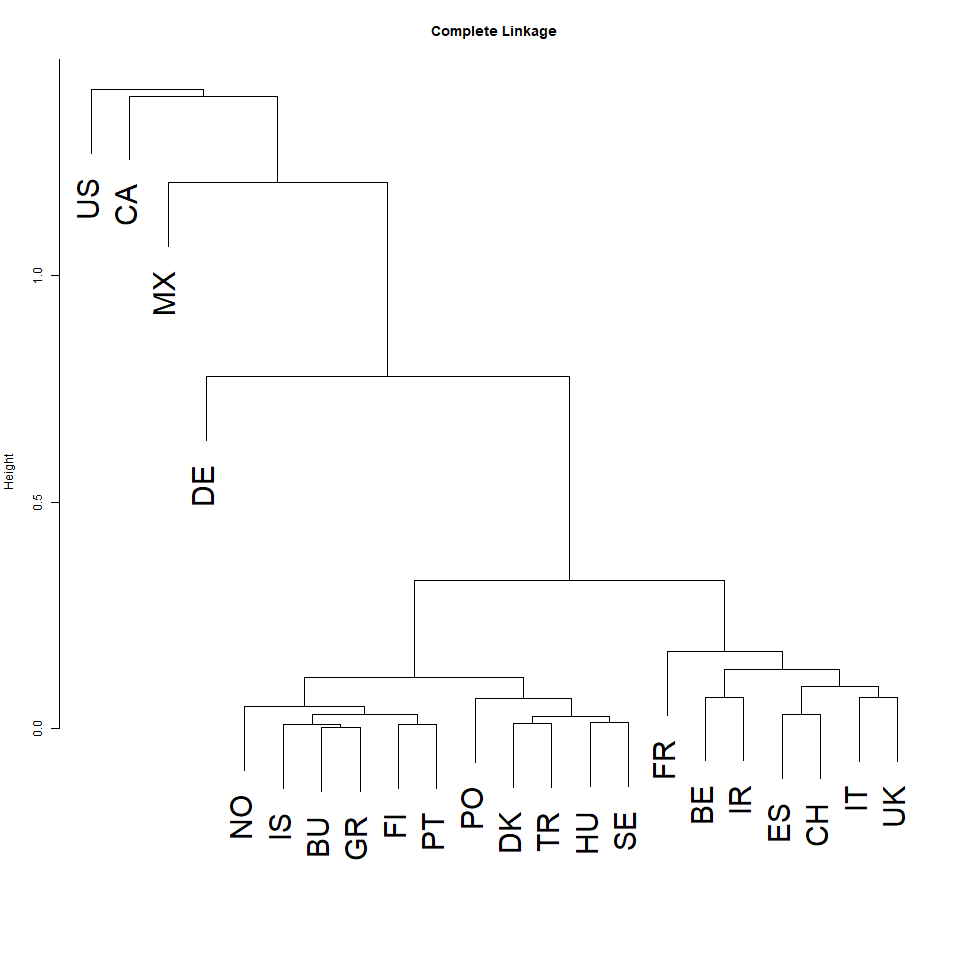}}
  \caption{The results for IPmoPCA}
  \label{heatmap15}
\end{figure}

\begin{figure}[H]
\centering
\subfigure[Heatmap for PROJ.]
{\includegraphics[width=2.5in,height=2in]{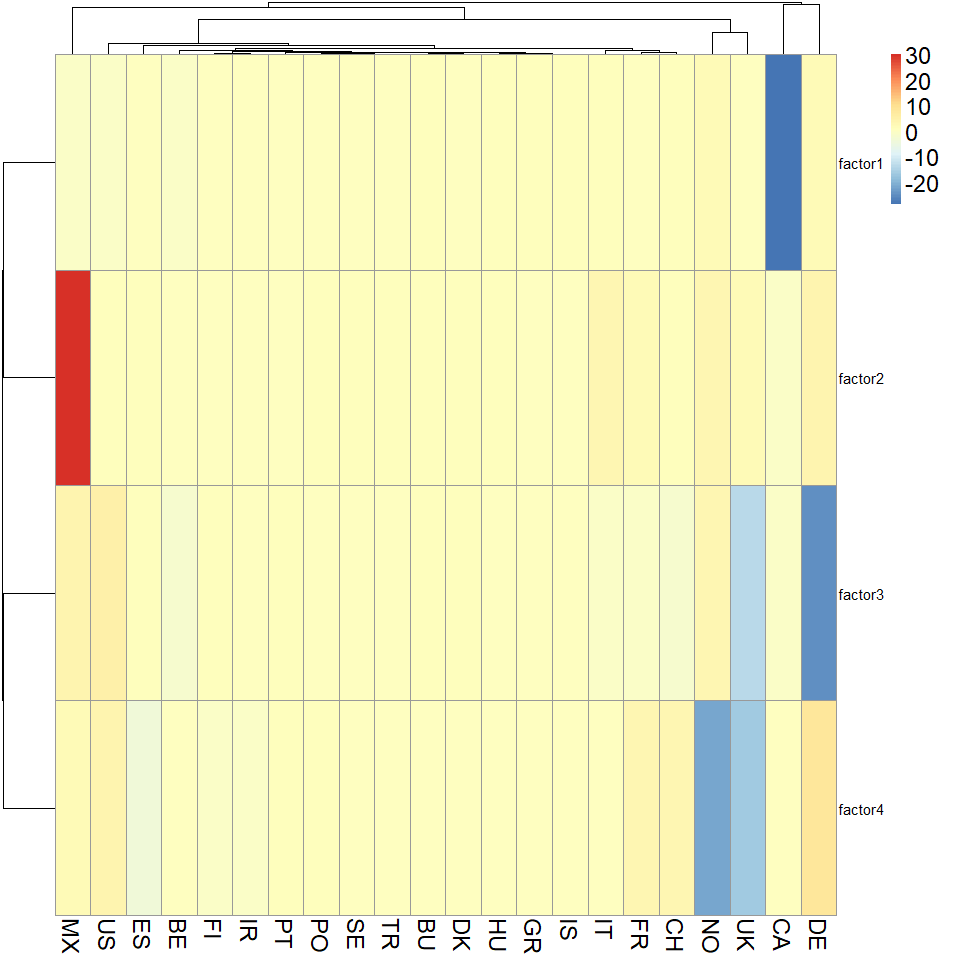}}
\quad
\subfigure[Hierarchical cluster for PROJ.]
{\includegraphics[width=2.5in,height=2in]{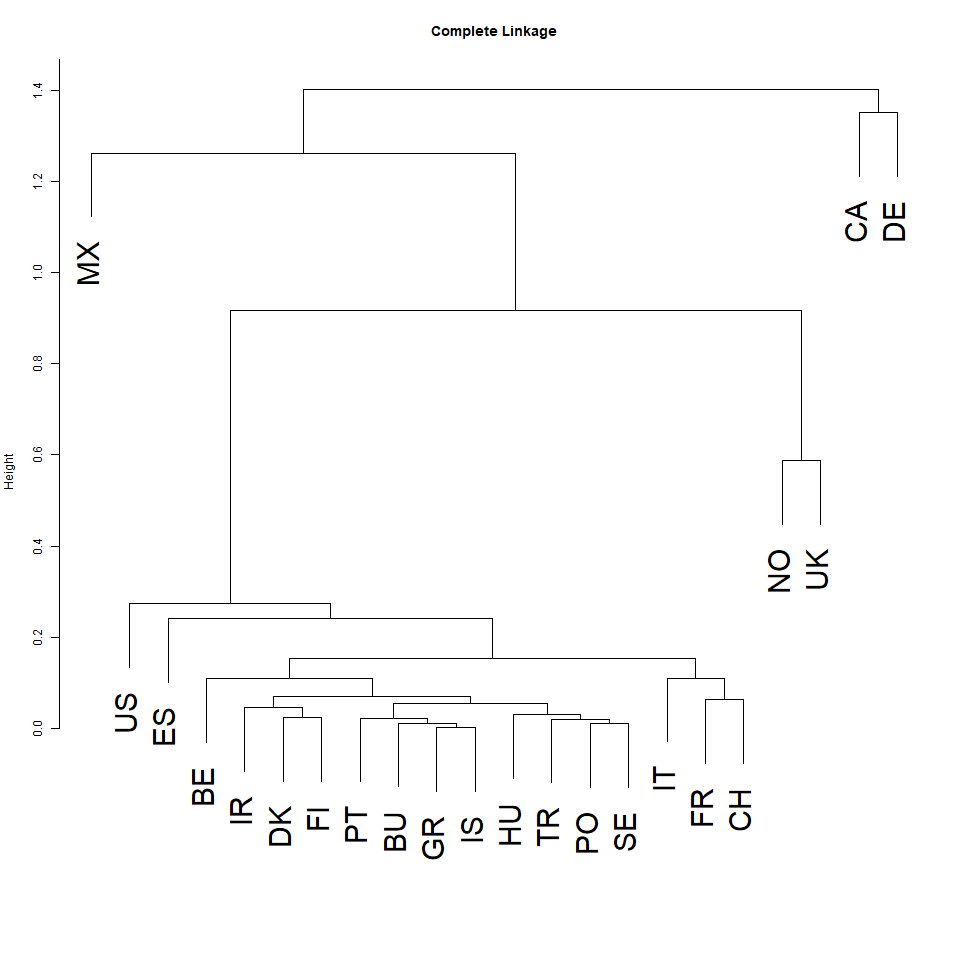}}
  \caption{The results for PROJ}
  \label{heatmap16}
\end{figure}

\subsubsection*{Hierarchical Clustering Results of Import Countries}

\begin{figure}[H]
\centering
\subfigure[Heatmap for iTIPUP(1).]
{\includegraphics[width=2.5in,height=2in]{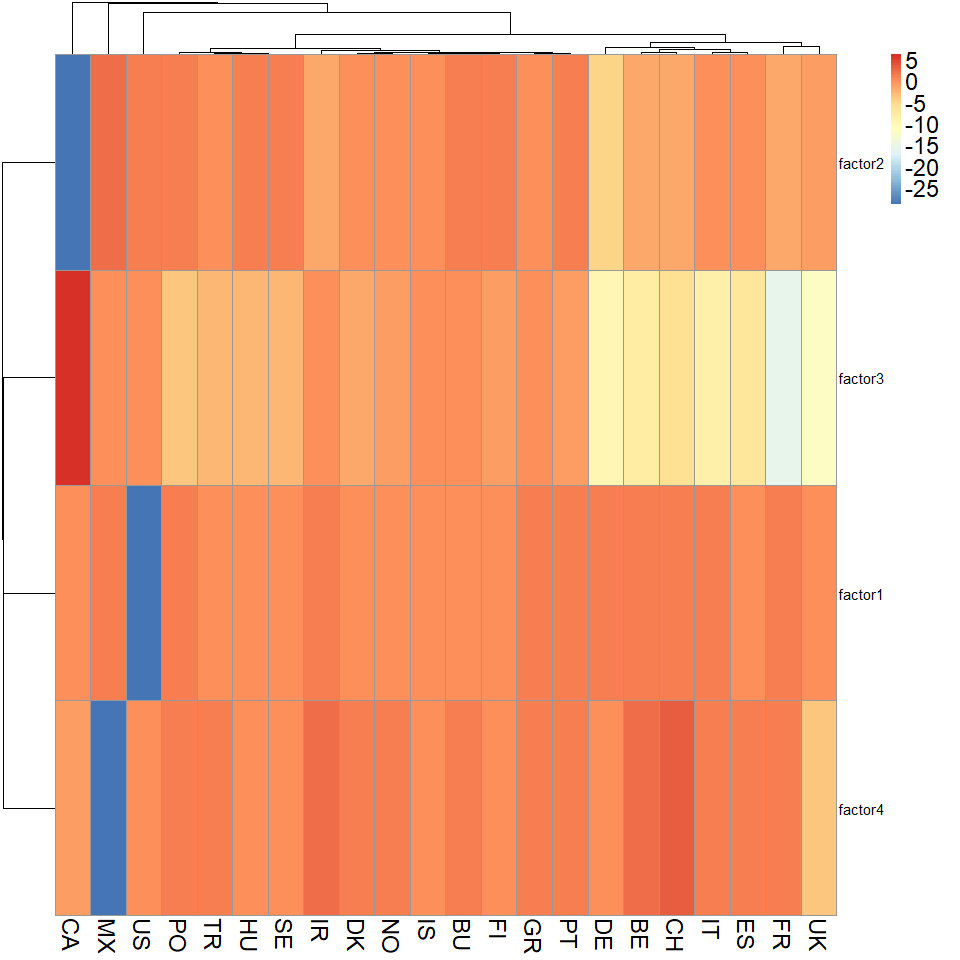}}
\quad
\subfigure[Hierarchical cluster for iTIPUP(1).]
{\includegraphics[width=2.5in,height=2in]{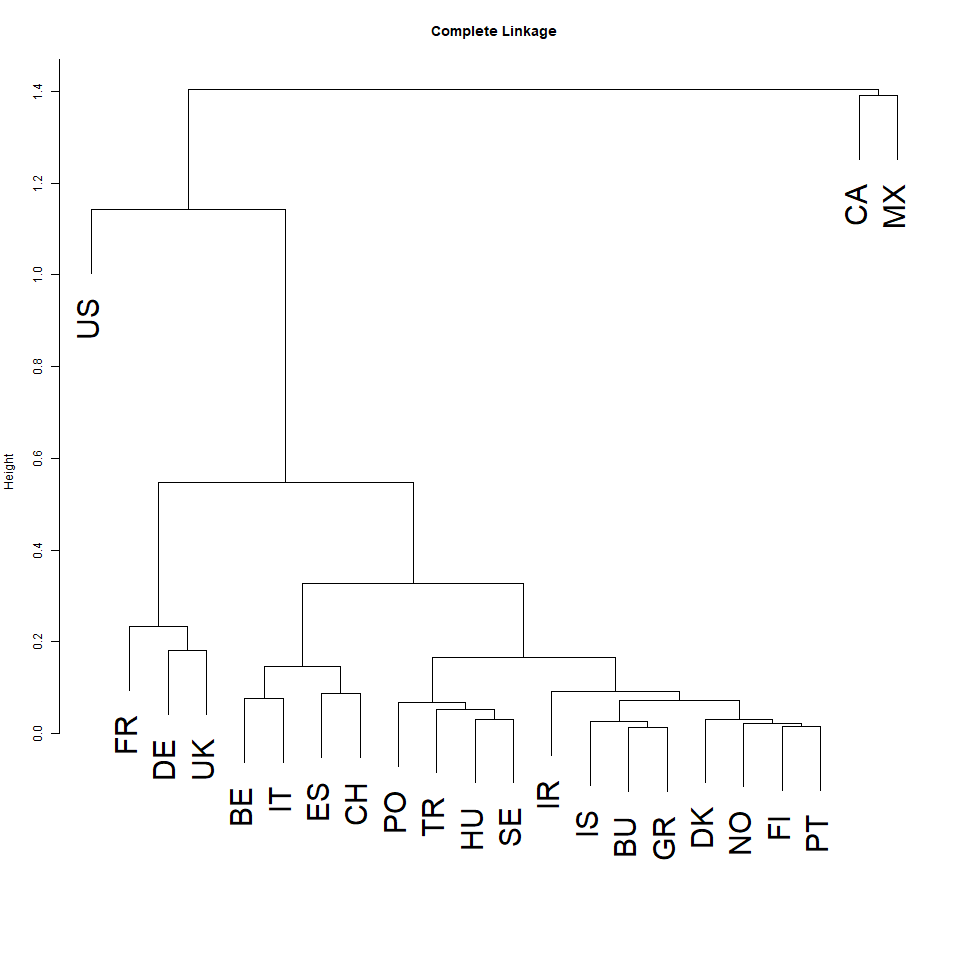}}
  \caption{The results for iTIPUP(1)}
  \label{heatmap21}
\end{figure}

\begin{figure}[H]
\centering
\subfigure[Heatmap for iTIPUP(2).]
{\includegraphics[width=2.5in,height=2in]{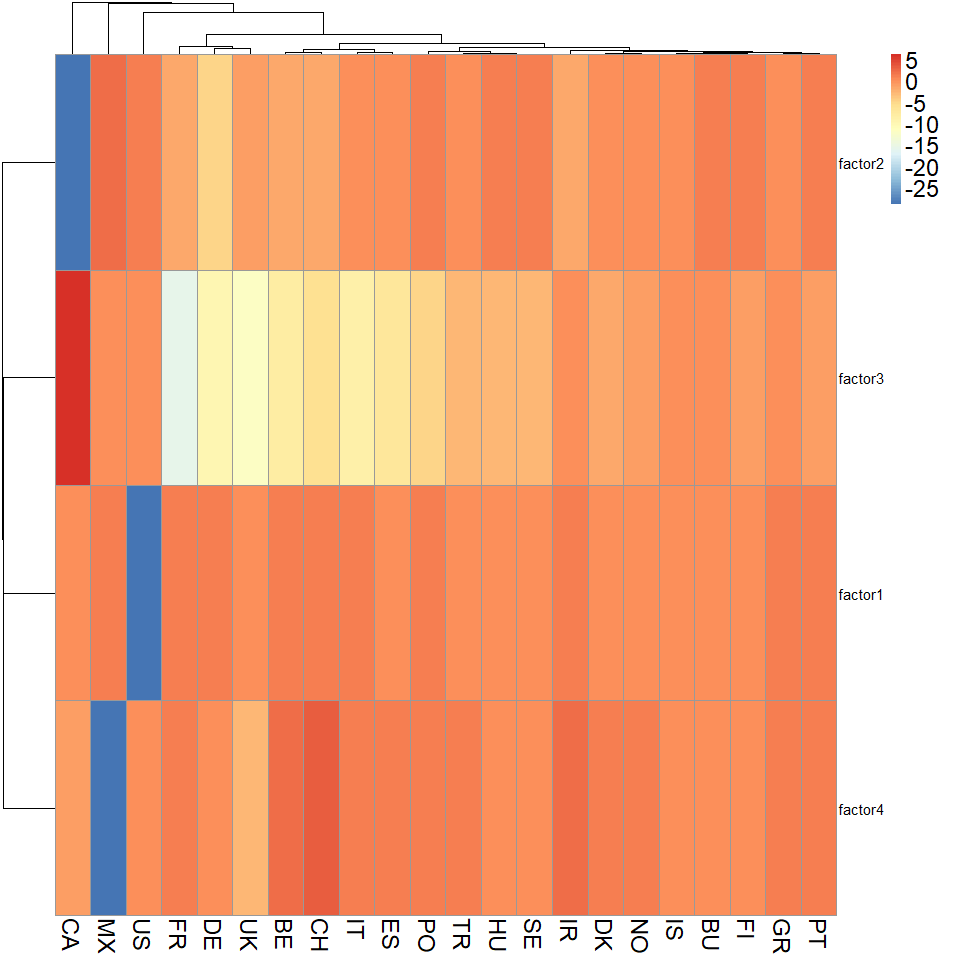}}
\quad
\subfigure[Hierarchical cluster for iTIPUP(2).]
{\includegraphics[width=2.5in,height=2in]{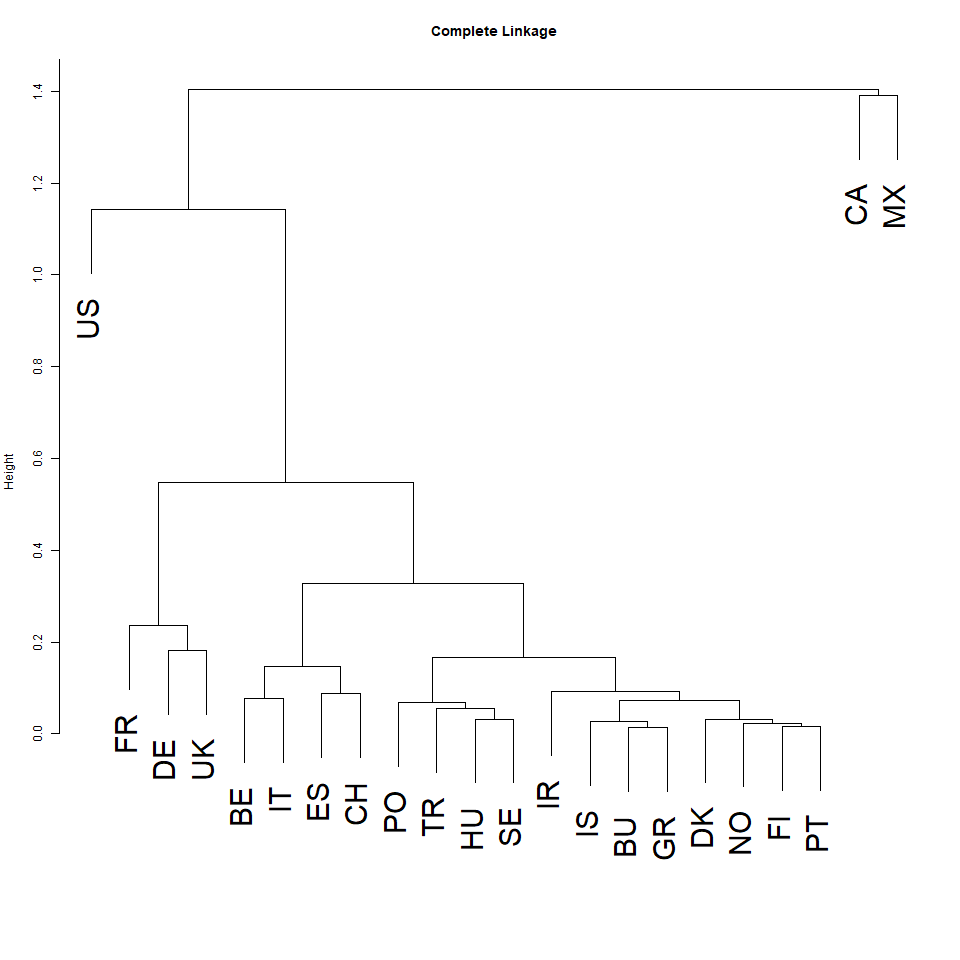}}
  \caption{The results for iTIPUP(2)}
  \label{heatmap22}
\end{figure}

\begin{figure}[H]
\centering
\subfigure[Heatmap for moPCA.]
{\includegraphics[width=2.5in,height=2in]{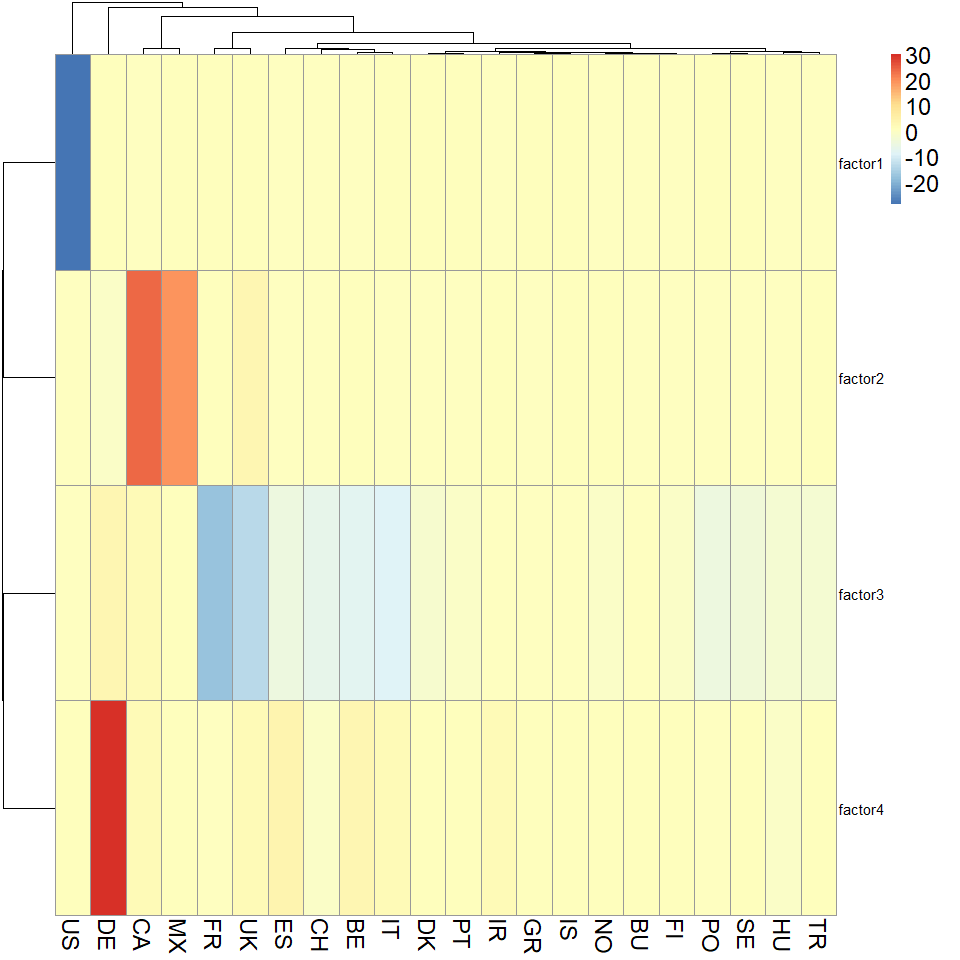}}
\quad
\subfigure[Hierarchical cluster for moPCA.]
{\includegraphics[width=2.5in,height=2in]{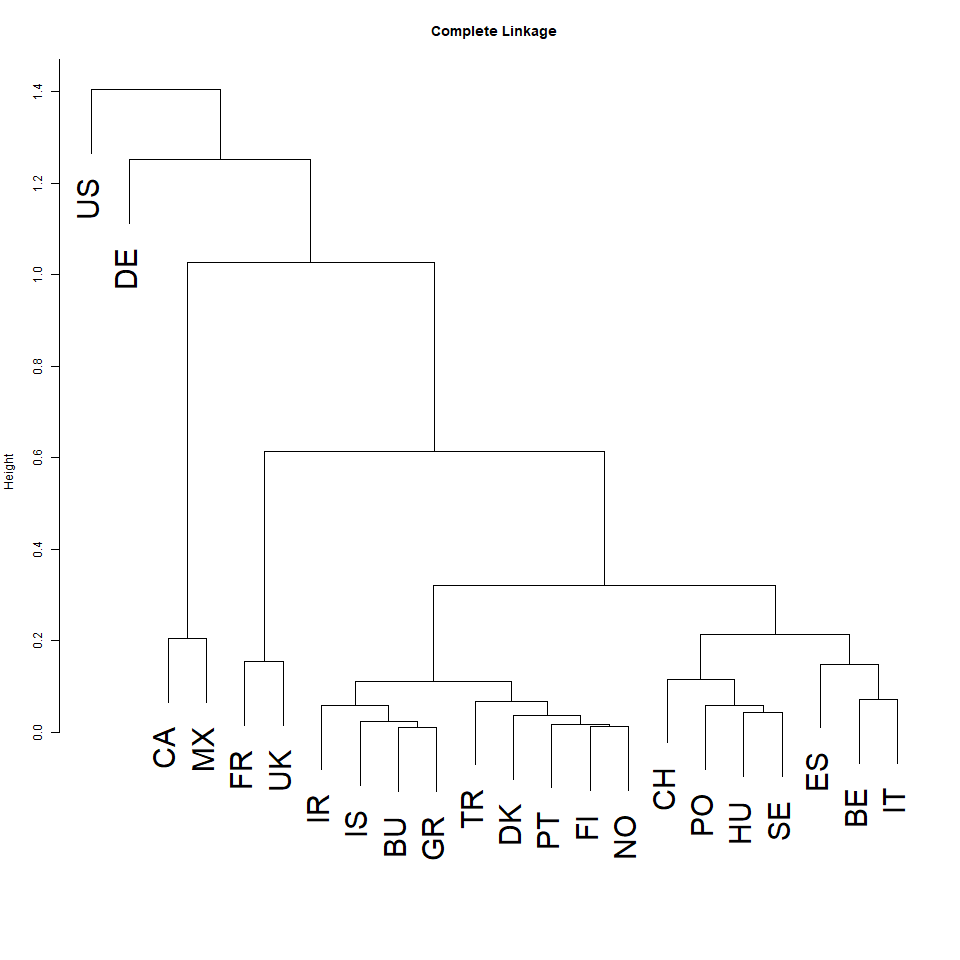}}
  \caption{The results for moPCA}
  \label{heatmap23}
\end{figure}

\begin{figure}[H]
\centering
\subfigure[Heatmap for PmoPCA.]
{\includegraphics[width=2.5in,height=2in]{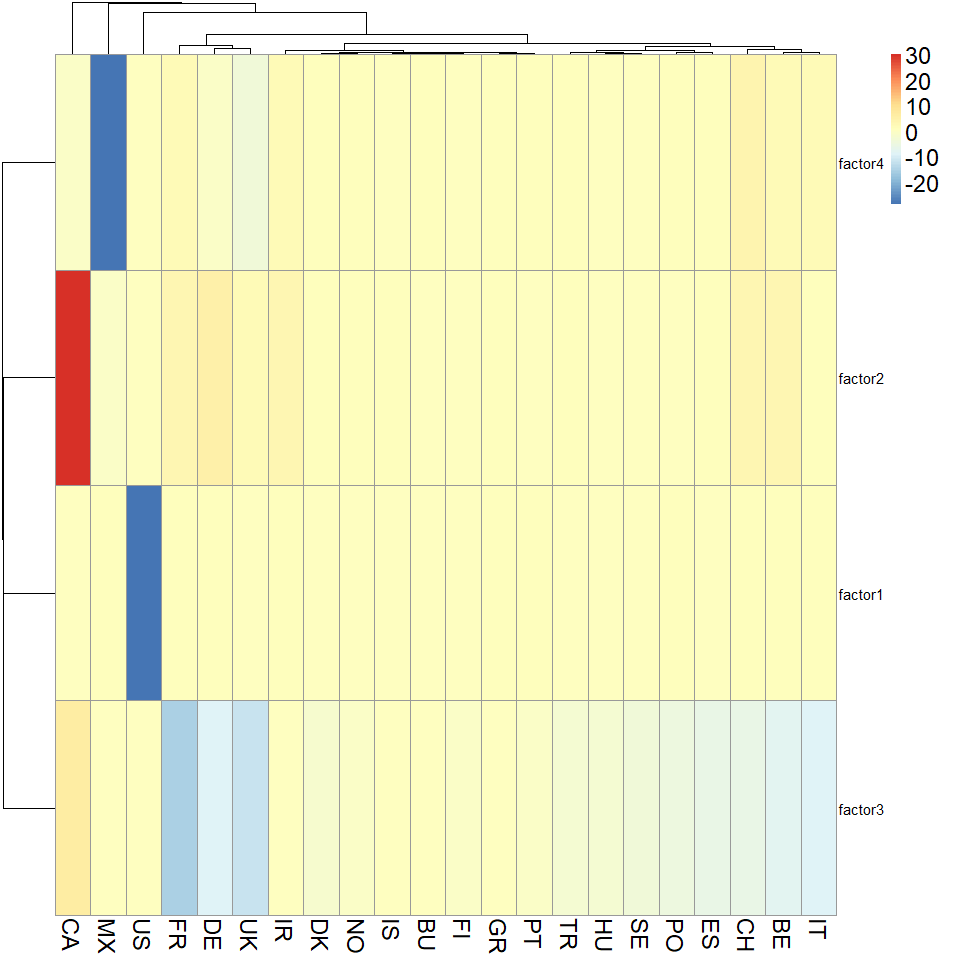}}
\quad
\subfigure[Hierarchical cluster for PmoPCA.]
{\includegraphics[width=2.5in,height=2in]{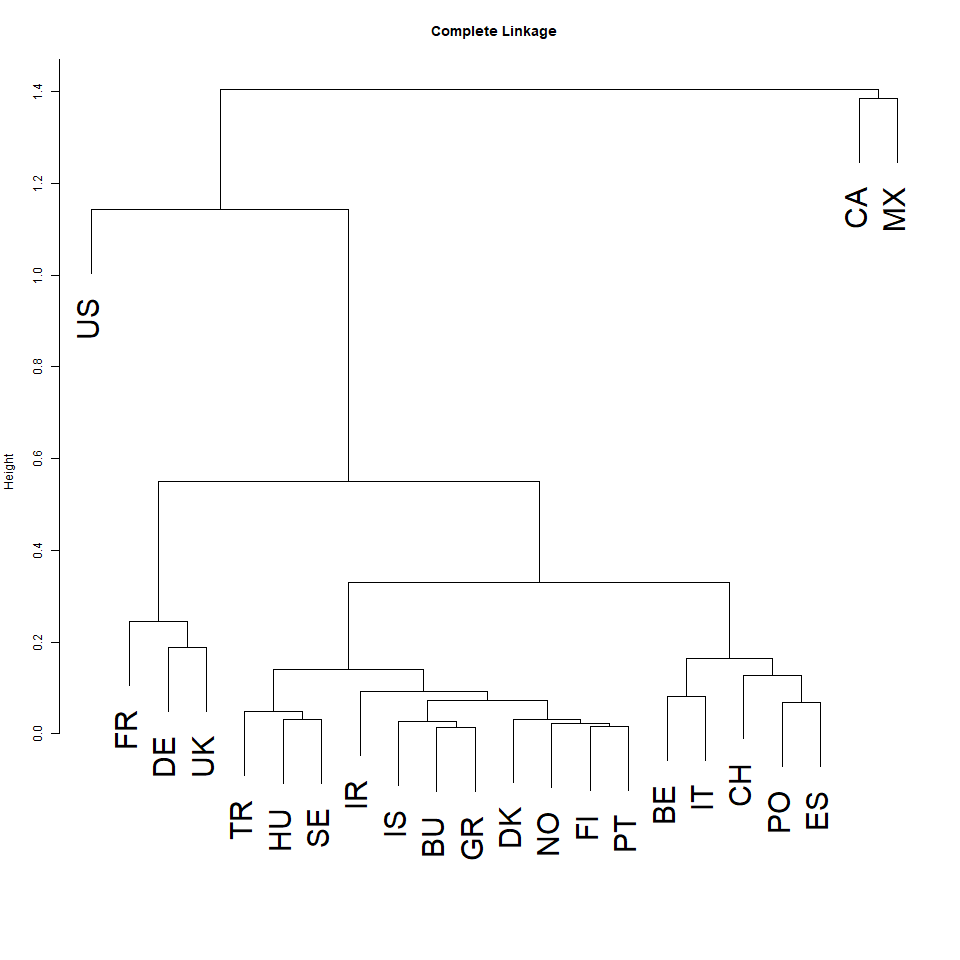}}
  \caption{The results for PmoPCA}
  \label{heatmap24}
\end{figure}

\begin{figure}[H]
\centering
\subfigure[Heatmap for IPmoPCA.]
{\includegraphics[width=2.5in,height=2in]{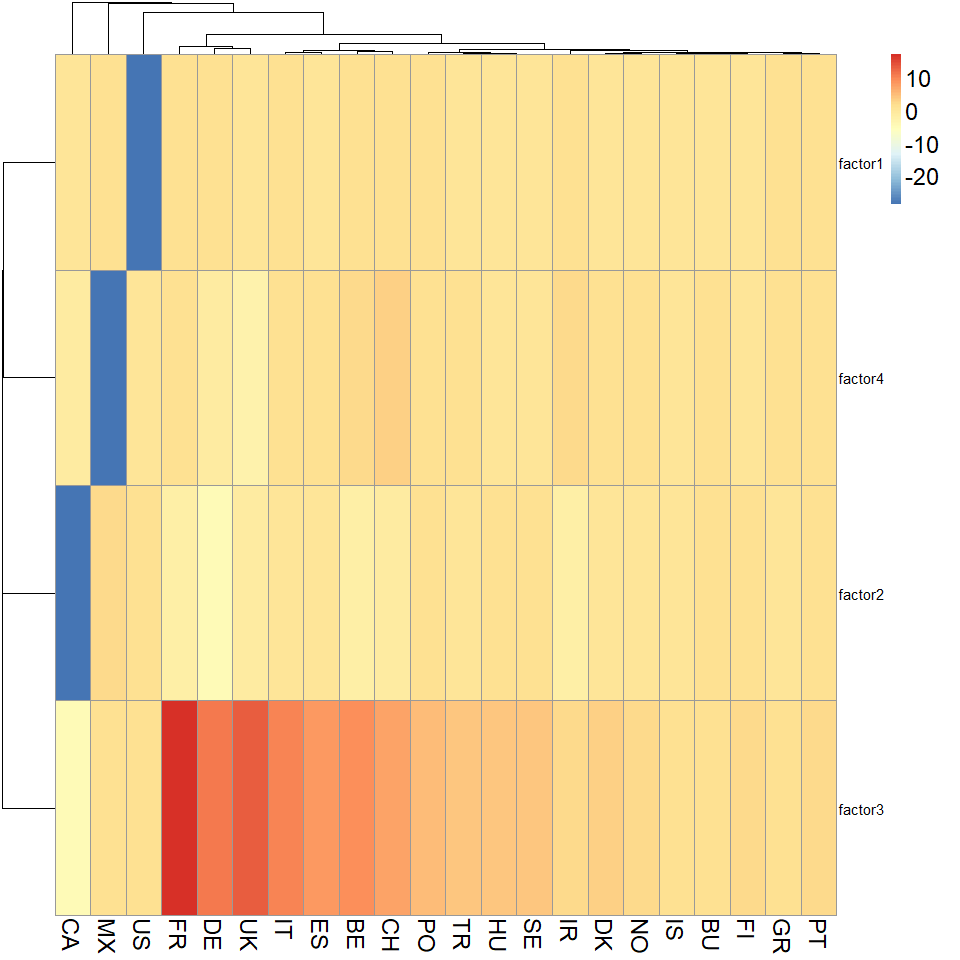}}
\quad
\subfigure[Hierarchical cluster for IPmoPCA.]
{\includegraphics[width=2.5in,height=2in]{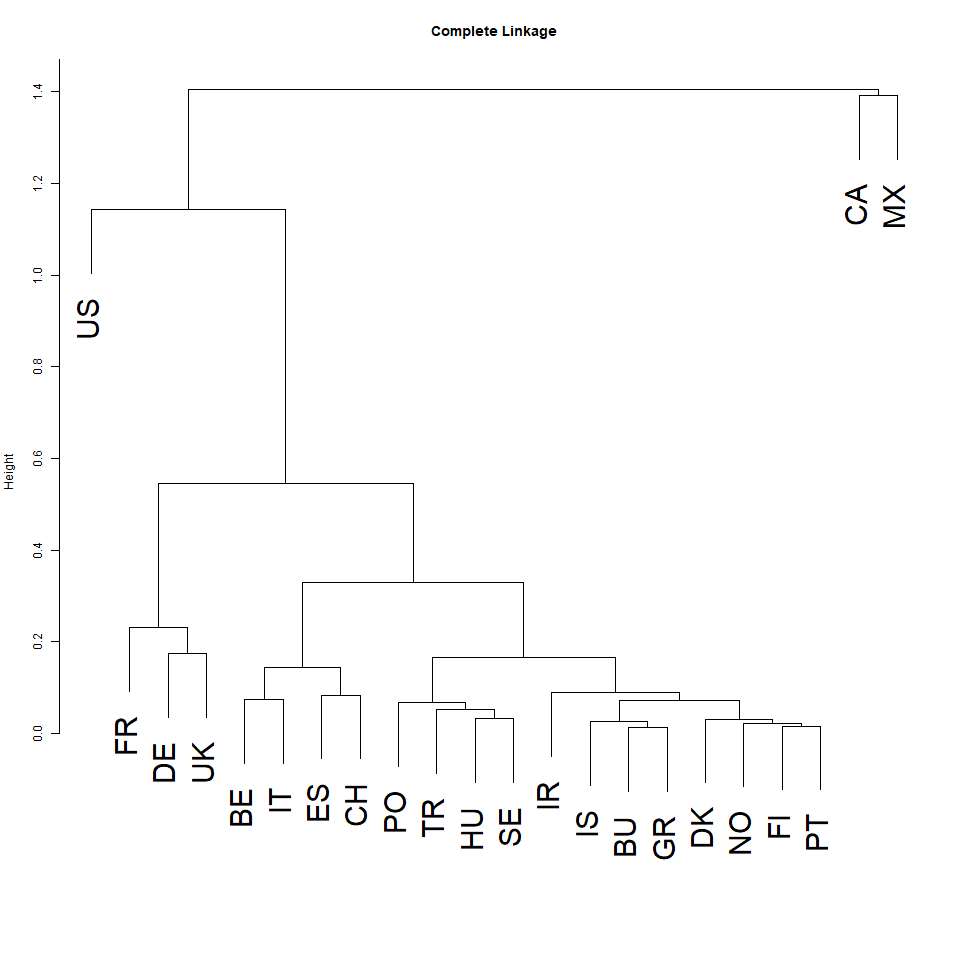}}
  \caption{The results for IPmoPCA}
  \label{heatmap25}
\end{figure}

\begin{figure}[H]
\centering
\subfigure[Heatmap for PROJ.]
{\includegraphics[width=2.5in,height=2in]{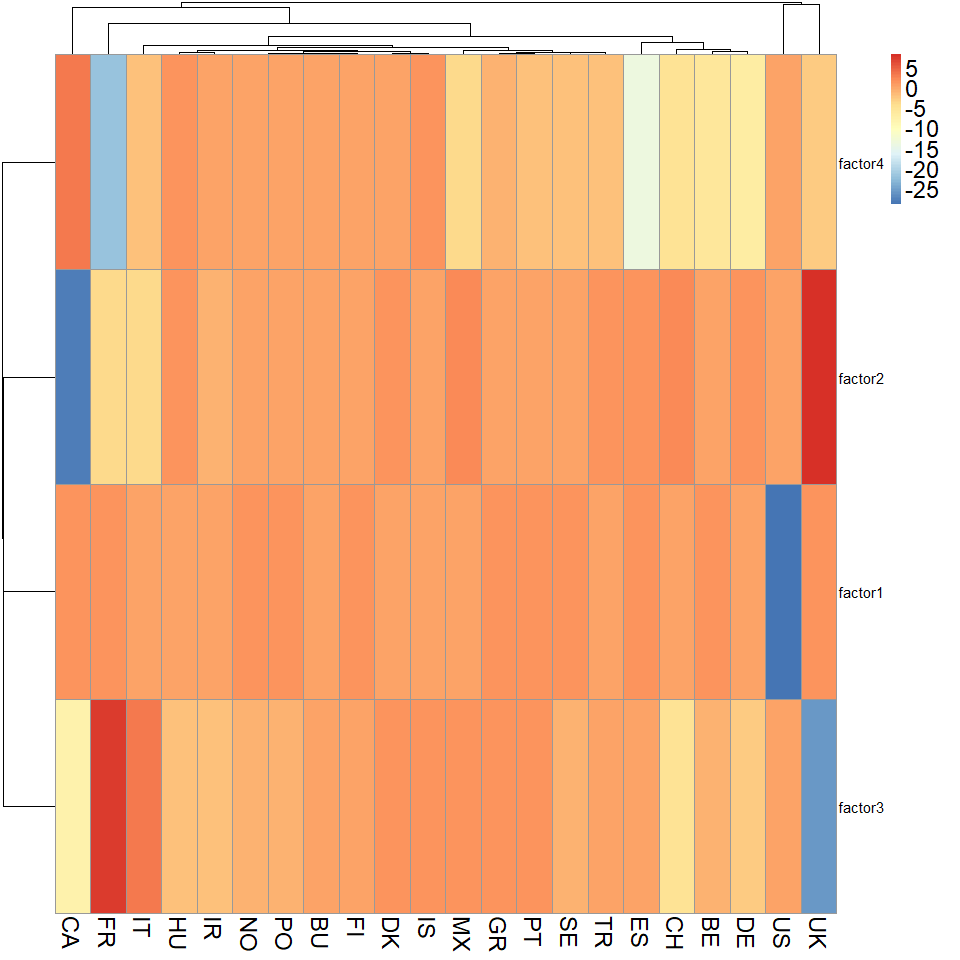}}
\quad
\subfigure[Hierarchical cluster for IPmoPCA.]
{\includegraphics[width=2.5in,height=2in]{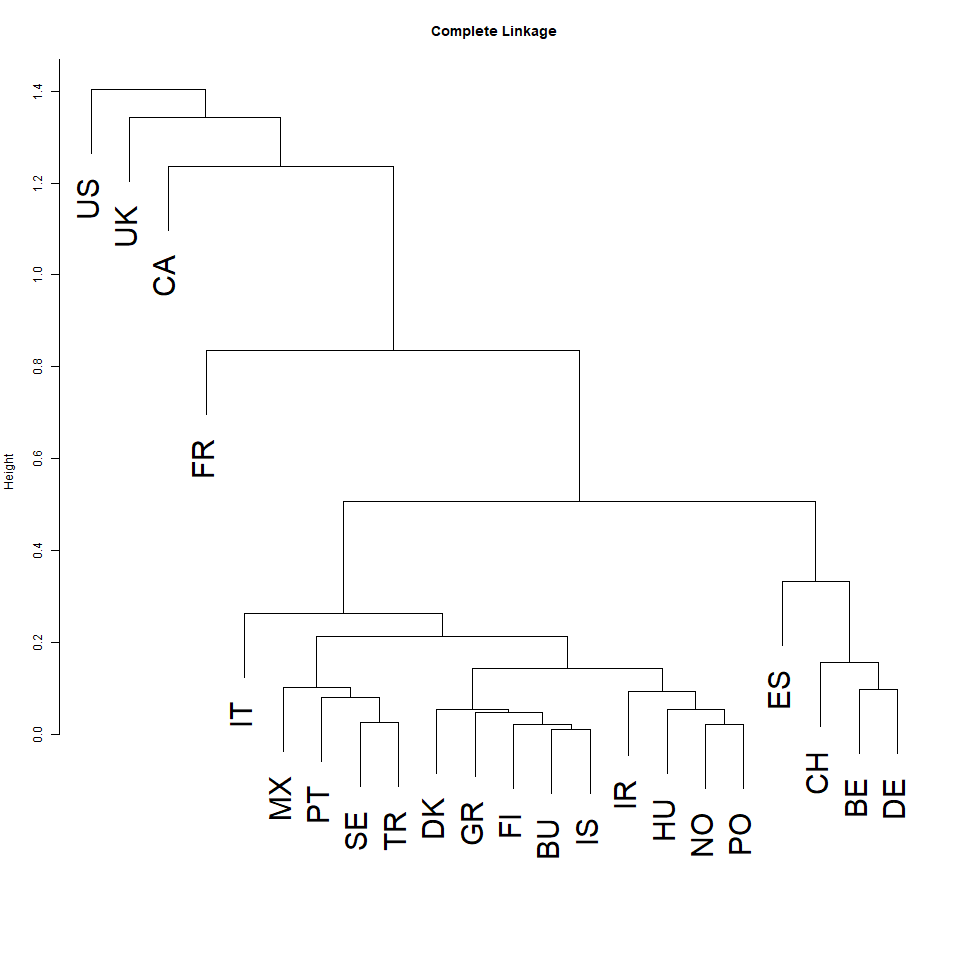}}
  \caption{The results for PROJ}
  \label{heatmap26}
\end{figure}

\subsubsection*{Hierarchical Clustering Results of Product Categories}

\begin{figure}[H]
\centering
\subfigure[Heatmap for iTIPUP(1).]
{\includegraphics[width=2.5in,height=2in]{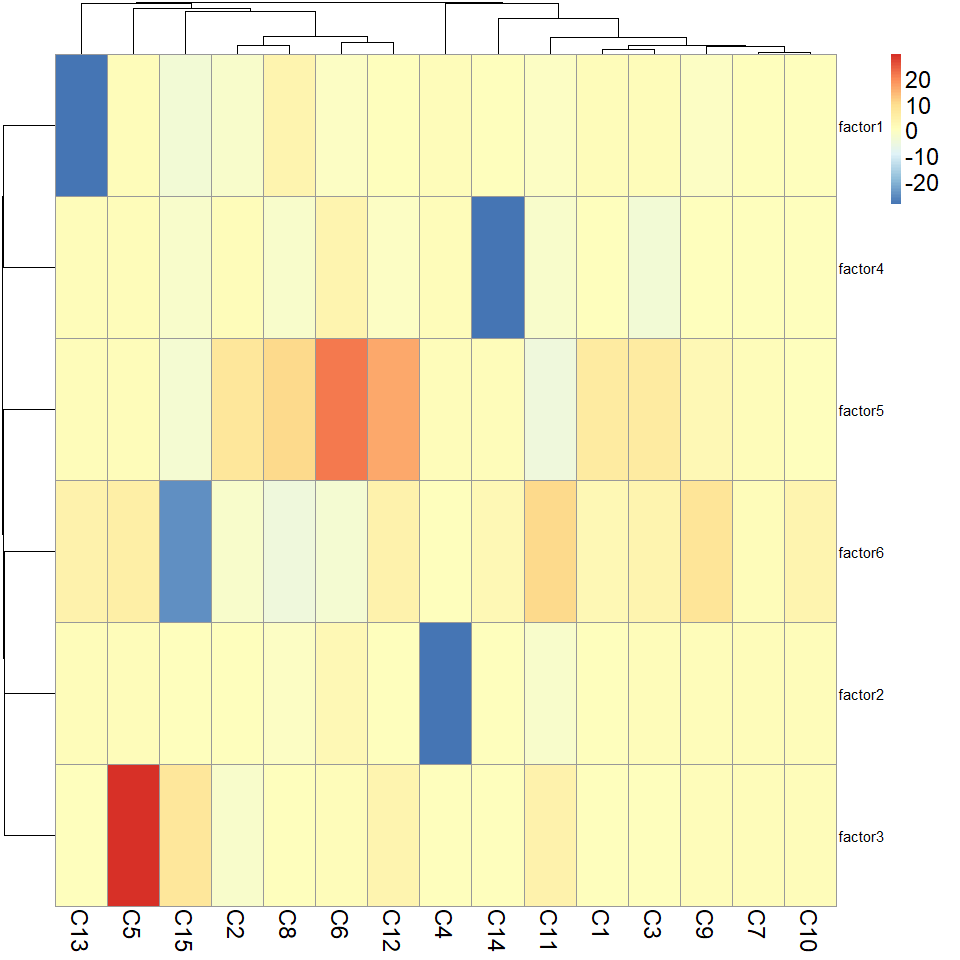}}
\quad
\subfigure[Hierarchical cluster for iTIPUP(1).]
{\includegraphics[width=2.5in,height=2in]{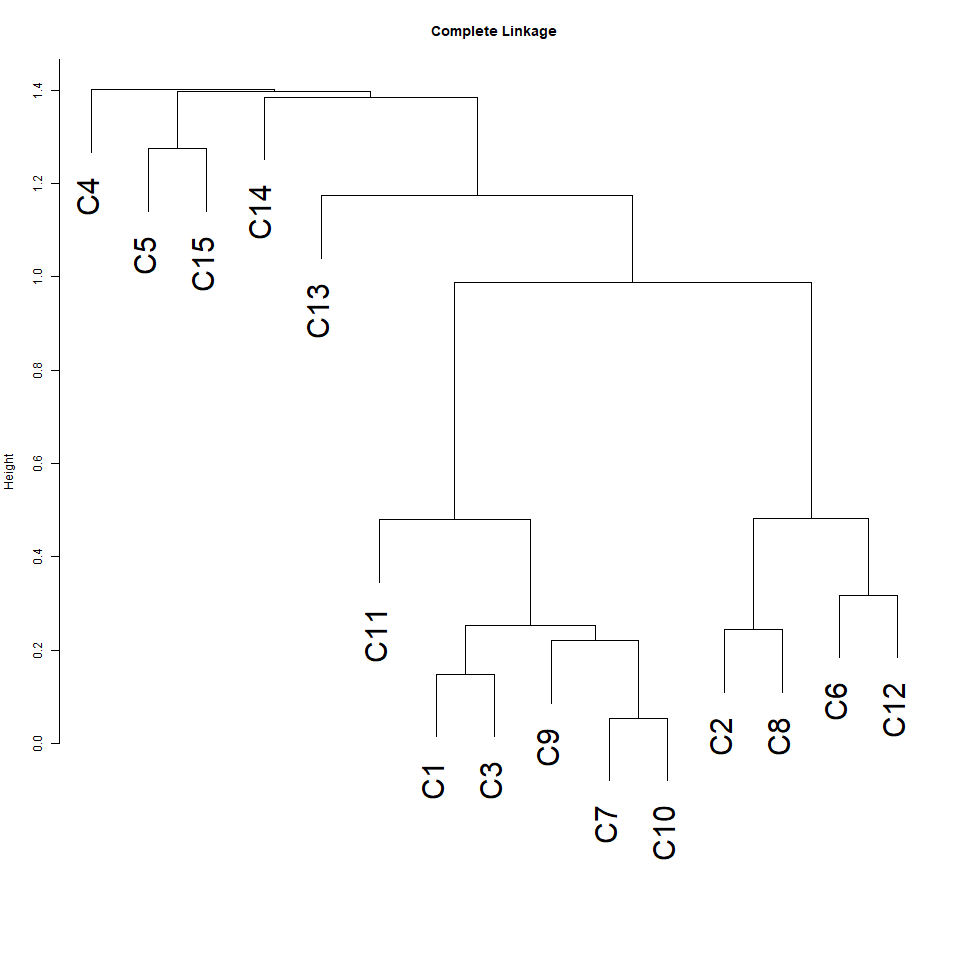}}
  \caption{The results for iTIPUP(1)}
  \label{heatmap31}
\end{figure}

\begin{figure}[H]
\centering
\subfigure[Heatmap for iTIPUP(2).]
{\includegraphics[width=2.5in,height=2in]{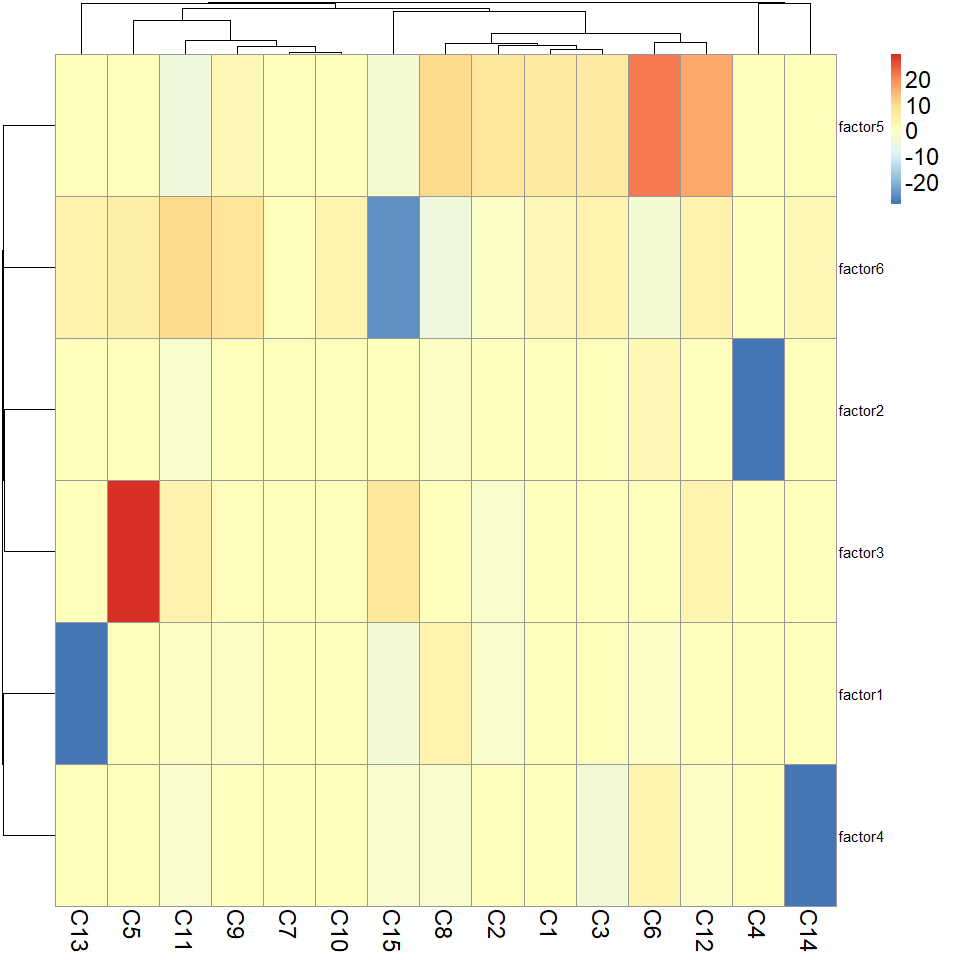}}
\quad
\subfigure[Hierarchical cluster for iTIPUP(2).]
{\includegraphics[width=2.5in,height=2in]{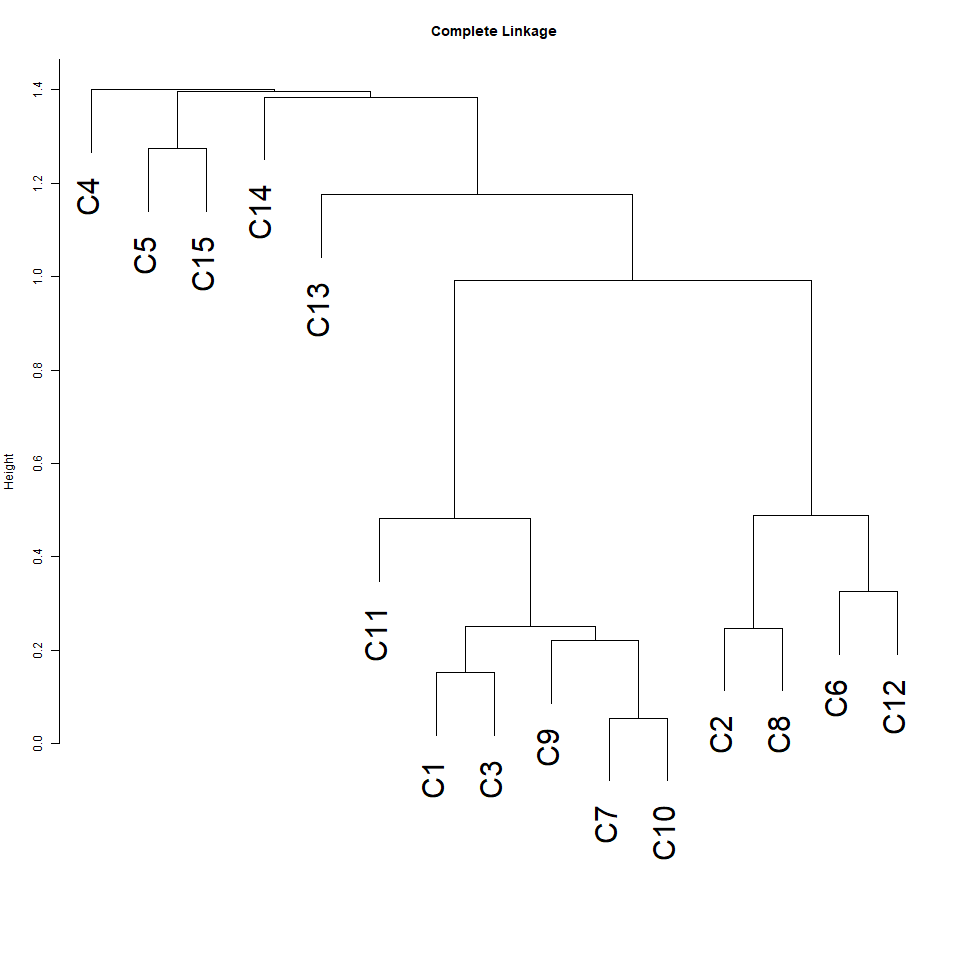}}
  \caption{The results for iTIPUP(2)}
  \label{heatmap32}
\end{figure}

\begin{figure}[H]
\centering
\subfigure[Heatmap for moPCA.]
{\includegraphics[width=2.5in,height=2in]{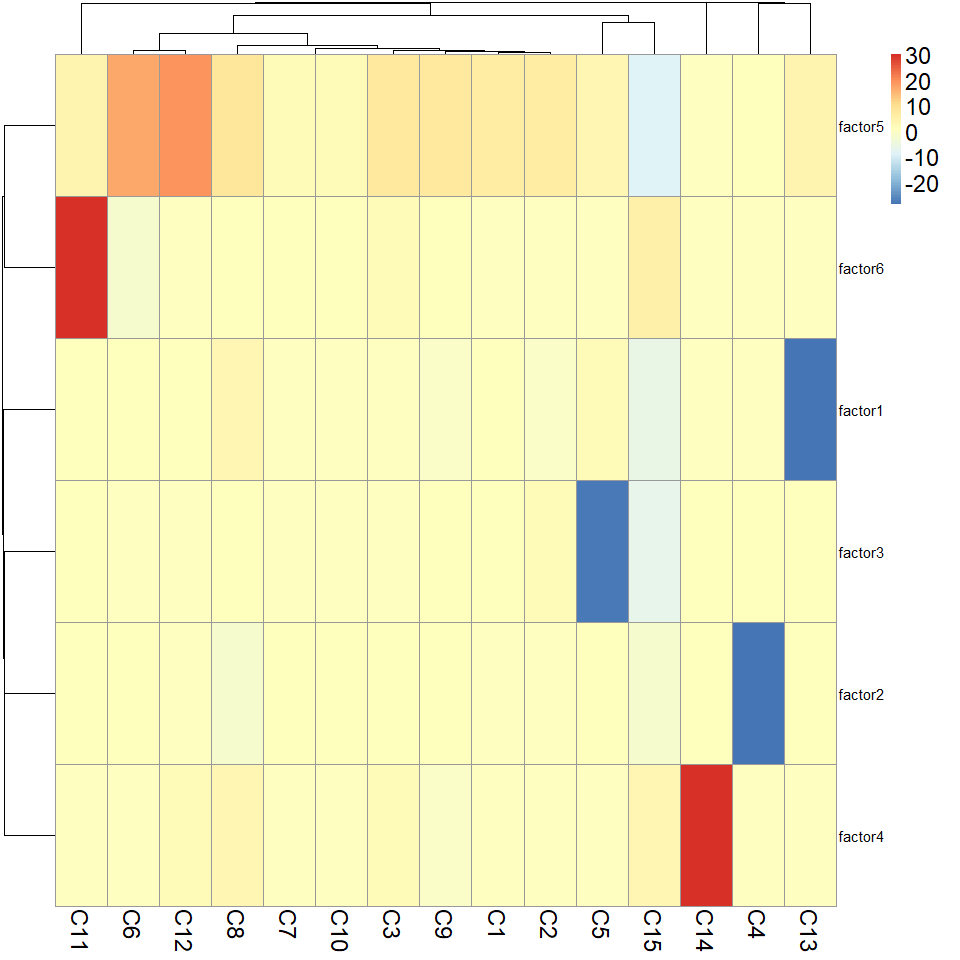}}
\quad
\subfigure[Hierarchical cluster for moPCA.]
{\includegraphics[width=2.5in,height=2in]{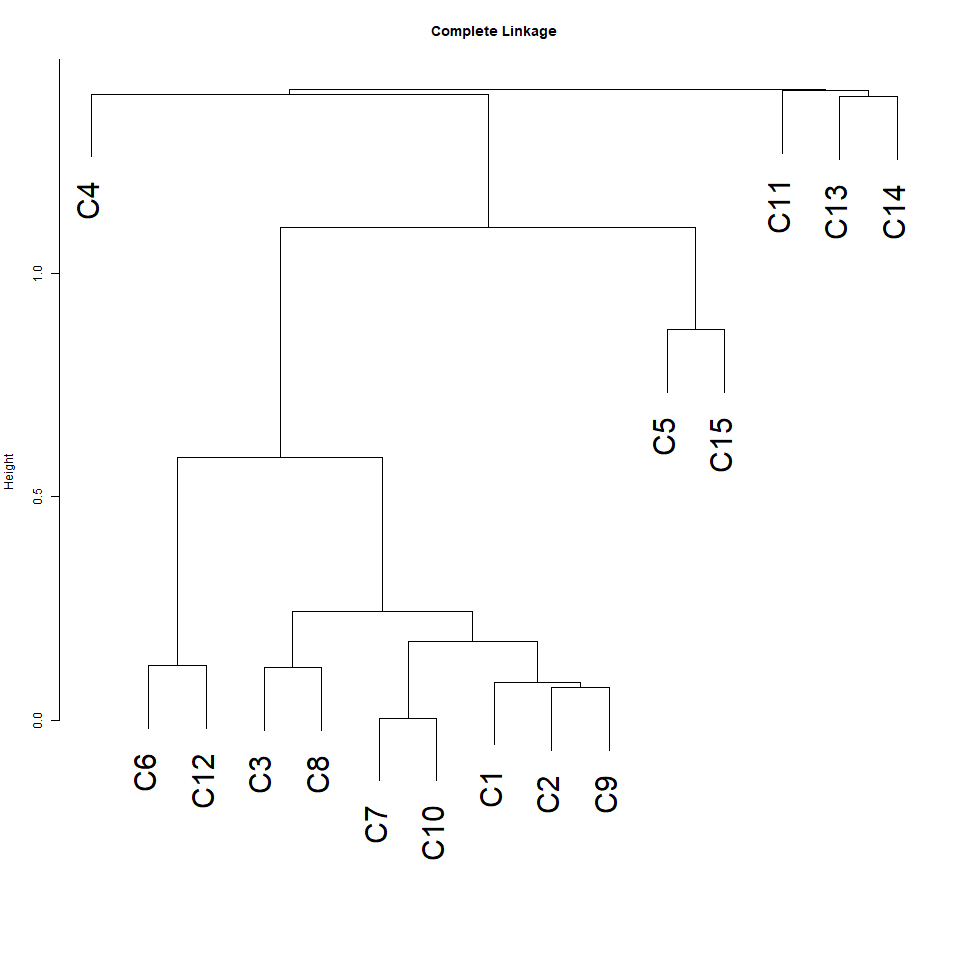}}
  \caption{The results for moPCA}
  \label{heatmap33}
\end{figure}

\begin{figure}[H]
\centering
\subfigure[Heatmap for PmoPCA.]
{\includegraphics[width=2.5in,height=2in]{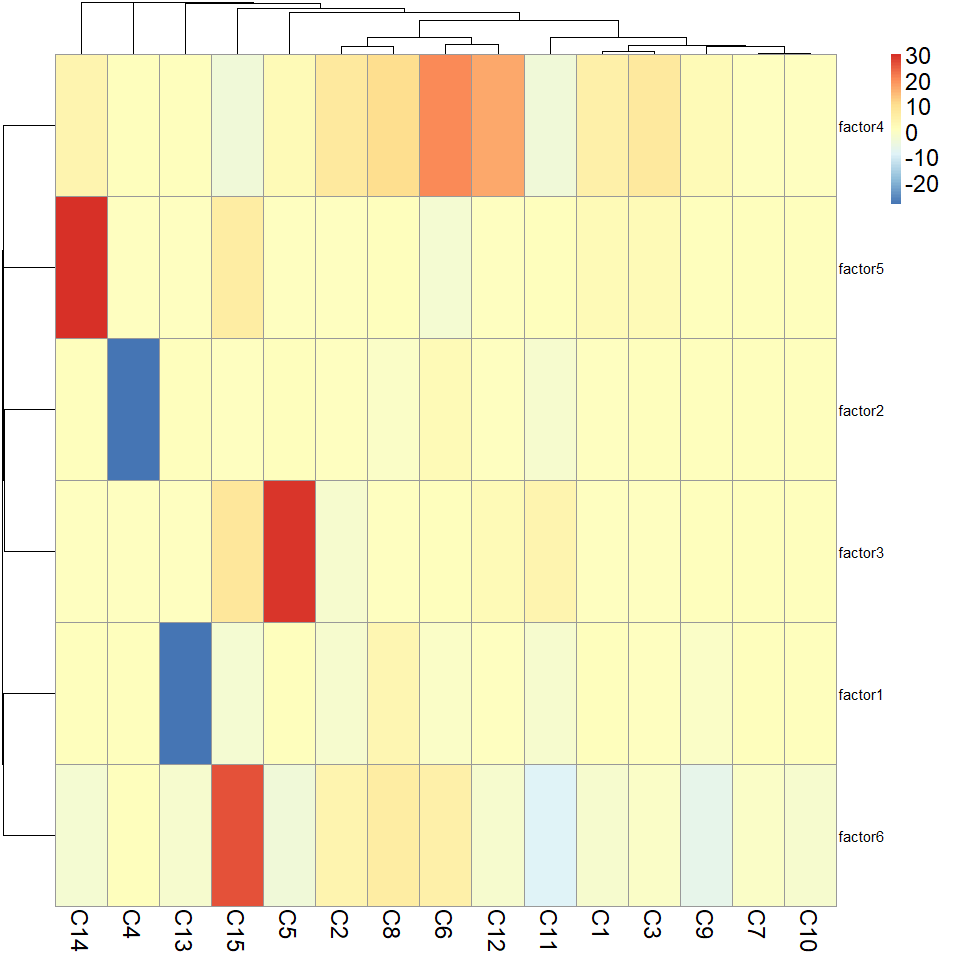}}
\quad
\subfigure[Hierarchical cluster for PmoPCA.]
{\includegraphics[width=2.5in,height=2in]{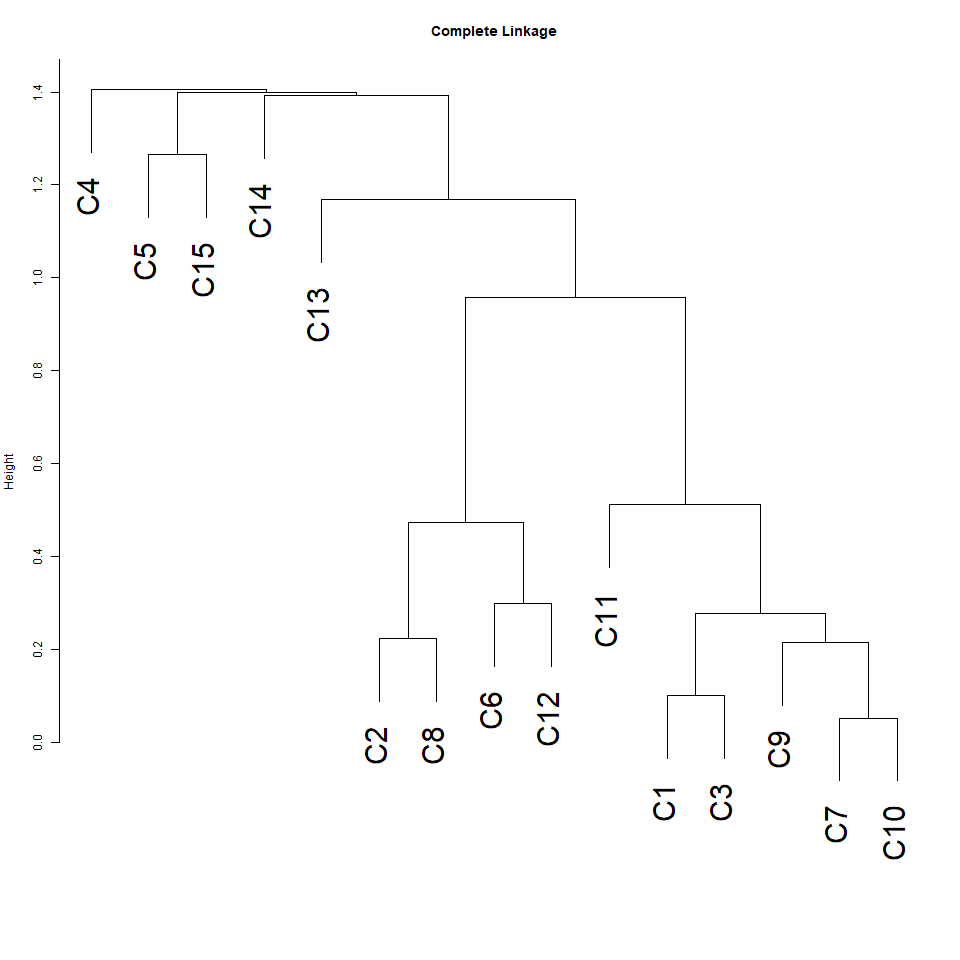}}
  \caption{The results for PmoPCA}
  \label{heatmap34}
\end{figure}

\begin{figure}[H]
\centering
\subfigure[Heatmap for IPmoPCA.]
{\includegraphics[width=2.5in,height=2in]{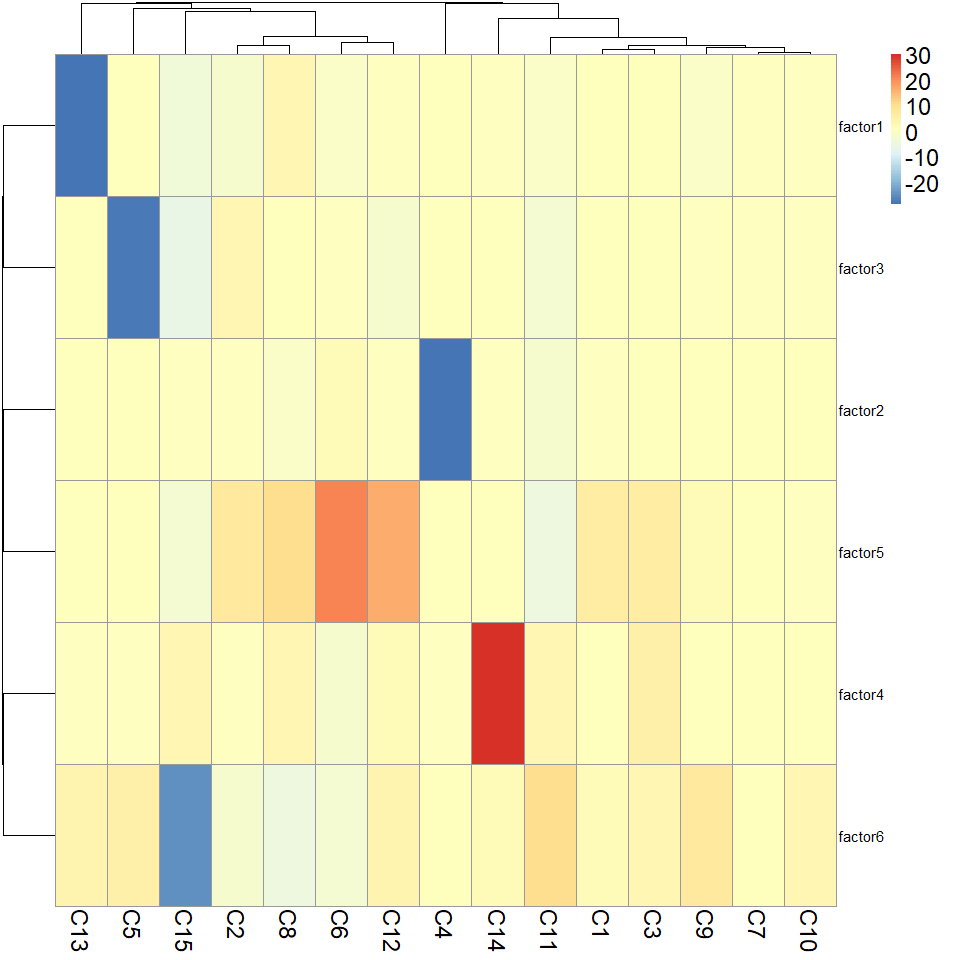}}
\quad
\subfigure[Hierarchical cluster for IPmoPCA.]
{\includegraphics[width=2.5in,height=2in]{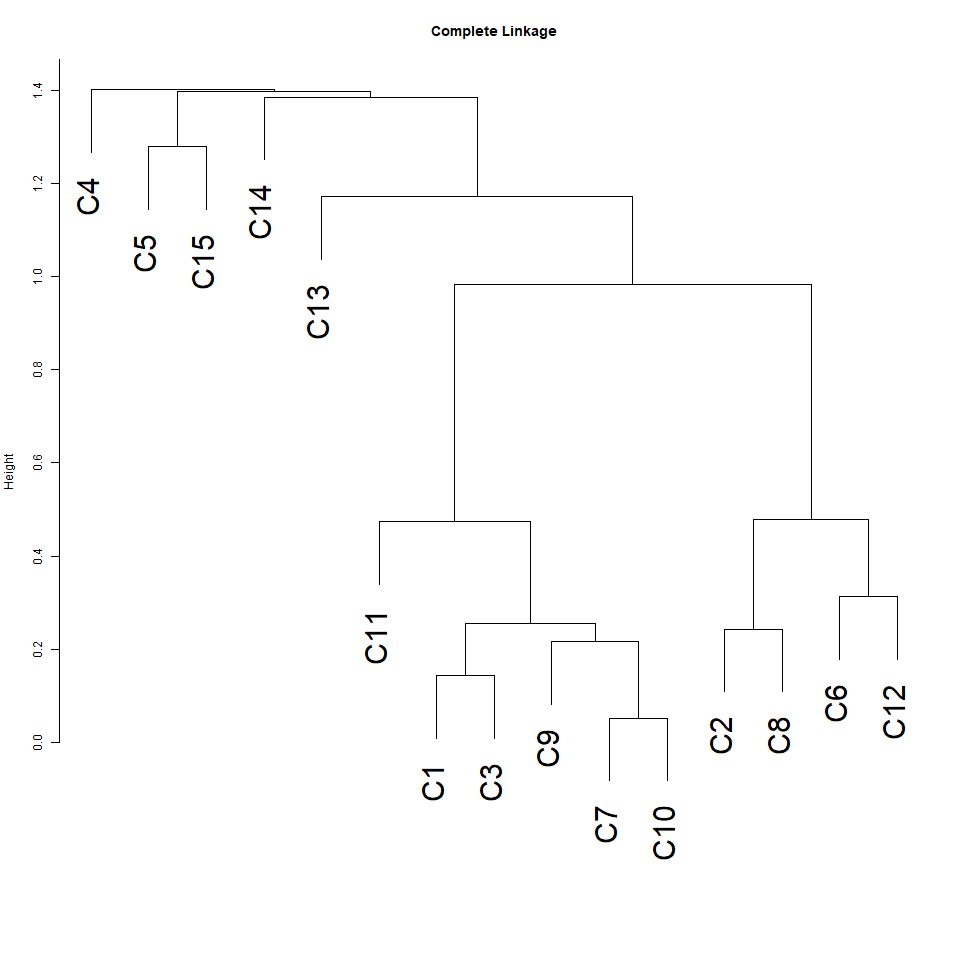}}
  \caption{The results for IPmoPCA}
  \label{heatmap35}
\end{figure}

\begin{figure}[H]
\centering
\subfigure[Heatmap for PROJ.]
{\includegraphics[width=2.5in,height=2in]{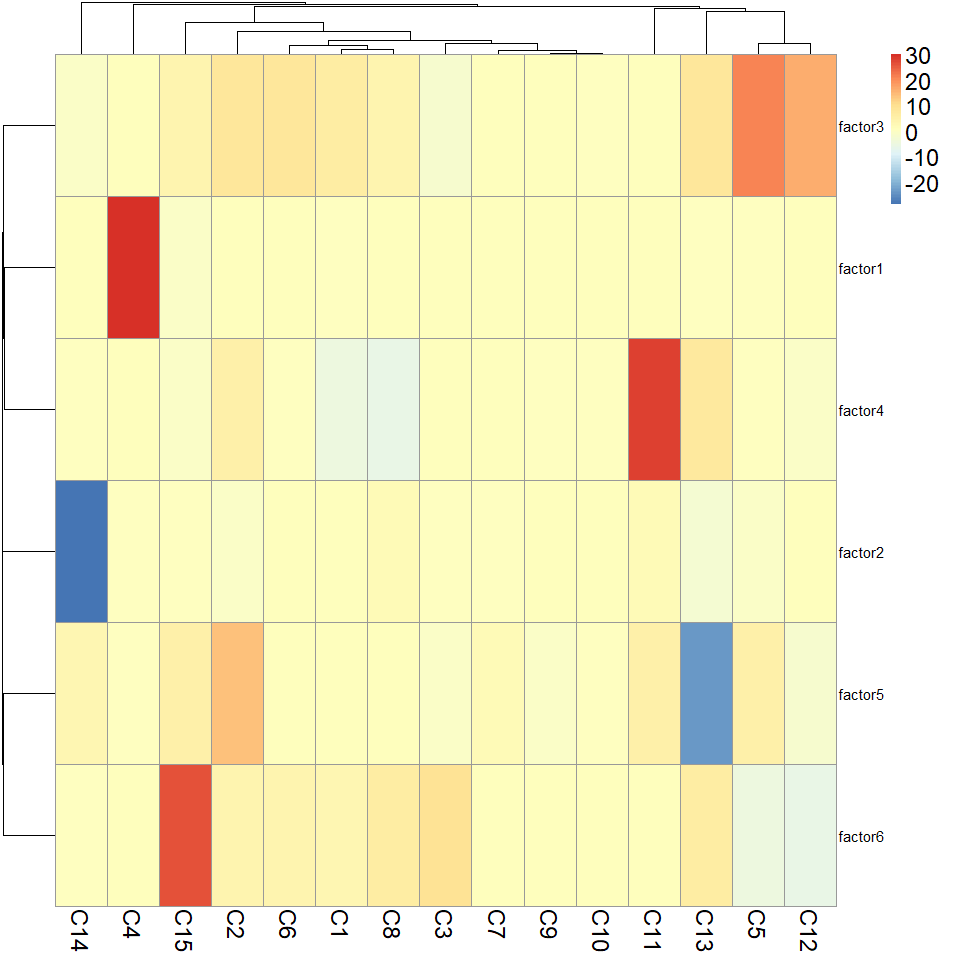}}
\quad
\subfigure[Hierarchical cluster for PROJ.]
{\includegraphics[width=2.5in,height=2in]{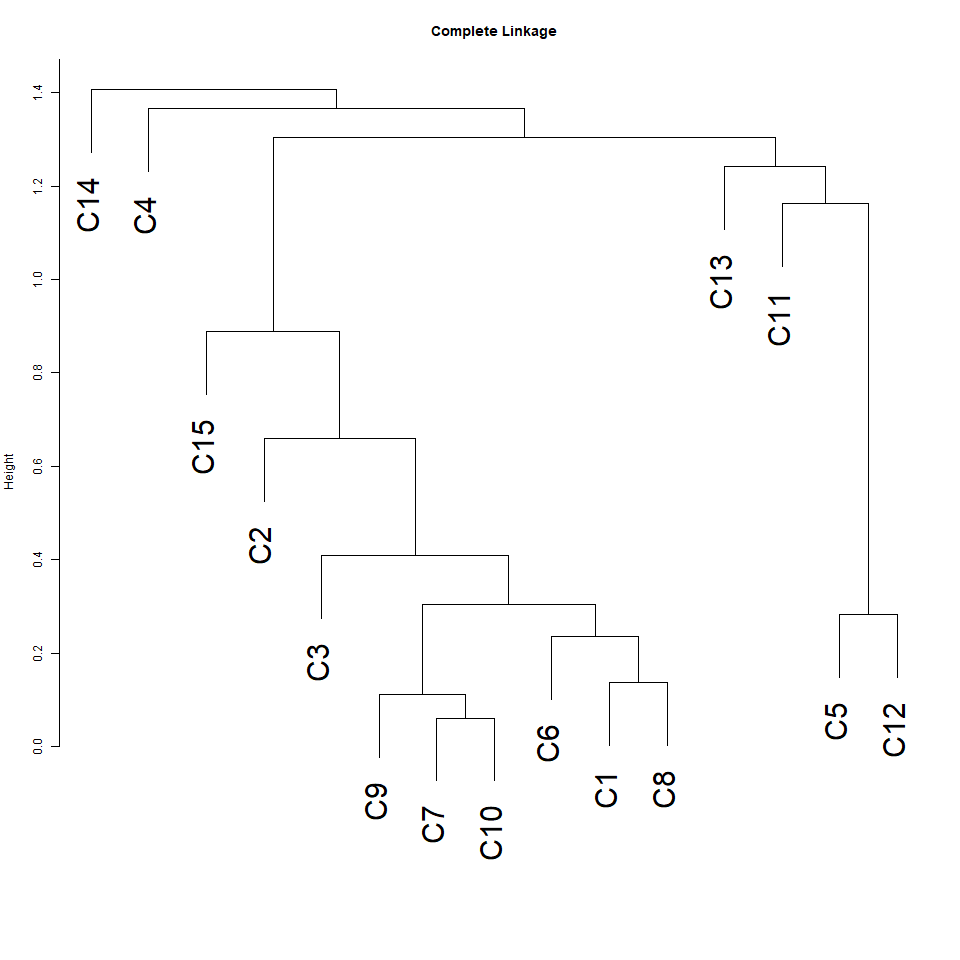}}
  \caption{The results for PROJ}
  \label{heatmap36}
\end{figure}

\section{More Information about moPCA}

\subsection{moPCA: variance-covariance tensor learning perspective}

Without loss of generality, one may allow decentering by assuming $\mE(\mathcal{F}_t)=\mathbf{0}$, $\mE(\mathcal{E}_t)=\mathbf{0}$  for tensor data $\mathcal{X}_t, t \in [T]$.
The population variance-covariance tensor is defined as
{\small
\beqrs
\mathcal{C}&=&\mE(\mathcal{X}_t\circ\mathcal{X}_t)= \mE(\mathcal{F}_t\circ\mathcal{F}_t)\times_1\mathbf{A}_1\times_2\cdots\times_D\mathbf{A}_D
\times_{D+1}\mathbf{A}_1\times_{D+2}\cdots\times_{2D}\mathbf{A}_D\\
&&+\mE(\mathcal{F}_t\circ\mathcal{E}_t)\times_1\mathbf{A}_1\times_2\cdots\times_D\mathbf{A}_D+\mE(\mathcal{E}_t\circ\mathcal{F}_t)\times_{D+1}\mathbf{A}_1\times_{D+2}\cdots\times_{2D}\mathbf{A}_D\\
&&+\mE(\mathcal{E}_t\circ\mathcal{E}_t),
\eeqrs}\noindent
where $\mathcal{C}\in\mR^{p_1\times \cdots\times p_D\times p_1\times \cdots\times p_D}$ is a $(2D)$th-order tensor with the element $c_{i_1\cdots i_D j_1\cdots j_D}=\mE(x_{t,i_1\cdots i_D}x_{t,j_1\cdots j_D})$ measuring the interrelation between entries.

Analog to the fact that several top eigenvectors of the variance-covariance matrix in the vector factor model span the low-dimensional feature subspace, the eigentensors of the variance-covariance tensor  $\mathcal{C}$ will span the low-dimensional Tucker tensor subspace $\mathcal{S}_{Tucker}$ in Section 2.
Here we introduce a tensor product operation, called \textit{tensor contraction} of $\mathcal{U} \in \mathbb{R}^{I_{1} \times \cdots \times I_{N} \times J_{1} \times \cdots \times J_{L}}$ and $\mathcal{V} \in \mathbb{R}^{J_{1} \times \cdots \times J_{L} \times K_{1} \times \cdots \times K_{M}}$, denoted by
$\langle\mathcal{U}, \mathcal{V}\rangle_{L}$, where the resulted tensor $\mathcal{D}\in
\mathbb{R}^{I_{1} \times \cdots \times I_{N} \times K_{1} \times \cdots \times K_{M}}$ has the entry $d_{i_{1}\cdots i_{N}k_{1}\cdots k_{M}}$, which is calculated by $\sum_{j_{1}, \cdots, j_{L}} u_{i_{1}\cdots i_{N}j_{1}\cdots j_{L}} v_{j_{1} \cdots j_{L} k_{1} \cdots k_{M}}.$
Then we may represent eigen equation of the $(2D)$th-order tensor $\mathcal{C}$ as $\langle\mathcal{C},\mathcal{U}\rangle_L=\lambda \mathcal{U}$, where $\lambda$ is the eigenvalue of the corresponding eigentensor $\mathcal{U}\in\mR^{p_{1} \times \cdots \times p_{D}}$; and $\mathcal{U}$ should be the approximation of tensor basis $\left\{\mathbf{A}_{1,i_{1}} \circ \cdots \circ \mathbf{A}_{D,i_{D}}\right\}_{i_1,\cdots,i_D}$ with cardinality $k=\Pi_{d=1}^Dk_d$. 
However, we can not recover $\{\mathbf{A}_d\}_{d=1}^D$ based on the outer-product type solution for the above eigen-decomposition system.

Comparatively, the mode-wise PCA is tractable by using partial structural information of  $\mathcal{C}$.
Take the subtensor $\mathcal{C}_{:i_2\cdots i_D:i_2\cdots i_D}=\mE(\mathcal{X}_{t,:i_2\cdots i_D}\mathcal{X}_{t,:i_2\cdots i_D}^{\top})$ for instance.
Under the pervasive assumption and the spike eigenvalues property (Assumption 3 in Section 3), aggregation across all modes except the $1$st and the $(D+1)$th modes yields
{\small$$\sum\limits_{i_2=1}^{p_2}\cdots\sum\limits_{i_D=1}^{p_D}\mathcal{C}_{:i_2\cdots i_D:i_2\cdots i_D}\approx p_{-1}\mathbf{A}_1\mE\left\{\mathcal{F}_{t}^{(1)}\mathcal{F}_{t}^{(1)\top}\right\}\mathbf{A}_1^{\top}.$$}\noindent
Hence $\mathbf{A}_1$ can be estimated by conducting truncated eigen decomposition on the sample subtensor $T^{-1}\sum_{t=1}^T$\\$\sum_{i_2=1}^{p_2}\cdots\sum_{i_D=1}^{p_D}
\mathcal{X}_{t,:i_2\cdots i_D}\mathcal{X}_{t,:i_2\cdots i_D}^{\top}$
or equivalently by
$T^{-1}\sum_{t=1}^T\mathcal{X}_t^{(1)}\mathcal{X}_t^{(1)\top}$, which is same as $\widehat{\mathbf{M}}_1$ up to a constant.

\subsection{moPCA: Streamline from Vectors to Tensors}

We provide the following Figure \ref{PCA-comparison} to show the unified streamline of the principal components method for tensor factor models of order $1$, $2$, and $3$, respectively.

\begin{figure}[H]
\centering
\begin{tikzpicture}
\node (1) at (-1,9.5) {$\mathbf{x}_t\in\mR^{p_1}$};
\draw[very thick] (-1.5,7)coordinate (A)
     --node[left] {$p_1$}
     (-1.5,9)coordinate (B)
     -- cycle;
\node[] (1) at (5.7,8) {$\left.\begin{array}{l}
    \widehat{\mathbf{M}}=\frac{1}{T p_1} \sum\limits_{t=1}^{T}\mathbf{x}_t\mathbf{x}_t^{\top}=
    \frac{1}{T p_1}\mathbf{X} \mathbf{X}^{\top}\\
    \mathbf{X}=(\mathbf{x}_{1},\cdots,\mathbf{x}_{T})\in\mR^{p_1\times T}\\ \widehat{\mathbf{A}}=\sqrt{p_1}~\mbox{eig}(\widehat{\mathbf{M}},r_1)
    \end{array}\right.$};

\draw[dashed] (-3,6)--(12,6);

\node (1) at (-0.5,5.5) {$\mathbf{X}_t\in\mR^{p_1\times p_2}$};
\draw[very thick] (-1.5,5)
-- node[left] {$p_1$}(-1.5,3)
-- node[below] {$p_2$}(0,3)
-- (0,5) -- cycle ;

\node[] (1) at (7,4) {$\left.\begin{array}{l}
    \widehat{\mathbf{M}}_{d}=\frac{1}{T p_1p_2} \sum\limits_{t=1}^{T} \mathbf{X}_{t}^{(d)} \mathbf{X}_{t}^{(d)\top}
    =\frac{1}{T p_1p_2}
    \mathbf{X}^{(d)}\mathbf{X}^{(d)\top}\\
    \mathbf{X}^{(d)}=(\mathbf{X}_{1}^{(d)},\cdots,\mathbf{X}_{T}^{(d)})\in\mR^{p_d\times (Tp_{-d})}\\
    \widehat{\mathbf{A}}_d=\sqrt{p_d}~\mbox{eig}(\widehat{\mathbf{M}}_d,r_d),~d=1,2
    \end{array}\right.$};

\draw[dashed] (-3,2)--(12,2);

\node (1) at (-0.5,1,0) {$\mathcal{X}_t\in\mR^{p_1\times p_2\times p_3}$};
\pgfmathsetmacro{\cubex}{1.5}
\pgfmathsetmacro{\cubey}{2}
\pgfmathsetmacro{\cubez}{1}
\draw[very thick] (0,0,0)
-- ++(-\cubex,0,0)
-- node[left] {$p_1$}++(0,-\cubey,0)
-- node[below] {$p_2$}++(\cubex,0,0) -- cycle
-- (0,0,0)
--++(0,0,-\cubez)
-- ++(0,-\cubey,0)
-- node[below,right] {$p_3$}++(0,0,\cubez) -- cycle
-- (0,0,0)
-- ++(-\cubex,0,0)
-- ++(0,0,-\cubez)
-- ++(\cubex,0,0) -- cycle;
\node[] (1) at (7,-1,-1) {$\left.\begin{array}{l}
    \widehat{\mathbf{M}}_{d}=\frac{1}{T p_1p_2p_3} \sum\limits_{t=1}^{T} \mathcal{X}_{t}^{(d)} \mathcal{X}_{t}^{(d)\top}
    =\frac{1}{T p_1p_2p_3}\mathbf{X}^{(d)}\mathbf{X}^{(d)\top}\\
    \mathbf{X}^{(d)}=(\mathcal{X}_{1}^{(d)},\cdots,\mathcal{X}_{T}^{(d)})\in\mR^{p_d\times (Tp_{-d})}\\
    \widehat{\mathbf{A}}_d=\sqrt{p_d}~\mbox{eig}(\widehat{\mathbf{M}}_d,r_d),~d=1,2,3
    \end{array}\right.$};
\end{tikzpicture}
\caption{Comparison among factor models.}
\label{PCA-comparison}
\end{figure}

\section{Symbol Table}

\begin{table}[H]
\caption{Symbols}
\label{Notation}
\begin{center}
\renewcommand\arraystretch{0.6}
{
\begin{tabular}{|cc|}
\hline
Symbol & Meaning \\
\hline
$a,b,c$ & scalars\\
$\mathbf{a},\mathbf{b},\mathbf{c}$ & vectors\\
$\mathbf{A},\mathbf{B},\mathbf{C}$ & matrices\\
$\mathbf{A}_{i\cdot}$ & the $i$th row of $\mathbf{A}$\\
$\mathbf{A}_{j}$ & the $j$th column of $\mathbf{A}$\\
$a_{ij}$ & the $(i,j)$th element of $\mathbf{A}$\\
$\mathcal{A},\mathcal{B},\mathcal{C}\in\mR^{p_1\times \cdots\times p_D}$ & the $D$th-order tensors\\
$p$ & $\Pi_{d=1}^Dp_d$\\
$p_{_{-d}}$ & $\Pi_{_{i \neq d}} ^D p_i$\\
$\mathcal{A}^{(d)}\in\mR^{p_d\times p_{_{-d}}}$ & the mode-$d$ matricization of $\mathcal{A}$\\
$\mathcal{A}_{i_1:\cdots :}\in\mR^{p_2\times \cdots\times p_D}$ & the $(D-1)$th-order sub-tensor with mode $1$\\
$\mathcal{A}_{i_{1}\cdots i_{d-1}: i_{d+1}\cdots i_{D}}$ & mode-$d$ fiber\\
$a_{i_1\cdots i_{_D}}$ & the $(i_1,\cdots,i_{_D})$th element of $\mathcal{A}$\\
$\otimes$ & Kronecker product\\
$\circ$ & outer product\\
$\times_d$ & tensor $d$-mode product\\
$\vec$ & vectorization operation\\
$\|\cdot\|$ & $L_2$-norm\\
$\|\cdot\|_F$ & Frobenius norm of matrix/tensor\\
$\|\cdot\|_{max}$ & maximum of absolute value of all elements\\
$\langle\cdot,\cdot\rangle_{L}$ & tensor contraction operation\\
$[n]$ & $\{1,\cdots,n\}$\\
$\mathcal{S}_{Trivial}\subset\mR^{p_1\times \cdots\times p_{_D}}$ & trivial tensor subspace\\
$\mathcal{S}_{Tucker}\subset\mR^{p_1\times \cdots\times p_{_D}}$ & Tucker tensor subspace\\
$\mbox{eig}(\mathbf{U}, r)\in\mR^{n\times r}$ & matrix with the top $r$ eigenvectors of $\mathbf{U}$\\
$\mathbf{A}_{[D]/\{d\}}$ & $\mathbf{A}_D\otimes\cdots\otimes \mathbf{A}_{d+1}\otimes \mathbf{A}_{d-1}\otimes\cdots\otimes \mathbf{A}_1$\\
$\mathbf{I}_n$ & $n$-dimensional identity matrix\\
$\mathbf{1}_{n\times n}$ & $n$-dimensional matrix of elements $1$\\
\hline
\end{tabular}}
\end{center}
\end{table}

\section{Reproducibility}
\setcounter{equation}{0}
\label{sec:Reproducibility}

The simulation studies and real data analysis were run under Windows 11 x64, 12th Gen Intel(R) Core(TM) i7-12700   2.10 GHz by R version 4.2.2.
